\pdfoutput=1
\documentclass[10pt, twoside,enabledeprecatedfontcommands]{scrbook}	

\usepackage[spanish,es-noshorthands,english]{babel} 						

\usepackage[protrusion=true,expansion=true]{microtype}			
\usepackage{amsmath,amsfonts,amsthm}					
\usepackage[pdftex]{graphicx}						
\usepackage[svgnames]{xcolor}						
\usepackage[hang, small,labelfont=bf,up,textfont=it,up]{caption}	
\usepackage{epstopdf}							
\usepackage{subfigure}							
\usepackage{booktabs}							
\usepackage{fix-cm}	
\usepackage[final]{pdfpages}
\usepackage{amssymb}
\usepackage{rotating}
\usepackage{multirow}
\usepackage{paralist}

\usepackage[colorlinks, citecolor=blue, linkcolor=black, urlcolor=blue]{hyperref}

\usepackage{dsfont}
\usepackage[normalem]{ulem}
\usepackage[square,numbers,comma,sort&compress]{natbib}
\usepackage{bm}
\usepackage{amssymb}

\usepackage[paperwidth=170mm, 
				   paperheight=244mm,
				   bindingoffset=80mm]{geometry}

\usepackage{arabtex}
\usepackage{utf8}

\usepackage[utf8]{inputenc} 

\usepackage{CJKutf8} 

\usepackage{setspace}
\usepackage{fancyhdr}

\usepackage{enumitem}
\usepackage{graphicx}
\usepackage{dcolumn}
\usepackage{bm}

\usepackage{braket}
\usepackage{color}
\usepackage{xcolor}

\renewcommand{\chaptermark}[1]{\markboth{\textit{Chapter \thechapter. #1}}{}}

\fancypagestyle{plain}{ 
     \fancyhead{} 
      
     } 
 
\fancypagestyle{chapter}{ 
     \fancyhead{} 

     \fancyfoot[LE, RO]{\footnotesize  \thepage}} 

\setkomafont{chapter}{cmr}

\addtokomafont{disposition}{\rmfamily}
\addtocontents{toc}{\normalfont}

\usepackage[explicit]{titlesec}
\usepackage{sectsty}				
\usepackage{lmodern}

\newlength\chapnumb
\titleformat{\chapter}[block]
{\doublespacing\normalfont\bfseries}{}{0pt}
{\parbox[b]{\chapnumb}{%
   \fontsize{65}{65}\selectfont\thechapter}%
  \parbox[b]{\dimexpr\textwidth-\chapnumb\relax}{%
    \raggedleft%
    \hfill{\doublespacing\bfseries\LARGE#1}\\
    \vspace{-15pt}\rule{\dimexpr\textwidth-\chapnumb\relax}{0.4pt}}}  

\titleformat{name=\chapter,numberless}[block]
{\doublespacing\normalfont\bfseries}{}{0pt}
{\parbox[b]{\chapnumb}{%
   \mbox{}}%
  \parbox[b]{\dimexpr\textwidth-\chapnumb\relax}{%
    \raggedleft%
    \hfill{\doublespacing\normalfont\bfseries\huge#1}\\
    \vspace{-15pt}\rule{\dimexpr\textwidth-\chapnumb\relax}{0.4pt}}}

\titleclass{\part}{top} 
\titleformat{\part}
[display]
{\vspace{15pc}\centering\normalfont\Huge\bfseries}
{\MakeUppercase{\partname} \thepart}
{0pt}
{\vspace{1pc}\huge\MakeUppercase}

\titlespacing*{\part}{0pt}{0pt}{20pt}

\usepackage{lettrine}

\newfont{\gwpfont}{cmssq8 scaled 1000}

\DeclareOldFontCommand{\bf}{\normalfont\bfseries}{\mathbf}
\DeclareOldFontCommand{\it}{\normalfont\itshape}{\mathit}

\newcommand{\msf}[1]{\mathsf{#1}}
\newcommand{\mcl}[1]{\mathcal{#1}}
\newcommand{\bsb}[1]{\bm{#1}}
\newcommand{\mbf}[1]{\mathbf{#1}}
\newcommand{\mathbbm}[1]{\text{\usefont{U}{bbm}{m}{n}#1}}
\newcommand{\mymathbb}[1]{\text{\usefont{U}{bbold}{m}{n}#1}}

\newcommand{\bone}{\bsb{1}}
\newcommand{\mone}{\mathbbm{1}}

\DeclareMathOperator{\ba}{\mathbf{a}}
\DeclareMathOperator{\bb}{\mathbf{b}}
\DeclareMathOperator{\bx}{\mathbf{x}}
\DeclareMathOperator{\bq}{\mathbf{q}}
\DeclareMathOperator{\by}{\mathbf{y}}
\DeclareMathOperator{\bz}{\mathbf{z}}
\DeclareMathOperator{\bff}{\mathbf{f}}
\DeclareMathOperator{\bn}{\mathbf{n}}
\DeclareMathOperator{\be}{\mathbf{e}}
\DeclareMathOperator{\bp}{\mathbf{p}}
\DeclareMathOperator{\bg}{\mathbf{g}}

\DeclareMathOperator{\bE}{\mathbf{E}}
\DeclareMathOperator{\bQ}{\mathbf{Q}}
\DeclareMathOperator{\bP}{\mathbf{P}}

\DeclareMathOperator{\bX}{\mathbf{X}}

\DeclareMathOperator{\br}{\mathbf{r}}
\DeclareMathOperator{\bc}{\mathbf{c}}
\DeclareMathOperator{\bd}{\mathbf{d}}
\DeclareMathOperator{\bU}{\mathbf{U}}
\DeclareMathOperator{\bV}{\mathbf{V}}
\DeclareMathOperator{\bW}{\mathbf{W}}
\DeclareMathOperator{\bu}{\mathbf{u}}
\DeclareMathOperator{\bv}{\mathbf{v}}
\DeclareMathOperator{\bw}{\mathbf{w}}

\DeclareMathOperator{\mcd}{\msf{c}_{\delta}}
\DeclareMathOperator{\mld}{\msf{l}_{\delta}}
\DeclareMathOperator{\mA}{\msf{A}}
\DeclareMathOperator{\mB}{\msf{B}}
\DeclareMathOperator{\mC}{\msf{C}}
\DeclareMathOperator{\mY}{\msf{Y}}
\DeclareMathOperator{\mZ}{\msf{Z}}
\DeclareMathOperator{\mS}{\msf{S}}
\DeclareMathOperator{\mP}{\msf{P}}
\DeclareMathOperator{\mQ}{\msf{Q}}

\DeclareMathOperator{\blambda}{\bsb{\lambda}}
\DeclareMathOperator{\bxi}{\bsb{\xi}}
\DeclareMathOperator{\bfeta}{\bsb{\eta}}
\DeclareMathOperator{\bphi}{\bsb{\phi}}
\DeclareMathOperator{\bpsi}{\bsb{\psi}}
\DeclareMathOperator{\bpi}{\bsb{\pi}}
\DeclareMathOperator{\brho}{\bsb{\rho}}
\DeclareMathOperator{\bvarphi}{\bsb{\varphi}}

\DeclareMathOperator{\bPi}{\mathbf{\Pi}}
\DeclareMathOperator{\bPhi}{\mathbf{\Phi}}
\DeclareMathOperator{\bPsi}{\mathbf{\Psi}}

\DeclareMathOperator{\mL}{\mcl{L}}
\DeclareMathOperator{\mT}{\mcl{T}}
\DeclareMathOperator{\mK}{\mcl{K}}
\DeclareMathOperator{\mD}{\mcl{D}}
\DeclareMathOperator{\mLU}{\mcl{L}_\mcl{U}}
\DeclareMathOperator{\mLV}{\mcl{L}_\mcl{V}}

\DeclareMathOperator{\mW}{\mcl{W}}

\DeclareMathOperator{\mt}{\msf{t}}
\DeclareMathOperator{\mk}{\msf{k}}

\DeclareMathOperator{\mcM}{\mcl{M}}
\DeclareMathOperator{\mM}{\msf{M}}
\DeclareMathOperator{\mN}{\msf{N}}
\DeclareMathOperator{\mm}{\msf{m}}
\DeclareMathOperator{\mO}{\msf{O}}
\DeclareMathOperator{\mH}{\msf{H}}
\DeclareMathOperator{\mI}{\msf{I}}

\DeclareMathOperator{\msA}{\msf{A}}
\DeclareMathOperator{\msB}{\msf{B}}
\DeclareMathOperator{\msC}{\msf{C}}
\DeclareMathOperator{\msD}{\msf{D}}
\DeclareMathOperator{\msE}{\msf{E}}
\DeclareMathOperator{\msF}{\msf{F}}
\DeclareMathOperator{\msG}{\msf{G}}
\DeclareMathOperator{\msJ}{\msf{J}}

\DeclareMathOperator{\msL}{\msf{L}}
\DeclareMathOperator{\msM}{\msf{M}}
\DeclareMathOperator{\msN}{\msf{N}}
\DeclareMathOperator{\msO}{\msf{O}}
\DeclareMathOperator{\msT}{\msf{T}}

\DeclareMathOperator{\msR}{\msf{R}}
\DeclareMathOperator{\msU}{\msf{U}}

\DeclareMathOperator{\msZ}{\msf{Z}}

\DeclareMathOperator{\msGamma}{\msf{\Gamma}}
\DeclareMathOperator{\msSigma}{\msf{\Sigma}}
\DeclareMathOperator{\msDelta}{\msf{\Delta}}
\DeclareMathOperator{\msLambda}{\msf{\Lambda}}
\DeclareMathOperator{\msOmega}{\msf{\Omega}}

\DeclareMathOperator{\msXi}{\msf{\Xi}}
\DeclareMathOperator{\msTheta}{\msf{\Theta}}

\DeclareMathOperator{\bI}{\mbf{I}}
\DeclareMathOperator{\0}{\mbf{0}}

\newcommand{\mR}{\mathbbm{R}}
\newcommand{\ome}{{\omega\epsilon}}
\newcommand{\omel}{{\omega\lambda}}
\newcommand{\omelp}{{\omega'\lambda'}}
\newcommand{\omep}{{\omega^{\prime}\epsilon^{\prime}}}

\begin{document}
\includepdf[pages=- , offset=72 -72,landscape=true,scale=0.97]{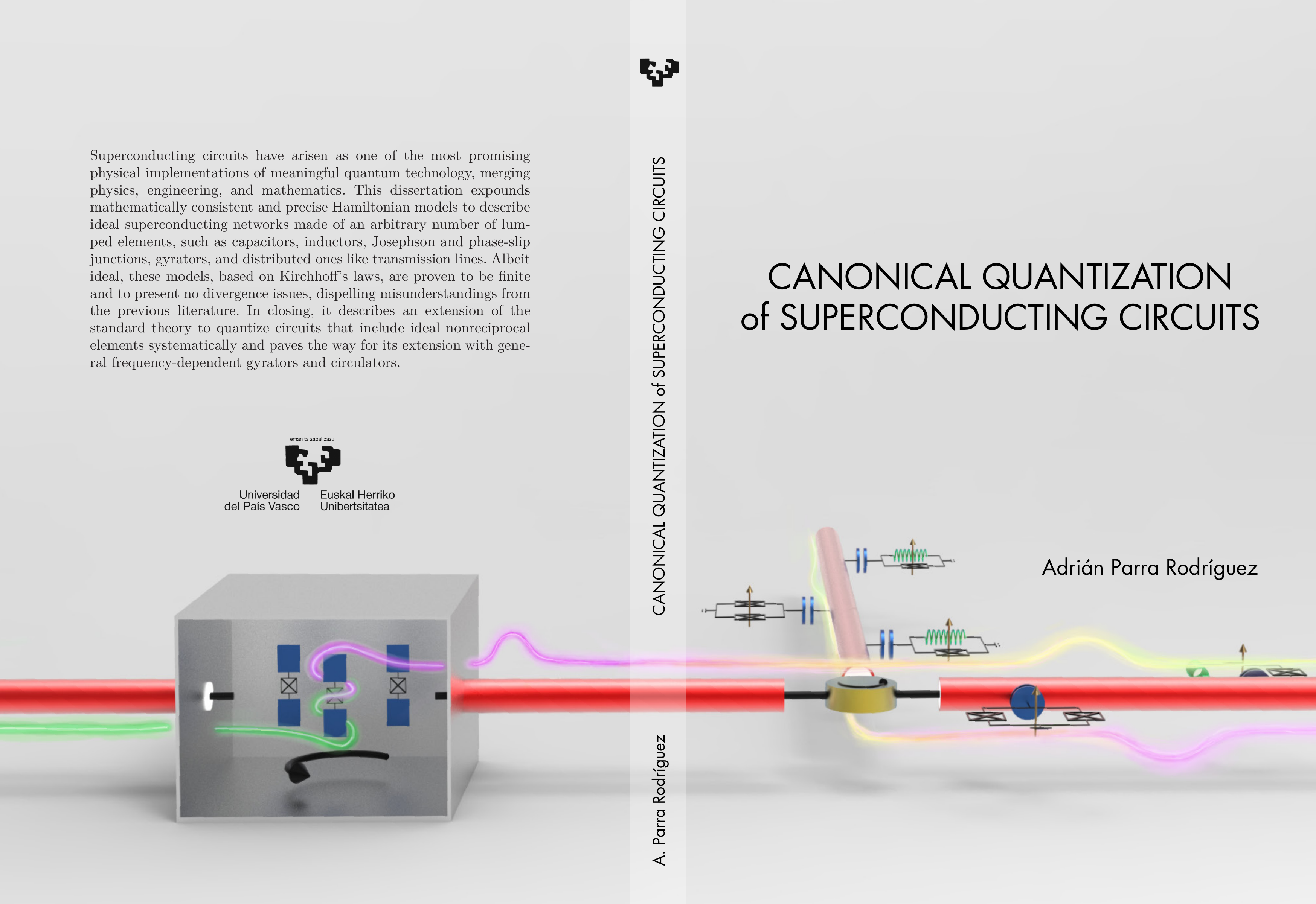}
\cleardoublepage

\begin{titlepage}
\thispagestyle{empty} 
\begin{figure}[h]
{\centering
{\includegraphics[width=0.45\textwidth]{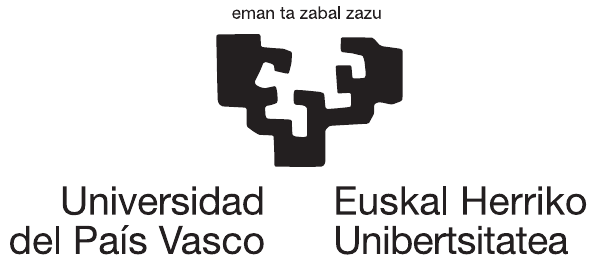}}\par}
\end{figure}

\begin{center}
\vspace{.5cm} 
{\Large Department of Physics}\\ 

\vspace{3.5cm} 

{\LARGE\textbf{Canonical Quantization}}\\
\vspace{0.5cm}
{\LARGE \textbf{of Superconducting Circuits}}\\

\vspace{3.5cm}

{\Large \textbf{Adri\'an Parra Rodr\'iguez}}\\

\vfill

{\Large PhD Thesis}\\
\vspace{1cm}
{\normalsize Leioa 2021}\\
\end{center}

\end{titlepage}

\thispagestyle{empty} 

\begin{center}
  Department of Physics\break
  University of the Basque Country (UPV/EHU)\break
  Postal Box 644, 48080 Bilbao, Spain
  
  \vspace{5pt}
  
%
%
  
%
%
%
\end{center}

\vfill

\begin{flushleft}

\vspace{3pt}

\noindent
Cover image credit: Irene Parra Rodr\'iguez.
\break
This document was generated with the 2020 \LaTeX~distribution.\break
The plots and figures of this Thesis were generated with Python and Illustrator.

\vspace{5pt}

\noindent
This work was funded by Basque Government PhD Grant PRE-2016-1-0284.

\vspace{5pt}

\noindent
\includegraphics[height=20pt]{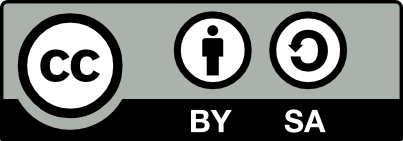}\break
2016-2021 Adri\'an Parra Rodr\'iguez.
This work is licensed under the Creative Commons Attribution-ShareAlike 4.0 International License.
To view a copy of this license, visit 
\href{http://creativecommons.org/licenses/by-sa/4.0/deed.en_US}
{http://creativecommons.org/licenses/by-sa/4.0/deed.en\_US}.
\end{flushleft}

\clearpage

\begin{titlepage}
\thispagestyle{empty} 
\begin{figure}[h]
{\centering
{\includegraphics[width=0.45\textwidth]{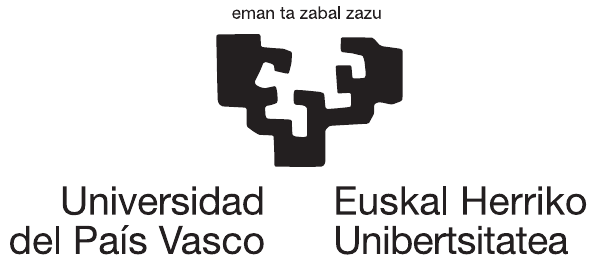}}\par}
\end{figure}

\begin{center}
{\Large Department of Physics}\\ 

\vspace{1.5cm} 

{\LARGE\textbf{Canonical Quantization}}\\
\vspace{0.5cm}
{\LARGE \textbf{of Superconducting Circuits}}\\
\end{center}

\vspace{3.75cm}

\begin{center}
\large{{\it Supervisor}:\\
\vspace{0.3cm}
Dr. I\~nigo Lu\'is Egusquiza Egusquiza}
\end{center}

\vfill

\begin{center}
\large{Submitted by Adri\'an Parra Rodr\'iguez\\
for the degree of Doctor of Philosophy in Physics}
\end{center}

\end{titlepage}

\newpage\null\thispagestyle{empty}\newpage

\thispagestyle{empty}

\vspace*{4cm}
\begin{flushright}
{\emph{{
A mi gran familia\\
}}}
\end{flushright}
\vspace*{8cm}
\begin{flushright}
	{\emph{{To those who inspired my quest 
	}}}
\end{flushright}
\newpage\null\thispagestyle{empty}\newpage


\thispagestyle{empty}

\vspace*{2cm}
\begin{flushright}
	\emph{\\
		Obsession is a double-edged sword with the power of driving one into madness\\
		and letting one thrive to conquer their desires}
\end{flushright}

\newpage\null\thispagestyle{empty}\newpage


{\pagestyle{plain}
\cleardoublepage
\pagenumbering{gobble}
\tableofcontents
\cleardoublepage}
    

\frontmatter

{\pagestyle{plain}
\chapter*{Acknowledgements / Agradecimientos}
\addcontentsline{toc}{section}{Acknowledgements / Agradecimientos}
Finally :).

This intends to be a chronological, but definitely not complete narrative of past events, naming people playing a significant role in my life in my pursuit of Physics (while myself being aware of it). I apologise to those which have felt left out and I thank you twice for my bad memory. This is supposed to be the hardest section to write in such a long-term work, and surely the most read. I guess one could say that it ``sets the bar" for the rest of the content. 

I have had a lot of luck in my life, many times and for many reasons. In summary, I have enjoyed a happy childhood with a great and safe environment around me that has helped me overcome my fears, and build the path to my dreams. Only after I left Madrid did I fully realize how special were my family, friends, and education. Let's start from the last.

I am not completely sure what was the first spark that engaged me into science. I definitely remember studying animals, burning materials with lenses (the fireplace at my grandparents was enthralling), playing (and kicking asses at :) chess, and being fascinated by numbers in my primary school years (first six of education in Spain). I remember feeling especially proud of myself whenever I beat Irene, the best student in my class and way-before-me a doctor in Nanoscience herself, in speed arithmetic. I took that ego kick not once but many times. Luckily, I did not get addicted to it, but I guess I lost that race in the end :).

I changed to Montserrat school for the final 4+2 years of education, a very open-minded institution which played a strong part in building my character. There I had the chance to receive maths lectures from Julian ``el Juli"; his rectitude singing polynomials and keeping the class quiet was both mesmerizing, and the aim of many jokes, right David? ``He" had the privilege of ``giving" me my first 10/10 in an exam (interestingly, Spanish students tend to report their marks in passive and not take full responsability of their failures or triumphs), and (Jesús) Braña, whose beautiful and clear blackboards were an outstanding effort given his back issue. I warmly remember two teachers of Spanish language and literature, María and Pilar, who not only taught me, but guided my teenage spirit in those difficult years. María pushed me to be more ambitious after almost failing her subject. Among the many anecdotes with Pilar, I still remember how bad I felt for not receiving the same public ``punishment" as another student who had not done (he rarely did) his homework. I could do nothing but privately report myself afterwards in my little fight for justice :). Her words were supportive and intelligent, and so far from what  one can listen from politicians or bosses nowadays. It was Javi(er Holgado), the first pure physicist teaching me (16-17). I can only now appreciate his efforts trying to answer to my many questions. Easy ones like: {\it how can one prove the relation between centripetal acceleration and linear velocity (with triangles)?}
\begin{equation}
	|\bsb{a}_c|=\lim_{\Delta t\rightarrow 0}\frac{\Delta \bsb{v}}{\Delta t}=\lim_{\Delta t\rightarrow 0}\frac{|\bsb{v}|}{|\bsb{r}|}\frac{\Delta \bsb{r}}{\Delta t}=\frac{|\bsb{v}|^2}{|\bsb{r}|},
\end{equation} 
and others rather more complex like: {\it how the hell are electrons and photons moving?}...a bit of the answer can be implicitly found in the following 200+ pages. There is bit more in my head, and I will always be glad to share it with anyone one who asks me (especially if the listener offers me a cup of hot chocolate). Javi's comedian style is still a classic reference in friends' gatherings. Unfortunately, rather few appreciated his manners in this difficult subject, which required being {\it fluent} in maths, spatial perception, language comprehension, and imagination. Curiously, I failed only one trimester exam before Uni, and that was Physics with Javi. I think I took up the challenge. I couldn't forget other teachers and tutors like Carmen (my interest for Chemistry did not last longer than her presence), and Eva, with whom I discussed my options after school, and in particular the ``Engineering vs Physics paradox". 

Those who have traveled with me long, have listened enough of that. In summary, it is still a question for me why I chose to study  Telecommunications Engineering instead of Physics, while both where in my list of choices: ``1. Telecom at UPM, 2. Telecom at Carlos III, 3. Telecom at Alcalá Uni, 4. Physics at UCM, ...". It looks like back then I was pretty sure of it. I guess the influence of many people, as for example my grandfather, who was an enthusiastic admirer of the telecom engineer in his village, played a role over the years. At the end of the day, many people in society (even several physicist colleagues) still think that applied sciences are more useful for finding a job. But, does that really matter after all?

Be as it may, I registered for the engineering school at ETSIT-UPM hoping that it would satisfy my eagerness for knowledge, while ending with a useful degree for my future career. First year passed quickly, and little did I engage in the learning (all of a sudden failing most of the exams), while mostly I enjoyed socializing. However, I remember coming back home some days disappointed with the courses and telling my pretty alarmed parents that I would study Physics after finishing Engineering. Their answer was pretty straightforward, and the words of my father still sound crystal clear in my head: ``you have always thought much of yourself, but you may not be capable of finishing the one you started. Stop dreaming". Hard and true words. But they had their effect. After a year and a half, I stopped wandering about, I put on the overalls and began a very strict work ethic. Results were immediate, and I remember with great joy all the library hours with two student colleagues and friends from my neighbourhood, ``Tanke" (Miguel) and ``Viti" (Victor). I am pretty sure that I owe them that degree. The breaks in-between long studying sessions, the long discussions with beers in the park, the basketball games, nights in Madrid... Gracias chavales. I want to thank two very important women during my time at ETSIT, Eloisa and Sara. I want to thank you both for all the great times we had together, and for having made me grow as a person in so many aspects. You showed true care for me, and I will be in debt all my life. 

The end of my years at ETSIT was crucially marked by my Electrodynamics teacher, Professor Jesús M. Rebollar. My interest in the subject, as you may now anticipate, was huge. He was an old-school teacher, former brilliant student, with a reputation of being very strict. I found in him many positive answers, but sometimes rather disappointing ones. Mainly, whenever I was going beyond the official syllabus, and coming back to my question... {\it how the hell do the the electrons and photons} \textbf{\textit{really}} {\it move?} I took my chance in the Microelectronics course, and I ended up me doing a research project on spin memories. The job required reading a lot of weird greek symbols in the Wikipedia, getting incredibly bothered by my lack of understanding of the word Hamiltonian, and discussing again and again with Jesús (unconsciously becoming the first plotter in this story). One day he told me about a famous spanish physicist called Ignacio Cirac working in quantum computation. When I could understand the goals of the topic, I knew I had finally found my door into Physics. I earned an Erasmus scholarship, left many people that I truly loved, and arrived in Aachen (Germany).

After one week, I visited a Physics tutor who pointed me towards the experts in quantum information and computation at the RWTH. I arranged my {\it learning agreement} to do all my engineering courses in one semester except for one, quantum info, which I would do in the second, being taught by Pr. Barbara Terhal. I am most thankful for all the great people I got the chance to meet on my Monday night breaks at ``Zuhause bar", and specially the Erasmus family that have accompanied me ever since in trips: Malik (merci, maître docteur musicien pour être un exemple du personne), Olaia (grazas por esas interminables conversas e vémonos en Santiago), Giulio (il maestro della pasta carbonara e le bombe...), Jose/Josean/Nene y Jelena (con sus sonrisas estridentes, sus birras al sol, sus tetes, sus chinos y sus ojos), Stefan (Herr Doktor Meisterengländer :), and Dario (il dottore trapanese in matematica e in dolci). That first year was pretty exhausting, working infinite night hours..., first programming with Adel while listening to Shore's music and eating at Chicken Pont, \setcode {utf8}\<شكرا لك سيد البرمجة>, and later working on problems sheets of quantum info with the online help of Niko, vielen vielen Dank noch mal daf\"ur! 

My performance in the physics lecture apparently was good enough to convince Professors Barbara Terhal and David DiVincenzo for letting me do a Masters project (with him). He introduced me to the topic of superconducting quantum circuits, which at the end has become my PhD Thesis topic. Let me thank you two again, hartelijk bedankt. There, I got the chance to taste a bit of what experts in quantum physics do. I want to specially thank that very special team of Postdocs and PhD students: Giovanni (grazie per i tuoi insegnamenti di fisica e la tua calma), François (sage et bon vivant), Shabir (\<ممنون معلم!>), Manu (ein guter und intelligenter Mann, danke!), Jascha (experto en campos, circuitos y lenguas :), Niko (der unermüdliche Arbeiter) and little Susanne (cuya bondad es inversamente proporcional a su tamaño mezclando todos los atributos de los machos Beta anteriores :). Thank you Firat, specially for showing to the world that one can (and should) do science in a honest way, süperiletken devrelerin ustası, teşekkürler.

After the german experience in a theoretical physics institute, I had the chance to spend a year in a quantum physics laboratory at Sussex University, next to Brighton. Although it did not quite meet my expectations, I learned a lot of things that year, both in the lab and (late at night) in my room. I want to thank few people who made my stay so much lighter than what it could have been otherwise. David(e) Francesco, thanks for being such a nice bloke and staying so grounded, that call with your grandma is still a classic (sto aspettando un invito per poter conoscere tua figlia e per avere un'altra opportunità di stracciarti a ping pong:)), Ethan (hilarious wizard at night, incredibly fit man in the morning, gg), Anton (huge doctor master engineer), Fan (as good philosopher as terrible badminton player, \begin{CJK}{UTF8}{gbsn}
 祝你好运，博士!), Tomas (almost a good basketball player) and Joe (virtuoso hands for music and experiments).
\end{CJK}Outside of the lab, I had great times with the italian team (sempre italiani :), Luigi (ci vediamo presto a Madrid?...o forse sulla neve) ed il grandissssimo Fabrizio (amante del vino y las {\it partículas}, di notte nel Mesmerist :). Equally important was the physicist's iberian support from Lucia (mil gracias por ser tan comprensiva y por ese trozo de suelo), Pedro (gracias por esas sabias y honestas palabras), and Samuel (gracias por esas discusiones y por tu ayuda). Gracias a los tres por todo lo que me enseñasteis. Gracias Carlos por aquellos momentos (en Seven Sisters, en Falmer bar, ...), espero que nuestros caminos vuelvan a juntarse pronto. ¡Tu apoyo fue muy importante!

After a mentally tough summer in 2015, I arrived in Bilbao with the hope of starting a master's in theoretical physics,...which eventually became a reality (cheers Iñigo for those waiting weeks biting my nails ;). Special thanks go to all the lecturers who taught me so much. Gracias Manu y Juan Luís por esas clases de temas tan complejos. Quiero pensar que {\it algún} poso ha quedado. Grazie a Michele, Gonzalo y JJ por vuestro silencioso apoyo. I want to thank the master's team and soon-docs Hodei (txirrindulari multikulturala, eskerrik asko), Ander (anaia ingeniari fisikaria) and Miguel (després de totes les converses que vam tenir, de vegades m'en recordo només de coses com Els Pets...) for the great and unselfish support they gave me while learning together.

I want to thank the members of QUTIS group. In particular, I want to thank Urtzi(las), por esas risas diarias, Unai (una mente maravillosa, y una persona aun más), Ryan (tough climber of rocks and situations, your roof is the sky, happily jealous of your language and maths mastery :), Mateo (el da Vinci del siglo XXI), Rodrigo (prisa mata muchacho), and Julen (the biggest bluffer who became the puniest biker). I want to thank Enrique, Lucas, Mikel and Jorge for their logistic support during my stay in the group. I cannot forget to mention Mario, Sal and Gary, with whom I got published (for the first time) a research article just right after the master's. I hope we can collaborate in the future, and enjoy some drinks, as we did in Shibuya with big Marios :). 

During the four years being a PhD student, I have had the chance to visit many places, and meet many interesting people. I have great memories of visiting Karlsruhe, vielen Dank an Jochen (ich hoffe, dass wir uns in der Zukunft wieder ruhig treffen können), Kollegen und Alexei, as well as Grenoble, merci beaucoup Nico, Javi, Serge, \& co pour tout, ce fut une très belle expérience et j'espère que nous pourrons collaborer à l'avenir. I want to thank great colleagues and friends during my four months in Shanghai\begin{CJK}{UTF8}{gbsn}
	(上海),
\end{CJK}and Beijing\begin{CJK}{UTF8}{gbsn}
(北京).
\end{CJK}In particular, the Chilean team composed of Pancho, Panchito and Gabo, los tres sois unos grandes y estoy seguro que volveremos a vernos en algún lugar, and PhD student Lijuan Dong\begin{CJK}{UTF8}{gbsn}
(谢谢 :).
\end{CJK}I want to specially thank Yu Jing, with whom I have learnt so much in these two years while teaching her some of the things I know about circuits. \begin{CJK}{UTF8}{gbsn}
我叫你一声小师妹，我一直希望把最好的都给你。你永远要记住，对我来说，你是最棒的!
\end{CJK}I want to thank Professor Kihwan Kim for his nice hospitality in our visit to his group at Tsinghua. Overall, I think I have never felt so well treated as in my trips in that country. The West would do well in learning some manners towards immigrants. 

I had to fly in the other direction around the globe to visit, first the Yale Quantum Institute, and then the EQuS group at MIT. I want to thank all the members in both groups that made my visits so enjoyable, full of physics conversations and a huge amount of learning. I want to thank Michel Devoret and Steve Girvin for the really interesting discussions we had on circuits, merci professeurs pour votre temps précieux. I want to extend my gratitude to Philippe, Steven, Clarke, Vlad, Jérémy, and Shruti at Yale, and Joel, Roni, Morten, Daniel, Barath, Ben, Gabriel, Yanjie, Terry and Kevin at MIT for all the designs and experiments you explained to me, and to Mirabella for her help. It was an amazing experience to see what world leading experts in this field do. Noch einmal vielen Dank an Jochen und Christina für die schöne Einladung und Unterbringung. 

Back in Europe, I had the great opportunity of collaborating with the lab led by Rudolf Gross and Frank Deppe at the Walther-Meißner Institut in Garching. I want to thank them all, together with Stefan, Kirill and the rest of the group members. Tatsächlich hoffe ich, dass wir uns wieder treffen können um mehr über supraleitende Schaltungen zu reden. Vielen vielen Dank für die Gastfreundschaft. Wie immer war mein Aufenthalt in Deutschland eine Freude. 

Unfortunately, my last research visit (to Konstanz), got too messed up due to covid-19 to be scientifically profitable. Still, I want to thank Guido Burkard and his group members for having let me discuss with them as one more during those three months. Gracias Monica por todas esas conversaciones sobre la física y la vida, y por haberte preocupado por mí tanto ;). Obrigado Thiago, mestre de línguas impossíveis, und vielen Dank Matthias, Vlad, Jonas, Amin, Benedikt, Florian, und Philipp. Noch vielen Dank an Susanne für die Hilfe. 

Between those big scientific trips, I had the opportunity to meet a rather large amount of scientists which I admire ``on and off the court" and with whom I have had great discussions, sometimes with a drink to cheer it up. That'd be the case of Alexandre Dauphin, AlejandroS González Tudela y Bermúdez, Diego Porras, Tomás Ramos, Angel Rivas, Pol Forn Díaz, etc. Apparently, many of them have bunched in {\it my} Madrid. 

I want to thank Laura Ortiz for all I learnt with her during some crazy months at the beginning of the PhD. I definitely met my match in the stubborn and passionate defense of ideas, and honestly, it still amazes me :).

Before I use the last lines for some very notable people in this story, I would like to thank Pablo and I\~nigo for the \LaTeX\ templates upon which this thesis is based on, and Sof\'ia for the style of the references (gracias por el apoyo y cariño que me has dado en esos momentos difíciles). 

Five years and a half remaining in a city are both a long or a short period of time depending on whether you look in front or behind. In any case, I am pretty sure that it would not have been the same without my flatmates Miguel and Ornella. Grazie a tutti e due, per avermi sopportato anche quando non smettevo di lamentarmi :), per le birrete, le notti insieme, etc. Alla fine, senza pianificarlo, la mia partenza ha signifigato l'arrivo di due persone molto importanti. Cecilia, la cuarta della famiglia, e a la bonne heure!! Tampoco me olvido de Giuseppe, Gael, Nuria, Raúl, Simone, Javi, Aroa y Jesús, por esas tardes/noches de calle ``arreglando" el mundo desde Goienkale!

I want to thank few scientists and friends with which I have had the special pleasure to share some years in Bilbao. 

Gracias a Enrique Rico, mi némesis adulta, por todas esas discusiones que hemos tenido. Te confieso que a veces no han sido fáciles, pero en todo caso, siempre me han parecido ricas en contenido. Mientras dure la pandemia, podrás seguir soñando que algún día me ganarás al basket, pero tranquilo, que probaremos empíricamente que habrá sido solo eso...un sueño (tal vez compartido por el gran Jose Andrés, el pitu...quiero decir Vilas, Iñaki y demás cuadrilla).

Gracias a Laura García Álvarez, por ser una buena compañera, mejor investigadora, y excelente amiga. Hemos compartido suficientes desgracias...disfrutemos de mejores momentos a partir de ahora :), y que se una el doctor Reina (gracias por las risas y la calma que me aportaste este año)!

Merci Dani, Camille et Juliette, por los ratos de ciencia, bares, monte y playa. Algún día miraremos atrás y nos reiremos a gusto. Ahora toca ponerse el mono y dar el callo, pero recordad que con un buen cocido en la mesa todo es más fácil ;). 

Eskerrik asko I\~nigo Arrazola, por ser gran compañero de fatigas en esta no trivial hazaña de terminar un doctorado. Menos aún con la mente sana. Allá donde vaya, me llevo muchas enseñanzas de esta tierra y de su gente. Arrakastaren sekretua zintzotasunean datza. Dena den, gogoratu hori: Euskadi bakarra da munduan. Bakarra da, bai, baina Euskadi ez da mundua :).

Back in Aachen, I remember well explaining to my german family eight years ago that I wanted to do a PhD in Physics. Their reactions where pretty funny, worthy of being recorded. Ich hatte keine Ahnung, dass es so lange dauern würde, aber ich kann versprechen, dass es sich den ganzen Weg gelohnt hat. Vielen Dank Judith und Christina, für eure Liebe. Vielen Dank noch mal für den Saeed, der sehr erwartete Bruder :).

\begin{figure}[h]
	\centering
	\includegraphics[width=1\textwidth]{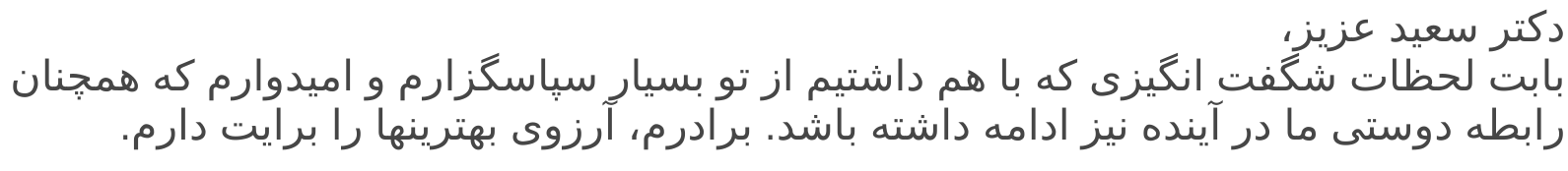}
\end{figure}

In this long story, I have always felt supported from friends back in Madrid. En especial, gracias Serg, Cris y Mart por vuestra paciencia conmigo, y con {\it mis movidas}. Es un placer saber que, no importa cuan lejos marche, siempre podré volver cerca de vosotr@s y ser feliz. 

Grazie Chiara, per essere diventata una persona molto importante per me e per aver portato la tua luce su questa ultima e più oscura parte del viaggio.\\

I want to thank my huge family, for being always an example of how much more people can achieve together rather than alone. Gracias a tod@s. I specially want to thank my grandad Higinio and my grandma Rosalía for their outstanding fight in life, and for having taught me the most important values that I (try to) have. Ojalá hubierais tenido la oportunidad que yo si tuve de estudiar, y así haber podido explicaros, cuan irrelevante es lo que aquí dejo escrito en comparación con lo que vosotros conseguisteis.

Gracias Ire, Gini y Nati, auténticos doctores de la vida, por el apoyo constante los 365,24... días del año. No sé si alguna vez podré devolveros tanto como me dais, pero lo intentaré. Os quiero.\\

Iñigo Luís Egusquiza Egusquiza, gratias multas tibi ago, virtute infinita, ad infinitum. Vous avez assez bien joué le rôle de l'abbé Faria.\\

{\raggedleft\vfill{
		\begin{flushright}
				{\it Libertas inaestimabilis res est}\\
			Imperator {\bf Iustinianus} edictum\\
			Corpus iuris civilis	
		\end{flushright}
}}

\cleardoublepage
\chapter*{Abstract}
\addcontentsline{toc}{section}{Abstract}
In the quest to produce quantum technology, superconducting networks, working at temperatures just above absolute zero, have arisen as one of the most promising physical implementations. The precise analysis and synthesis of such circuits have required merging the fields of physics, engineering, and mathematics.

In this dissertation, we develop mathematically consistent and precise Hamiltonian models to describe ideal superconducting networks made of an arbitrary number of lumped elements, such as capacitors, inductors, Josephson and phase-slip junctions, gyrators, etc., and distributed ones like transmission lines. We give formal proofs for the decoupling at high and low frequencies of lumped degrees of freedom from infinite-dimensional systems in different coupling configurations in models based on the effective Kirchhoff's laws. We extend the standard theory to quantize circuits that include ideal nonreciprocal elements all the way to their Hamiltonian descriptions in a systematic way.  Finally, we pave the way on how to quantize general frequency-dependent gyrators and circulators coupled to both transmission lines and other lumped-element networks. 

We have explicitly shown, that these models, albeit ideal, are finite and present no divergence issues. We explain and dispel misunderstandings from the previous literature. Furthermore, we have demonstrated the usefulness of a redundant basis for performing separation of variables of the transmission line (1D) fields in the presence of point-like (lumped-element) couplings by time-reversal symmetry-breaking terms, i.e. nonreciprocal elements.
\cleardoublepage
\chapter*{Resumen}
\addcontentsline{toc}{section}{Resumen}
\begin{otherlanguage}{spanish}
Durante miles de años, los seres humanos han desarrollado métodos sofisticados para contar o calcular. Entre el ábaco sumerio y las computadoras electrónicas programables, se han dado pasos fundamentales para llegar a una expresión mínima de información, el bit; una variable binaria con los valores de ``0" y ``1". Las primeras y más lentas computadoras mecánicas manejadas por el ser humano sentaron las bases de los sistemas autónomos electromecánicos utilizados en las guerras de principios del siglo pasado. En los últimos años, y después de la invención y posterior miniaturización del transistor, las máquinas eléctronicas han alcanzado el récord de velocidad de la computación. Sin embargo, durante décadas se ha sugerido que ni el paradigma computacional clásico formalizado por Turing, ni el vehículo clásico de la electrodinámica utilizado para el transporte de información, son límites fundamentales para la velocidad de un cálculo.

De hecho, a principios de los 80, Paul Benioff introdujo por primera vez la versión mecánico cuántica de la máquina de Turing basada en una descripción clásica de Charles Bennett, mientras que Yuri Manin y Richard Feynman imaginaron paradigmas de simulación, un problema computacional específico en el que se imita una descripción parcial de la realidad, basados en las reglas de la mecánica cuántica. Estos primeros esfuerzos fueron resumidos brillantemente por el último con la frase: 

{\it La naturaleza no es clásica, maldita sea, y si quieres hacer una simulación de la naturaleza, será mejor que la hagas con la mecánica cuántica, y en verdad es un maravilloso problema, porque no parece tan fácil}. 

El comienzo de la era de la computación cuántica se fundó sobre la base de que la unidad mínima de información es el bit cuántico o {\it qubit}. Este se puede codificar en un estado cuántico $\ket{\psi}$ que vive en un espacio de Hilbert bidimensional $\mcl{H}$, que comúnmente se expande en la base de estados $\ket{0}$ y $\ket{1}$. Sin embargo, a diferencia de su homóloga clásica, se permite una combinación ponderada de cero y uno, es decir, $\ket{\psi}=a \ket{0}+ b\ket{1}$ con $\{a,\,b \}\in \mathbbm{C}$, con la única restricción de que $|a|^2+|b|^2=1$. De hecho, las representaciones más generales de información cuántica requieren el uso de una matriz, conocida como {\it matriz densidad}, $\rho=\sum_i p_i \ket{\psi_i}\bra{\psi_i}$, donde $p_i$ son coeficientes de ponderación con la propiedad de que $\mathrm{Tr} (\rho) = 1$, siendo $\ket{\psi_i}$ cualquier vector bidimensional y donde $\ket{\cdot}\bra{\cdot}$ representa el producto externo de dos vectores. Curiosamente, la matriz densidad permite representar no solo estados {\it puros}, e.g. $\rho = \frac{1}{2} (\ket{0} + \ket{1}) (\bra {0} + \bra{1})$, sino también estados {\it mixtos}, como $\tilde{\rho} = \frac{1}{2}(\ket{0}\bra{0} + \ket{1}\bra{1})$. Así mismo, la matriz densidad $\rho$ de un registro de $N$ qubits actua en un producto tensorial de espacios de Hilbert de un solo qubit $\mcl {H} = \mcl{H} _1\otimes\mcl{H} _2\otimes\dots\otimes\mcl {H} _N $. Dichos espacios cuánticos de múltiples qubits permiten la descripción de estados {\it  entrelazados} más generales que no tienen equivalente clásico, por ejemplo $\rho = \ket {\psi}\bra{\psi}$ y $\ket{\psi} = (\ket{0}\ket {0} + \ket{1}\ket{1})/2$, donde tras conocer el resultado de una primera medición del primer qubit, por ejemplo en el estado cero, se sabe con certeza que el resultado de otra medición en el segundo qubit también será cero, y viceversa.

Provistos con los fundamentos de los sistemas mecánicos cuánticos controlables de luz-materia\footnote{Serge Haroche y David J. Wineland recibieron el premio Nobel en 2012 ``por métodos experimentales innovadores que permiten medir y manipular sistemas cuánticos individuales"}, a principios de los noventa comenzó una carrera para hacer realidad el sueño de Feynman y construir máquinas cuánticas capaces de calcular a voluntad. Cirac y Zoller fueron pioneros en un método para implementar puertas de dos qubits, el bloque fundamental para realizar un estado entrelazado a partir de uno separable (puro) con iones atrapados. En ese caso, el qubit estaría codificado en el espín del electrón más externo de un ion suspendido en el aire por un potencial electromagnético lento. Aquella brillante propuesta se implementó rápidamente en el laboratorio de David Wineland.

El campo de la información cuántica ganó gran interés después del descubrimiento, por parte de Peter Shor, de un algoritmo cuántico para factorizar grandes números primos basado en el entrelazamiento de un registro de qubits que superaría exponencialmente al mejor algoritmo clásico conocido. Desde entonces, se han descubierto otros algoritmos cuánticos con una aceleración potencial sobre los clásicos como el algoritmo de Grover para la búsqueda de bases de datos o HHL para resolver sistemas lineales. Mientras tanto, también se han diseñado otras implementaciones físicas que actualmente se encuentran en desarrollo en laboratorios como por ejemplo, circuitos de óptica lineal, puntos cuánticos, resonancia magnética nuclear, enrejados ópticos, defectos cristalográficos en diamante, etc. Debemos otorgar una mención especial a los circuitos superconductores trabajando apenas por encima de la temperatura del cero absoluto, que muy recientemente han superado el llamado umbral de {\it ventaja cuántica} dentro del paradigma de dispositivos ruidosos de escala intermedia (NISQ). En esencia, el laboratorio de Google afirmó en 2019 que el algoritmo cuántico programable que se ejecuta en su chip superconductor para calcular probabilidades de estados cuánticos tiene una aceleración muy sustancial con respecto a cualquier clásico.

La tecnología cuántica superconductora se basa en materiales que por debajo de un cierto umbral de temperatura y energía (``gap" superconductor) tienen una resistencia insignificante. En este régimen de trabajo, las ecuaciones fenomenológicas de London-Maxwell  capturan correctamente la física relevante cuando la carga fundamental de la teoría se toma como dos electrones acotados, conocidos como pares de Cooper. Más allá de que los pares de Cooper fluyan alrededor de materiales superconductores en ausencia de resistencia, el material compacto se comporta como un diamagneto perfecto, es decir, expulsa perfectamente el campo magnético en su interior. En otras palabras, un superconductor es más que un conductor perfecto, como lo demostraron por primera vez Meissner y Ochsenfeld en 1933. Más tarde, Bardeen, Cooper y Schrieffer introdujeron una refinada teoría microscópica de la superconductividad en su influyente artículo, donde explicaron que cualquier interacción potencial negativa entre electrones sería suficiente para que se formaran pares de electrones por debajo de cierta temperatura\footnote{John Bardeen, Leon N. Cooper y John R. Schrieffer ganaron el premio Nobel de física en 1972 ``por su trabajo conjunto en el desarrollo de la teoría de la superconductividad, comúnmente conocida como teoría BCS".}.

Más allá de su origen cuántico microscópico, se puede formular una teoría efectiva por debajo del ``gap" superconductor en términos de grados de libertad colectivos, en particular, de la fase macroscópica $\varphi $ de la función de onda en cada isla superconductora. En otras palabras, una teoría clásica efectiva (ecuaciones de London-Maxwell) surge de la naturaleza cuántica de los superconductores. Sin embargo, al trabajar en regímenes de energía suficientemente bajos, donde la población térmica es casi insignificante, esta fase macroscópica se comporta cuánticamente. De hecho, la fase superconductora no disipativa, junto con su respuesta lineal al campo electromagnético permitió construir circuitos superconductores que se comportan como átomos artificiales, dando lugar al campo de la electrodinámica cuántica de circuitos (cQED). Esto no es más que el análogo del histórico campo de la electrodinámica cuántica en cavidades (CQED), donde los átomos reales se ``colocan" (se atrapan en campos electromagnéticos lentos) en cavidades para aumentar la interacción electromagnética. A diferencia de los átomos reales, los artificiales hechos a partir de circuitos superconductores tienen la ventaja de no tener un límite fundamental para el parámetro de acoplo entre ellos y modos de luz. De hecho, este puede ser diseñado arbitrariamente dentro de unos limites. El procedimiento general para el estudio de circuitos que funcionan en el régimen cuántico es derivar sus hamiltonianos clásicos, cuyas ecuaciones de movimiento se escriben en términos de pares de variables conjugadas que se convierten en operadores cuánticos.

En 1962, se dió un gran paso en la comprensión de la superconductividad con el descubrimiento del efecto Josephson que lleva el nombre de su descubridor, Brian Josephson\footnote{Brian D. Josephson ganó el premio Nobel de Física en 1973 ``por sus predicciones teóricas de las propiedades de una supercorriente a través de una barrera de potencial, en particular aquellos fenómenos que generalmente se conocen como el efecto Josephson".}. La primera ecuación fenomenológica encontrada relaciona la corriente que pasa a través de una unión formada entre dos superconductores separados por una delgada barrera aislante y la diferencia de fase en la unión de forma no lineal $I_J=I_c \sin(\delta \varphi_J)$, donde $I_c$ es una corriente constante crítica, y $\delta\varphi_J = \varphi_1- \varphi_2$ es la diferencia de fases de la función de onda en los dos superconductores. La segunda ecuación dice que la diferencia de fase entre dos islas diferentes es linealmente proporcional a la caída de voltaje como $V_J=\frac{\Phi_q}{2\pi} \delta\dot{\varphi}_J$, donde $\Phi_q =h/2e $ es el cuanto de flujo magnético. Introduciendo la segunda relación en la primera, uno puede interpretar el efecto Josephson como un inductor no lineal. Años más tarde, este efecto no lineal se ha convertido en el dispositivo clave para hacer átomos artificiales con circuitos superconductores. Este modelo fenomenológico fue completado en 1968 por McCumber y Stewart, quienes mostraron la necesidad de una contribución resistiva (R) y capacitiva (C) a la respuesta de la unión cuando se somete a una fuente de corriente, acuñando el nombre de ``modelo de unión con resistencia y capacidad en paralelo" (RCSJ en inglés). Hoy en día, sin embargo, la resistencia intrínseca de la unión no es típicamente el fenómeno principal que induce la pérdida de coherencia en los circuitos cuánticos y, por lo tanto, comúnmente se desprecia frente a otros fenómenos. Dentro de la aproximación de disipación despreciable, los circuitos superconductores pueden ser bien descritos por una dinámica hamiltoniana. No obstante, en lugar de utilizar las ecuaciones diferenciales microscópicas que gobiernan la electrodinámica de los superconductores, es posible capturar la física esencial dentro de la denominada aproximación de elementos concentrados.

La aproximación de elementos concentrados, válida para longitudes de onda mayores que la longitud característica del circuito, divide el problema electromagnético en una red en dos subproblemas: el topológico y el geométrico. Para ello, se requiere de una aproximación crucial: los cables que conectan los elementos concentrados son conductores perfectos. Además, esta suposición implica que todos los campos eléctricos y magnéticos presentes viven dentro de los elementos concentrados (de ahí su nombre) de manera que no interactúan fuertemente con otros elementos de la malla. Para el problema geométrico, es posible resolver para cada elemento concentrado (por separado) su respuesta a los campos eléctricos dentro de los elementos. Por ejemplo, las relaciones lineales (simétricas bajo inversión temporal) entre las fuentes electromagnéticas y las ecuaciones de Maxwell se pueden capturar en los coeficientes de capacitancia $C$ e inductancia $L$, en lo que se conoce como ecuaciones constitutivas de los elementos concentrados. La dinámica lineal colectiva que rompe de manera efectiva la simetría de inversión temporal puede ser capturada por el elemento lineal no recíproco fundamental, el girador, que tiene cuatro terminales (dos puertos) y se describe mediante un parámetro de resitancia $R$ y un matriz antisimétrica dos por dos. Para un conjunto dado de coeficientes geométricos, el problema de la red eléctrica se reduce a un problema topológico de conexiones de cableado que representan un conjunto de ecuaciones diferenciales para voltajes y corrientes, o flujos y cargas, conocidas como leyes de Kirchhoff.

Es bien sabido que la aproximación de elementos concentrados impone una frecuencia de corte ultravioleta demasiado fuerte para encontrar mapeos uno a uno entre elementos concentrados ideales aislados y volúmenes 3D de un chip real. Por ejemplo, una sección de una guía de ondas superconductora coplanar puede contener idealmente un número infinito de modos electromagnéticos cuasi-transversales (quasi-TEM) con una frecuencia que aumenta monótonamente. En ese caso, las guías se describen mejor con un modelo de línea de transmisión de condensadores e inductores diferenciales ideales cuya dinámica está anclada a las leyes  de Kirchhoff diferenciales, también conocidas como {\it ecuaciones del telegrafista}. De esta manera podemos reducir el problema clásico de 3 + 1 dimensiones a uno efectivo de 1 + 1 dimensiones, dado que los campos electromagnéticos están restringidos a un volumen pequeño y tienen ciertas simetrías, e.g. los modos TEM en el cable coaxial.

La capacidad de las ecuaciones de Kirchhoff y la aproximación de elementos concentrados se observa cuando se describen sistemas electromagnéticos lineales multipuerto. Los ingenieros del siglo pasado demostraron que se puede ajustar sistemáticamente la respuesta de ``scattering" lineal $\mS(\omega)$ de un entorno electromagnético lineal 3D multipuerto (sin pérdidas), ahora conocido normalmente en la comunidad de circuitos QED como {\it caja negra}, a un circuito que contiene etapas (infinitas) de condensadores, inductores, giradores y transformadores ideales de elementos concentrados. Vale la pena recordar que ningún circuito equivalente, cuando se utiliza como herramienta de modelado, no captura la física microscópica del interior de la caja electromagnética pero si que describe información útil para el observador externo.

Hay dos configuraciones principales de circuitos superconductores, los que contienen cavidades 3D con chips planos cuasi-2D posiblemente incrustados donde las cavidades juegan un papel en el procesamiento de información cuántica, por ejemplo, aumentando los acoplos efectivos entre qubits, y aquellos en los que estas cavidades actúan como filtros de ruido, encapsulando toda la zona criogénica. En ambos casos, líneas de transmisión pueden conectarse a los puertos de una cavidad o directamente a los chips planares. En ese sentido, es útil ver el chip dividido en cajas negras lineales o no lineales donde puede necesitarse una descripción más o menos precisa para capturar la dinámica esencial. Las cavidades 3D se pueden incorporar fácilmente en el análisis completo de Kirchhoff después de hacer uso de simulaciones de las ecuaciones de Maxwell con programas de ordenador o mediciones de ``scattering" directas. Como se ha mencionado anteriormente, dado que los chips superconductores están configurados para trabajar en temperaturas criogénicas (alrededor de $10$ mK), los modelos efectivos de Kirchhoff deben entenderse desde un punto de vista mecánico cuántico.

Por tanto, el problema se traduce a cuantizar un conjunto de ecuaciones diferenciales escritas en términos de variables conjugadas (canónicamente), es decir, encontrar un hamiltoniano que contenga un conjunto (mínimo) de pares de variables conjugadas, típicamente flujos y cargas, y un corchete de Poisson (el cual pasa a ser un conmutador) que determinará la evolución en el tiempo. Como es bien sabido, las variables conjugadas con un corchete canónico facilitan el proceso, pero no es estrictamente obligatorio. De hecho, ni siquiera se requiere un término de energía cinética de segundo orden en un lagrangiano para encontrar un hamiltoniano útil sin necesidad de invocar el procedimiento de Dirac para eliminar ligaduras. 

El trabajo de esta tesis responde a una sencilla pregunta. ¿Es posible encontrar sistemáticamente una teoría cuántica convergente de chips superconductores a partir de las ecuaciones de Kirchhoff de elementos concentrados y distribuidos? La respuesta es sí. De hecho, un análisis simple muestra que el desacoplo de grados de libertad a frecuencia cero e infinita debe ocurrir ya que los condensadores y los inductores se comportan como terminales abiertos y a tierra (a tierra y abiertos) para frecuencias infinitas (cero) respectivamente. El punto clave para demostrar esta afirmación es describir correctamente el acoplo entre subsistemas multimodo (teóricamente infinitos) y subsistemas de dimensión finita. Este resultado contrasta con los modelos fenomenológicos hamiltonianos históricos de luz-materia en electrodinámica cuántica de cavidades con niveles de energía de átomos vestidos por los modos electromagnéticos, donde se invocaban frecuencias de corte ultravioleta o técnicas de renormalización para hacer predicciones finitas de cantidades observables.

Aquí deducimos que, sin importar si algunos elementos concentrados en el circuito son una representación válida de los volúmenes de chips 3D para frecuencia infinita, los circuitos de Kirchhoff tienen desacoplos naturales que pueden hacerse explícitos en el hamiltoniano, lo que hace que los programas de teoría cuántica de campos como la renormalización sean  innecesarios y superfluos en este contexto. En otras palabras, la aproximación de elementos concentrados, incluso cuando es combinada con la forma diferencial de una línea de transmisión, introduce una escala de longitud. Esto resulta especialmente práctico cuando se utilizan modelos hamiltonianos para predecir los Lamb ``shifts" o desplazamientos multimodo (acoplos efectivos) de (entre) qubits. Un segundo resultado importante demostrado aquí es la adecuación de una descripción de espacio de configuraciones doblado con variables de carga {\it y} de flujo  para cuantizar circuitos con líneas de transmisión y elementos no recíprocos, véase circuladores o giradores genéricos, elementos que rompen la simetría de inversión temporal. La redundancia aparentemente prescindible introducida en el análisis resulta ser el punto de partida correcto para derivar el hamiltoniano, y puede eliminarse sistemáticamente haciendo uso de la simetría de dualidad electromagnética en el espacio de fases.
	
En esta tesis, hemos desarrollado herramientas analíticas para obtener modelos cuánticos canónicos de circuitos superconductores dentro del contexto de las leyes de Kirchhoff mostrando explícitamente la ausencia de divergencias cuyo origen reside en la teoría clásica macroscópica. En resumen, hemos estudiado los diferentes problemas de divergencia que aparecen en una configuración QED de circuito mínima e ilustrativa que contiene un resonador de línea de transmisión multimodo acoplado capacitivamente a una unión Josephson. Hemos ampliado este análisis a un catálogo de múltiples sistemas dimensionales infinitos acoplados linealmente a grados de libertad finitos no armónicos. Además, hemos introducido los elementos no recíprocos ideales de una manera exacta en las descripciones hamiltonianas efectivas de los circuitos de elementos concentrados. Finalmente, hemos introducido la descripción más genérica en un espacio doblado para derivar el hamiltoniano exacto de líneas de transmisión acopladas a través de sistemas lineales no recíprocos a grados de libertad no armónicos. Más específicamente:

En el capítulo \ref{chapter:chapter_2}, hemos analizado un modelo de Rabi cuántico multimodo en cQED a partir de un circuito equivalente macroscópico de elementos concentrados. Hemos mostrado explícitamente la convergencia del desplazamiento de Lamb en ausencia de cualquier frequencia de corte fenomenológica extra, que surge de una renormalización natural de los parámetros hamiltonianos con un número creciente de modos armónicos. También hemos estudiado las implicaciones de una capacitancia de la unión de Josephson de valor finito, que introduce una frequencia de corte asociada a una longitud eléctrica natural en el acoplo con los modos de alta frecuencia. Hemos demostrado que al construir un modelo de Rabi cuántico a partir de circuitos de elementos concentrados acoplados capacitivamente, es crucial incluir la renormalización natural de una transformación de Legendre exacta para obtener los parámetros hamiltonianos correctos a partir de los valores macroscópicos de los elementos del circuito. Además, hemos mostrado una conexión entre los modelos hamiltonianos con un número truncado de modos y aproximaciones de baja energía del modelo dimensional infinito del resonador de la línea de transmisión. Hemos señalado la utilidad de este enfoque en el contexto de experimentos de régimen de acoplo ultrafuerte (USC) donde es obligatorio tener en cuenta los efectos multimodo. Recalcamos que estos modelos han sido posteriormente utilizados por el grupo dirigido por el Profesor Gary Steele para ajustar experimentos con qubits tipo transmon acoplados a un resonador multimodo en régimen de acoplo USC.

En el capítulo \ref{chapter:chapter_3}, hemos analizado críticamente una serie de enfoques para la cuantización de circuitos superconductores con un entorno infinito dimensional, principalmente líneas de transmisión y cajas negras genéricas de inmitancia multipuerto, con particular interés en el tema de divergencias en los desplazamientos de Lamb o acoplos efectivos (adiabáticos) predichos por las constantes de acoplo. Con respecto a las líneas de transmisión, hemos hecho uso de construcciones matemáticas sólidas, problemas de autovalores para operadores diferenciales de segundo orden, con ecuaciones de frontera que incluyen el autovalor, para describir correctamente los parámetros de acoplo capacitivos (e inductivos) libres de divergencia para redes no lineales (elementos concentrados). Al hacerlo, hemos identificado las longitudes eléctricas e inductivas fundamentales que definen los límites de dichos parámetros, que hemos establecido de manera óptima en unos valores para que el hamiltoniano final no tenga acoplos modo-modo en el sector armónico. Curiosamente, el parámetro de acoplo para longitud capacitiva o inductiva pura es de tipo Lorentz-Drude, con una caída suave $g_n \sim\omega_n^{- 1/2}$. Al transformar de nuevo a la descripción de campos en el hamiltoniano, uno puede leer directamente el término diamagnético $A^2$, que depende del parámetro de longitud asociado. Se ha realizado un análisis análogo con cajas negras recíprocas lineales multipuerto con un conjunto infinito de modos. Aquí, un modelo de elementos concentrados se ha truncado a un número de modos de $N$ antes de tomar el límite infinito. Bajo el criterio la descripción hamiltoniana final sea la de un conjunto infinito de modos armónicos independientes acoplados a un conjunto finito de variables, hemos mostrado la convergencia del límite infinito en la descripción hamiltoniana final usando la transformación canónica pertinente. Hemos realizado este análisis en un catálogo de configuraciones de acoplo lineal y hemos demostrado el mismo comportamiento. De cara al futuro, sería interesante comprobar experimentalmente la predicción del acoplo máximo alcanzable con una línea de transmisión que se puede mejorar con un modelo de condensador de acoplo de longitud finita. 

A partir de los análisis exhaustivos de los capítulos \ref{chapter:chapter_2} y \ref{chapter:chapter_3}, con la formulación de modelos cuánticos de Rabi multimodo en el contexto de las ecuaciones de Kirchhoff que presentan un desacoplo natural de la luz y la materia, uno debería ser capaz de lograr los modelos hamiltonianos sin divergencia similares en otras configuraciones de mecánica cuántica (no relativistas). Un ejemplo interesante sería con átomos acoplados a modos de guías de ondas (CQED) en la aproximación dipolar donde normalmente se invoca  una longitud de átomo efectiva de manera fenomenológica, como por ejemplo el radio de Bohr.

En el capítulo \ref{chapter:chapter_4}, hemos descrito un procedimiento para incluir elementos ideales no recíprocos a las descripciones hamiltonianas exactas de redes de elementos concentrados en una generalización de las técnicas estándar basadas en la teoría de grafos de redes y la elección de variables de flujo como grados de libertad. Hemos usado esta técnica en dos ejemplos de circuitos. En primer lugar, un girador Viola-DiVincenzo de dos puertos conectado a las uniones Josephson. En segundo lugar, hemos cuantizado el circuito de elementos concentrados equivalente para una impedancia no recíproca de dos puertos genérica. También hemos discutido un problema técnico con respecto a la introducción de elementos ideales no recíprocos en una descripción de variable de flujo que carecen de descripción de admitancia, y demostramos que el problema se reduce a eliminar ligaduras, con una reducción de variables independientes. Finalmente, hemos discutido el método de cuantización dual en términos de cargas de ``loop" que podrían ser particularmente útiles en futuras tecnologías superconductoras basadas en uniones tipo ``phase-slip" y elementos no recíprocos. En la misma dirección que en el capítulo posterior \ref{chapter:chapter_5}, se pueden buscar generalizaciones de descripciones hamiltonianas basadas en un espacio de configuración doblado o mixto a partir de una descripción lagrangiana redundante en términos de flujos de nodo y cargas de bucle. De esta manera, se podrá cuantizar trivialmente el circuito de la admitancia multipuerto de elementos concentrados, dual a la matriz de impedancia tratada en este capítulo.

En el capítulo \ref{chapter:chapter_5}, hemos presentado una técnica de cuantización canónica más general para circuitos descritos en términos de una descripción de carga de flujo redundante en el espacio de configuración. En lugar de eliminar la redundancia en las ecuaciones de movimiento de Euler-Lagrange (lagrangiano), lo hacemos en el espacio de fase (hamiltoniano) haciendo uso de una simetría de dualidad. Esta base en el espacio doblado se convierte en el punto de partida más eficaz para derivar hamiltonianos exactos de redes que contienen un número arbitrario de líneas de transmisión acopladas puntualmente por elementos ideales no recíprocos debido al término que rompe la simetría de inversión temporal, que mezcla los campos del espacio de configuración de una manera no trivial. Hemos encontrado una base completa del operador diferencial de Sturm-Liouville en el espacio duplicado que corresponde a la base de modos normales que diagonaliza exactamente al hamiltoniano. En una generalización del capítulo anterior \ref{chapter:chapter_3}, ampliamos la técnica para describir conexiones lineales puntuales (capacitivas/inductivas) a la línea de transmisión y encontramos el hamiltoniano exacto de un circuito que contiene una unión de Josephson acoplada capacitivamente a una línea de transmisión, que a su vez está conectada a otras dos por medio de un circulador. Naturalmente, el modelo presenta las mismas propiedades de convergencia de los capítulos anteriores ya que ampliamos el espacio de Hilbert en el que viven las funciones que usamos para desarrollar los campos de flujo y carga. A su vez, los acoplos modo-modo de la línea también pueden eliminarse exactamente en el hamiltoniano bajo el mismo criterio de optimalidad del capítulo \ref{chapter:chapter_3}. Finalmente, hemos realizado por primera vez un análisis sobre cómo extender la teoría para cuantizar los circuitos que contienen líneas de transmisión acopladas por cajas negras no recíprocas cuya respuesta depende de la frecuencia. El problema se reduce al análisis de circuitos con condiciones de contorno generales con condensadores, inductores y elementos ideales no recíprocos. Trabajando en la base doblada, hemos demostrado la eliminación de variables no dinámicas redundantes para casos particulares sin inductores o sin circuladores en la frontera, siendo este último caso también tratable con la base reducida. Hemos proporcionado una prueba de la dimensión máxima no trivial del espacio de fases para el hamiltoniano para lineas con acoplo lineal genérico. Sin embargo, advertimos que será necesario más trabajo para encontrar la transformación simpléctica que lleve al hamiltoniano completo a su base diagonal.

En general, esta tesis expande la teoría para encontrar modelos canónicos hamiltonianos para circuitos basados en las leyes de Kirchhoff, una cuestión de especial relevancia para derivar modelos para tecnologías cuánticas superconductoras. Esperamos que los resultados presentados aquí ayuden a analizar y sintetizar nuevos circuitos superconductores teniendo en cuenta la compleja naturaleza infinito-dimensional de los sistemas acoplados de luz-materia, así como nuevos dispositivos que rompen de manera efectiva la simetría de inversión temporal. Intrínsecamente, estos circuitos tienen el potencial de revelar misterios del universo aún sin resolver y de dar un salto tecnológico sustancial a la humanidad. Además, suponemos que la teoría de los operadores autoadjuntos en la que se asientan muchos resultados de esta tesis será de interés para la comunidad matemática, en particular, el uso de un espacio de un Hilbert doblado para eludir problemas de valores propios en la frontera de segundo orden con la raíz cuadrada de los autovalores en la propia condición de frontera.
\end{otherlanguage}
\cleardoublepage
\chapter*{List of publications}
\addcontentsline{toc}{section}{List of publications}

This Thesis is based in the following publications and preprint:
\\

{\bf Chapter 2: \nameref{chapter:chapter_2}}

\begin{enumerate}
	
\item 
{M. F. Gely$^*$, A. Parra-Rodriguez$^*$, D. Bothner, Y. M. Blanter, S. J. Bosman, E. Solano, and G. A. Steele, {\it Convergence of the multimode quantum Rabi model of circuit quantum electrodynamics},  \href{https://journals.aps.org/prb/abstract/10.1103/PhysRevB.95.245115}{Phys. Rev. B {\bf 95}, 245115 (2017)}}.
\item[]$^*$ These authors contributed equally to the article.
\end{enumerate}

{\bf Chapter 3: \nameref{chapter:chapter_3}}

\begin{enumerate}[resume]
	
\item 
{A. Parra-Rodriguez, E. Rico, E. Solano and I. L. Egusquiza, {\it Quantum networks in divergence-free circuit QED}, \href{https://iopscience.iop.org/article/10.1088/2058-9565/aab1ba/meta}{Quantum Sci. Technol. {\bf 3}, 024012 (2018)}}.
\end{enumerate}

{\bf Chapter 4: \nameref{chapter:chapter_4}}

\begin{enumerate}[resume]
	
	\item
{A. Parra-Rodriguez, I. L. Egusquiza, D. P. DiVincenzo and E. Solano, {\it Canonical circuit quantization with linear nonreciprocal devices}, \href{https://journals.aps.org/prb/abstract/10.1103/PhysRevB.99.014514}{Phys. Rev. B {\bf 99}, 014514 (2019)}}.
	
\end{enumerate}

{\bf Chapter 5: \nameref{chapter:chapter_5}}

\begin{enumerate}[resume]

\item
{A. Parra-Rodriguez, I. L. Egusquiza, {\it Canonical quantization of telegrapher's equations coupled by ideal circulators}, \href{https://arxiv.org/abs/2010.12572}{arxiv:2010.12572 (2020)}}.

\end{enumerate}

\newpage
\vspace{1cm}
Other articles published in the course of this thesis yet not included in it are:

\begin{enumerate}[resume]
		\item
	{L. Lamata, A. Parra-Rodriguez, M. Sanz, E. Solano, {\it Digital-analog quantum simulations with superconducting circuits}, \href{tandfonline.com/doi/full/10.1080/23746149.2018.1457981}{Adv. Phys: X {\bf 3}, 1457981 (2018)}}.
		\item 
	{S. Pogorzalek, K. G. Fedorov, M. Xu, A. Parra-Rodriguez, M. Sanz, M. Fischer, E. Xie, K. Inomata, Y. Nakamura, E. Solano, A. Marx, F. Deppe, and R. Gross, {\it Secure quantum remote state preparation of squeezed microwave states}, \href{https://www.nature.com/articles/s41467-019-10727-7}{Nat. Comm. {\bf10}, 2604 (2019)}}.
	\item
	{A. Parra-Rodriguez, P. Lougovski, L. Lamata, E. Solano, and M. Sanz, {\it Digital-analog quantum computation}, \href{https://journals.aps.org/pra/abstract/10.1103/PhysRevA.101.022305}{Phys. Rev. A {\bf101}, 022305 (2020)}}.
	
\end{enumerate}

Preprints submitted:
\begin{enumerate}[resume]
	\item
	{R. Asensio-Perea, A. Parra-Rodriguez, G. Kirchmair, E. Solano, E. Rico, {\it Chiral states and nonreciprocal phases in a Josephson junction ring
		}, \href{https://arxiv.org/abs/2009.11254}{arxiv:2009.11254 (2020)}}.
	
\end{enumerate}
\cleardoublepage}


\mainmatter
\pagestyle{fancy}

\chapter{Introduction}
\label{chapter:chapter_1}
\thispagestyle{chapter}
\hfill\begin{minipage}{0.85\linewidth}
	{\emph{La verdad adelgaza y no quiebra, y siempre nada sobre la mentira como el aceite sobre el agua\\\\
	(The truth may be stretched thin, but it never breaks, and it always surfaces above lies, as oil floats on water)}}
\end{minipage}
\begin{flushright}
	\textbf{Miguel de Cervantes Saavedra} \\
	{Don Quijote de la Mancha}
\end{flushright}
\vspace*{1cm}
For thousands of years, humans have developed sophisticated methods to count or compute. Between the Sumerian abacus and the programmable electronic computers~\cite{Ifrah:2000}, fundamental steps have been taken to arrive to a minimal expression of information, the bit, i.e. a binary variable with the values of ``0" and ``1". The earliest and slowest humanly-controlled mechanical computers laid the foundations to electro-mechanical autonomous systems used at the wars of the beginning of last century. In recent years, and after the invention and posterior miniaturization of the transistor\footnote{The Nobel Prize in Physics 1956 was awarded jointly to William B. Shockley, John Bardeen and Walter H. Brattain “for their researches on semiconductors and their discovery of the transistor effect”.}~\cite{Lilienfeld:1928,Bardeen:1948,Shockley:1948}, pure electrical machines have achieved the speed record of computation. However, it has been suggested for decades already that neither the classical computational paradigm formalized by Turing~\cite{Turing:1937} nor the classical electrodynamics~\cite{Jackson:1999} vehicle to carry the information are fundamental limits for the speed of a calculation. 

In fact, in the early 80's, Paul Benioff~\cite{Benioff:1982} introduced for the first time the quantum mechanical version of the Turing's machine based on a classical description by Charles Bennett~\cite{Bennett:1973}, while Yuri Manin~\cite{Manin:1980} and Richard Feynman~\cite{Feynman:1982} envisioned paradigms of simulation, a specific computational problem in which a partial description of reality is imitated based on the rules of quantum mechanics. These early efforts were brilliantly summarized by the latter in the sentence: 

{\it Nature isn't classical, dammit, and if you want to make a simulation of nature, you'd better make it quantum mechanical, and by golly it's a wonderful problem, because it doesn't look so easy}. 

The beginning of the quantum computation era was founded on the basis that the minimal information unit is the quantum bit or {\it qubit}, which can be encoded in a quantum state $\ket{\psi}$ living in a two-dimensional Hilbert space $\mcl{H}$ with a basis commonly denoted as $\ket{0}$, and $\ket{1}$, see Nielsen and Chuang~\cite{Nielsen:2010} for an introduction to the topic. Classically, a pure state corresponds to being either in zero or in one. On the other hand, a quantum mechanical pure state is a weighted combination of zero and one, i.e. 
\begin{equation}
	\ket{\psi}=a \ket{0}+ b\ket{1},\quad\{a,\,b \}\in \mathbbm{C},
\end{equation}
with the only restriction that $|a|^2+|b|^2=1$, and $\mathbbm{C}$ denoting the set of complex numbers. In fact, more general representations of quantum information require the use of a matrix, known as {\it density matrix}, 
\begin{equation}
	\rho=\sum_i p_i \ket{\psi_i}\bra{\psi_i},
\end{equation}
where $p_i$ are weighting coefficients with the property that $\mathrm{Tr}(\rho)=1$, $\ket{\psi_i}$ any two-dimensional vector, and  $\ket{\cdot}\bra{\cdot}$ representing outer product of vectors. Interestingly, the density matrix allows the representation not only of {\it pure} states, e.g. $\rho=\frac{1}{2}(\ket{0}+\ket{1})(\bra{0}+\bra{1})$ but also non-coherent {\it mixed} states, e.g. $\tilde{\rho}=\frac{1}{2}(\ket{0}\bra{0}+\ket{1}\bra{1})$. Furthermore, the density matrix $\rho$ of a register of $N$ qubits acts on a tensor product of single-qubit Hilbert spaces  $\mcl{H}=\mcl{H}_1\otimes\mcl{H}_2\otimes\dots\otimes\mcl{H}_N$. Such multi-qubit quantum spaces permit the description of more general {\it entangled} states without classical counterpart, e.g. $\rho=\ket{\psi}\bra{\psi}$ and $\ket{\psi}=(\ket{0}\ket{0}+\ket{1}\ket{1})/2$, where upon a first measurement of the qubit-1 in the state zero, one knows for certain that the outcome of an other measurement  in qubit-2 is also zero, and reversely with state one. 

Armed with the fundamentals of controllable light-matter quantum mechanical systems\footnote{Serge Haroche and David J. Wineland received in 2012 the Nobel Prize “for ground-breaking experimental methods that enable measuring and manipulation of individual quantum systems”}~\cite{Wineland:1975,Hansch:1975,Paul:1990,Haroche:1989,Phillips:1998}, in the early 90's a race began to make Feynman's dream come true and build quantum machines able to compute at will. Cirac and Zoller~\cite{Cirac:1995} pioneered a method to implement two-qubit gates, the building blocks to produce an entangled state from a separable one, with trapped ions, where in that case, the qubit would be encoded in the spin of the most-outer electron of an ion suspended in mid-air by a slow electromagnetic potential. Such proposal was quickly implemented in the laboratory of David Wineland~\cite{Monroe:1995}.

The field of quantum information gained strong interest after the discovery by Peter Shor of a quantum algorithm to factorize large prime numbers \cite{Shor:1994,Shor:1997} based on the entanglement of a register of qubits which would exponentially outperform the best known classical one. Other quantum algorithms have been envisaged with potential speed-up over classical ones, Grover's algorithm \cite{Grover:1996} for data-base search or HHL's for solving linear systems~\cite{Harrow:2009} and its extension~\cite{Wossnig:2018}. Meanwhile, other physical implementations have also been engineered, and are currently under strong development in laboratories, such as linear optics~\cite{Chuang:1995,Knill:2001}, quantum dots~\cite{Loss:1998}, nuclear magnetic resonance (NMR)~\cite{Cory:1997,Gershenfeld:1997}, optical lattices~\cite{Brennen:1999}, crystallographic defects in diamond~\cite{Cappellaro:2009}, etc. Special mention must be given to superconducting circuits~\cite{Nakamura:1999} working barely above absolute zero temperature, which have very recently beaten the so-called quantum {\it advantage} threshold within the paradigm of noisy intermediate-scale devices (NISQ)~\cite{Preskill:2018}. In essence, the Google team claimed that the programmable quantum algorithm run in their superconducting chip to compute quamtum-state probabilities has a substantial speed-up with respect to any classical one \cite{Arute:2019}.

Superconducting quantum technology is based on materials that below a certain threshold temperature, and energy (the superconducting gap) have negligible resistance~\cite{Tinkham:2004}. In this working regime, phenomenological London-Maxwell~\cite{London:1935} equations correctly capture the relevant physics when the fundamental charge of the theory is taken to be two bounded electrons, known as Cooper pairs~\cite{Cooper:1956}. Beyond the fact that Cooper pairs flow about it superconducting materials in absence of resistance, the bulk material behaves as a perfect diamagnet, i.e. it perfectly expels the magnetic field in its interior. In other words, a superconductor is more than just a perfect conductor, as first proved by Meissner and Ochsenfeld in 1933~\cite{Meissner:1933}. A refined microscopic (quantum) theory of superconductivity was later introduced by Bardeen, Cooper and Schrieffer in their seminal paper~\cite{Bardeen:1957}, where they explained that any negative potential interaction between electrons would suffice for electron pairs to be formed below certain temperature\footnote{John Bardeen, Leon N. Cooper and John R. Schrieffer won the 1972 Nobel Price in Physics ``for their jointly developed theory of superconductivity, usually called the BCS-theory".}. 

Interestingly, an effective and very efficient theory below the superconducting gap can be framed in terms of collective degrees of freedom, i.e. the macroscopic phase $\varphi$ of the aggregate wavefunction in each superconducting island~\cite{Chaikin:1995}. In other words, an effective classical theory (London-Maxwell) emerges from the fundamentally quantum bosonic nature of superconductors, and symmetry breaking~\cite{Tinkham:2004,Bardeen:1957}. Crucially however, by working at sufficiently low energy regimes where the thermal population is almost negligible, this macroscopic phase behaves quantum mechanically~\cite{Martinis:1987,Clarke:1988}. In fact, the non-dissipative superconducting phase, together with its linear response to the electromagnetic field allowed the construction of superconducting circuits behaving as artificial atoms, giving birth to the field of circuit quantum electrodynamics (cQED)~\cite{Blais:2004}. This is nothing but the analogue to the historic field of cavity quantum electrodynamics (CQED)~\cite{Walther:2006} where real atoms are placed in cavities for the enhancement of electromagnetic interaction. As opposed to real atoms, artificial ones made of superconducting circuits have the advantage of no fundamental limit for the coupling parameter between each other or with light modes, and in fact can be properly engineered. The general procedure for the study of circuits working in the quantum regime will be to derive their classical Hamiltonians, whose equations of motion are written in terms of pairs of conjugated variables that are promoted to quantum operators. 

A big step was taken in the understanding of superconductivity with the discovery of the Josephson effect~\cite{Josephson:1962}, named after its discoverer, Brian Josephson\footnote{Brian D. Josephson won the Nobel prize in Physics in 1973 ``for his theoretical predictions of the properties of a supercurrent through a tunnel barrier, in particular those phenomena which are generally known as the Josephson effect".}. The first phenomenological equation found relates the current passing through a junction made of two superconductors separated by a thin insulating barrier and the difference of phase in the junction in a nonlinear fashion, i.e. 
\begin{equation}
	I_J=I_c \sin(\delta \varphi_J),\label{eq:ac_JJ_relation}
\end{equation}
where $I_c$ is a critical constant current, and $\delta\varphi_J=\varphi_1-\varphi_2$ is the difference of phases of the wave function on the two superconductors. The second one tells us that the phase difference  between two different islands is linearly proportional to the voltage drop as~\cite{Josephson:1974}
\begin{equation}
	V_J=\frac{\Phi_q}{2\pi} \delta\dot{\varphi}_J,\label{eq:dc_JJ_relation}
\end{equation}
where $\Phi_q=h/2e$ is the quantum of flux, $h$ Planck's constant, and $e$ the electron's charge. Note that, upon the use of the (\ref{eq:dc_JJ_relation}) in (\ref{eq:ac_JJ_relation}) one may interpret the Josephson effect as a nonlinear inductor. Years later, this nonlinear effect has become the key ingredient to make artificial atoms with superconducting circuits. This phenomenological picture was completed in 1968 by McCumber~\cite{McCumber:1968} and Stewart~\cite{Stewart:1968}, who showed the requirement of a resistive (R) and a capacitive (C) contribution to the response of the junction when subjected to a current source, coining the name resistively and capacitively shunted junction (RCSJ) model. Nowadays, however, the intrinsic resistance of the junction is not typically the leading phenomenon inducing the loss of coherence in quantum circuits, and thus is commonly neglected. Within the approximate absence of dissipation, superconducting circuitry may well be described by Hamiltonian dynamics. However, instead of using the microscopic differential equations governing the electrodynamics of superconductors, it is sometimes possible to capture the essential physics within the so-called lumped element approximation.

\begin{figure}[h]
	\centering
	\includegraphics[width=.95\textwidth]{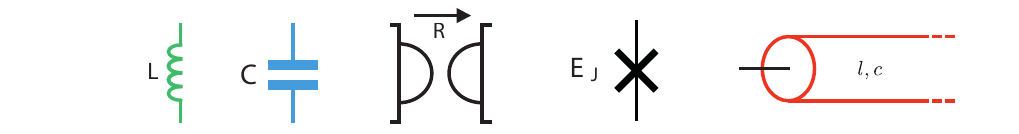}
	\caption{Circuital representation of the lumped element inductor $(L)$, capacitor $(C)$, gyrator $(R)$ and pure Josephson element $(E_J)$, and the distributed transmission line, described by its capacitance $(c)$ and indutance $(l)$ per unit paramenters.}
	\label{fig:L_C_G_JJ_TL}
\end{figure}

The lumped element approximation, valid for wavelengths bigger than the characteristic length of the circuit, divides the electromagnetic problem in a network into two sub-problems~\cite{Devoret:1997}; the topological one, and the geometric one. A crucial assumption however is required, namely that the wires connecting lumped elements are perfectly conducting. Furthermore, the approximation implies that all electric and magnetic fields live inside of the lumped elements~\cite{Feynman:2010}, thus the name,  in such a way that they do not interact strongly with other lumped elements. Regarding the geometrical problem, it is possible to solve for each lumped element separately its response to the electric fields. For instance, linear time-reversal symmetric relations between electromagnetic sources and Maxwell equations can be captured in capacitance $C$ and inductance $L$ coefficients, in what are known as constitutive equations of the lumped elements. Collective linear dynamics effectively breaking time-reversal symmetry can be captured by the fundamental nonreciprocal linear element, the gyrator~\cite{Tellegen:1948}, which has four terminals (two ports) and is described by a resistance parameter $R$ and a two by two matrix, see Fig. \ref{fig:L_C_G_JJ_TL}. For a given set of geometrical coefficients, the electrical network problem is reduced to a topological problem of wiring connections representing a set of differential equations for voltages and currents, or fluxes and charges, known as Kirchhoff's laws. 

General as it is, a finite network of ideal lumped elements cannot capture physics of real-chip 3D volumes above certain frequencies. For instance, a section of a co-planar superconducting waveguide can ideally hold an infinite number of quasi-transversal electromagnetic (quasi-TEM) modes with monotonically-increasing frequency, and thus an infinite lumped element network is required to match its impedance response. Alternatively, it is possible in that case to reduce the 3+1 dimensional classical problem to an effective 1+1 dimensional one, given that the electromagnetic fields are constrained to a small volume and have certain symmetries~\cite{Pozar:2009} (like a co-axial cable). In fact, one can describe the infinite-dimensional waveguides directly with a transmission line model (see sketch in Fig. \ref{fig:L_C_G_JJ_TL}) of differential ideal capacitors and inductors whose dynamics are anchored to differential Kirchhoff's laws also known as telegrapher's equations~\cite{Pozar:2009}.

The power of Kirchhoff's equations and the lumped element approximation is illustrated when describing linear multi-port electromagnetic systems without attending to their geometrical structure. Engineers of the past century proved that it is systematically possible to fit the scattering response $\mS(\omega)$ of a multi-port 3D (lossless) linear electromagnetic environment, now known in the circuit QED community as a {\it black-box}, to a circuit containing (infinite) stages of lumped element capacitors, inductors, gyrators and ideal transformers. For example, a general lossless impedance $\msZ(\omega)$ can be decomposed in lumped-element circuit as we will see further in chapters. A very complete reference covering the synthesis problem of lumped circuits may be found in \cite{Newcomb:1966}, showing what they are known as multi-port immitance Foster/Cauer (lossless) and Brune (lossy) expansions, or the more generic scattering matrix expansion from Belevitch and Oono--Yasuura. It may be worth recalling that no such equivalent circuits, when used as a modelling tool, can capture the microscopic physics from inside of the electromagnetic box, although it does provide useful information for the outside observer. 

\begin{figure}[h]
	\centering
	\includegraphics[width=0.5\textwidth]{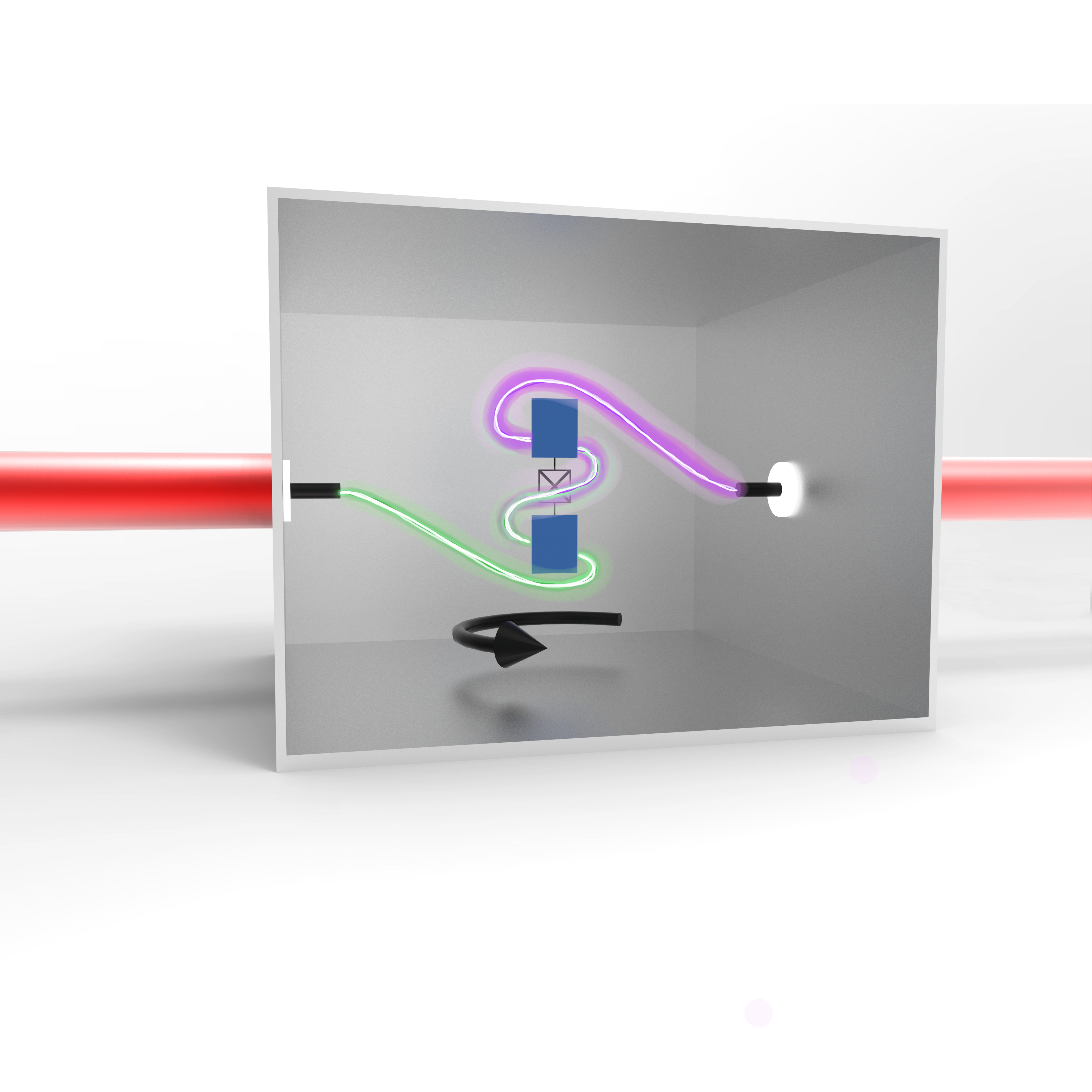}
	\caption{Artistic representation of a nonreciprocal electromagnetic box connected throught output ports to transmission lines, and coupled by a capacitive antenna through the electric field to a Josephson junction. Designed by Irene Parra Rodr\'iguez.}
	\label{fig:NR_blackbox_2TL_1JJ}
\end{figure}

There are two main superconducting circuit set-ups, those containing 3D cavities with possibly embedded quasi-2D planar chips where the cavities play a role in the quantum information processing, for example by enhancing effective couplings between qubits, and those where these cavities act as noise isolators encapsulating the whole cryogenic area. In both cases, transmission lines may be connected either to the ports of a cavity or directly to the planar chips as in Fig. \ref{fig:NR_blackbox_2TL_1JJ}. In that sense, it is useful to see the chip divided in linear or nonlinear black-boxes where a finer or coarser point of view may be required for capturing the essential dynamics. The 3D cavities can be easily incorporated in the full Kirchhoff's analysis after making use of multi-physics simulations of Maxwell equations or direct scattering measurements, as first proposed by Nigg et al.~\cite{Nigg:2012}. As previously mentioned, given that the superconducting chips are set to work in cryogenic temperatures (around $10$ mK), the effective Kirchhoff models must be understood from a quantum mechanical point of view. 

The problem thus simplifies to quantizing a set of differential equations written in terms of (canonically) conjugate variables, i.e.  finding a Hamiltonian containing a (minimal) set of conjugate variable pairs~\cite{Solgun:2014,Solgun:2015,Minev:2020}, typically fluxes and charges, and a Poisson bracket to be promoted to a commutator which will determine the time evolution. Actually, conjugate variables with a canonical bracket facilitate the process but it is not strictly compulsory, see for instance~\cite{Faddeev:1988}. In fact, not even a second order kinetic energy term in a Lagrangian is required for finding a correct Hamiltonian without needing to invoke Dirac's procedure~\cite{Dirac:1950,Dirac:1959} for removing constrains~\cite{Faddeev:1988,Jackiw:1994}.

\begin{figure}[h!]
	\centering
	\includegraphics[width=1\textwidth]{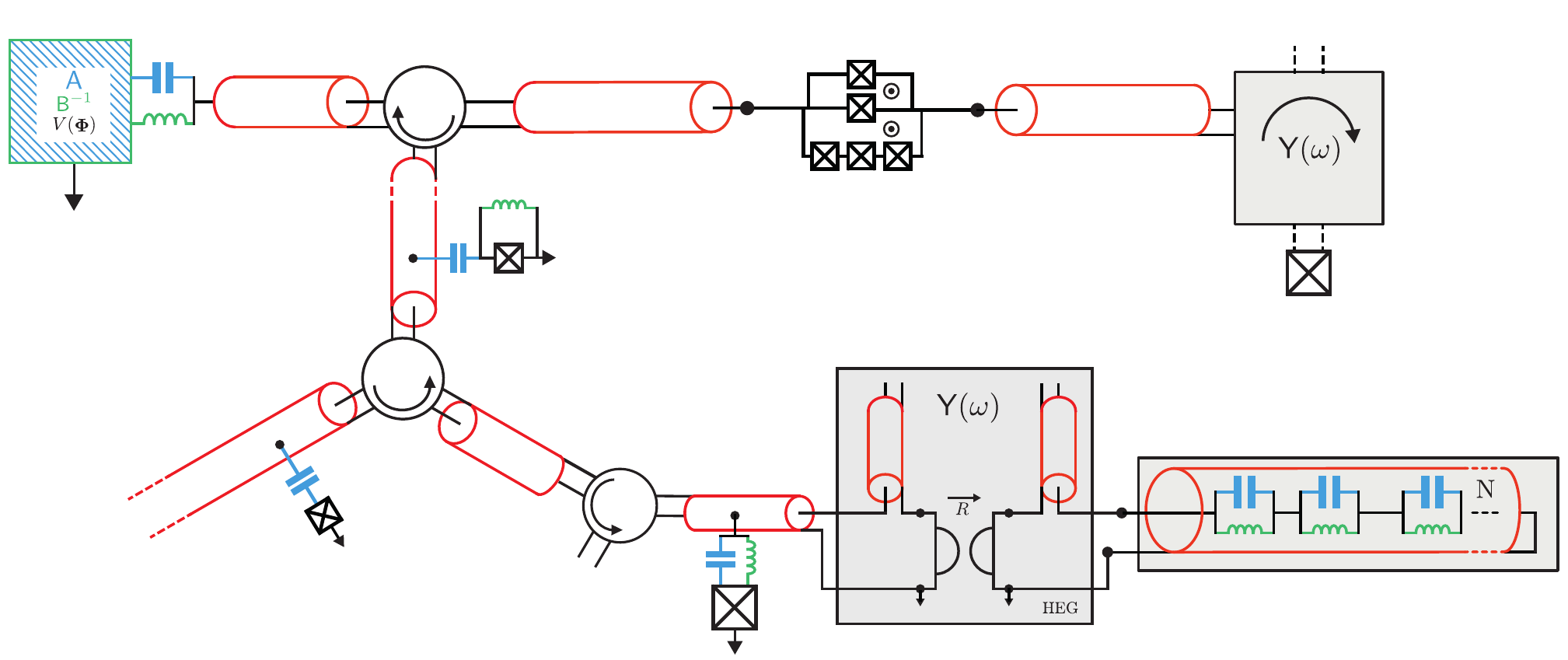}
	\caption{Superconducting network containing lumped and distributed elements.}
	\label{fig:QNetwork}
\end{figure}

The work in this Thesis responds to a simple question: 

{\it Is it possible to find systematically a convergent quantum theory of superconducting chips from lumped and distributed Kirchhoff's equations (like that in Fig. \ref{fig:QNetwork})?} 

The answer is yes. In fact, a simple analysis shows that the decoupling of the lumped degrees of freedom (e.g. Josephson junctions) at zero and infinite frequency must occur as capacitors and inductors behave as open and short (short and open) terminals for infinite (zero) frequencies respectively. The key point to prove the statement is to correctly describe the coupling between unbounded multi-mode subsystems and finite-dimensional ones. This result is in contrast with the phenomenological and historical nonrelativistic light-matter models in cavity QED. For example, in the multi-mode quantum Rabi models in the dipole approximation, with energiy levels of atoms dressed by the electromagnetic modes, ultraviolet frequency cutoffs or renormalization techniques have been commonly invoked to make finite predictions of observable quantities~\cite{Bethe:1947}.

Here we argue that, no matter whether partial lumped elements in the circuit are a valid representation of 3D chip volumes for infinite frequency, Kirchhoff's circuit models have natural cutoffs which can be made explicit in the Hamiltonian, making quantum field theory programs such as renormalization unnecessary and superfluous in this context. In other words, the lumped element approximation, even when combined with the differential form for a transmission line, introduces a length scale. This comes in as specially practical when Hamiltonian models are used for predicting or designing~\cite{Solgun:2019,Ku:2020} the multi-mode dispersive Lamb shifts (effective couplings) of (between) qubits. The above statement can be simply understood in the context of a spin-boson model described by a multi-mode quantum Rabi Hamiltonian
\begin{equation}
	H/\hbar=\sum_n^N \omega_n a^\dag_n a_n +\sum_{n,i} g_n^{(i)}(a_n+a_n^\dag) \sigma_x^{(i)} +  \sum_i \omega_{q}^{(i)} \sigma_z^{(i)},
\end{equation}
where $\omega_n$ are the frequencies of a set of harmonic modes, $g_n^{(i)}$ is the coupling parameter of mode $n$ to spin $i$, $\omega_{q}^{(i)}$ is the frequency of its corresponding spin, $a_n$ and $a_n^\dag$ are annihilation and creation operators of mode $n$, and $\sigma_x$ and $\sigma_z$ are the first and third Pauli matrices. Under an adiabatic elimination~\cite{Blais:2004}, an effective Hamiltonian $\tilde{H}/\hbar\approx\left(\omega_{q}^{(i)}+\chi^{(i)}\right)\sigma_z^{(i)}+\tilde{g}_{i,j}(\sigma_+^{(i)}\sigma_-^{(j)} + \text{h.c.})$ for the spins can be found, where their frequencies are shifted by 
\begin{equation}
	\chi^{(i)}\propto \sum_n^N \frac{g_n^2}{\Delta_n^{(i)}},
\end{equation}
with $\Delta_n^{(i)}\equiv \omega_n-\omega_q^{(i)}$, and $\sigma_+^{(i)}$ the raising operator of the spin, and the qubits are coupled by the parameters 
\begin{equation}
	\tilde{g}_{i,j}\propto \sum_n^N g_n^{(i)}g_n^{(j)}\left(\frac{1}{\Delta_n^{(i)}}+\frac{1}{\Delta_n^{(j)}}\right).
\end{equation}

It is easy to realize that in the limit of monotonically increasing frequencies and coupling parameters, e.g. $\omega_n\rightarrow\infty$ and $g_n\propto\sqrt{\omega_n}$ with $n\rightarrow\infty$ (as it is usually considered~\cite{Blais:2004,Gardiner:2004}), the above sums diverge for an infinite number of harmonic modes~\cite{Filipp:2011}. In the following chapters, we show that the lumped-element coupling of any infinite dimensional electromagnetic systems to other (localized) degrees of freedom contains a natural cutoff, leading to the convergence of these sums (or integrals in the continuum limit). Naturally, this entails $g_n$ falling off as $\omega_n^{-\epsilon}$ for $n\rightarrow\infty$ with $\epsilon>0$.

A second important result demonstrated here is the adequacy of a doubled description in terms of flux {\it and} charge variables (both in the configuration space) for quantizing circuits with transmission lines and nonreciprocal elements, i.e. circulators or gyrators, elements that break time-reversal symmetry. The apparently dispensable redundancy introduced in the analysis turns out to be the correct starting point to derive exact Hamiltonians, and it can be systematically eliminated by making use of the electromagnetic duality symmetry in the phase space.

\section{What you will find in this thesis}
In this Thesis\footnote{The notation is homogeneous within each chapter, but not necessarily throughout the Thesis.}, we provide canonically quantized Hamiltonian models of reciprocal and nonreciprocal superconducting circuits based on Kirchhoff's laws within a mixed lumped-differential approximation containing no divergence issues. For that, we first review in {\bf chapter \ref{chapter:chapter_1}} the historical quantization procedure of the LC harmonic oscillator from a more general and pedagogical perspective.

In {\bf chapter \ref{chapter:chapter_2}}, we introduce a simple multi-mode model in circuit QED containing the fictitious divergence issue in the point-like capacitive (kinetic energy) coupling, readily, a Josephson junction coupled to a transmission line resonator described by the one-port Foster expansion. We show explicitly the convergence of the model even with the removal of the Josephson capacitance for an increasing but finite number of modes, by computing the Lamb shift. Finally, we make explicit the connection between the low-lying energy spectrum and a truncated Hamiltonian model. A numerically exact convergent coupling parameter is first presented. 

In {\bf chapter \ref{chapter:chapter_3}}, we present the core theory for the exact Hamiltonian description of networks with transmission lines linearly coupled to nonlinear reciprocal networks using pure flux variables. We explain the mathematical details missing in the literature that permit the exact expansion of the flux field with capacitive and inductive coupling in the Lagrangian. An exact Legendre transformation is presented for a catalogue of coupling configurations and we show the relevant length scale for the decoupling of dressed harmonic modes to the networks. We show the connection between the methods described in this chapter and those of the seminal reference by Devoret~\cite{Devoret:1997} in the computation of quantum fluctuations of dressed infinite dimensional systems. We further review the discrete infinite limit of the black-box quantization procedure with its associated inner product.   

In {\bf chapter \ref{chapter:chapter_4}}, we introduce the nonreciprocal ideal elements, the gyrator and the circulator, in Lagrangian and Hamiltonian descriptions of superconducting circuits. We extend the generic rules for node-flux quantization of Burkard {\it et al.} \cite{Burkard:2004} and Solgun {\it et al.} \cite{Solgun:2015}, and apply them to the quantization of pedagogical and pathological examples, i.e. the generic two-port nonreciprocal lossless impedance capacitively coupled to Josephson junctions, and the Hall effect Viola-DiVincenzo gyrator \cite{Viola:2014}. We quantize dual circuits containing Josephson (phase-slip) junctions, parallel capacitors (series inductors) and admittance (impedance) described circulator in terms of node-flux (loop-charge) variables. 

In {\bf chapter \ref{chapter:chapter_5}},  we study the quantization of circuits with transmission lines coupled by ideal nonreciprocal elements. We introduce the double configuration-space description as an optimal method for quantizing such systems. We show how to remove the redundant degrees of freedom making use of the electromagnetic duality symmetry in the phase space instead of the configuration space. We exemplify the above procedure extending the quantization of a circuit well treated within the context of chapter \ref{chapter:chapter_2}, a Josephson junction  capacitively coupled to a transmission line, and connect it through a circulator to two other lines. Finally, we set the grounds for the quantization of realistic frequency-dependent nonreciprocal devices coupled to transmission lines and Josephson junctions, which remains an open problem.

\section{Quantization of the LC oscillator revisited}
\begin{figure}[h!]
	\centering
	\includegraphics[width=0.6\textwidth]{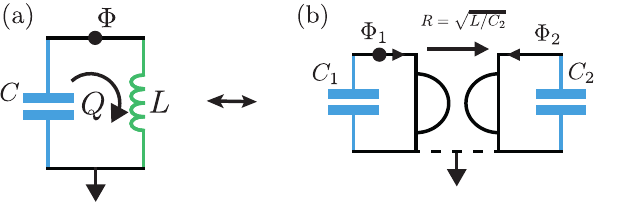}
	\caption{Superconducting network containing lumped and distributed elements.}
	\label{fig:LC_CGC_circuit}
\end{figure}

As an introduction to the general topic of the Thesis, let us review the canonical quantization of the LC resonator, the ubiquitous example of circuit QED.  Following  standard analysis~\cite{Devoret:1997,Burkard:2004,Ulrich:2016}, the Lagrangian of circuit (a) in Fig.~\ref{fig:LC_CGC_circuit}(a) can be described in terms of a single node-flux variable $\Phi(t)=\int_{-\infty}^t dt' \, V(t')$, or a loop-charge variable $Q(t)=\int_{-\infty}^t  dt'\, I(t')$ variables, i.e. 
\begin{align}
	L_\Phi=\frac{C\dot{\Phi}^2}{2}-\frac{\Phi^2}{2L},\quad \text{and}\,\quad
	L_Q=\frac{L\dot{Q}^2}{2}-\frac{Q^2}{2C},\label{eq:Lag_Phi_Q_LC_resonator}
\end{align}
which correspond to the equations of motion (EOMs)
\begin{align}
	C\ddot{\Phi}+\frac{\Phi}{L}&=0,\quad \text{and}\,\quad
	L\ddot{Q}+\frac{Q}{C}=0,\nonumber
\end{align}
respectively. Please note that the equation in node-flux (loop-charge) variable is the Kirchhoff current $I_C=-I_L$ (voltage $V_C=V_L$) law. In order to generalize the procedure to more complex circuits with a pure node-flux or loop-charge description, one must deal with a bigger set of constraints imposed by the opposite Kirchhoff law to the EOM. This reduction of nondynamical variables is typically performed directly (i) in the Lagrangian, or equivalently (ii) in the EOMs upon which one induces a Lagrangian, but can also be done (iii) in the phase space as we show later. Extended discussions on how to perform this variable reduction using graph theory can be found in~\cite{Devoret:1997,Burkard:2004,Burkard:2005,Solgun:2014,Solgun:2015,Ulrich:2016}. Arbitrary connections of lumped-elements may bring in free-particle dynamics, e.g. connecting two capacitors in series, which are typically eliminated using graph theory reduction techniques, but could also be dealt with in the Hamiltonian (phase space). We make our contribution to this topic later in chapter~\ref{chapter:chapter_4} with the inclusion of nonreciprocal ideal lumped elements within the Burkard node-flux analysis~\cite{Burkard:2005}, and an example of free-particle embedded in a lumped-element non-reducible circuit. 

Naturally, the Hamiltonian for both descriptions (\ref{eq:Lag_Phi_Q_LC_resonator}) can be easily derived by a Legendre transformation involving the definition of the conjugate variables $\Pi=\partial L_\Phi/\partial \dot{\Phi}$ and $\pi=\partial L_Q/\partial \dot{Q}$ with the canonical Poisson brackets $\{\Phi,\Pi\}=\{Q,\pi\}=1$, readily
\begin{align}
	H_\Phi&=\frac{\Pi^2}{2C}+\frac{\Phi^2}{2L},\quad \text{and}\,\quad
	H_Q=\frac{\pi^2}{2L}+\frac{Q^2}{2C}.
\end{align}
The canonical quantization of the above Hamiltonians ends with the promotion of the conjugate variables to quantum operators, and the exchange of Poisson brackets by the commutators, e.g.  $\{\Phi,\Pi\}\rightarrow[\hat{\Phi},\hat{\Pi}]=i\hbar$, and analogously for $Q$ and $\pi$. Upon the definition of annihilation and creation operators, e.g. $\hat{\Phi}=\sqrt{\hbar/2C}(a+a^\dag)$ and $\hat{\Pi}=-i\sqrt{\hbar C/2}(a-a^\dag)$ and analogously for $\hat{Q}$ and $\hat{\pi}$, such that $[a,a^\dag]=1$, both Hamiltonians are equivalent to that of the harmonic oscillator
\begin{equation}
	H=\hbar \omega_R a^\dag a,
\end{equation}
where the frequency is $\omega_R=1/\sqrt{LC}$. Typical quantum optics analyses can follow from here.
\subsubsection*{Doubled configuration space}
It is instructive to derive the above result from a redundant Lagrangian description, see \cite{Ulrich:2016} for more examples. The main idea is to use the duality symmetry present in electrodynamics, here represented by the loop-charge and node-flux descriptions. For instance, a set of redundant Euler-Lagrange (E-L) equations of motion for the same LC circuit in Fig.~\ref{fig:LC_CGC_circuit}(a) is 
\begin{align}
\label{eq:LC_EOMs_doubled_space}
\begin{split}
C\ddot{\Phi}+\dot{Q}&=0,\\
L\ddot{Q}-\dot{\Phi}&=0,
\end{split}
\end{align}
derivable from the Lagrangian
\begin{equation}
L=\frac{1}{2}\left(C\dot{\Phi}^2+L\dot{Q}^2-\dot{Q}\Phi+\dot{\Phi}Q\right).\label{eq:Lagrangian_LC_ds}
\end{equation}

In contrast to Lagrangian (\ref{eq:Lag_Phi_Q_LC_resonator}), now we have two coupled variables in the Lagrangian. However, we still have a full rank kinetic matrix readily at hand, allowing us to perform again a Legendre transformation $\Pi=\partial L/\partial \dot{\Phi}$ and $\pi=\partial L/\partial \dot{Q}$ to derive the Hamiltonian 
\begin{equation}
\tilde{H}=\frac{(\Pi-Q/2)^2}{2C}+\frac{\left(\pi+\Phi/2\right)^2}{2L}.
\end{equation}
A canonical transformation of the two pairs of conjugate variables 
\begin{align}
	\tilde{\Pi}&=\Pi-Q/2, \quad\tilde{\pi}=\pi-\Phi/2,\nonumber\\
	\tilde{\Phi}&=\Phi/2+\pi,\quad\tilde{Q}=Q/2+\Pi,\nonumber
\end{align}
reveals its one-oscillator nature 
\begin{equation}
	H=\frac{\tilde{\Pi}^2}{2C}+\frac{\tilde{\Phi}^2}{2L}\equiv_{\text{quant.}}\hbar \omega a^\dag a,\label{eq:H_LC_doubled_space}
\end{equation}
where we have made implicit the quantization procedure and the transformation to annihilation and creation operators. It is worth highlighting that the two nondynamical phase-space variables which do not appear in (\ref{eq:H_LC_doubled_space}) correspond to two of the four free parameters in the solution of the classical differential equations (\ref{eq:LC_EOMs_doubled_space}), i.e. the overall constant charge and flux in the circuit. We will generalize this procedure for the canonical quantization of transmission lines coupled to nonreciprocal linear systems in later chapter~\ref{chapter:chapter_5}.

As a brief introduction to the quantization of circuits with nonreciprocal elements, realize that (\ref{eq:Lagrangian_LC_ds}) is equivalent to that of two capacitors coupled by an ideal gyrator, see Fig.~\ref{fig:LC_CGC_circuit}(b),

\begin{equation}
L_{\text{C-G-C}}=\frac{1}{2}\left(C_1\dot{\Phi}_1^2+C_2\dot{\Phi}_2^2-\frac{1}{R}\left[\dot{\Phi}_2\Phi_1-\dot{\Phi}_1\Phi_2\right]\right),\label{eq:Lagrangian_CC_gyrator}
\end{equation}
when $R=\sqrt{L/C_2}$, and identifying $\frac{\Phi_2}{R}\leftrightarrow Q$. It is worth noticing that although time-reversal symmetry would be broken in the  circuit in Fig.~\ref{fig:LC_CGC_circuit}(b) and in the double configuration-space description of the LC circuit in Fig.~\ref{fig:LC_CGC_circuit}(a), their dynamics are equivalent to the reduced and time-reversal symmetric description of the harmonic oscillator~\cite{Rymarz:2018}. 

The analysis performed in this section shows the typical pathway towards a canonical quantization of circuits. In essence, one must be able to (i) write Lagrangians with non-singular kinetic energy terms, and (ii) perform a Legendre transformation. Afterwards, extra canonical transformations may be pertinent for the reduction of nondynamical variables or free-particle dynamics. It is the main study of this Thesis the extension of this idea to harmonic infinite dimensional systems, with either discrete or continuous spectra, coupled point-wise (locally in space) through lumped elements to additional nonlinear degrees of freedom.

\chapter{Convergence of the Multimode QRM in cQED}
\label{chapter:chapter_2}
\begin{figure}[h]
	\centering
	\includegraphics[width=0.5\textwidth]{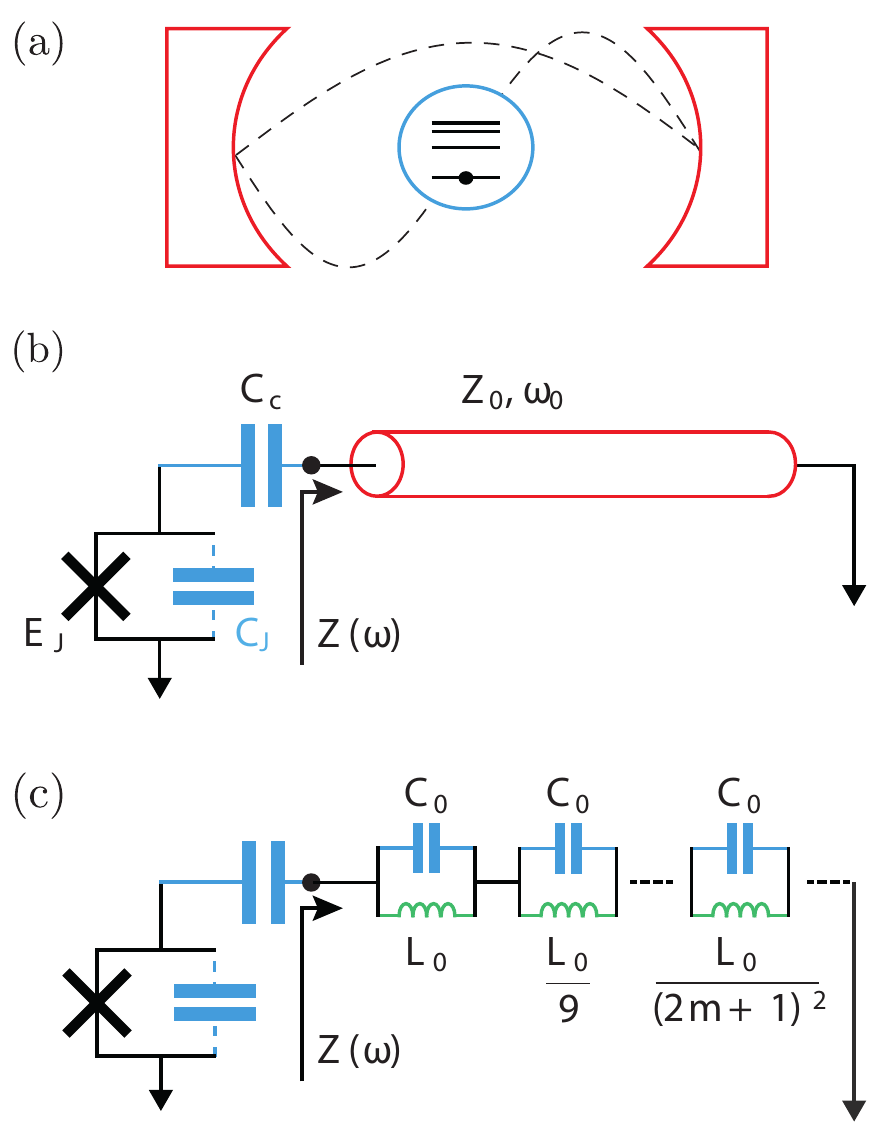}
	
	\caption{(a) Schematic representation of cavity QED: A multilevel atom (in blue) coupled to the electromagnetic cavity modes (dashed black lines). (b) Circuit QED example covered in this work: A Josephson junction anharmonic LC oscillator, or ``artificial atom", coupled to modes of a transmission line. (c) Lumped element equivalent of (b).}
	\label{fig:circuit}
\end{figure}

\thispagestyle{chapter}
\hfill\begin{minipage}{0.85\linewidth}
		{{\emph{Wahrlich es ist nicht das Wissen, sondern das Lernen, nicht das Besitzen, sondern das Erwerben, nicht das Da-Seyn, sondern das Hinkommen, was den grössten Genuss gewährt. Wenn ich eine Sache ganz ins Klare gebracht und erschöpft habe, so wende ich mich davon weg, um wieder ins Dunkle zu gehe; so sonderbar ist der nimmersatte Mensch, hat er ein Gebäude vollendet so ist es nicht um nun ruhig darin zu wohnen, sondern um ein andres anzufangen
			\\\\
			(Surely it is not the knowing but the learning, not the possessing but the acquiring, not the being-there but the getting there that afford the greatest satisfaction. If I have exhausted something, I leave it in order to go again into the dark. Thus is that insatiable man so strange: when he has completed a structure it is not in order to dwell in it comfortably, but to start another)
}}}
\end{minipage}
\begin{flushright}
	\textbf{Carl Friedrich Gauss}\\
	Briefwechsel mit Wolfgang Bolyai
\end{flushright}
\vspace*{1cm}
Cavity quantum electrodynamics (CQED) studies the fundamental interaction between light and matter, coupling individual atoms and electric field of cavity modes through the dipole moment, as described by the Rabi model~\cite{Rabi:1936} and depicted in Fig.~\ref{fig:circuit}(a). The mere confinement of the electromagnetic field to a finite region enhances this interaction and the atomic transition frequencies are Lamb-shifted~\cite{Lamb:1947,Heinzen:1987}. The first attempts at calculating these energy shifts made apparent the first shortcomings of QED theory, mainly that the transition energies of the atom diverge as the infinite number of electromagnetic modes are considered. There were several efforts to address this mathematical issue, and the concept of energy {\it renormalization} was born~\cite{Bethe:1947}. Akin to cavity QED is the field of circuit QED~\cite{Wallraff:2004}, where artificial atoms such as anharmonic superconducting circuits made of Josephson junctions couple to the modes of a waveguide resonator or an open transmission line. Such systems allow the study of plenty of quantum effects~\cite{Schuster:2007,Bishop:2009}, and are one of the most promising platforms for the realization of quantum processors~\cite{Takita:2016,Kelly:2015,Riste:2015,Arute:2019}. Despite experimental successes, ``ad hoc'' multi-mode extensions of the Rabi model suffered from divergences when considering the limit of infinite modes in a waveguide resonator~\cite{Houck:2008,Filipp:2011,Kockum:2014}.

Aware of this problem, Nigg \textit{et al.}~\cite{Nigg:2012} developed the method of black-box quantization to obtain Hamiltonians with higher predictive power~\cite{Girvin:2014}. This method proved very practical for treating weakly anharmonic systems, and indeed cure the divergence problem. In a nutshell, the method consists on inserting all the linear phenomena of the Josephson junctions inside of the electromagnetic box, whose response is approximated by a Foster lumped-element circuit, and expanding the nonlinear part of it in the normal mode basis. However, this method was not designed for systems with strong anharmonicity, such as a Cooper pair box~\cite{Buettiker:1987,Bouchiat:1998,Nakamura:1999,Koch:2007}. Furthermore, in applying the black-box procedure with the Foster circuit, the form of the quantum Rabi Hamiltonian is not preserved, and while it gives a convergent energy spectrum, it is not clear how to identify and connect it to a coupling rate between the two bipartite systems. In other words, it is no longer possible to directly identify which parts of the Rabi interaction lead to certain energy shifts of bare quantities, e.g. Bloch-Siegert shifts, highly relevant for studying the physics of ultra-strongly coupled (USC) systems~\cite{FornDiaz:2010,Andersen:2017,Bosman:2017,PuertasMartinez:2019}. Further improvements of the black-box quantization methodology were introduced by Solgun et al. \cite{Solgun:2014,Solgun:2015} where instead of the Foster decomposition, the more accurate Brune expansion was used for the exact treatment of lossy phenomena. 

In this chapter, we derive a first-principle Hamiltonian model addressing these issues in a minimal set-up, which motivates the search for canonical models for more general circuits in the following chapters. The approach used here is intimately related to the correct treatment presented by Paladino et al. \cite{Paladino:2003} where a convergent Hamiltonian model for a general environment capacitively coupled to a Josephson junction was derived but not further analysed, see below in chapter \ref{chapter:chapter_3}. This Hamiltonian, expressed in the basis of the uncoupled resonator modes and the artificial atom, is valid for arbitrary atomic anharmonicities and allows us to understand why previous attempts at extending the Rabi Hamiltonian failed. The presence or not of a Josephson capacitance $C_J$ in our study leads to two important results. First, in the limit $C_J\rightarrow 0$, the coupling rates follow a square root increase up to an infinite number of modes ($\propto\sqrt{n}$). Without introducing a cutoff in the number of coupled modes, we show that a first principle analysis of the quantum circuit leads to convergence of the energy spectrum. The $C_J=0$ limit also highlights a natural renormalization of Hamiltonian parameters, arising from the circuit analysis, which is essential to understanding how to reach correct multi-mode extensions of the Rabi model. Secondly, we study the experimentally relevant case $C_J>0$, which introduces a cutoff that suppresses the coupling to higher modes~\cite{DeLiberato:2014,Garcia-Ripoll:2015,Malekakhlagh:2016}. In particular, we provide an analysis of this regime and discuss the physics of this cutoff in the context of a lumped element circuit model. This results in a useful tool for studying multi-mode circuit QED in the framework of the Rabi Hamiltonian, or for studying strongly anharmonic regimes, out of reach of the black-box quantization method. This model was indispensable in extracting the Bloch-Siegert shift in the experiment of Ref.~\cite{Bosman:2017}, where a na\"ive extension of the Rabi model would predict a Lamb shift of more than three times the atomic frequency due to 35 participating modes before any physically-motivated cutoff, such as the qubit's physical size or the Junction capacitance, becomes relevant.

The circuit QED system studied here is an artificial atom (AA) formed from an anharmonic oscillator~\cite{Koch:2007}, capacitively coupled to a quarter-wave ($\lambda/4$) transmission line resonator~\cite{Pozar:2009} as depicted in Fig.~\ref{fig:circuit}(b). The AA is a superconducting island connected to ground through a Josephson junction characterized by its Josephson energy $E_J$. It has a capacitance to ground $C_J$ and is coupled to the voltage anti-node of the resonator, with characteristic impedance $Z_0$ and fundamental mode frequency $\omega_0/2\pi$, through a capacitance $C_c$. The Josephson junction acts as a non-linear inductor, providing a source of single-photon anharmonicity in the oscillations of current flowing through it. In order to clearly illustrate the most novel aspect of the model, the renormalization of the charging energy, we first consider the case $C_J=0$. Despite the absence of a cutoff in the number of coupled modes in this case, we find that the energy spectrum still converges. The case of $C_J>0$ is discussed at the end of this chapter and in detail in the Appendix~\ref{appendix_a}.

\section{Circuit Hamiltonian} We consider a Hamiltonian in which each uncoupled harmonic mode of the resonator, with resonance frequency $\omega_m$ and annihilation operator $\hat{a}_m$, is coupled to the transition between the bare atomic states $\ket{i}$, $\ket{j}$ with energies $\hbar\epsilon_i$, $\hbar\epsilon_j$  through a coupling strength $\hbar g_{m,i,j}$~\cite{Zhu:2013,Sundaresan:2015}. We derive such a Hamiltonian by constructing a lumped element equivalent circuit, or Foster decomposition, of the transmission line resonator as represented in Fig.~\ref{fig:circuit}(c). The input impedance of a shorted transmission line, at a distance $\lambda/4$ from the short, $Z(\omega) = iZ_0\tan (\pi\omega/(2\omega_0))$, is equal to that of an infinite number of parallel LC resonators with capacitances $C_0 = \pi/(4\omega_0 Z_0)$ and inductances $L_m = 4Z_0/((2m+1)^2\pi \omega_0)$. In order to consider a finite number of modes $M$ in the model, one replaces the $m\ge M$ LC circuits in Fig.~\ref{fig:circuit}(c) by a short circuit to ground. This removes the $m\ge M$ resonances in the resonator input impedance $Z(\omega)$ with little effect on $Z(\omega)$ for $\omega \ll \omega_M$. The focus of this chapter is on the evolution of the Hamiltonian parameters as a function of this system size $M$, and the consequences on the energy spectrum. Using the tools of circuit quantization~\cite{Devoret:1997}, we obtain as Hamiltonian of the system, (see Appendix~\ref{appendix_a} for the calculation)

\begin{figure*}[t]
	\centering
	\includegraphics[width=0.9\textwidth]{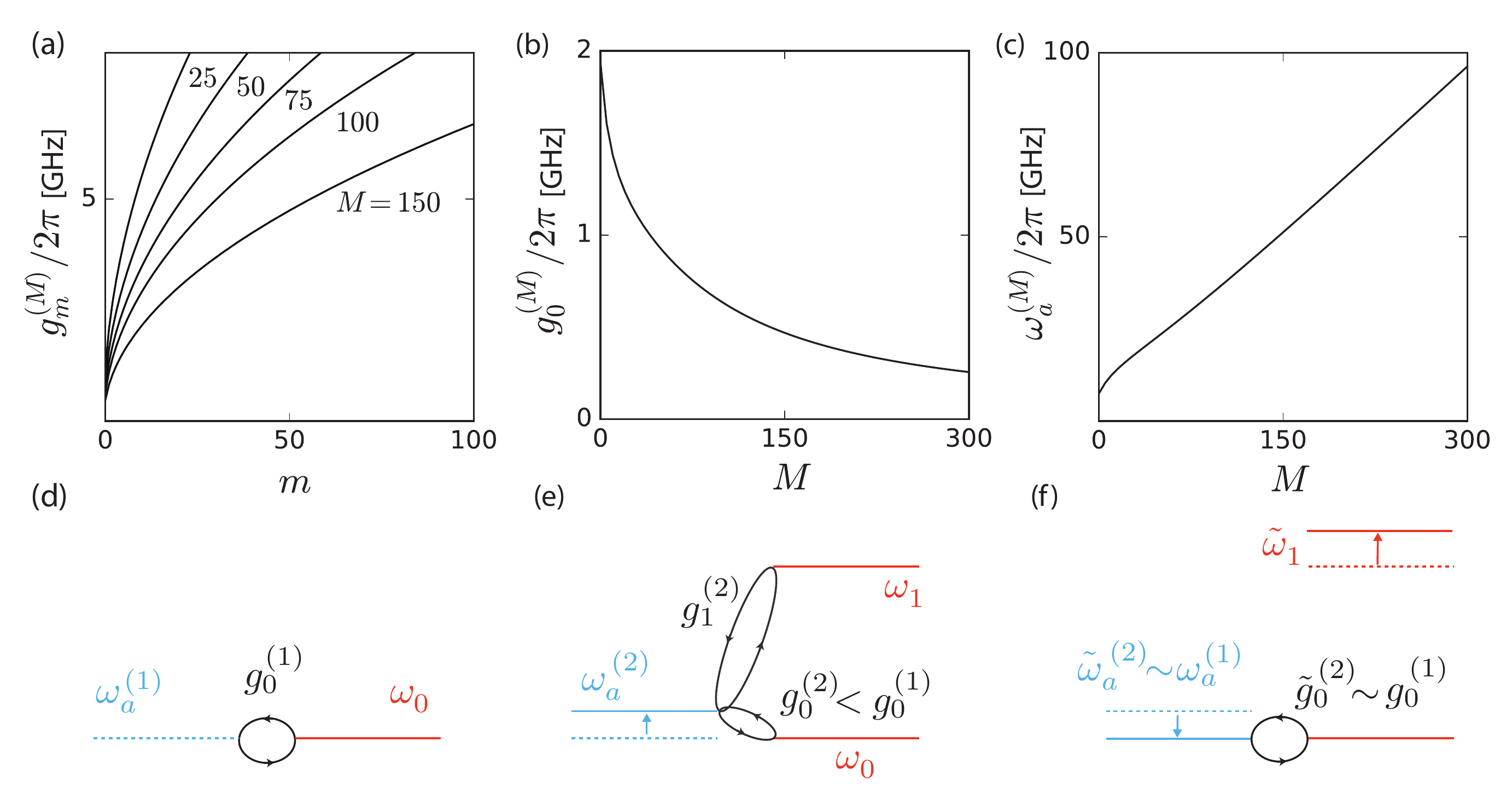}
	\caption{Renormalization of the Hamiltonian parameters and dressing of the atom by higher modes. For the plots, we choose $\omega_0/2\pi = 10$ GHz, $Z_0=50\ \Omega$, $C_c=50$ fF and $E_J/h=20$ GHz. Although this places the AA in the transmon limit, we note that the scaling shown in the figure are exact for all regimes, including the CPB limit. (a) Coupling of the ground to first excited state transition of the atom $\ket{g}\rightarrow\ket{e}$ to the resonator modes as a function of the mode number $m$: $\hbar g_{m}^{(M)} = \sqrt{2m+1}V_{\text{zpf,0}}2e\bra{g}^{(M)}\hat{n}_J\ket{e}^{(M)}$ for different values of $M$. The $M$-dependence of the coupling is detailed in (b). (b),(c) Renormalization of the coupling strength and the atomic frequency through the change in charging energy $E_C^{(M)}$ as a function of $M$. For large $M$, the coupling diminishes with $1/M$ and the atomic frequency increases linearly with $M$. (d-f) Schematic energy diagrams of the renormalization procedure in the case of the atom and fundamental mode at resonance. (a) $M=1$ (b) $M=2$.  Adding a mode shifts the bare atomic energy upwards and changes the values of the couplings. (f) 
		Dressing the atom with the second mode results in a dressed atomic state with atomic frequency and coupling to the $m=1$ mode close to that of the $M=1$ model.}
	\label{fig:renormalization}
\end{figure*}

\begin{align}
\begin{split}
\hat{H}^{(M)} =&\, \sum_i\hbar\epsilon_i^{(M)} \ket{i}^{(M)}\bra{i}^{(M)}  + \sum_{m=0}^{m<M}\hbar\omega_m\hat{a}_m^\dagger \hat{a}_m \\
&+ \sum_{i,j}\sum_{m=0}^{m<M}\hbar g^{(M)}_{m,i,j}\ket{i}^{(M)}\bra{j}^{(M)}(\hat{a}_m+\hat{a}_m^\dagger)\ .
\end{split}
\label{eq:circuit_hamiltonian}
\end{align}
The eigenfrequencies of the higher resonator modes $\omega_m$ are related to that of the fundamental mode through $\omega_m = (2m+1)\omega_0$. The coupling strength $\hbar g^{(M)}_{m,i,j} = 2e V_{\text{zpf},m}\bra{i}^{(M)}\hat{n}_J\ket{j}^{(M)}$ scales with the square root of the mode number $m$ through the zero-point voltage fluctuations of the $m$-th mode $V_{\text{zpf},m}=\sqrt{2m+1}\sqrt{\hbar\omega_0/2C_0}$.
Since we will concentrate on the frequency and coupling of the first atomic transition $\ket{g}\rightarrow\ket{e}$, we use the shorthand $\omega_a^{(M)} = \epsilon_e^{(M)}-\epsilon_g^{(M)}$ and $ g_{m}^{(M)} = g_{m,g,e}^{(M)}$ throughout this chapter.

The (bare) AA eigenstates $\ket{i}^{(M)}$ and energies $\hbar \epsilon_i^{(M)}$ in Eq.\ (\ref{eq:circuit_hamiltonian}) are those that diagonalize the Hamiltonian
\begin{equation}
\hat{H}^{(M)}_{\text{AA}} = 4E_C^{(M)} \hat{n}_J -E_J \cos(\hat{\varphi}_J)\ .
\label{eq:CPB_hamiltonian}
\end{equation}
Here $\hat{n}_J$ is the quantum number of Cooper-pairs on the island conjugate to $\hat{\varphi}_J$ the superconducting phase difference across the junction, and $E_C^{(M)}$ is the charging energy of the island. This choice of the decomposition of the Hamiltonian is one in which the bare atom corresponds to purely anharmonic degrees of freedom (currents flowing only through the junction) and the bare cavity to purely harmonic degrees of freedom (currents flowing only through the linear cavity inductors). 

The crucial consequence of this model is a renormalization of the parameters of the Hamiltonian as modes are added. In particular, the charging energy $E^{(M)}_C$ of the (bare) AA depends explicitly on the number of modes included in the equivalent circuit
\begin{equation}
E^{(M)}_{C} = \frac{e^2}{2}\frac{C_0+MC_c}{C_0C_c}\ ,
\label{eq:Ec}
\end{equation}
as reported previously in~\cite{Paladino:2003,Bergenfeldt:2012,Jaako:2016}. For the case of $M \rightarrow \infty$ with $C_J = 0 $, the charging energy of the bare atom diverges. This divergence arises from the definition of the bare atom as current oscillations flowing only through the junction. As $M \rightarrow \infty$, the impedance path through only the series capacitors of the resonator equivalent circuit diverges. Charge from currents through the junction can no longer oscillate on $C_c$ and $\omega_a^{(M)}$ diverges. For the case of $M=1$ and $C_c\ll C_0$, Eq.~(\ref{eq:Ec}) simplifies to the standard definition of the charging energy $E_C = e^2/2C_c$~\cite{Koch:2007}. With $M > 1$, we will see that a more complex picture emerges.

\section{Renormalization of the atomic parameters}
In Fig.~\ref{fig:renormalization}, we explore the renormalization of the parameters of this model as the number of modes $M$ is increased. Through the change in charging energy, both the eigenstates $\ket{i}^{(M)}$ and coupling strengths $g_m^{(M)}$ depend on $M$. For a fixed number of modes $M$, the coupling $g_m^{(M)}$ of the atom to mode $m$ scales with the square root of the mode number $m$: $g_m^{(M)} = g_0^{(M)}\sqrt{2m+1}$. From this coupling, each mode will induce a Lamb shift of the atomic energy $\chi_m\simeq -2 \big(g_m^{(M)}\big)^2/\omega_m = -2 \big(g_0^{(M)}\big)^2/\omega_0$, a formula valid in the transmon regime when $\omega_m$ is much larger than the bare atomic frequency. With the typical assumption of coupling and bare atomic frequency independent of $M$, summing the Lamb shifts of every mode would lead to diverging values of the dressed atomic frequencies. This leads to the divergences found in typical multi-mode extensions of the quantum Rabi model.

In the model presented here, however, we find that the full quantization of the lumped-element circuit leads to a Hamiltonian in which both the bare atomic frequency $\omega_a$ and the couplings to the modes $g_m$ are explicitly dependent on the number of modes $M$ included in the model. As the number of modes $M$ in the model increases, the bare atomic couplings $g_m^{(M)}$ are suppressed (Fig.~\ref{fig:renormalization}(a,b)), and the bare atomic frequency $\omega_a^{(M)}$ increases (Fig.~\ref{fig:renormalization}(c)), diverging for an infinite number of modes. As we will see, however, convergence is obtained in the dressed transition energy of the atom when including the Lamb-shift from higher modes of the resonator.

As an illustration of how renormalization in this model leads to convergence of the spectrum, let us consider the case shown in Fig.~\ref{fig:renormalization}(d--e) in which the fundamental mode is resonant with the atomic frequency $\omega_a^{(1)}$ when $M=1$ (Fig.~\ref{fig:renormalization}(d)).  Including an additional mode with frequency $\omega_1$ will lead to an upwards shift of the bare atomic transition $\omega_{a}^{(1)}\rightarrow\omega_{a}^{(2)} > \omega_{a}^{(1)}$ and a change of the coupling $g^{(1)}_{0}\rightarrow g^{(2)}_{0}<g^{(1)}_{0}$ through the renormalization of the charging energy (Fig.~\ref{fig:renormalization}(e)). Diagonalizing the subsystem of the atom and mode 1 in this model, the transition energy of the atom is shifted down again near resonance with the fundamental mode $\omega_{a}^{(2)}\rightarrow \tilde{\omega}_{a}^{(2)}\sim \omega_{a}^{(1)}$ by the dispersive shift, and the coupling of the atomic transition to the fundamental mode is increased $g_{0}^{(2)}\rightarrow\tilde{g}_{0}^{(2)}\sim g^{(1)}_{0}$ (Fig.~\ref{fig:renormalization}(f)). In this way, the resulting vacuum Rabi splitting of the fundamental mode is found to be similar to that of the $M=1$ model, despite the decrease in the bare coupling rates $g_0^{(2)}$.

Note that in this model, the value of $E_J/E_C^{(M)}$ of the bare atom, which determines its anharmonicity~\cite{Koch:2007}, is also a function of $M$. It would seem that in the limit $M \rightarrow \infty$, the bare atom would be deep in the Cooper pair box limit. However, including the hybridization with the cavity, the low energy sector of $\hat{H}^{(M+1)}$ is well approximated by a model with $M$ modes where the charging energy is not $E_C^{(M)}$ but
\begin{equation}
\tilde{E}_C^{(M)} = E_C^{(M+1)} - \hbar(\bar{g}^{(M+1)}_{M})^{2}/4\omega_{M}\ ,
\label{eq_EC_tilde_M}
\end{equation}
where $\bar{g}^{(M)}_m$ is the coupling constant without the dipole moment $\bra{i}^M\hat{n}_J\ket{j}^M$. For this to hold, $\hbar \omega_M$ must be larger than the characteristic energy of the low energy sector of $\hat{H}^{(M)}$. In this case, the vacuum of the $(M+1)$-th mode, shifted by $\left(\bar{g}^{(M+1)}_{M}/\omega_{M}\right)\hat{n}_J$, is a good variational choice for the low lying energy sector of $\hat{H}^{(M+1)}$. In this subspace, the effective Hamiltonian is of the same form as $\hat{H}^{(M)}$, but with charging energy $\tilde{E}_C^{(M)}$, see Appendix~\ref{appendix_a}. This result matches with the zero-th order of a Schrieffer-Wolff approximation~\cite{Schrieffer:1966,Bravyi:2011}. We can iterate this procedure to a mode $L$. For $M\rightarrow\infty$, an effective Hamiltonian with $L$ modes will have a finite charging energy 
\begin{equation}
\tilde{E}_C^{(L)} = \lim_{M\rightarrow\infty}E_C^{(M)} - \sum_{m\geq L}^{M}\hbar(\bar{g}^{(M)}_{m})^{2}/4\omega_{m} .\label{eq_EC_tilde_LM}
\end{equation}

The interaction with higher modes therefore modifies the charging energy of the dressed atom, leading to a convergence of the atomic anharmonicity as well. This formula applies for all values of $C_J$, but for $C_J=0$, we have $\tilde{E}_C^{(M)} = E_C^{(M)}$, i.e., the dressing from higher modes exactly compensates the renormalization of the charging energy.

\begin{figure}[t]
	\centering
	\includegraphics[width=0.65\textwidth]{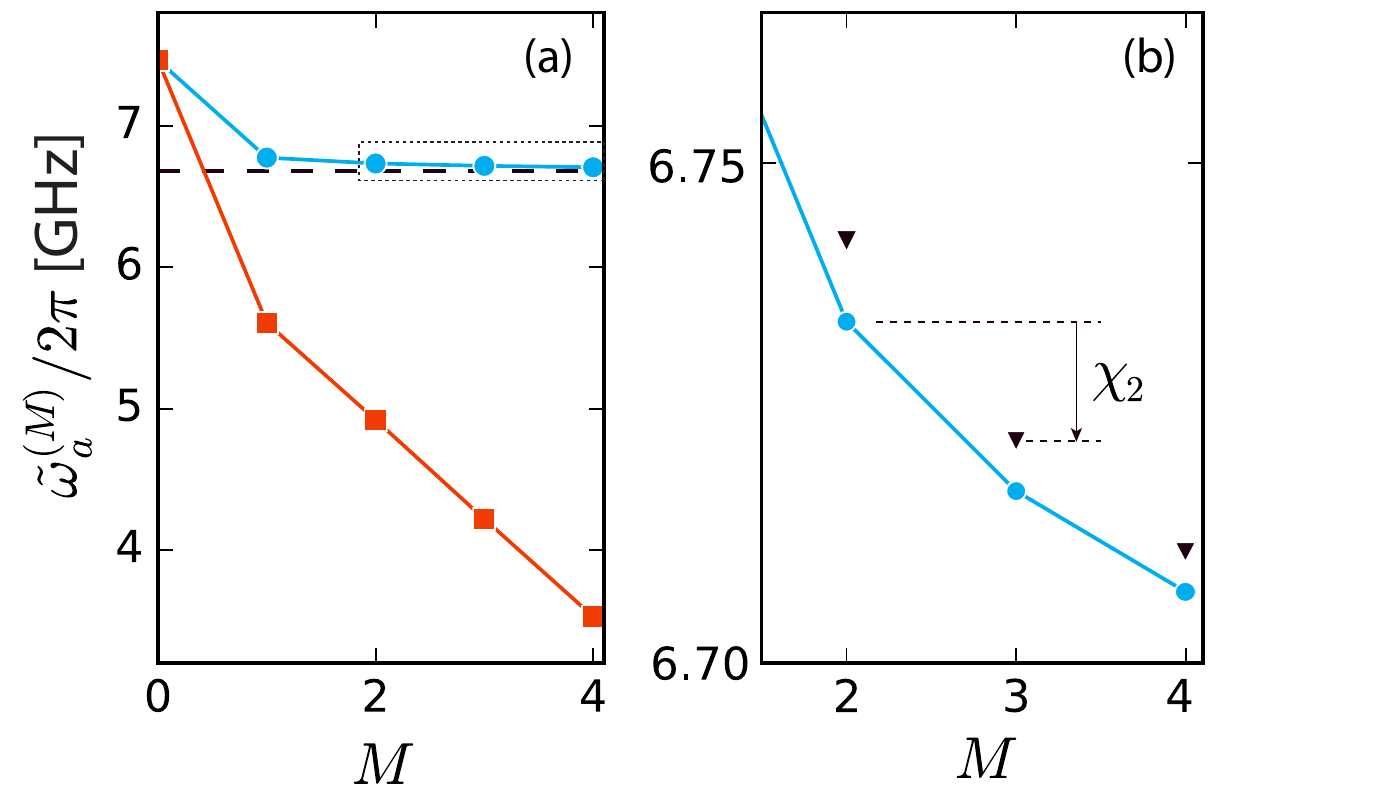}
	\caption{Computed spectrum as a function of the number of modes $M$ included in the model.  (a) Dots (squares) correspond to a diagonalization of the circuit (non-)renormalized extended-Rabi model. The frequency obtained by black-box quantization of the circuit Fig.~\ref{fig:circuit}(b) (dashed line) provides a point of reference corresponding to the case when all modes are included for the linearized system, and quartic anharmonicities are added for a truncated 3-mode model. (b) Zoom of dashed box in (a). Triangles show the prediction based on a first order classical approximation of the Lamb shift: $\chi_M = \tilde{\omega}_{a}^{(M+1)}-\tilde{\omega}_{a}^{(M)} \simeq -2 ((g_{M}^{(0)})^2/\omega_M)((\omega_{a}^{(0)})^{2}/\omega_M^2)$, see Appendix~\ref{appendix_a} for details.}
	\label{fig:lambshift}
\end{figure}
In order to illustrate the effectiveness of this renormalization, in Fig.~\ref{fig:lambshift} we compare a diagonalization of Hamiltonian (\ref{eq:circuit_hamiltonian}) to a non-renormalized multi-mode extension of the quantum Rabi model, implemented by removing the $M$-dependence of the charging energy $E_C^{(M)}\rightarrow e^2/(2C_c)$. The dashed line indicates the result of the black-box quantization (BB) method~\cite{Nigg:2012} as a point of reference, with a quartic anharmonic term, consisting of 3 harmonic modes (and 5 states per mode). That entails, first, numerically solving the frequencies of the linear system through the transcendental equation $Z_{TL}(\omega)=Z_{L_J}(\omega)+Z_{C_c}(\omega)$, i.e. replacing the junction by its linear term, to get the diagonal Hamiltonian $H_0=\sum_m \omega_m a_m^\dag a_m$. Second, adding the quartic term of the nonlinearity $H_1\approx\frac{E_J}{4!}\varphi_J^4$ in terms of the normal modes $\hat{\varphi}_J\propto \sum_m\sqrt{\frac{\hbar}{2\omega_m}} (a_m+a_m^\dag)$, for a truncated number of them, and numerical diagonalization. The calculations are performed using the same physical parameters as in Fig.~\ref{fig:renormalization}. Compared to the non-renormalized model, which diverges linearly, a diagonalization of the first-principle Hamiltonian (\ref{eq:circuit_hamiltonian}) converges towards the value expected from BB.

It is also interesting to note that the corrections from this model are non-perturbative: perturbation theory fails to give a value for the Lamb shift resulting from including an extra mode. Using a circuit analysis of coupled LC oscillators (see Appendix~\ref{appendix_a}), in the transmon regime, $E_J \gg E_C^{(M)}$, we find an estimate of the shift in the dressed AA energy when including an additional mode in the model given by $\chi_m\simeq -2(g_m^2/\omega_m)(\tilde{\omega}_{a}^2/\omega_m^2)$. This formula can be used to estimate the number of relevant modes to include in a simulation, and can be though of as a replacement of the usual expression for the Lamb shift $\chi_m^{\text{Lamb}}\simeq -2 g_m^2/\omega_m$. Along these lines, let us mention that Hassler et al.~\cite{Hassler:2019} further demonstrated the usefulness of effective open Jaynes/Cummings models for the description of the relevant physics in a meaningful frequency range in the strong coupling limit. 

\section{A high-frequency cutoff}
\begin{figure}[h!]
	\centering
	\includegraphics[width=0.55\textwidth]{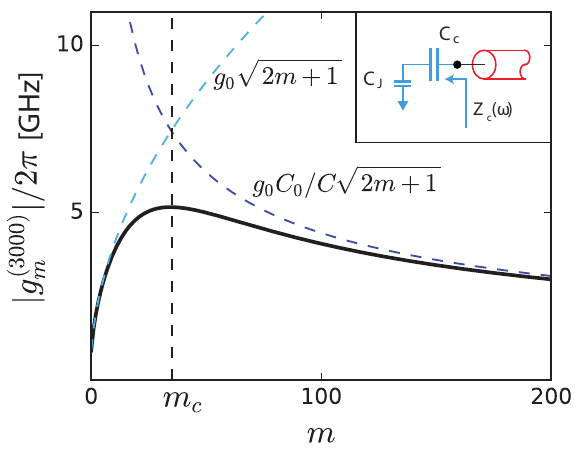}
	\caption{High-frequency cutoff for $C_J \neq 0$. The capacitive loading at the left boundary of the resonator shown in the inset transforms this point from a voltage anti-node to a voltage node for higher modes. The mode $m_c\simeq 35$ marks this transition. The solid line corresponds to the (envelope of the) coupling strength as a function of the number of modes. With $C_J \neq 0$ the parameter converges to a non-zero value for large $M$ hence the choice $M=3000$. Dashed lines: asymptotic values of the coupling, with $g_0 = g_0^{(M=3000)}$ and $C = C_cC_J/(C_c+C_J)$.}
	\label{fig:cutoff}
\end{figure}

In a realistic system, higher modes will tend to decouple from the atom due to several coexisting physical mechanisms~\cite{Filipp:2011}. One such mechanism is the capacitance of the Josephson junction $C_J$. In particular, the capacitive loading of the cavity from the AA illustrated in the inset of Fig.~\ref{fig:cutoff} leads to a decreasing impedance to ground $Z_c(\omega) \simeq i(C_J+C_c)/\omega C_J C_c$ at the end of the resonator when $\omega\gg\omega_{a}$. When the mode frequencies become such that this impedance is lower that the characteristic impedance of the resonator $|Z_c(\omega)| \ll |Z_0|$, this voltage anti-node of the resonator, to which the AA couples, becomes a voltage node, and the coupling vanishes. Additionally, the eigen frequencies will span from those of a $\lambda/4$ resonator for the lower modes to those of a $\lambda/2$ resonator $\omega_m \rightarrow 2m\omega_0$ for the higher lying modes.

This effect can be captured with the same quantization procedure applied to the circuit in Fig.~\ref{fig:circuit}(c) with $C_J \ne 0$ and is detailed in the Appendix~\ref{appendix_a}. Mathematically, the cutoff in the coupling is due to a mode-mode coupling term of the form $\sum_{m=0}^{m<M}\sum_{m'=m+1}^{m<M}G^{(M)}_{m,m'}(\hat{a}_m+\hat{a}_m^\dagger)(a_{m'}+a_{m'}^\dagger)$, which arises naturally from the circuit quantization. This is the equivalent of the $A^2$ term discussed in Refs.~\cite{DeLiberato:2014,Garcia-Ripoll:2015,Malekakhlagh:2016}. Diagonalizing the Hamiltonian of coupled resonator modes leads to decreasing zero-point voltage fluctuations of the modes at the coupling node. As shown in Fig.~\ref{fig:cutoff}, with a capacitance to ground $C_J = 5$ fF close to the experimental parameters of Ref.~\cite{Bosman:2017}, the expected cutoff occurs when $|Z_c(\omega_{m_c})| \simeq |Z_0|$, or equivalently at the mode number $m_c \simeq (C_J+C_c)/2\omega_0 Z_0C_JC_c$. This mechanism is accompanied by the appearance of an upper bound in the renormalized charging energy, such that Eq.~(\ref{eq:Ec}) becomes
\begin{equation}
E^{(M)}_{C} = \frac{e^2}{2}\frac{C_0+MC_c}{MC_cC_J+C_0(C_c+C_J)},
\end{equation}
and $E^{(M)}_{C}\rightarrow e^2/2C_J$ for $M\rightarrow \infty$. We emphasize however, that this cutoff is not a necessary condition for the convergence of the energy spectrum: the model described above with $C_\textrm{J} = 0$ converges even in the absence of such a cutoff. This is to be contrasted with typical models of (natural) atoms coupled to cavity modes where high frequency cutoffs must be imposed to obtain finite predictions~\cite{Mandel:1995,Gardiner:2004}. It would be interesting to study if the ideas developed in this chapter and Thesis apply to such systems. A first but unfinished inquiry in this line has been done in~\cite{Malekakhlagh:2016}. 

Summarizing, in this chapter we have fully analysed a first-principles multi-mode quantum Rabi model of circuit QED from a compact lumped element equivalent circuit. We have made explicit the convergence of quantities such as the Lamb shift without the need of any extra phenomenological high frequency cutoff, arising from a natural renormalization of the Hamiltonian parameters as more modes are added in the model. We have also studied the implications of a finite junction capacitance, which introduces a cutoff in the coupling to high frequency modes, but does not change the renormalization that occurs when additional modes are included in the circuit. We have shown the crucial role played by the exact Legendre transformation in order to have a correctly renormalized capacitively coupled Hamiltonian model, an issue not previously encountered in inductively coupled systems in the node-flux fundamental variable description. This framework is specially important for an intuitive understanding and modeling of experiments in the multi-mode ultra-strong coupling regime. This formulation of the multi-mode quantum Rabi model in the context of circuits hints at an intuitive picture on how this renormalization can arise physically.

\chapter{Distributed and Lumped Reciprocal Networks}
\label{chapter:chapter_3}
\thispagestyle{chapter}
\hfill\begin{minipage}{0.85\linewidth}
{{\emph{Tous les effets de la nature ne sont que résultats mathématiques d'un petit noinbre de lois immuables
\\\\
(All the effects of Nature are nothing but the mathematical consequences of a small number of immutable laws)}}}
\end{minipage}
\begin{flushright}
	\textbf{Pierre-Simon Laplace}\\
	Théorie des probabilités 
\end{flushright}
\vspace*{1cm}
When dealing with networks of superconducting qubits, several methods have been used to derive first-principle quantum Hamiltonian models to describe the effective dynamics and statistical properties observed in the experiments. The two seminal works in this field correspond to the ``Quantum Network Theory'' derived by Yurke and Denker \cite{YurkeDenker:1984} and the systematic Hamiltonian description of Devoret ``Quantum Fluctuations in Superconducting Circuits'' \cite{Devoret:1997}. In the former, the basic rules for first principles circuit quantization of linear and non-linear elements were presented in order to derive input-ouput relations of charge operators. However, no general Hamiltonian description was derived and thus the dynamics of the conjugate flux operators were not shown. On the other hand, dissipative elements were introduced in an analog manner to the Caldeira-Leggett model \cite{CaldeiraLeggett:1983} with semi-infinite transmission lines, and equivalent results to those of Caves \cite{Caves:1980} on the noise-amplification relation were obtained. The second crucial reference \cite{Devoret:1997} provided us with general rules to derive Hamiltonians of lumped electrical elements based on node-flux variables, and made use of the Caldeira-Leggett model to describe a closed Hamiltonian for an LC-oscillator inductively coupled to the impedance environment, which was replaced by infinite harmonic oscillators. These general rules for quantizing lumped-element circuits were later extended with systematic approaches for commonly-used classes of circuits \cite{Burkard:2004,Burkard:2005}.

In \cite{Chakravarty:1986}, Chakravarty and Schmid presented for the first time the action of a transmission line in the form that is used nowadays, when describing the input-output impedance of an open transmission line with a Josephson impurity. They used the path integral formulation to derive the spectral density, akin to the results  of Leggett \cite{Leggett:1984}, and obtained for that case a quadratic behaviour for small frequencies and linear growth for large frequency. 

In 2004, Blais et al. \cite{Blais:2004} quantized the transmission line following the canonical quantization procedure of Devoret \cite{Devoret:1997} for the continuum of harmonic oscillators. They considered multiple modes, and invoked a physical ultraviolet cutoff because the system cannot be exactly one dimensional. It was implicit in their work that the coupling of the modes to the qubit scale in the form $g_k\sim\sqrt{\omega_k}$. This approximate approach to quantize circuits with transmission lines that requires frequency cutoffs for the multi-mode Rabi Hamiltonian in cQED has routinely been used in theory \cite{Bourassa:2009} and experiments \cite{Houck:2008,Filipp:2011,Sundaresan:2015}. Without the phenomenological introduction of several types of mode truncations, such models, in which the scaling  $g_k\sim\sqrt{\omega_k}$ is present, would have predicted divergent Lamb-shifts \cite{Lamb:1947} or effective qubit-qubit couplings in the dispersive approximation. It must be remarked that mode truncation can be circumvented for estimating finite qubit decay constants, by carrying out a Markov approximation, as in the result of Wigner and Weisskopf \cite{Weisskopf:1930}. In this regard, see \cite{Malekakhlagh:2017} for an explicit presentation. Yet again, one needs adjustments and further approximations in order to recover finite predictions for these physical quantities.

The study of different classes of circuits resulted in other models that would not present divergent predictions for observables. That is the case in the work of Bourassa et al. \cite{Bourassa:2012}, where a transmission line with an inline Josephson junction was studied. Although not explicitly shown, the modified normal modes of that system would be coupled with monotonically decaying constants above certain saturation frequency. There was however no strict separation of anharmonic and harmonic degrees of freedom and thus a simple multi-mode quantum Rabi model could not be recovered. Interestingly, Bourassa himself introduced the finite length of the capacitor to provide an approximate but non-divergent Hamiltonian model for the capacitive coupling multi-mode model in his PhD thesis \cite{BourassaThesis:2012}. 

As commented in previous chapter, similar approaches have been used to describe linear electromagnetic environments with impedance black-boxes connected to non-linear elements at their ports \cite{Nigg:2012,Solgun:2014,Solgun:2015} in what is now known as ``black-box quantization''. Again, Nigg et al. \cite{Nigg:2012} generalized the concept introduced in \cite{Bourassa:2012} where the degree of freedom corresponding to the flux/phase differences across Josephson junctions could effectively interact with general electromagnetic normal modes. This method has proved very efficient to describe the physics of Josephson junctions in the transmon regime inside 3D cavities, see e.g. \cite{Kirchmair:2013,Shankar:2013,Vlastakis:2013}, due to the fact that the interaction between the harmonic modes and the non-linear element cannot be assumed to be local. The black-box techniques do also take into account that the electromagnetic modes have finite bandwidth, because the cavity has open ports from which energy flows away. Such procedures require finding a discrete equivalent lumped element circuit to simulate the linear response of the cavity from the point of the junctions. In \cite{Nigg:2012}, the Josephson fluxes were included in the linear description of the whole system, while nonlinear couplings appeared on expanding the cosine potentials in the normal mode basis. However, in \cite{Solgun:2014,Solgun:2015}, such Josephson variables where kept independent, thus reaching Hamiltonian models with both linear capacitive and inductive couplings between the normal modes of the environment and the anharmonic variables.

Although the black-box methods above have been very successful in describing experiments, it has not been hitherto clear how such linear systems can be later coupled to other systems. Thus, in this chapter we focus on techniques that involve coupling linear systems with infinite modes to reciprocal lumped-element quantum circuits. As is only to be expected, this entails some dressing of the infinite modes, that are ineluctably modified by the coupling. 

The first quantization of a general impedance, modelled as an infinite series of harmonic oscillators and capacitively coupled to Josephson junctions, was derived by Paladino et al. \cite{Paladino:2003}. A complete Hamiltonian without the diamagnetic $A^2$-term was derived by using the correct basis of harmonic variables. It was also noted that there is no need to add counterterms to the Hamiltonian, because in the experiment there is only access to renormalized parameters. In fact, the model developed there allows the engineering of the system, as the Hamiltonian is written in terms of the bare parameters. It was however not shown that the coupling of the anharmonic to the harmonic degrees of freedom would decay above certain frequency. Using different techniques but aware of \cite{Paladino:2003}, Bergenfeldt and Samuelsson \cite{Bergenfeldt:2012} quantized canonically a system with a 1D transmission line resonator capacitively coupled to a quantum dot. They noticed that the Hamiltonian could be bipartite-diagonalized and that the capacitive linear coupling constants would monotonically decay for high-frequency modes. Other methods to derive non-divergent quantum-Rabi Hamiltonians for tranmission line resonators coupled to Josephson junctions where also developed by Bamba et al.~\cite{Bamba:2014}, Malekakhlagh and T\"ureci in \cite{Malekakhlagh:2016} and by Mortensen at al. \cite{Mortensen:2016}, although the non-divergent characteristic of the coupling constants was not then explored and explained. Recently, two works \cite{Gely:2017,Malekakhlagh:2017} have independently been able to explain the mechanisms by which the infinite degree of freedom in a transmission line resonator decouple above certain mode when they are capacitively connected to a Josephson junction, without making any assumptions on the circuit parameters. The first method has been reviewed in previous chapter \ref{chapter:chapter_2}, whereas the second used the previous results achieved in \cite{Malekakhlagh:2016}. Finally, we remark that non-divergent but approximate methods to describe Josephson junctions capacitively coupled to transmission lines have been studied by Koch et al. \cite{Koch:2010} and Peropadre et al. \cite{Peropadre:2013} among others.

In this chapter, we generalize the ideas introduced by Paladino et al. \cite{Paladino:2003}, Bourassa et al. \cite{Bourassa:2012}, Bamba et al.~\cite{Bamba:2014}, and Malekakhlagh et al. in \cite{Malekakhlagh:2016,Malekakhlagh:2017}, following and extending well based mathematical machinery \cite{Walter:1973} previously used in \cite{ParraRodriguez:2016}. This is done to describe general networks of superconducting circuits that include circuital elements supporting infinite modes, such as transmission lines of finite, semi-infinite and infinite length, and general impedances, coupled to lumped-element networks capacitively, inductively and galvanically. Using the theory of eigenvalue problems with eigenvalue boundary conditions, for whose expansion theorems we developed a new proof, we recover the results achieved in \cite{Gely:2017} with the Foster-decomposition method and in \cite{Malekakhlagh:2016,Malekakhlagh:2017} with a regularization technique on the space-local interactions. We identify and solve the sources of the technical problems by those presentations.

As was surmised in most related works, the main source of complications in the quantization of such systems lies  in the need to invert an infinite dimensional kinetic matrix, in different guises and origins.

For systems modeled directly from Lagrangian densities for transmission lines, the usual functional techniques available for continuous linear fields are suspect in the present context because their couplings to discrete variables complicate the issue.  It is therefore imperative to use an alternative approach. One such is to perform the Legendre transformation for the discrete system and then take the continuous limit in the Hamiltonian, such as in the approaches of Bamba et al.~\cite{Bamba:2014} and Malekaklagh et al. \cite{Malekakhlagh:2016}. An alternative (which we follow in this work) is to expand in modes and then obtain a canonical Hamiltonian for the whole system, as has been done, for instance, by \cite{Bourassa:2009} and many others. In this second approach, we signal and clarify the issues involved in the choice of modes, and explicitly compute the intrinsic cutoff for the coupling constants. The crucial point is that the separate identification of lumped element network, on the one hand, and transmission line, on the other, that is used in the Lagrangian presentation, cannot persist when passing on to the required Hamiltonian formalism, and proper dressing of the infinite continuous modes with the discrete modes is necessary. This also requires the correct identification of the degrees of freedom. For instance, when several transmission lines are present and coupled the same network, it is not always possible to separate modes as pertaining only to one transmission line: the presence of the network forces the modes to be distributed on several transmission lines.

Modelling a system with infinite degrees of freedom coupled to a network with a finite number of modes can be done in a number of ways. A Lagrangian density is not the only possibility, quite evidently. In the context of linear passive non dissipative electrical circuits an alternative is given by an analysis of immittances, be they impedances or admittances, with infinite poles, which are then translated into lumped element circuits, with infinite capacitances and inductances. In order to write down a Hamiltonian it is again necessary to invert an infinite dimensional kinetic matrix. A possible approach is to consider a truncation in the number of modes associated with the impedance to a finite number $N$, to proceed with an inversion and then to the limit $N\to \infty$. In many cases of interest, this procedure leads to the uncoupling of the impedance modes and the finite network. This comes about because some coupling vector in the Lagrangian  has infinite norm in that limit. More precisely, because that coupling vector has infinite norm with respect to a specific inner product, determined by the inverse capacitance matrix of the infinite dimensional system. We give two solutions to this problem, and then show their equivalence. The first one, in parallel to the presentation for transmission lines, consists in a canonical transformation in the line of that presented by Paladino et al. \cite{Paladino:2003} and in previous chapter \ref{chapter:chapter_2}. That is, a rearrangement of the degrees of freedom, dressing the impedance modes with network modes. We present this formalism for the first Foster expansion of an impedance coupled capacitively to a network. The second solution comes from the identification of the proper normal modes of the impedance in the Hamiltonian, by the standard canonical transformation only in impedance modes. This is first done for finite $N$ and then taking the limit $N\to\infty$. We extend this analysis to multiport impedances.

After this introductory section, we present a catalogue of configurations with transmission lines in the formalism of Lagrangian density, in which we study systematically mode expansions, counting of degrees of freedom, and separability of modes. We defer to Appendix \ref{appendix_b} the relevant mathematical apparatus used in this chapter. In the following section we turn to the coupling of networks to canonical impedances. We first study the reassignment of modes, dressing the impedance  with network modes, and then the diagonalization of impedance modes in the Hamiltonian to avoid the uncoupling in the infinite mode limit. In the fourth section we retake transmission lines. We use the previous analysis to provide explicit analytically computable examples, after a general discussion on spin-boson models as derived from these capacitive couplings. We finish the chapter with a summary of conclusions and proposals for future works.

\section[Networks with transmission lines]{Networks with transmission lines}
\label{sec:netw-with-transm}
Quantum networks of superconducting qubits make use of transmission lines to either carry information away from the computational system with open boundaries or to store and manipulate it in the form of resonators. As any conducting box does, superconducting transmission lines theoretically support an infinite number of electromagnetic modes as bosonic degrees of freedom. Typically, the dynamics of the whole system is well described in terms of controlled anharmonic subsystems, e.g. qubits, interacting with a countable, possibly infinite, number of bosonic modes, i.e. harmonic oscillators. A typical requirement for such effective models to be valid is that the coupling strength between subsystems is small compared to the energy defining the subsystems themselves. However, even in this small-coupling limit, multi-mode effects can have crucial effects on the predicted effective coupling between two separated information units \cite{Filipp:2011}. 

In this section, we develop the tools required to write exact quantum Hamiltonians of systems of general anharmonic subsystems linearly coupled to transmission lines with closed or open boundary conditions, keeping the multi-mode feature of the lines, and verifying that ultraviolet divergenceless predictions are a natural consequence of the canonical quantization procedure.

Our viewpoint   issues from a Lagrangian description. We then  write the transmission lines in terms of an infinite set of modes, and  carefully proceed to a Hamiltonian formulation in which to perform canonical quantization. As usual, the description in terms of modes is not unique. We make explicit use of this freedom to identify the most adequate choice, where the criterion is that the network modes are uncoupled from each other in the Hamiltonian formalism. In particular, given the capacitive coupling scheme we study, $\lambda/4$ or $\lambda/2$ resonator modes expansions are seen to be inadequate: they would lead to a description in which network and transmission line are uncoupled. The correct set of modes necessarily is dressed by the parameters of the coupling. This dressing means that the mode form functions are \emph{not} eigenfunctions of a Sturm--Liouville operator. In fact, carrying out a na\"{\i}ve separation of variables, we see that the mode form functions are determined by a boundary condition differential equation singular value problem, in which the singular value also enters the boundary condition. This kind of singular value problem is hugely different from the Sturm--Liouville case, and standard textbook material does not cover it. We provide mathematical details and a new proof of the expansion theorems one requires in order for these functions to be indeed mode functions in Appendix \ref{Walter_appendix}. Applying these techniques we obtain definite predictions for the couplings, with a natural intrinsic cutoff frequency. This cutoff frequency comes about because the dressing of the transmission lines requires a length parameter, $\alpha$, that provides us with a natural ultraviolet cutoff. It is important to stress that this cutoff is intrinsic to the model, with no need to argue about the validity of the model itself for it to appear.
\subsection{Linear coupling to lumped-element networks}
\label{subsec:chap2_linear_cupling}

In this subsection, we study common linear coupling configurations between a finite set of degrees of freedom and transmission lines, namely mixed inductive and capacitive ``point-like'' coupling to (i) a network or to (ii) multiple networks, and (iii) mixed inductive and capacitive galvanic coupling.

\subsubsection{Mixed linear coupling}
\label{subsubsec:chap2_mixed_linear_coupling}

\begin{figure*}[]
	\centering{\includegraphics[width=0.5\textwidth]{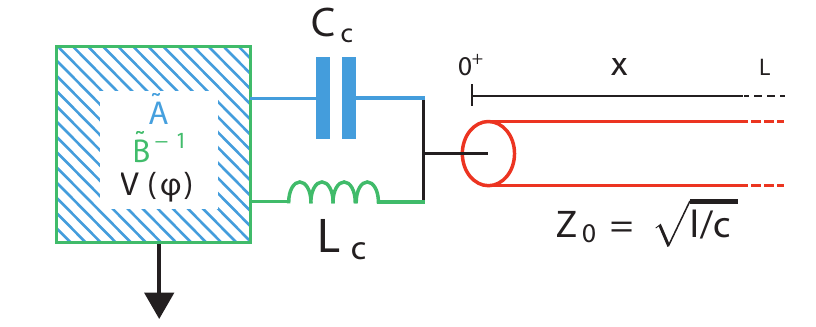}}
	\caption{\label{fig:TL_LCcoup_Network} Transmission line inductively and capacitively coupled to a finite network. The network has internal flux degrees of freedom $\bphi_{i}$, with capacitance $\msA$ and inductive $\msB^{-1}$ matrices and general non-linear potential $V(\bphi)$.}
\end{figure*}

We consider the circuit in Fig. \ref{fig:TL_LCcoup_Network} with a network of degrees of freedom $\bphi$ linearly coupled through capacitor $C_c$ and inductor $L_c$ to a transmission line at one end. Given that our non-linear network has a nonlinear potential in flux variables, as it is the case when there are Josephson junctions, it facilitates the analysis to choose flux variables as our set of position-like coordinates. 

Following standard microwave theory \cite{Blais:2004,Bourassa:2009,Pozar:2009}, the Lagrangian of this circuit can be written in terms of a discrete set of flux variables describing the network, collected in the column vector $\bphi$, and a flux field $\Phi(x,t)$,
\begin{align}
L=&\, \frac{1}{2}\dot{\bphi}^T \msA\dot{\bphi}-\frac{1}{2}\bphi^T \msB^{-1}\bphi-V(\bphi)+\int_{0}^{L}dx\,\left[\frac{c}{2}\dot{\Phi}(x,t)^2-\frac{1}{2l}(\Phi'(x,t))^2\right]\nonumber\\
&+C_c\left(\frac{\dot{\Phi}(0,t)^2}{2}-\dot{\bphi}^T\ba \dot{\Phi}(0,t)\right)-\frac{1}{L_c}\left(\frac{\Phi(0,t)^2}{2}-\bphi^T\bb\Phi(0,t)\right),\label{eq:Lag_TL_LCcoup_Network}
\end{align}
where $\msA=\tilde{\msA}+C_c \ba\ba^T$ and $\msB^{-1}=\tilde{\msB}^{-1}+\bb\bb^T/L_c$ are the capacitance and inductance submatrices of the network respectively, and $\ba$ and $\bb$ are coupling vectors to the network from the transmission line with finite norm. Notice here that we do not assume any specific description of the network in terms of branch or node flux variables. We do nonetheless emphasize that the network has to be connected non-trivially to the common ground in order for current to circulate through $C_c$ and $L_c$. We remark that in the whole analysis we can take the limits of $C_c\rightarrow0$ and $L_c\rightarrow\infty$ to disconnect the transmission line from the network through its corresponding element. The classical equations of motion of this system read
\begin{align}
 c\ddot{\Phi}(x,t)&=\frac{\Phi''(x,t)}{l}\label{eq:EulLag_TL_LCcoup_Network11},\\
 \frac{\Phi'(0,t)}{l}&=C_c\left(\ddot{\Phi}(0,t)-\ba^T\ddot{\bphi}\right)+\frac{1}{L_c}\left({\Phi}(0,t)-\bb^T{\bphi}\right),\label{eq:EulLag_TL_LCcoup_Network12}\\
 \msA\ddot{\bphi}+\msB^{-1}{\bphi}&=C_c\ba\ddot{\Phi}(0,t)+\frac{1}{L_c}\bb\Phi(0,t)-\frac{\partial V(\bphi)}{\partial \bphi}\label{eq:EulLag_TL_LCcoup_Network13}.
\end{align}

Let us first assume, for simplicity, that the transmission line has finite length $L$ (see Appendix \ref{subsec:infin-length-transm} for infinite length transmission lines, and explicit computation in \ref{sec:an-example:-half} and \ref{sec:an-example-galvanic}). A textbook analysis would carry out separation of variables, that is, it would introduce a decomposition of the flux field in a countable basis of functions $\Phi(x,t)=\sum_n \Phi_n(t)u_n(x)$, justified physically as normal modes. There is an issue in this case, however, in that there is a coupling at the endpoint $x=0$ with the network that involves the second derivative with respect to time of the flux field. Even if all the network variables were set to zero, we would still have, from Eq. (\ref{eq:EulLag_TL_LCcoup_Network12}), a boundary condition that would involve the separation constant ($-\omega_n^2$ in (\ref{eq:EVP_TL_LCcoup_Network_eq0}) below). Furthermore, setting the network variables to zero would not be consistent, since the transmission line sources the network in equation (\ref{eq:EulLag_TL_LCcoup_Network13}).

Here we will take the following approach: we shall retain the dependence of the boundary condition on the separation constant, by introducing a length parameter $\alpha$ that will later be set to an optimal value, according to a precise optimality criterion. Namely, that in the Hamiltonian presentation there be no coupling amongst the transmission line modes.

In this manner, the  field equations for the line yield the following homogeneous eigenvalue problem
\begin{align}
 \ddot{\Phi}_n(t)&=-\omega_n^2\Phi_n(t),\label{eq:EVP_TL_LCcoup_Network_eq0}\\
 u_n''(x)&=-k_n^2u_n(x),\label{eq:EVP_TL_LCcoup_Network_eq1}\\
 u_{n}'(0)&=-k_n^2\alpha u_n(0)+\frac{1}{\beta}u_n(0),\label{eq:EVP_TL_LCcoup_Network_eq2}\\
u_n(L)&=0,\label{eq:EVP_TL_LCcoup_Network_eq3}
\end{align}
where the frequencies are related to the wavenumbers through $\omega_n^2=k_n^2/lc$, and we have assumed for concreteness a short to ground boundary condition at $x=L$. Notice that this choice is not a restriction of our method, and other boundary conditions can be considered at $x=L$, i.e. the general case as at the other end $x=0$.

As we have already pointed, this form of Eq. (\ref{eq:EVP_TL_LCcoup_Network_eq2}) can be derived by setting to zero the network fluxes in (\ref{eq:EulLag_TL_LCcoup_Network12}), in which case the parameter $\alpha$ would be given by $C_c/c$. It can also be obtained by solving $\ddot{\bphi}$ in (\ref{eq:EulLag_TL_LCcoup_Network13}), substituting it in (\ref{eq:EulLag_TL_LCcoup_Network12}) and consistently imposing  $\bphi=-\msB\partial_{\bphi}V(\bphi)$. In this case the parameter $\alpha$ would be given as $(C_c/c)\left(1-C_c\ba^T\msA^{-1}\ba\right)$, which, as we will see, is  optimal from our point of view.  Indeed, and as previously envisaged in \cite{Malekakhlagh:2016,Bamba:2014}, the second approach uses the information about the network capacitance matrix $\msA$ and its coupling vector $\ba$ to derive a Hamiltonian without mode-mode coupling in the purely harmonic sector. Please see Appendix~\ref{appendix_b} for a detailed analysis of how and why the two procedures give matching Hamiltonians. The physical reason for this choice is that in this manner the inhomogeneous source term  corresponds to the current through the anharmonic potential.

So far we have concentrated on the more crucial parameter $\alpha$. The second length parameter, $\beta$, is more easily determined. Nonetheless we also allow it to be free, and its value will also be fixed a posteriori. We remark that Dirichlet,  Neumann, and Robin homogeneous boundary conditions are included in the analysis, with corresponding sets of parameters $\beta_D=0$ for Dirichlet (line ended in open circuit), $(\alpha_N,\beta_N)=(0,\infty)$ for Neumann (line ended in short circuit), and $\alpha_R=0$ for Robin (pure inductive coupling). Analogously, the pure capacitive coupling boundary condition corresponds to $\beta=\infty$.

For fixed parameters $\alpha$ and $\beta$, the system of equations
from 
Eq.
(\ref{eq:EVP_TL_LCcoup_Network_eq1}) through to (\ref{eq:EVP_TL_LCcoup_Network_eq3}) define a generalized eigenvalue problem, with an easily determined secular equation and generalized eigenfunctions.
Furthermore, those eigenfunctions satisfy the following orthogonality conditions
\begin{align}
\langle u_n,u_m\rangle_{\alpha}&=c \left(\int_{0}^{L}dx\, u_n(x) u_m(x)+  \alpha u_n(0) u_m(0)\right)=N_{\alpha}\delta_{nm},\label{eq:TL_LCcoup_Network_ortho_1}\\
\langle u_n,u_m\rangle_{1/\beta}&=\frac{1}{l}\left(\int_{0}^{L}dx\, u_n'(x) u_m'(x)+ \frac{1}{\beta} u_n(0) u_m(0)\right)=\omega_n^2 N_{\alpha}\delta_{nm},\label{eq:TL_LCcoup_Network_ortho_2}
\end{align}
where $N_{\alpha}$ is a free normalization constant in capacitance units.

From these considerations, a number of authors have used these generalised eigenfunctions and orthogonality for an expansion in modes. We should note however that the possiblility of expanding a function in these eigenfunctions, i.e. that they form a basis in a suitable space of functions, is by no means deducible from standard Sturm--Liouville theory. Fortunately, the topic has been examined in the mathematical literature (see, inter alia, \cite{Walter:1973}), and it is indeed the case that an expansion theorem does hold. We provide more mathematical details, and a new proof of the expansion theorem, in Appendix \ref{Walter_appendix}.

Now, knowing that we can expand in this generalized eigenbasis, we write the Lagrangian (\ref{eq:Lag_TL_LCcoup_Network}) as 
\begin{equation}	L=\frac{1}{2}\dot{\bX}^T\msC\dot{\bX}-\frac{1}{2}\bX^T \msL^{-1}\bX -V(\bphi),
\end{equation}
where we have defined the vector of fluxes
\begin{equation}
\label{eq:xvector}
\bX=
\begin{pmatrix}
\bphi\\
\bPhi
\end{pmatrix},
\end{equation}
and the capacitance and inductance matrices
\begin{align}
\,\, \msC=&\begin{pmatrix}
\msA& -C_c \ba\bu^T\\
-C_c \bu\ba^{T} & N_{\alpha} \mone + d\bu \bu^T
\end{pmatrix},\label{eq:C_LCcoup_Network}\\
 \msL^{-1}=&\begin{pmatrix}
\msB^{-1}& - \bb\bu^T/L_c\\
- \bu\bb^{T}/L_c & N_{\alpha} (\omega_n^2) + e\bu \bu^T
\end{pmatrix}.\label{eq:Linv_LCcoup_Network}
\end{align}
with $\bu\equiv (u_0(0), u_1(0),...u_n(0),\ldots)^T$ being the coupling vector (of infinite dimensionality), the parameters $d \equiv C_c-c \alpha$ and $e\equiv 1/L_c - 1/\beta l$, $\mone$ the infinite-dimensional identity matrix and $(\omega_n^2)=\mathrm{diag}(\omega_0^2,\omega_1^2,...)$ the diagonal matrix of squared frequencies of the eigenvalue problem. Notice that $\bu$ is generically normalizable. Even more importantly, the quantity $\bu^T\left[N_\alpha \mone+d\bu\bu^T\right]^{-1}\bu=1/C_c$ is finite unless $C_c$ is zero.  The vector $\bu$ is in fact an element of the $l^2$ sequence Hilbert space, by the construction of Appendix \ref{sec:finite-length-transm}, and its norm depends directly on the parameter $\alpha$, namely $\left|\bu\right|^2=\bu^T\bu=N_\alpha/\alpha c$. The dimensionful parameter $N_\alpha$ was introduced so that this norm be adimensional.

We can now invert the capacitance matrix and derive the Hamiltonian 
\begin{equation}
H=\frac{1}{2}{\bP}^T\msC^{-1}{\bP}+\frac{1}{2}\bX^T \msL^{-1}\bX+V(\bphi),\label{eq:Ham_LCcoup_Network}
\end{equation}
where the conjugate charge variables to the fluxes are $\bP=\partial L/\partial \bX=(\bq^T,\bQ^T)^T$, and the inverse capacitance matrix is 
\begin{equation}
\msC^{-1}= \begin{pmatrix}
\msA^{-1}+ \frac{C_c^2 \left|\bu\right|^2}{D} \msA^{-1}\ba\ba^T \msA^{-1}&   \frac{C_c}{D} \msA^{-1} \ba\bu^T\\
\frac{C_c}{D} \bu\ba^T\msA^{-1}& \frac{1}{N_{\alpha}}\mone+ \frac{1}{\left|\bu\right|^2}\left( \frac{1}{D}- \frac{1}{N_{\alpha}}\right) \bu \bu^T
\end{pmatrix},\label{eq:Cinv_LCcoup_Network}
\end{equation}
with $D= N_{\alpha}+\left|\bu\right|^2(d- C_c^2 \ba^T \msA^{-1}\ba)$. It now behoves us to insert the requirement that there be no mode-mode coupling in the description of the transmission line.  Recalling that $d$ and $e$ depend on the parameters $\alpha$ and $\beta$, which we have so far left undetermined, we can choose these parameters $\alpha$ and $\beta$ to satisfy the equations $D=N_{\alpha}$ and $e=0$, thus removing the harmonic mode-mode couplings, with the result 
\begin{align}
\alpha&=\frac{C_c(1-C_c\ba^T \msA^{-1}\ba)}{c},\label{eq:TL_LCcoup_Network_alpha_fix}\\
\beta&=L_c/l.\label{eq:TL_LCcoup_Network_beta_fix}
\end{align}
Next, in order to find the frequencies $\omega_n$, we have to solve the eigenvalue problem (\ref{eq:EVP_TL_LCcoup_Network_eq1}-\ref{eq:EVP_TL_LCcoup_Network_eq3}) with the values of $\alpha$ and $\beta$ presented in (\ref{eq:TL_LCcoup_Network_alpha_fix}) and (\ref{eq:TL_LCcoup_Network_beta_fix}), and the final Hamiltonian will be
\begin{align}
H=&\, \frac{1}{2}\bq^T(\msA^{-1}+ \frac{C_c^2}{\alpha c} \msA^{-1}\ba\ba^T \msA^{-1})\bq+\frac{1}{2}\bphi^T\msB^{-1}\bphi+V(\bphi)+\sum_n \frac{Q_n^2}{2N_{\alpha}}+\frac{N_{\alpha}\omega_n^2\Phi_n^2}{2}\nonumber\\
&+\frac{C_c}{N_{\alpha}} (\bq^T\msA^{-1} \ba)\sum_n Q_n u_n(0)-\frac{1}{L_c} (\bphi^T\bb)\sum_n \Phi_n u_n(0),\label{eq:Ham_LCcoup_Network2}
\end{align}
where we have used the normalization equality $\left|\bu\right|^2=N_{\alpha}/\alpha c$. 

To complete the process of quantization, we promote the conjugate variables to operators with the commutator $[\hat{X}_i,\hat{P}_j]=i\hbar \delta_{ij}$.  Finally the quantum Hamiltonian in terms of annihilation and creation operators, related to flux and charge variables by $\hat{\Phi}_n=i\sqrt{\hbar/2\omega_n N_\alpha}(a_n-a_n^{\dagger})$ and $\hat{Q}_n=\sqrt{\hbar\omega_n N_\alpha/2}(a_n+a_n^{\dagger})$,
\begin{align}
\tilde{H}=& \frac{1}{2}\hat{\bq}^T(\msA^{-1}+ \frac{C_c^2}{\alpha c} \msA^{-1}\ba\ba^T \msA^{-1})\hat{\bq}+\frac{1}{2}\hat{\bphi}^T\msB^{-1}\hat{\bphi}+V(\hat{\bphi})+\sum_n \omega_n a_n^\dagger a_n\nonumber\\
&+C_c\sqrt{\frac{\hbar}{2 N_\alpha}} (\hat{\bq}^T\msA^{-1} \ba)\sum_n (a_n+a_n^\dagger)\sqrt{\omega_n} u_n(0) \nonumber\\
&-\frac{i}{L_c}\sqrt{\frac{\hbar}{2 N_\alpha}} (\hat{\bphi}^T\bb)\sum_n (a_n-a_n^{\dagger}) \frac{u_n(0)}{\sqrt{\omega_n}}.\nonumber 
\end{align}

This Hamiltonian is as exact as the starting point, the Lagrangian (\ref{eq:Lag_TL_LCcoup_Network}), and here we can see a first result: the (capacitive) coupling constants $g_n\propto \sqrt{\omega_n}u_n(0)$ do not grow without bound. As we discuss in detail in  \ref{sec:generic-behaviour}, the large $n$ behaviour of $u_n(0)$ is $1/n$, while that for $\omega_n$ is $n$. It follows that $g_n\sim n^{-1/2}$. There is no need for an ultraviolet cutoff extrinsic to the model (\ref{eq:Lag_TL_LCcoup_Network}); rather,  the correct choice of modes to expand in has provided us with a natural length scale, intrinsic to the model, that translates into  an intrinsic ultraviolet cutoff.
\subsubsection{Linearized galvanic coupling}
\label{sec:line-galv-coupl}
Another very common circuit configuration that has been used in cQED is the so called {\it galvanic} coupling between harmonic modes and non-harmonic variables, see \cite{Bourassa:2009}. Indeed, such a configuration has proved to be the most efficient way thus far to reach the ultrastrong coupling regime in light-matter interactions \cite{Niemczyk:2010,FornDiaz:2017,Yoshihara:2017}.
\begin{figure*}[h]
	\centering{\includegraphics[width=0.65\textwidth]{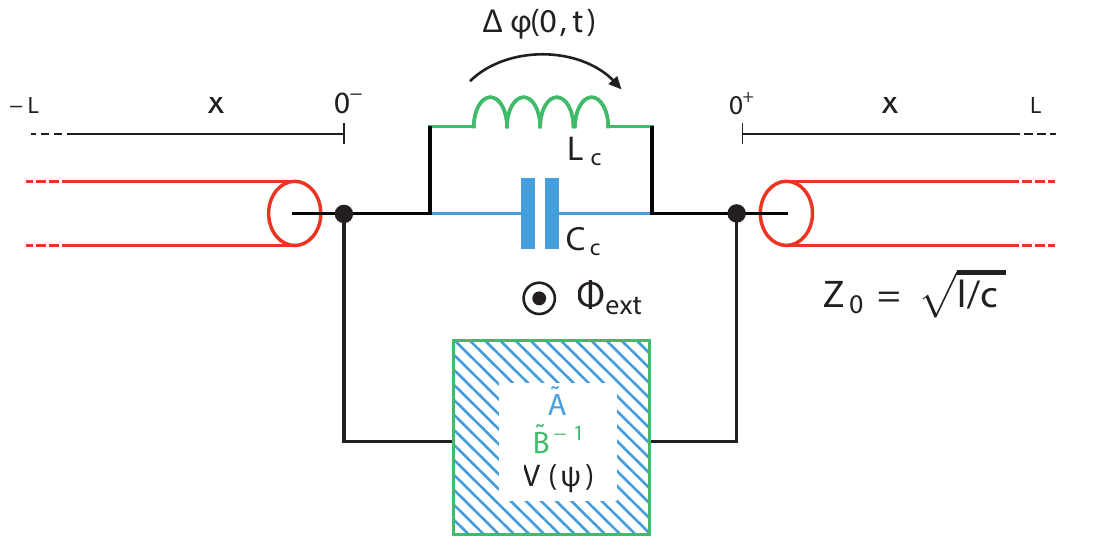}}
	\caption{\label{fig:TL_LCgalvcoup_Network} Transmission line galvanically coupled to a finite network. The network has internal flux degrees of freedom $\bphi_{i}$, with capacitance $\msA$ and inductive $\msB^{-1}$ matrices and general non-linear potential $V(\bphi_{i})$.}
\end{figure*}

The Lagrangian describing a generalized galvanic configuration, see Fig. \ref{fig:TL_LCgalvcoup_Network}, can be written as 
\begin{align}
L=& \frac{1}{2}\dot{\boldsymbol{\psi}}^T \tilde{\msA}\dot{\boldsymbol{\psi}}-\frac{1}{2}\boldsymbol{\psi}^T \tilde{\msB}^{-1}\boldsymbol{\psi}-V(\boldsymbol{\psi})+\int_{-L}^{L}dx\,\left[\frac{c}{2}\dot{\Phi}(x,t)^2-\frac{1}{2l}(\Phi'(x,t))^2\right]\nonumber\\
&+\frac{C_c}{2}\Delta\dot{\Phi}(0,t)^2-\frac{1}{2L_c}\Delta\Phi(0,t)^2,\label{eq:Lag_TL_LCgalvcoup_Networks}
\end{align}
where the set of internal variables is collected in a column vector $\boldsymbol{\psi}=(\psi_1,...,\psi_N)^T$, and $\Delta\Phi(0,t)$ is the flux difference in the line. The first order of business is  to identify a good set of independent variables. In order to achieve that,  we  impose  Kirchoff's laws in the connection, a constraint that fixes at least one of the degrees of freedom in the network,
\begin{equation}
\psi_N(t)=\Delta \Phi(0,t)+\Phi_{\mathrm{ext}}+ \bg^T\bphi(t),\label{eq:TL_LCgalvcoup_Networks_Vtransf}
\end{equation}
where the new truncated set of variables is $\bphi\equiv(\psi_1, \psi_2,...\psi_{N-1})^T$, and $\bg$ is a constant vector on that reduced subspace. We reduce the number of variables and find that the Lagrangian has both capacitive and inductive coupling to the flux difference in the line $\Delta \Phi(0,t)$
\begin{align}
L=& \frac{1}{2}\dot{\bphi}^T \msA\dot{\bphi}-\frac{1}{2}\bphi^T \msB^{-1}\bphi-V(\bphi,\Phi_{\mathrm{ext}})+\int_{-L}^{L}dx\,\left[\frac{c}{2}\dot{\Phi}(x,t)^2-\frac{1}{2l}(\Phi'(x,t))^2\right]\nonumber\\
&+\frac{C_{cA}}{2}\Delta\dot{\Phi}(0,t)^2-\frac{1}{2L_{cB}}\Delta\Phi(0,t)^2+\frac{1}{2L_B}(2 \Delta \Phi(0,t)\Phi_{\mathrm{ext}}+\Phi_{\mathrm{ext}}^2)\nonumber\\
&-C_A(\dot{\bphi}^T\ba)\Delta\dot{\Phi}(0,t)+\frac{1}{L_B}(\bphi^T\bb)(\Delta\Phi(0,t)+\Phi_{\mathrm{ext}}),\label{eq:Lag_TL_LCgalvcoup_Networks_2}
\end{align}
where $C_{cA}=(C_c+C_A)$ and $L_{cB}=L_c L_B/(L_c + L_B)$, with $C_A$ and $L_B$ the coupling capacitance and inductance parameters coming out of the transformation (\ref{eq:TL_LCgalvcoup_Networks_Vtransf}) in (\ref{eq:Lag_TL_LCgalvcoup_Networks}). In this reduction, we decompose the matrix $\tilde{\msA}$ as
\begin{equation}
\tilde{\msA}=\begin{pmatrix}
\msA_1 & \ba_1\\
\ba_1^T & C_A
\end{pmatrix}.
\end{equation}
It follows that $\msA=\msA_1+\ba_1\bg^T+\bg\ba_1^T+C_A\bg\bg^T$ and $\ba=-\bg-\ba_1/C_A$ in (\ref{eq:Lag_TL_LCgalvcoup_Networks_2}). An analogous procedure provides us with matrix $\msB$ and coupling vector $\bb$. The equations of motion for this Lagrangian at the boundary (in the lines we still have the wave equation) are 
\begin{align}
\frac{\Phi'(0^-,t)}{l}=&\frac{\Phi'(0^+,t)}{l},\\
C_A\ba^T\ddot{\bphi} +\frac{1}{L_B}\left(\bb^T{\bphi} + \Phi_{\mathrm{ext}}\right)=&\frac{\Phi'(0^-,t)}{l}+C_{cA}\Delta\ddot{\Phi}(0,t) + \frac{1}{L_{cB}}\Delta{\Phi}(0,t),\label{eq:EulLag_TL_LCgalvcoup_Networks12}\\
\msA\ddot{\bphi}+\msB^{-1}{\bphi}+\frac{\partial V(\bphi,\Phi_{\mathrm{ext}})}{\partial \bphi}=&C_A\ba\Delta\ddot{\Phi}(0,t)+\frac{1}{L_B}\bb\left(\Delta\Phi(0,t)+\Phi_{\mathrm{ext}}\right)\label{eq:EulLag_TL_LCgalvcoup_Networks13}.
\end{align}

We decompose again the flux field in a countable basis of functions $\Phi(x,t)=\sum_n \Phi_n(t)u_n(x)$ (given that we assumed the line of finite length) and the field equations for the line yield the following homogeneous eigenvalue problem
\begin{align}
u_n''(x)&=-k_n^2u_n(x),\label{eq:EVP_TL_LCgalvcoup_Networks_eq1}\\
u_{n}'(0^-)&=u_{n}'(0^+)=k_n^2\alpha \Delta u_n(0)-\frac{1}{\beta}\Delta u_n(0),\label{eq:EVP_TL_LCgalvcoup_Networks_eq2}\\
u_n(-L)&=u_n(L)=0,\label{eq:EVP_TL_LCgalvcoup_Networks_eq3}
\end{align}
where $\Delta u_n(0)=u_n(0^-)-u_n(0^+)$. Again, Eqs. (\ref{eq:EVP_TL_LCgalvcoup_Networks_eq1}-\ref{eq:EVP_TL_LCgalvcoup_Networks_eq3}) define a generalized eigenvalue problem with eigenvalue-dependent boundary conditions, see Sec. \ref{sec:galvanic-coupling}, whose eigenfunctions satisfy the following orthogonality conditions
\begin{align}
  \langle u_n,u_m\rangle_{\alpha}&=c \left(\int_{0}^{L}dx\, u_n(x) u_m(x)+  \alpha \Delta u_n(0) \Delta u_m(0)\right)=N_{\alpha}\delta_{nm},\label{eq:TL_LCgalvcoup_Networks_ortho_1}\\
 \langle u_n,u_m\rangle_{1/\beta}&=\frac{1}{l}\left(\int_{0}^{L}dx\, u_n'(x) u_m'(x)+ \frac{1}{\beta} \Delta u_n(0) \Delta u_m(0)\right)=\omega_n^2 N_{\alpha}\delta_{nm},\label{eq:TL_TL_LCgalvcoup_Networks_ortho_2}
\end{align}
where $N_{\alpha}$ is a free normalization constant in capacitance units. Notice that in this case we can choose real eigenfunctions, and we have done so. Making use of the above equations we can rewrite the Lagrangian (\ref{eq:Lag_TL_LCgalvcoup_Networks_2}) as 
\begin{equation}
L=\frac{1}{2}\dot{\bX}^T\msC\dot{\bX}^T-\frac{1}{2}\bX^T \msL^{-1}\bX -V(\bphi,\Phi_{\mathrm{ext}}),
\end{equation}
with $\bX=({\bphi}^T,
{\bPhi}^T)^T$ and 
\begin{align}
\msC=&\begin{pmatrix}
\msA& -C_A \ba\Delta\bu^T\\
-C_A \Delta\bu\ba^{T} & N_{\alpha} \mone + d\Delta\bu \Delta\bu^T
\end{pmatrix},\label{eq:C_TL_LCgalvcoup_Networks}\\
\msL^{-1}=&\begin{pmatrix}
\msB^{-1}& - \bb\Delta\bu^T/L_B\\
- \Delta\bu\bb^{T}/L_B & N_{\alpha} (\omega_n^2) + e\Delta\bu \Delta\bu^T
\end{pmatrix},\label{eq:Linv_TL_LCgalvcoup_Networks}
\end{align}
where we have defined the coupling vector $\Delta\bu\equiv (\Delta u_0(0), \Delta u_1(0),...)^T$ and the parameters $d \equiv C_{cA}-c \alpha$ and $e\equiv 1/L_{cB} - 1/\beta l$. As usual, $\mone$ stands for the infinite-dimensional identity matrix, and $(\omega_n^2)$ is the diagonal matrix of squared frequencies. Following the same steps as in the previous section we derive the Hamiltonian
\begin{align}
H=&\, \frac{1}{2}\bq^T(\msA^{-1}+ \frac{C_A^2}{\alpha c} \msA^{-1}\ba\ba^T \msA^{-1})\bq+\frac{1}{2}\bphi^T\msB^{-1}\bphi+V(\bphi)+\sum_n \frac{Q_n^2}{2N_{\alpha}}+\frac{N_{\alpha}\omega_n^2\Phi_n^2}{2}\nonumber\\
&+\frac{C_A}{N_{\alpha}} (\bq^T\msA^{-1} \ba)\sum_n Q_n \Delta u_n(0)-\frac{1}{L_B} (\bphi^T\bb)\sum_n \Phi_n \Delta u_n(0),\,\label{eq:Ham_TL_LCgalvcoup_Networks}
\end{align}
from which canonical quantization can be done. Again the criterion has been the elimination of the harmonic mode-mode couplings, and the solution for the parameters reads
\begin{align}
\alpha&=\frac{C_{cA}-C_A^2\ba^T \msA^{-1}\ba}{c},\quad
\beta=L_{cB}/l.\nonumber
\end{align}
\subsubsection{Multiple networks coupled to line}
\label{sec:mult-netw-coupl}
\begin{figure*}[]	\centering{\includegraphics[width=0.65\textwidth]{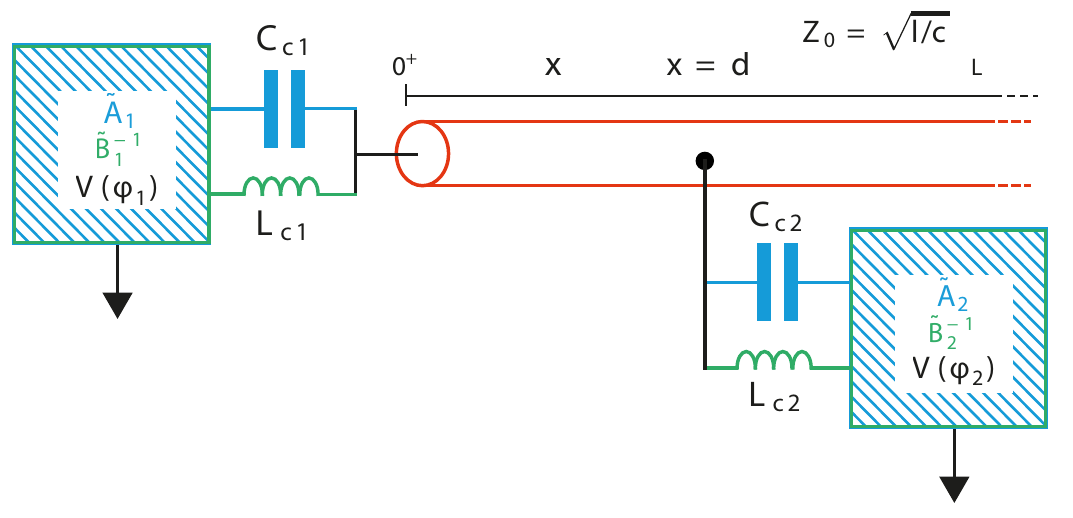}}
	\caption{\label{fig:TL_LCcoup_2Networks} Transmission line linearly coupled to two finite networks. The networks have internal flux degrees of freedom $\bphi_{i}$, with capacitance $\tilde{\msA}_{i}$ and inductive $\tilde{\msB}_i^{-1}$ matrices and general non-linear potential $V(\bphi_{i})$, with $i\in \{1,2\}$.}
\end{figure*}

We consider now the generalization of \ref{sec:line-galv-coupl} with a number $M$ of networks linearly coupled to a common transmission line, e.g. the circuit in Fig. \ref{fig:TL_LCcoup_2Networks} has two networks of degrees of freedom $\bphi_i$ coupled through capacitors $C_{ci}$ and inductors $L_{ci}$ to a transmission line at positions $\vec{x}=({0,d})$.

A Lagrangian of the generalized circuit, with $M$ networks, can be written as 
\begin{align}
 L= \sum_i^M\frac{1}{2}\dot{\bphi}_i^T \msA_i\dot{\bphi}_i-\frac{1}{2}\bphi_i^T \msB_i^{-1}\bphi_i-V(\bphi_i)
+ C_{ci}\left(-(\dot{\bphi}_i^T \ba_i)\dot{\Phi}(x_i,t)+\frac{\dot{\Phi}(x_i,t)^2}{2}\right)\nonumber\\
 \,\, -\sum_i^M \frac{1}{L_{ci}}\left(-(\bphi_i^T \bb_i)\Phi(x_i,t)+
\frac{\Phi(x_i,t)^2}{2}\right)
+\int_{0}^{L}dx\,\left[\frac{c}{2}\dot{\Phi}(x,t)^2-\frac{1}{2l}(\Phi'(x,t))^2\right],\nonumber 
\end{align}
where $\msA_i=\tilde{\msA}_i+C_{ci} \ba_i\ba_i^T$ and $\msB_i^{-1}=\tilde{\msB}_i^{-1}+\bb_i\bb_i^T/L_{ci}$ are the capacitance and inductance submatrices of the network respectively and $\ba_i$ and $\bb_i$ are coupling vectors to the finite networks from the transmission line. Following the same procedure as in last section,  we expand the flux field in an eigenbasis $\Phi(x,t)=\sum_n \Phi_n(t)u_n(x)$ and derive the wave equations and a number of boundary conditions of two possible forms, namely 
\begin{align}
u_{n}'(x_i)=&-k_n^2\alpha_i u_n(x_i)+\frac{1}{\beta_i}u_n(x_i),\,\,\,\,\,\,\forall x_i \in \{0,L\}\quad\mathrm{and}\label{eq:EVP_TL_LCcoup_2Networks_eq22}\\
\Delta u_{n}'(x_i)=&-k_n^2\alpha_i u_n(x_i)+\frac{1}{\beta_i}u_n(x_i),\,\,\,\,\,\,\forall x_i \notin \{0,L\},\label{eq:EVP_TL_LCcoup_2Networks_eq23}
\end{align}
depending on whether the $i^{th}$-network is connected at one end of the line or inbetween, respectively. Here $\Delta u_{n}'(x_i)\equiv u_{n}'(x_i^+)-u_{n}'(x_i^-)$, and for networks connected with boundary conditions of Eq. (\ref{eq:EVP_TL_LCcoup_2Networks_eq23}), we further require continuity of $u_{n}(x)$ at $x_i$ . Regardless of the position of connection, the new inner products for the eigenfunctions are
\begin{align}
 \langle u_n,u_m\rangle_{\{\alpha_i\}}&=c \left(\int_{0}^{L}dx\, u_n(x) u_m(x)+  \sum_{x_i}  \alpha_i u_n(x_i) u_m(x_i)\right)=N_{\alpha}\delta_{nm},\nonumber\\ 
 \langle u_n,u_m\rangle_{\{1/\beta_i\}}&=\frac{1}{l}\left(\int_{0}^{L}dx\, u_n'(x) u_m'(x)+ \sum_{x_i}\frac{1}{\beta_i} u_n(x_i) u_m(x_i)\right)=\omega_n^2 N_{\alpha}\delta_{nm},\nonumber 
\end{align}

The Lagrangian can thus be rewritten as 
\begin{equation}
L=\frac{1}{2}\dot{\bX}^T\msC\dot{\bX}^T-\frac{1}{2}\bX^T \msL^{-1}\bX -V(\boldsymbol{X}),\nonumber
\end{equation}
where $\bX=(\bphi_1^T,\bphi_2^T,...,\bphi_M^T,
{\bPhi}^T)^T$ with the new capacitance and inductance matrices  
\begin{align}
\msC&=\begin{pmatrix}
\msA_1&0&\dots&-C_{c1} \ba_1\bu_1^T\\
0&\msA_2&&-C_{c2} \ba_2\bu_2^T\\
\vdots&&\ddots&\vdots\\
-C_{c1} \bu_1\ba_1^{T} & -C_{c2} \bu_2\ba_2^{T} &\dots& N_{\alpha} \mone + \sum_i^M d_i\bu_i \bu_i^T
\end{pmatrix},\label{eq:C_LCcoup_2Networks}\\
\msL^{-1}&=\begin{pmatrix}
\msB_1^{-1}&0&\dots&-\bb_1\bu_1^T/L_{c1}\\
0&\msB_2^{-1}&&-\bb_2\bu_2^T/L_{c2}\\
\vdots&&\ddots&\vdots\\
-\bb_1\bu_1^T/L_{c1} & -\bb_2\bu_2^T/L_{c2} &\dots&  N_{\alpha} (\omega_n^2) + \sum_i^M e_i\bu_i \bu_i^T
\end{pmatrix},\label{eq:Linv_LCcoup_2Networks}
\end{align}
where we have defined the coupling vectors to the $i^{th}$ network as  $\bu_i \equiv\bu(x_i)=(u_0(x_i),u_1(x_i),...)^T$.

Let us for now assume the invertibility of the capacitance matrix $\msC$ (we examine this assumption critically in the next subsection, \ref{subsec:invertibility_variable_counting}).
Using the property that the coupling vectors are orthogonal $\langle \bu_i,\bu_j\rangle=\delta_{ij}N_{\alpha}/\alpha_i c$, see Eq. (\ref{eq:genericsumrule}) of Appendix \ref{Walter_appendix}, we determine
\begin{align}
 &\msC^{-1}=\nonumber\\
&\begin{pmatrix}
\msA_1^{-1}+ s_1 \msA_1^{-1}\ba_1\ba_1^T \msA_1^{-1}&0&\dots&t_1\msA_1^{-1}\ba_1\bu_1^T\\
0&\msA_2^{-1}+ s_2 \msA_2^{-1}\ba_2\ba_2^T \msA_2^{-1}&&t_2 \msA_2^{-1}\ba_2\bu_2^T\\
\vdots&&\ddots&\vdots\\
t_1 \bu_1\ba_1^{T}\msA_1^{-1} & t_2 \bu_2\ba_2^{T}\msA_2^{-1} &\dots& \frac{1}{N_{\alpha}} \mone + \sum_i^M r_i\bu_i \bu_i^T
\end{pmatrix},\nonumber
\end{align}
where we have defined  paramaters $s_i \equiv -C_{ci}^2 |\bu_i|^2/D_i$, $t_i= C_{ci}/D_i$,  $r_i=1/|\bu_i|^2(1/D_i-1/N_{\alpha})$ and $D_i\equiv N_{\alpha}+ |\bu_i|^2(d_i-C_{ci}^2\ba_i^T \msA_i^{-1} \ba_i)$. Finally, we can choose the relevant coefficients of the eigenvalue problem $(\alpha_i, \beta_i)$ such that $r_i=e_i=0$, $\forall i$. That is, we solve the equations $D_i = N_\alpha$ for $\alpha_i$ and $\beta_i = L_{ci}/l$, in order to arrive to a Hamiltonian with a well defined infinite harmonic set
\begin{align}
H&= \frac{1}{2}\sum_i\hat{\bq}_i^T(\msA_i^{-1}+ \frac{C_{ci}^2}{\alpha_i c} \msA_i^{-1}\ba_i\ba_i^T \msA_i^{-1})\hat{\bq}_i+\frac{1}{2}\hat{\bphi}_i\msB_i^{-1}\hat{\bphi}_i+V(\hat{\bphi}_i)+\sum_n\hbar \omega_n a_n^\dagger a_n\nonumber\\
&+\sum_i C_{ci}\sqrt{\frac{\hbar}{2 N_\alpha}} (\hat{\bq}_i^T\msA_i^{-1} \ba_i)\sum_n (a_n+a_n^\dagger)\sqrt{\omega_n} u_n(x_i) \nonumber\\
&-\sum_i\frac{i}{L_{ci}}\sqrt{\frac{\hbar}{2 N_\alpha}} (\hat{\bphi}_i^T \bb_i)\sum_n (a_n-a_n^{\dagger})\frac{u_n(x_i)}{\sqrt{\omega_n}},\label{eq:Ham_TL_LCcoup_2Networks3}
\end{align}
where we have promoted conjugate variables to operators as in previous sections. Again, the coupling coefficients of the capacity part are governed by $\sqrt{\omega_n}u_n(x_i)$, and thus have a large $n$ behaviour of the form $n^{-1/2}$.
\subsection{Invertibility and variable counting}
\label{subsec:invertibility_variable_counting}
In the previous section we have assumed that the capacitance matrix $\msC$ has inverse, and thus there is no overcounting of velocity degrees of freedom. However, this assumption does not always hold. Fortunately, it can be easily checked, by determining the conditions for the existence of a zero eigenvalue. Let us first examine the simple case of the network connected to the transmission line, under the assumption that the capacitance submatrix $\msA$ is invertible. The condition for the invertibility of $\msC$ in (\ref{eq:C_LCcoup_Network}) is determined by analyzing the possible existence of a zero eigenvalue, for which
\begin{equation}
\msC\begin{pmatrix}
\by\\
\bz
\end{pmatrix}=0.\nonumber
\end{equation}

The above matrix equation reduces to 
\begin{align}
\quad\msA\by&=C_c \ba (\bu^T \bz),\label{eq:C_LCcoup_Network_invert_1}\\
C_c \bu (\ba^T \by)&=N_\alpha \bz+d\bu(\bu^T \bz).\label{eq:C_LCcoup_Network_invert_2}
\end{align}
Solving $\by$ in (\ref{eq:C_LCcoup_Network_invert_1}) and substituting in (\ref{eq:C_LCcoup_Network_invert_2}) we can derive the following equation
\begin{equation}
(\bu^T \bz)\frac{C_c}{\alpha c}\left(1- C_c \ba^T \msA^{-1}\ba\right)=0,\nonumber
\end{equation}
where we have used the sum rule $\left|\bu\right|^2=N_{\alpha}/\alpha c$ and $d=C_c-\alpha c$.  Notice furthermore that if we were to assume $\bu^T\bz$ were zero, equation (\ref{eq:C_LCcoup_Network_invert_2}) would tell us that $\bu$ and $\bz$ are parallel, and we would be forced to have $\bz$ of zero norm, so we can conclude that  $\bu^T\bz\neq0$ if the eigenvector is not trivial. It follows  that, unless $\alpha$ is zero or infinity (in which case there is no  capacitive connection), a non-trivial solution can appear only when $\left(1- C_c \ba^T \msA^{-1}\ba\right)=0$. Thus, unless  $\left(1- C_c \ba^T \msA^{-1}\ba\right)$ is zero, $\msC$ is invertible.

Having a non invertible capacitance matrix means that at least one combination of the initial variables will not be dynamical, and will be frozen in a value determined by the potential part. For our purposes, namely the provision of quantum mechanical models, this is a complication that can readily be eliminated by a good choice of variables, in which this frozen variable is discarded.

A different analysis corresponds to the inductance matrix. In this case the question at hand is the presence of zero modes. For a linear network where the potential $V(\bphi)=0$, the condition for the invertibility of the inductance matrix $\msL^{-1}$ can also be examined. In particular, consider $\msL^{-1}$  given by Eq. (\ref{eq:Linv_LCcoup_Network}). Solving the equation $\msL^{-1} (\by, \bz)^T=0$, and using the second sum rule  $\bu^T (N_\alpha (\omega_n^2))^{-1} \bu=\sum_n u_n(0)^2/N_\alpha \omega_n^2=\beta l$, see equation (\ref{eq:secondsumrule}) in Appendix \ref{Walter_appendix}, we can derive 
\begin{equation}
(\bu^T \bz)\frac{\beta l}{L_c}\left(1- \frac{1}{L_c} \bb^T \msB\bb\right)=0.\nonumber
\end{equation}
Similarly to the capacitance coupling case, for $\beta \neq \{0,\infty\}$ the inductance matrix $\msL^{-1}$ is  not invertible when $\left(1- \frac{1}{L_c} \bb^T \msB\bb\right)=0$. In contrast to the capacitive case, given that there is no general potential,  $V(\bphi)=0$, the description with such set of degrees of freedom can be used but a zero-mode will appear. 

The generalization to the $M$-networks connected to the transmission line is straightforward. In order for the capacitance matrix $\msC$ and the inductance matrix $\msL^{-1}$ in Eqs. (\ref{eq:C_LCcoup_2Networks}, \ref{eq:Linv_LCcoup_2Networks}) respectively to be non-invertible we require that 
\begin{align}
 \frac{C_{ci}}{\alpha_i c}\left(1- C_{ci} \ba_i^T \msA_i^{-1}\ba_i\right)&=0,\nonumber\\
 \frac{\beta_i l}{L_{ci}}\left(1- \frac{1}{L_{ci}} \bb_i^T \msB_i\bb_i\right)&=0,\nonumber
\end{align}
for all the networks, i.e. $ \forall i\in \{1,...,M\}$.

As we will see presently,  a frequent approach in the field of superconducting circuits is to truncate the number of modes to a finite quantity $N$. In so doing, the possibility exists that in the large $N$ limit the model presents non-dynamical modes, even if it is not the case for finite $N$, and some computations can present inadequate behaviours in that limit.

There is an additional reason, which will be exemplified in the next subsection, and is directly in line with the results presented above for transmission lines. In transmission lines capacitively coupled to networks we have seen that the choice of expansion modes is not free if we demand that the Hamiltonian description of the complete system  be understood as  being given by an infinite number of independent modes with no coupling among themselves, together with a network Hamiltonian, and coupling of this network to the independent modes. This basic idea of coupling of otherwise independent subsystems is essential in phenomenological model building. It is the case, however, that in some circumstances a na\"{\i}ve separation of subsystems will lead to non invertibility of the capacitance operator. That is, the separation in subsystems has led to overcounting of variables. One has to identify the structure of modes that precisely accounts for the proper amount of independent variables, by assessing how the couplings restrict our freedom in the choice of expansion.

In the next section, we are going to discuss this problem for the  particular case of the connection of two transmission lines via a very simple network. We will also provide an alternative solution and explanation.

\subsection{Linear lumped-element coupling between lines}
\label{subsec:TL_LCcoup_TL}

\begin{figure*}[]
	\centering{\includegraphics[width=0.7\textwidth]{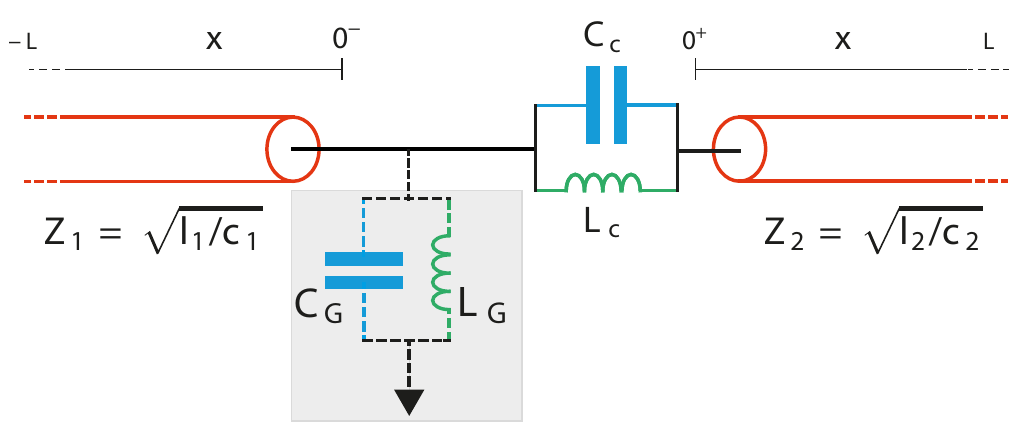}}
	\caption{\label{fig:TL_LCcoup_TL}Transmission line inductively and capacitively coupled to a finite network. Two transmission lines characterized by their capacitance $c_i$ and inductance $l_i$ per unit length are coupled and grounded through $LC$-resonators. Given that the flux field is described in partite bases, ground capacitor $C_G$ (inductor $L_G$) becomes necessary when the lines are capacitively (inductively) coupled for the $\msC$ $(\msL)$ matrix to be invertible.}
\end{figure*}

Let us now consider the circuit in Fig. \ref{fig:TL_LCcoup_TL}, in which two transmission lines are coupled via a simple network. It is apparent that there are two subsystems, namely the left and right transmission lines. We shall see, however, that the description in those terms would be wrong if  either of the capacities $C_G$ and $C_c$ were absent. In such a situation, $C_GC_c=0$, there is a need for considering the whole system to identify the proper expansion in modes. Notice the difference with respect to the example of galvanic coupling, subsection \ref{sec:line-galv-coupl}, in that here the two endpoints are only connected by one oscillator. 

In this example we see that this essentially means that for there to be a description in terms of separate subsystems  there needs to be an endpoint variable for each transmission line that suitably dresses the transmission line modes, and that those endpoint variables be independent among themselves.

The Lagrangian for the circuit  reads
\begin{align} 
L =& \int_{-L}^{0^-}dx\,\left[\frac{c_1}{2}\dot{\Phi}_1(x,t)^2 -\frac{1}{2l_1}(\Phi_1'(x,t))^2\right]\nonumber\\
& +\int_{0^+}^{L}dx\,\left[\frac{c_2}{2}\dot{\Phi}_2(x,t)^2-\frac{1}{2l_2}(\Phi_2'(x,t))^2\right]\nonumber\\
&+\frac{C_{c}}{2}(\dot{\Phi}_1(0^-,t)-\dot{\Phi}_2(0^+,t))^2-\frac{1}{2L_{c}}({\Phi}_1(0^-,t)-{\Phi}_2(0^+,t))^2\nonumber\\
&+\frac{C_{G}}{2}\dot{\Phi}_1(0^-,t)^2-\frac{1}{2L_{G}}{\Phi}_1(0^-,t)^2, \label{eq:Lag_TL_LCcoup_TL}
\end{align}
giving rise to wave equations for each of the transmission lines,
and  Kirchhoff's equations
\begin{align}
  -\frac{1}{l}\Phi_1(0^-,t)&=C_{G}\ddot{\Phi}(0^-,t)+\frac{1}{L_{G}}{\Phi}(0^-,t)-\frac{1}{l}\Phi_2(0^+,t),\nonumber\\
  -\frac{1}{l}\Phi_2(0^+,t)&=	C_{c}(\ddot{\Phi}_1(0^-,t)-\ddot{\Phi}_2(0^+,t)) + \frac{1}{L_{c}}({\Phi}_1(0^-,t)-\Phi_2(0^+,t)).\nonumber 
\end{align}

As has been our approach all along, we now look for expansions  $\Phi_1(x,t)=\sum_n \Phi_n(t)u_n(x)$ and $\Phi_2(x,t)=\sum_n \Psi_n(t)v_n(x)$ that provide us with a good description of the system.  Following the arguments presented in previous examples and in Appendix \ref{Walter_appendix}, we introduce one pair of free parameters,  $\alpha_i$ and $\beta_i$, for each boundary condition equation (\ref{eq:EVP_TL_LCcoup_TL_eq2}, \ref{eq:EVP_TL_LCcoup_TL_eq3}), and achieve separation of variables with eigenvalue-dependent boundary conditions,
\begin{align}
\ddot{\Phi}_n(t)&=-\omega_n^2\Phi_n(t),\,\,\,
u_n''(x)=-k_n^2u_n(x),\label{eq:EVP_TL_LCcoup_TL_eq11}\\
\ddot{\Psi}_n(t)&=-\Omega_n^2\Psi_n(t),\,\,\,
v_n''(x)=-\chi_n^2v_n(x),\label{eq:EVP_TL_LCcoup_TL_eq12}\\
u_{n}'(0^-)&=k_n^2\alpha_1 u_n(0^-)-\frac{1}{\beta_1} u_n(0^-),\label{eq:EVP_TL_LCcoup_TL_eq2}\\
v_{n}'(0^+)&=-\chi_n^2\alpha_2 v_n(0^+)+\frac{1}{\beta_2} v_n(0^+).\label{eq:EVP_TL_LCcoup_TL_eq3}
\end{align}
These generalized singular value problems fall in the class studied in Appendix \ref{Walter_appendix}, and the expansion theorems are guarantee of our approach. 
Again, the eigenfunctions fulfill the orthogonality relations 
\begin{align}
  \langle u_n,u_m\rangle_{\alpha_1}&=c_1 \left(\int_{-L}^{0^-}dx\, u_n(x) u_m(x)+  \alpha_1 u_n(0^-)  u_m(0^-)\right)=N_{\alpha_1}\delta_{nm},\nonumber\\
  \langle u_n,u_m\rangle_{1/\beta_1}&=\frac{1}{l_1}\left(\int_{-L}^{0^-}dx\, u_n'(x) u_m'(x)+ \frac{1}{\beta_1}  u_n(0^-) u_m(0^-)\right)=\omega_n^2 N_{\alpha_1}\delta_{nm},\nonumber\\
  \langle v_n,v_m\rangle_{\alpha_2}&=c_2 \left(\int_{0^+}^{L}dx\, v_n(x) v_m(x)+  \alpha_2 v_n(0^+)  v_m(0^+)\right)=N_{\alpha_2}\delta_{nm},\nonumber\\
  \langle v_n,v_m\rangle_{1/\beta_2}&=\frac{1}{l_2}\left(\int_{0^+}^{L}dx\, v_n'(x) v_m'(x)+ \frac{1}{\beta_2}  v_n(0^+) v_m(0^+)\right)=\Omega_n^2 N_{\alpha_2}\delta_{nm},\nonumber
\end{align}
where $N_{\alpha_i}$ are free normalization constants with dimensions of capacitance. The Lagrangian (\ref{eq:Lag_TL_LCcoup_TL}) is now rewritten in terms of modes as 
\begin{equation}
L=\frac{1}{2}\dot{\bX}^T\msC\dot{\bX}^T-\frac{1}{2}\bX^T \msL^{-1}\bX\nonumber
\end{equation}
where the full flux vector is $\bX=(\bPhi,
{\bPsi})^T$ and the capacitance and inverse inductance matrices are
\begin{align}
\msC=&\begin{pmatrix}
N_{\alpha_1} \mone + d_1\bu \bu^T& -C_{c} \bu\bv^T\\
-C_{c} \bv\bu^{T} & N_{\alpha_2} \mone + d_2\bv \bv^T
\end{pmatrix},\label{eq:C_TL_LCcoup_TL}\\
\msL^{-1}=&\begin{pmatrix}
N_{\alpha_1} (\omega_n^2) + e_1\bu \bu^T& - \bu\bv^T/L_{c}\\
- \bv\bu^{T}/L_{c} & N_{\alpha_2} (\Omega_n^2) + e_2\bv \bv^T
\end{pmatrix}.\label{eq:Linv_TL_LCcoup_TL}
\end{align}
Following the notational conventions we have used previously, 
the coupling vectors are named  $\bu= ( u_0(0^-),  u_1(0^-),...)^T$ and  $\bv= ( v_0(0^+),  v_1(0^+),...)^T$. We introduce  parameters $d_1 = C_{\Sigma}-c_1 \alpha_1$ and $d_2 = C_{c}-c_2 \alpha_2$ where $C_{\Sigma}=C_{c}+C_{G}$ is the total capacitance,  and $e_1= 1/L_{\Sigma} - 1/\beta_1 l_1$ and $e_2= 1/L_{c} - 1/\beta_2 l_2$ with $L_{\Sigma }=L_{c}L_{G}/(L_{c}+L_{G})$ being  the equivalent parallel inductance.

\subsubsection{Derivation of the Hamiltonian}
\label{sec:deriv-hamilt}
It is easy to calculate the inverse of the capacitance matrix (Legendre transformation) in this basis of modes
\begin{equation}
\msC^{-1}=\begin{pmatrix}
\frac{1}{N_{\alpha_1}} \mone + \delta_1\bu \bu^T& \delta_3 \bu\bv^T\\
\delta_3 \bv\bu^{T} & \frac{1}{N_{\alpha_2}} \mone + \delta_2\bv \bv^T
\end{pmatrix},\nonumber
\end{equation}
Inserting the definitions of $d_1$ and $d_2$ and the normalization of vectors $\left|\bu\right|^2=N_{\alpha_1}/c_1 \alpha_1$ and $\left|\bv\right|^2=N_{\alpha_2}/c_2 \alpha_2$, we can check that the parameters are
\begin{align}
	\delta_1=\frac{c_1\alpha_1(c_1 \alpha_1 - C_G)}{C_G N_{\alpha_1}^2},&\quad
	\delta_2=\frac{c_2 \alpha_2(c_2\alpha_2 C_{\Sigma}-C_c C_G)}{C_cC_G N_{\alpha_2}^2},\nonumber\\
	\delta_3=&\frac{c_1 c_2 \alpha_1 \alpha_2}{C_G N_{\alpha_1}N_{\alpha_2}}.\nonumber
\end{align}

We enforce now the condition that the modes in a transmission line have no direct coupling among themselves, i.e. $\delta_1=\delta_2=0$. This criterion determines the coefficients $\alpha_1$ and $\alpha_2$ to be $C_G/c_1$ and $C_GC_c/c_2(C_G+C_c)$ respectively. These are clearly the natural capacity length scales for each of the transmission lines, since they are given in both cases by the ratio of the total capacity from the endpoint of the line to the reference zero potential divided by the capacity density of the line. The coupling strength $\delta_3$ simplifies then to $C_GC_c/N_{\alpha_1}N_{\alpha_2}(C_G+C_c)$.

Let us examine possible pathological situations. First of all, bear in mind that fixing the length scales $\alpha_1$ and $\alpha_2$ as above is necessary according the criterion we presented. Nonetheless, any value other than 0 or infinity would provide us with a description of the system, for general values of the parameters of the lumped elements.
The value zero is excluded because it would not provide us with the description of the coupling. Such a condition entails there being no current at the endpoint. As to infinity, this would fix the value of the potential at the endpoint, again inhibiting coupling.

There are two other pathological cases, given by $C_cC_G=0$. First, $C_c=0$. In this situation, $\delta_2$ blows up unless $\alpha_2$ is set to 0. But we are then in a case in which there is proper coupling of the transmission lines with our Hamiltonian  description. Let us therefore examine this case in more detail directly in the capacitance matrix itself, assuming that $\alpha_2\neq0$. In this  case with $C_c=0$, $d_2=-c_2 \alpha_2$ and, using the general result that $c_2\alpha_2=N_{\alpha_2}/\left|\bv\right|^2$, the bottom-right submatrix in (\ref{eq:C_TL_LCcoup_TL}) becomes proportional to the projector $\mone-\bv\bv^T/\left|\bv\right|^2$. This projector spans the vector space orthogonal to $\bv$, and gives 0 when acting on $\bv$. It follows that the column vector $(0^T,\bv^T)^T$ is an eigenvector of $\msC$ with eigenvalue 0. We have a nondynamical variable in our description.

Passing now to the case $C_G=0$, one can write the capacitance matrix in this situation in the form
\begin{equation}
\nonumber 
\msC\to
\begin{pmatrix}
N_{\alpha_1}\left(\mone-\frac{\bu\bu^T}{\left|\bu\right|^2}\right)&0\\ 0&  N_{\alpha_2}\left(\mone-\frac{\bv\bv^T}{\left|\bv\right|^2}\right)
\end{pmatrix}+
C_c\begin{pmatrix}
\bu\bu^T&  -\bu\bv^T\\ - \bv\bu^T & \bv\bv^T
\end{pmatrix},
\end{equation}
whence the zero eigenvalue vector $\left(\bu^T/\left|\bu\right|^2,\bv^T/\left|\bv\right|^2\right)^T$ is readily computed.
Again, (\ref{eq:C_TL_LCcoup_TL}) becomes singular and we have an overcounting of the degrees of freedom. 

This analysis has provided us with an understanding of the issue beyond the purely algebraic treatment, in that we can only partition usefully the subsystem into its two composants if indeed there are enough degrees of freedom. In particular, we must have enough kinetic terms. This also suggests that in systems for which the partitioning fails in terms of non-invertibility of the capacitance matrix, it is sensible to study the possibility that our modelling is lacking some additional (parasitic) capacitances, which would solve the problem.

Regarding the rank of the inductance matrix (\ref{eq:Linv_TL_LCcoup_TL}),  and following Sec. \ref{subsec:invertibility_variable_counting} above, we study the equation $\msL^{-1} (\by^T, \bz^T)^T=0$. Introducing the parameters $e_1$, $e_2$ and the sum rules
$\bu^T (N_{\alpha_1} (\omega_n^2))^{-1} \bu=\sum_n u_n(0)^2/N_{\alpha_1} \omega_n^2=\beta_1 l_1$ and $\bv^T (N_{\alpha_2} (\Omega_n^2))^{-1} \bv=\sum_n v_n(0)^2/N_{\alpha_2} \Omega_n^2=\beta_2 l_2$, we derive the equation
\begin{equation}
\frac{\beta_2 l_2}{L_c}\left(1- \frac{L_\Sigma}{L_c} \right)=0.\nonumber
\end{equation} 
Again, we can distinguish a few interesting cases. If there is inductive coupling between the lines, i.e. $L_c$ has a finite value, the limit of open ground inductor $L_G\rightarrow \infty$ makes the inductive matrix singular. On the other hand, if we disconnect the lines $L_c\rightarrow \infty$, we can find orthogonal complete bases as long as we develope the right field mode in a basis with $\beta_2 l_2\rightarrow \infty$.

\subsection{Exact and approximate quantization methods}
\label{subsec:ch2_approximate_quantization}

Up to here we have studied exact methods to derive quantum Hamiltonians of circuits with transmission lines linearly coupled to networks of finite variables or other transmission lines. Thus, insofar as the starting point, namely the Lagrangian, is a good description of the system under study, so is the Hamiltonian, with no further approximation. Additionally, there are no divergences intrinsic to these models, since there is a natural cutoff.

\begin{figure*}[h]
	\centering{\includegraphics[width=0.55\textwidth]{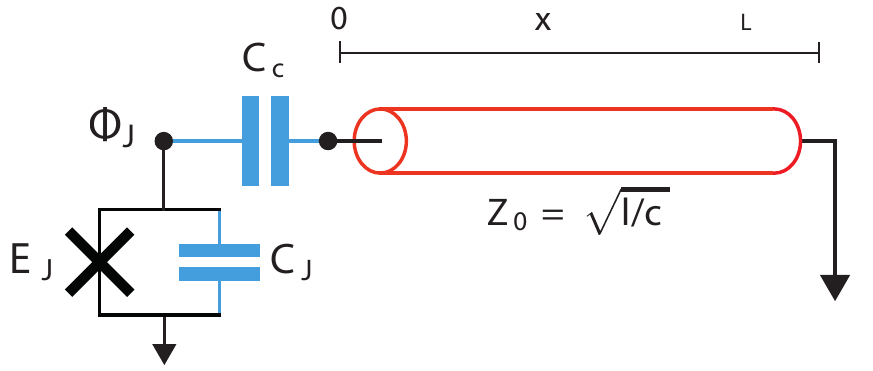}}
	\caption{\label{fig:TL_Ccoup_CQubit}Transmission line capacitively coupled to a network composed of a unique Josephson junction. The exact coupling parameters $g_n$ have a finite cutoff frequency due to the Josephson junction capacitance $C_J$, see Eqs. (\ref{eq:TL_Ccoup_CQubit_gn}) and (\ref{eq:TL_Ccoup_CQubit_gn_approx}), and Fig. \ref{fig:TL_Ccoup_CQubit_gn}.}
\end{figure*}

In this section we shall first use the previous techniques in a particularly simple example, for which we shall later portray some of the approximations present in the literature, with a view to clarifying how those approximations are the actual source of the divergences there encountered. The  example is that of a charge qubit capacitively coupled to a finite length transmission line resonator ended in a short to ground, see Fig. \ref{fig:TL_Ccoup_CQubit}.
Following \ref{subsubsec:chap2_mixed_linear_coupling}, the Lagrangian of the circuit in Fig. \ref{fig:TL_Ccoup_CQubit} can be written, once the field has been expanded in modes, as 
\begin{equation}
L=\frac{1}{2}\dot{\bX}^T\msC\dot{\bX}-\frac{1}{2}\bX^T \msL^{-1}\bX +E_J\cos(2\pi\phi_J/\Phi_q),\nonumber
\end{equation}
where we have defined the vector of fluxes $\bX=(\phi_J,\bPhi^T)^T$, $\Phi_q$ is the magnetic flux quantum, and the capacitance and inductance matrices are
\begin{align}
 \msC=&\begin{pmatrix}
C_J+C_c& -C_c \bu^T\\
-C_c \bu & N_{\alpha} \mone + d\bu \bu^T
\end{pmatrix},\nonumber\\ 
 \msL^{-1}=&\begin{pmatrix}
0& 0^T\\
0 & N_{\alpha} (\omega_n^2)
\end{pmatrix},\nonumber 
\end{align}
where $d=C_c-\alpha c$. For definiteness, we rewrite here the homogeneous eigenvalue problem, Eqs. (\ref{eq:EVP_TL_LCcoup_Network_eq1}-\ref{eq:EVP_TL_LCcoup_Network_eq3}), that must be solved in order to find the eigenfrequencies $\omega_n$ and generalized eigenfunctions $u_n(x)$,
\begin{align}
u_n''(x)=&-k_n^2u_n(x),\label{eq:EVP_App_Quant_eq1}\\
u_{n}'(0)=&-k_n^2\alpha u_n(0),\label{eq:EVP_App_Quant_eq2}\\
u_n(L)=&0.\label{eq:EVP_App_Quant_eq3}
\end{align}
\subsubsection{Exact Legendre transformation }
We can solve the eigenvalue problem with eigenfunctions normalized by $\langle u_n,u_m\rangle_{\alpha}$ as in  (\ref{eq:TL_LCcoup_Network_ortho_1}). The eigenfunctions are readily seen to be $u_n(x)=A_n \sin(k_n(L-x))$, where the amplitude for  the $n^{\mathrm{th}}$ mode is
\begin{equation}
A_n=\sqrt{\frac{2 N_\alpha}{c L}} \sqrt{\frac{1+(\alpha k_n)^2}{1+(\alpha/ L) + (\alpha k_n)^2}},\nonumber
\end{equation}
given the choice of normalization in (\ref{eq:TL_LCcoup_Network_ortho_1}). As to the wavenumbers of the modes, they are the positive non trivial solutions of  the transcendental equation
\begin{equation}
\alpha k = \cot (k L).\nonumber
\end{equation}
We can calculate the components of the coupling vector $\bu=(u_1(0),u_2(0),...)^T$,
\begin{equation}
u_n(0)=\sqrt{\frac{2 N_\alpha}{c L}} \sqrt{\frac{1}{1+(\alpha/ L) + (\alpha k_n)^2}},\nonumber
\end{equation}
from which one can directly infer the finiteness of the norm, since $k_n\sim n$, whence $u_n(0)\sim n^{-1}$. Thanks to the results of Appendix \ref{Walter_appendix}, we know that this finite norm is $\left|\bu\right|^2=N_\alpha/\alpha c$. Using Eq. (\ref{eq:TL_LCcoup_Network_alpha_fix}), we find that the choice $\alpha=C_cC_J/c(C_c+C_J)$  results in an exact Hamiltonian
\begin{equation}
H=\frac{1}{2}{\bP}^T\msC^{-1}{\bP}-\frac{1}{2}\bX^T \msL^{-1}\bX +E_J\cos(2\pi\phi_J/\Phi_q),\nonumber
\end{equation}
without mode-mode coupling, where the inverse of the capacitance matrix is
\begin{equation}
\msC^{-1}=\begin{pmatrix}
\frac{1}{C_J}& \frac{C_c}{N_\alpha C_\Sigma} \bu^T\\
\frac{C_c}{N_\alpha C_\Sigma} \bu & \frac{1}{N_{\alpha}} \mone
\end{pmatrix},\nonumber
\end{equation}
and the charge conjugate variables are $\bP=\partial L/\partial \dot{\bX}=(q_J,\bQ^T)^T$. The parameter $C_\Sigma$ is again $C_J+C_c$. Promoting the conjugate variables to operators and introducing annihilation and creation operators for the harmonic sector, we derive the Hamiltonian
\begin{equation}
H = 4 E_C \hat{n}_J^2-E_J \cos(2\pi\hat{\phi}_J/\Phi_q)+ \sum_n \hbar g_n \hat{n}_J(a_n+a_n^\dagger)+ \sum_n \hbar \omega_n a_n^\dagger a_n,\nonumber
\end{equation}
with the charge energy of the qubit $E_C =e^2/2 C_J$ and coupling defined as 
\begin{equation}
g_n= v_{p}\frac{C_c}{C_\Sigma} \sqrt{\frac{Z_0}{R_Q}}\sqrt{\frac{2\pi k_n}{L(1+(\alpha/L) + (\alpha k_n)^2)}}.\label{eq:TL_Ccoup_CQubit_gn}
\end{equation}
Here $R_Q= h/(2e)^2$ is the quantum of resistance and $v_{p}=1/\sqrt{lc}$ the velocity of propagation in the line. As  shown in \cite{Gely:2017,Malekakhlagh:2017} the coupling grows as $g_n\propto \sqrt{k_n}$ for low frequency modes, where the wavenumbers resemble those of an open line, i.e. $k_n\approx (2n+1)\pi/2L$, and decays as $g_n\propto 1/\sqrt{k_n}$ for large $n$, when the wavenumbers tend to those of a line ended in a short, $k_n\approx n \pi/L$.  Observe that $g_n\sim n^{-1/2}$ for large $n$.

\begin{figure*}[h]
	\centering{\includegraphics[width=0.6\textwidth]{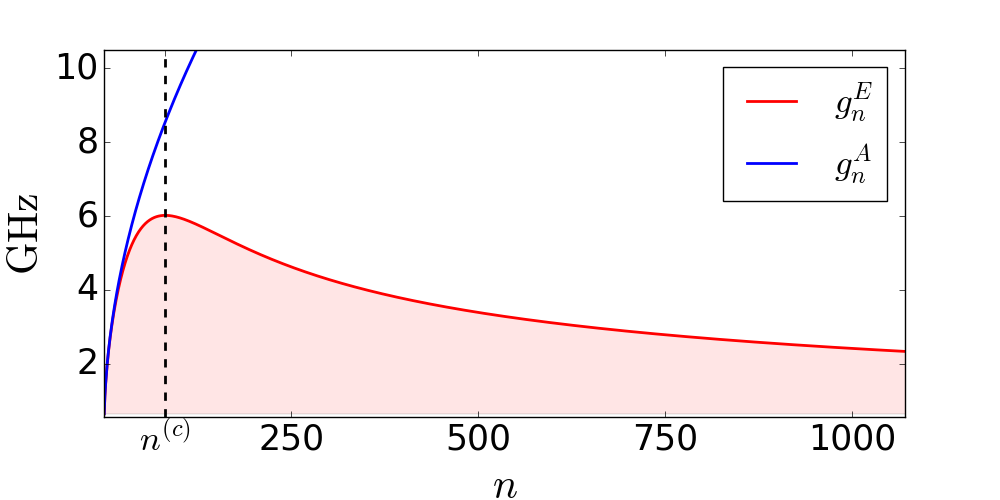}}
	\caption{\label{fig:TL_Ccoup_CQubit_gn}Capacitive coupling $g_n$ per mode for the circuit in Fig. \ref{fig:TL_Ccoup_CQubit}. We apply the exact and approximate formulae to the experiment of Device A in \cite{Bosman:2017}, with parameters $C_c=40.3\,\mathrm{fF}$, $C_J=5.13 \,\mathrm{fF}$, $c=249\,\mathrm{pF/m}$, $l=623\,\mathrm{nH/m}$ and $L=4.7\,\mathrm{mm}$. In red, $g_n$ for the exact derivation, Eq. (\ref{eq:TL_Ccoup_CQubit_gn}): a natural cutoff of the coupling constant appears at mode number $n^{(c)}=81$ with frequency $\omega_{81}=702.5\,\mathrm{GHz}$ (dashed line). In blue, the approximate divergent coupling constant from Eq. (\ref{eq:TL_Ccoup_CQubit_gn_approx}).}
\end{figure*}

Given that, for large $n$, $(k_{n+1}-k_n)/k_n\sim n^{-1}$, if the maximum of $g_n$ is given at high $n$ we can treat the wavenumber as a continuous variable, and predict at which mode number the coupling saturates by solving the equation $\partial g_n /\partial k_n=0$, which yields as a result $k_n^{(c)}\approx\sqrt{\frac{1+\alpha/L}{\alpha^2}}$. For typical experimental values where the capacitance of the network is smaller than the total capacitance of the transmission line $\alpha/L\ll1$, such a device has many modes with frequencies close to a $\lambda/4$-resonator and the saturation point appears in a high frequency mode, see Fig. \ref{fig:TL_Ccoup_CQubit_gn}.

Clearly, $1/\alpha$ is a natural ultraviolet smooth cutoff. The frequencies in the model go all the way to infinity, but the coupling is indeed moderated by this cutoff. In our search of the literature we have not identified situations in which the maximum of the coupling has been detected, since the modes under study have lain rather below it. We submit this as a prediction for future experiments. 

\subsubsection{Approximations and introduction of a cutoff}
\label{sec:appr-intr-cutoff}
Let  us now compare the exact results above with some presentations in the literature which rely on some widely used  approximations.

As stated above, for low lying modes, such that $\alpha k_n\ll 1$, the couplings $g_n$ will scale with $\sqrt{k_n}$. The issue is that for many of the experimental setups the wavenumber for maximal coupling is very large in comparison with accessible wavenumbers.

Let us examine the situation in which  the total capacitance external to the transmission line, $C_cC_J/(C_c+C_J)$, is much smaller than the total capacitance of the transmission line, $cL$. In such a case, $\alpha \ll L$. Then two consistent approximations can be made. In the first place, the secular equation is best rewritten as $(\alpha/L) (kL)=\cot(kL)$. For $\alpha\ll L$ and low-lying modes, the equation is approximately $\cot(kL)=0$. In fact, a perturbative analysis shows that this is consistent for $k_n$ as long as $\alpha\ll L/n$. Secondly, the boundary condition (\ref{eq:EVP_App_Quant_eq2}) in such a case is well approximated by $u_n'(0)$, again with the same restriction for $n$, namely $n\ll L/\alpha$.
This approximation $u_n'(0)\approx 0$ gives us eigenfunctions $u_n(x)=N_c \cos(k_n x)$, where $N_c$ is a normalization constant, and wavenumbers $k_n=(2n+1)\pi/2L$. In contrast with the exact Hamiltonian in the  section above, this approximation, equivalent to setting $\alpha$ to zero,  results in a coupling vector $\bu=(u_1(0),u_2(0),...)$ with infinite norm $\left|\bu\right|^2=\sum_{n=1}^{\infty} N_c^2\rightarrow\infty$. This basis of modes is orthogonal with respect to the inner product $\langle u_n,u_m\rangle_{\alpha=0}$ (\ref{eq:TL_LCcoup_Network_ortho_1}), such that the two normalization constants are related through $N_c=\sqrt{2N_\alpha/cL}$.

The issue now is that the capacitance matrix $\msC$ for the case of $\alpha=0$, while formally simple, presents vectors $\bu$ all of whose components equal $N_c$, and the formulae for inversion cannot be applied. Let us therefore introduce a truncation of modes of the transmission line to $N$. The vector $\bu$ has in this case norm squared $\left|\bu\right|^2= 2 N_\alpha N/cL$. Define the vector of unit length $\be=\bu/\left|\bu\right|$. The capacitance matrix reads
\begin{align}
\msC=&
\begin{pmatrix}
C_\Sigma&- C_c \sqrt{\frac{2N N_\alpha}{cL}}\be^T\\ - C_c \sqrt{\frac{2N N_\alpha}{cL}}\be& N_\alpha\left(\mone+\frac{2 C_c N}{cL}\be\be^T\right)
\end{pmatrix}\nonumber\\
=&\begin{pmatrix}
C_J&0\\0&N_\alpha
\end{pmatrix}+C_c
\begin{pmatrix}
1& - \sqrt{\frac{2N N_\alpha}{cL}}\be^T\\- \sqrt{\frac{2N N_\alpha}{cL}}\be& \frac{2N_\alpha N}{cL}\be\be^T
\end{pmatrix}
\,\nonumber 
\end{align}
Assume that $C_c\ll cL$ (which entails the smallness of the total external capacitance in comparison to the capacitance of the transmission line). As long as the number of modes under consideration, $N$, is not too large, we can consider that the terms with $C_c$ are perturbative with respect to the others. Given two matrices $\msA$ and $\msB$ such that $\msB$ can be understood as very small with respect to the invertible $\msA$, we have the approximate expression $\left(\msA+\msB\right)^{-1}\approx\msA^{-1}-\msA^{-1}\msB\msA^{-1}$. Thus, to order $\left(C_cN/cL\right)^{1/2}$ we have an approximate inverse capacitance matrix
\begin{equation}
\nonumber
\msC^{-1}\approx
\begin{pmatrix}
C_\Sigma^{-1}& \frac{C_c}{C_\Sigma}\sqrt{\frac{2N}{N_\alpha cL}}\be^T\\  \frac{C_c}{C_\Sigma}\sqrt{\frac{2N}{N_\alpha cL}}\be & N_\alpha^{-1}\mone
\end{pmatrix}\approx \begin{pmatrix}
C_\Sigma^{-1}& \frac{C_c}{C_\Sigma N_\alpha}\bu^T\\  \frac{C_c}{C_\Sigma N_\alpha}\bu & N_\alpha^{-1}\mone
\end{pmatrix}.
\end{equation}
It is important to insist that this approximation is only valid if the number of modes taken into account is such that indeed $C_c N/cL\ll1$.  Nonetheless, the rightmost expression does not portray explicitly the truncation in modes. 

We now use this approximate inverse capacitance matrix. Promoting the conjugate variables $\bP=\partial L/\partial \dot{\bX}=(q_J, \bQ^T)^T$ to operators and changing to annihilation and creation operators for the harmonic sector, we derive the Hamiltonian
\begin{equation}
H \approx 4 E_C \hat{n}_J^2-E_J \cos(2\pi \hat{\phi}_J/\Phi_q)+ \sum_n \hbar g_n \hat{n}_J(a_n+a_n^\dagger)+ \sum_n \hbar \omega_n a_n^\dagger a_n\nonumber
\end{equation}
with the charge energy of the qubit $E_C =e^2/2 C_\Sigma$ and the coupling defined as 
\begin{equation}
g_n= v_{p}\frac{C_c}{C_\Sigma} \sqrt{\frac{Z_0}{R_Q}}\sqrt{\frac{2\pi k_n}{L}},\label{eq:TL_Ccoup_CQubit_gn_approx}
\end{equation}
where $R_Q= h/(2e)^2$ is the quantum of resistance and $v_{p}=1/\sqrt{lc}$ the velocity of propagation in the line.
As repeatedly stated this Hamiltonian is only an approximation, valid for the low lying modes. The number of modes for which it applies can be large, if the experimental parameters are adequate. However, were we to take this Hamiltonian as the starting point, and $N\to\infty$, we would have divergences in the spectral function and other relevant quantities. Their origin is that the approximations that prominently feature in its derivation are incompatible with the ultraviolet limit, here represented by  $N\to\infty$.

Thus, in the literature, where frequently this Hamiltonian has indeed been taken as the phenomenological model to describe the system at hand, the introduction from outside of the model of an ultraviolet cutoff has been proposed almost systematically. From our point of view this is unnecessary, since this phenomenological Hamiltonian is only a good approximation to the lower modes, and there is a natural length that provides us with the cutoff, namely $\alpha$.

\section{Networks with canonical impedances}
The quantization techniques described above are useful to obtain Hamiltonian descriptions of circuit networks with transmission lines, starting from  first principles. More general passive environments, e.g. 3D superconducting cavities, have been used to design high-coherent qubits \cite{Paik:2011}. Lumped-element descriptions of the response function of such environments have been used within the {\it black-box} paradigm to derive Hamiltonians \cite{Nigg:2012,Solgun:2014,Solgun:2015}. This technique relies on a lumped-element description with numerable modes, such that its impedance response $Z(s)$ agrees with that of an electromagnetic environment either simulated with a computer solving Maxwell's equations or directly measured in an experiment.

The separation of system and environment degrees of freedom was not possible in \cite{Nigg:2012}, because the lumped-element circuit   expansion of the impedance was approximated with the first Foster form, and the linear part of the system was incorporated into the impedance. Thus, the Josephson-junction phase-drop degree of freedom had to be written in terms of all the harmonic variables, resulting in mode-mode couplings to all orders, see Eq. (6) in \cite{Nigg:2012}. On the other hand, other lumped-element descriptions, such as the second Foster expansion \cite{Devoret:1997} and the Brune expansion \cite{Solgun:2014,Solgun:2015}, presented in an in-built way separation of the environment degrees of freedom and the ones of the network it is attached to through its ports. As mentioned in \cite{Nigg:2012} and \cite{Solgun:2014,Solgun:2015}, such descriptions have intrinsic convergent properties.

For historical reasons, we first review the derivation of Paladino et al. \cite{Paladino:2003} where a flux variable is capacitively coupled to a one port general lossless passive and reciprocal impedance $Z(s)$ expanded in an infinite series of harmonic oscillators, i.e. the first Foster form. Recall that such description with a stage of a lone capacitor without inductor would correspond to a total impedance $Z_T(s)=\frac{1}{s C_B} + Z(s)$, as  seen by the anharmonic variable, with a pole at $s=0$. The rest of the expansion must be an electromagnetic environment whose impedance response at frequency $s=0$ has a zero, i.e. $Z(0)=0$. The generalization to the coupling of the general impedance to a more complex network can be easily done using the results of the sections above. We recall that the addition of lossy environments (represented by immittances in the linear case) requires a continuum description of nondissipative systems (e.g. harmonic oscillators)~\cite{CaldeiraLeggett:1983,Devoret:1997}.

In the second section, we extend the multi-network case that we studied in previous section \ref{subsec:TL_LCcoup_TL}, where the infinite dimensional subsystem was a transmission line, to a general multi-mode infinite-dimensional lossless passive and reciprocal environment that couples linearly to finite-dimensional networks. We also show how this analysis can be applied for example to simplify the quantization of the $1^{\mathrm{st}}$-Foster circuit done by Paladino et al. Mathematical details and other particular circuit cases as the $2^{\mathrm{nd}}$-Foster expansion are left for an Appendix section. In this section, we have restricted ourselves to the analysis of infinite-dimensional environments capacitively coupled to networks to lighten up the proofs, as the combined case with inductive coupling is an easy extension of this problem.

\subsection{$1^{\mathrm{st}}$ Foster-form impedance quantization}
\label{sec:1st-foster-expansion}
Equivalently to the analysis in  section \ref{sec:netw-with-transm}, the main goal is to find a Hamiltonian where the infinite set of canonical variables are coupled to a finite set of them without mode-mode couplings  and without divergence issues. In the literature these two points are frequently related by referring to the $A^2$-term. In the present context, as will become clearer in the next section, divergences are physical inasmuch as they impact either on an infinite mass renormalization for the finite set of variables or in quantities determined by the spectral density, that codifies the effect of the environment on the reduced dynamics of the network. In both cases, the divergences can be traced back to the normalizability of the coupling vectors/matrices in the block diagonal decomposition, with respect to the proper inner product. Similarly to the case of network lines (see in particular \ref{sec:appr-intr-cutoff}), it might well be the case that the root of the divergence is that an approximation that is valid for a truncation to a finite number $N$ of impedance modes is not valid in the limit $N\to\infty$. An alternative problem arises when some transformations are carried out in the finite $N$ case, and intermediate computations become invalid in the infinite limit.  This has caused difficulties in the literature that have led several authors to convoluted arguments to be able to discard such divergences.

Here we consider an example that can, with a special choice of parameters $C_n$ and $L_n$, also be used to describe a transmission line. This will prove convenient to relate  both approaches.

We first follow, with a number of simplifications and generalizations, an analysis proposed by Paladino et al. \cite{Paladino:2003}, and we prove that it corresponds to a particular canonical transformation to diagonalize impedance modes. The focus is to allow us some freedom in rescaling and reorganizing the impedance modes, with the criterion that the final Hamiltonian presents no mode-mode coupling. This freedom is the analogue to the freedom in $\alpha$ and $\beta$ parameters in the transmission line case.

\begin{figure*}[ht]
	\centering{\includegraphics[width=0.55\textwidth]{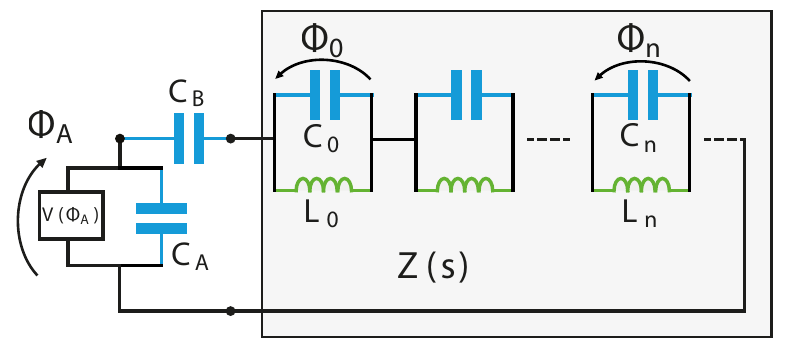}}
	\caption{\label{fig:1st_Foster_form}$1^{\mathrm{st}}$-Foster form capacitively coupled to an anharmonic flux variable. One port impedance $Z(s)$ modelled by a series of LC oscillators ($C_n,L_n$) is capacitively coupled to an anharmonic variable $\phi_A$ with capacitance $C_A$ and potential $V(\Phi_A)$ through capacitor $C_B$.}
\end{figure*}

Let us consider a family of circuits described by  Fig. \ref{fig:1st_Foster_form}. The corresponding Lagrangian     can be written choosing as variables the branch flux differences at the capacitors $C_n$ and $C_A$,
\begin{align}
L&= \frac{C_A}{2} \dot{\Phi}_A^2+\frac{C_B}{2} \left(\dot{\Phi}_A-\sum_{n}\dot{\Phi}_{n}\right)^2+\sum_{n}\left[\frac{C_{n}}{2}\dot{\Phi}_{n}^2-\frac{1}{2L_{n}}\Phi_{n}^2\right]-V(\Phi_A)\nonumber\\
&=\frac{1}{2}\dot{\bPhi}^T \msC\dot{\bPhi}-\frac{1}{2}\bPhi^T \msL^{-1}\bPhi - V(\Phi_A),\label{eq:Lag_Paladino_circuit}
\end{align}
where $\bPhi=(\Phi_A,\bPhi_{n}^T)^T=(\Phi_A, \Phi_0, \Phi_1, ...)^T$. The capacitance matrix reads
\begin{equation}
\msC=\begin{pmatrix}
C_{\Sigma} & -C_B\be_n^T\\
-C_B\be_n & \msC_{n}+C_B \be_n\be_n^T
\end{pmatrix},\label{eq:Paladino_Cmat}
\end{equation}
where we define $C_{\Sigma}=C_A+C_B$, the variable $\Phi_A$ couples equally  to all $\Phi_n$, and thus to $\bPhi_{n}$ through the vector of ones $\be_n=(1, 1, 1, ...)^T$, and the inductance matrix is 
\begin{equation}
\msL^{-1}=\begin{pmatrix}
0 & 0\\
0 & \msL_n^{-1}
\end{pmatrix}.\nonumber
\end{equation}

This coupling vector has an $l^2$ norm $\left|\be_n\right|^2=N$, with $N$ the number of stages, that diverges as $N$ tends to infinity. In this limit, then, one must be careful in assigning meaning while inverting the capacitance matrix when $N\rightarrow\infty$. In particular, and as signalled in Appendix \ref{App:1st_Foster}, we will have a pathological case if, in the limit $N\to\infty$, we have that $\be_n^T\msC_n^{-1}\be_n\to\infty$.  In what follows we are assuming $C_A\neq0$, such that the total capacitance matrix is full rank even in that limit, something that can be checked using the same arguments as in Sec.  \ref{subsec:invertibility_variable_counting} and taking the limit, see again  Appendix \ref{App:1st_Foster}.

Our objective is to carry out changes of variables that allow us to write a Hamiltonian with no mode-mode coupling that still mantains a coupling between the external variable and the impedance variables. In order to do so, let us start by dressing the external coordinate with mode coordinates, allowing ourselves some freedom in the amount of dressing. That is, we introduce a new coordinate $\eta$ as a linear combination of the old coordinates  $\Phi_A= \eta + t \be_n^T\bPhi_{n}$ with a free parameter $t$, and write the modified Lagrangian
\begin{equation}
L_I=\frac{1}{2}\dot{\bx}^T \msC_{x}\dot{\bx}-\frac{1}{2}\bx^T \msL^{-1}\bx - V(\eta + t  \be_n^T\bPhi_{n})\label{eq:Lag_Paladino_circuit_I}
\end{equation}
where $\bx=(\eta,\bx_{n}^T)^T$, with $\bx_{n}=\bPhi_n$ and
\begin{equation}
\msC_{x}=\begin{pmatrix}
a & b\be_n^T\\
b\be_n & \msM_{n}
\end{pmatrix}.\label{eq:Paladino_diag_C_I}
\end{equation}
The block capacitance matrix $\msM_{n}=\msC_n+d \be_n\be_n^T$, and we introduce parameters  $a=C_{\Sigma}$, $b=(tC_{\Sigma}-C_B)$ and $d = C_B - 2 C_B t + C_{\Sigma} t^2$.

In the second step, we rescale the coordinates to diagonalize the capacitance matrix $\msM_{n}$ as $\bx_{n}=M_0^{1/2} \msM_{n}^{-1/2}\by_{n}$, where $\by=(\eta, \by_{n}^T)^T$ and $M_0$ is a constant with dimensions of capacitance. That is, 
\begin{align}
\msC_{y}=\begin{pmatrix}
a & b\bff_n^T\\
b\bff_n & M_0\mone
\end{pmatrix},\quad \mathrm{and}\label{eq:Paladino_diag_C_II} \quad 
\msL_{y}^{-1}=\begin{pmatrix}
0 & 0\\
0 & (\msL_{n}^{y})^{-1}
\end{pmatrix}.
\end{align}
The inductance block submatrix reads $(\msL_{n}^{y})^{-1}=M_0 \msM_{n}^{-1/2}\msL_{n}^{-1}\msM_{n}^{-1/2}$, while (and this is the crucial point) the new coupling matrix is given by $\bff_n=M_0^{1/2}\msM_{n}^{-1/2}\be_n$. The Lagrangian in this second step is therefore
\begin{equation}
L_{II}=\frac{1}{2}\dot{\by}^T \msC_{y}\dot{\by}-\frac{1}{2}\by^T \msL_y^{-1}\by - V(\eta + t  \be_n^T\bPhi_{n}),\nonumber 
\end{equation}
where we have kept the old variables in the anharmonic potential for simplicity. It can be easily checked that the new coupling vectors $\bff_n$ have finite norm in the limit of infinite oscillators even if  $\lim\limits_{N\rightarrow\infty}\be_n^T \msC_n^{-1} \be_n=\infty$, see Appendix \ref{App:1st_Foster} for the complete proof. In this special but very common case (see for example  \cite{Gely:2017}), 
\begin{equation}
\lim\limits_{N \rightarrow\infty} M_0^{-1}\left|\bff_n\right|^2=\lim\limits_{N \rightarrow\infty} \be_n^T\msM_{n}^{-1}\be_n=1/d.\nonumber
\end{equation}
In the third step we undo the initial point transformation through $\eta = \Phi_A - t \be_n^T\bPhi_{n}$, in order to remove the interaction from the general potential $V(\Phi_A)$,
\begin{equation}
L_{III}=\frac{1}{2}\dot{\bz}^T \msC_{z}\dot{\bz}-\frac{1}{2}\bz^T \msL_{z}^{-1}\bz -V(\Phi_A)\nonumber
\end{equation}
with $\bz=(\Phi_A,\bz_n)$, $\msL_{z}^{-1}=\msL_{y}^{-1}$, and where the capacitance matrix has trasformed to
\begin{align}
\msC_{z}=&\begin{pmatrix}
C_{\Sigma} & (b+tC_{\Sigma})\bff_n^T\\
(b+tC_{\Sigma})\bff_n & M_0\mone+C_{\Sigma}t^2\bff_n\bff_n^T
\end{pmatrix}.\nonumber
\end{align}

Now that we have finite-norm coupling vectors, we can invert the capacitance matrix 
\begin{align}
\msC_{z}^{-1}=&\begin{pmatrix}
\bar{a} & \bar{b}\bff_n^T\\
\bar{b}\bff_n & M_0^{-1}\mone+\bar{d}\bff_n\bff_n^T
\end{pmatrix},\nonumber 
\end{align}
where we have defined the parameters
\begin{align}
\bar{a}=&\frac{M_0 + C_{\Sigma}t^2\left|\bff_n\right|^2}{D_z},\qquad
\bar{b} = \frac{-(b+C_{\Sigma}t)}{D_z},\nonumber\\
\bar{d} =&\frac{(b+C_{\Sigma}t)^2-C_{\Sigma}^{2}t^{2}} {M_0D_z},\qquad
D_z=M_0 C_{\Sigma}+((b+C_{\Sigma}t)^2-C_{\Sigma}^{2}t^{2})\left|\bff_n\right|^2,\nonumber
\end{align}
and derive the Hamiltonian
\begin{equation}
H=\frac{1}{2}\bp^T \msC_{z}^{-1}\bp+\frac{1}{2}\bz^T \msL_{z}^{-1}\bz+V(\Phi_A).\nonumber 
\end{equation}
We denote with $\bp=\partial L_{III}/\partial \dot{\bz}=(q_A,\bp_n^T)^T$ the charge variables conjugate to the $\bz$ fluxes. In order to simplify the Hamiltonian and remove the mode-mode coupling ($A^{2}$-like term) in the capacitance sector, we use one of the solutions to the equation $\bar{d}=0$, i.e. $t=C_B/C_{\Sigma}$ with the condition that $b=0$, and we obtain $d=C_s=C_A C_B/C_{\Sigma}$, which is the series capacitance seen by the impedance. Finally, we remove the mode-mode coupling in the inductance matrix with a canonical unitary transformation $\msU$, such that $(\bar{\msL}_n^{z})^{-1}=\msU(\msL_n^{z})^{-1}\msU^T$ be diagonal. The variables are then rotated through $\bar{\bp}_n=\msU \bp_n$ and $\bar{\bz}_n=\msU^T \bz_n$. All the expressions can be simplified in the limit of infinite oscillators, and for this specific case where $\lim\limits_{N\rightarrow \infty}\left|\bff_n\right|^2\rightarrow M_0/d$, the Hamiltonian reduces to
\begin{equation}
H=\frac{q_A^{2}}{2C_A}+V(\Phi_A)-\frac{C_B}{M_0 C_{\Sigma}}q_A\sum_{n}\bar{f}_{n}\bar{p}_{n}+\sum_{n}\frac{1}{2}\left[\frac{\bar{p}^2_{n}}{M_0}+M_0 \Omega_{n}^{2}\bar{z}_{n}^{2}\right].\nonumber 
\end{equation}
with the frequencies $\Omega_{n}\equiv (M_0 \bar{L}_{n}^{z})^{-1/2}$ and the rotated coupling vectors $\bar{\bff}_n=\msU\bff_n$, which preserve the same norm as the old ones $|\bar{\bff}_n|^2=\left|\bff_n\right|^2$. As previously commented, the canonical quantization procedure 
directly goes  through, by promoting the canonical variables to operators with canonical commutation relations  $[\Phi_A,q_A]=i\hbar$ and $[\bar{z}_n,\bar{p}_m]=i\hbar \delta_{nm}$.

This procedure has followed the structure of that presented in \cite{Paladino:2003}. We have been explicit about each step, in order to dispel some misconceptions that have arisen in the literature. We shall see in the next section, by extending it to the multiport case, that it can be replaced by the introduction of a single canonical transformation that only pertains to the impedance modes. The transformation, parametrized by an operator $\msM_n$, incorporates the freedom we gave us via the $t$ parameter, which disappears, and will be determined by the requirement of no mode-mode coupling.

\subsection{Multiport impedance quantization}
\label{sec:multi-port-impedance}
\begin{figure*}[]
	\centering{\includegraphics[width=0.5\textwidth]{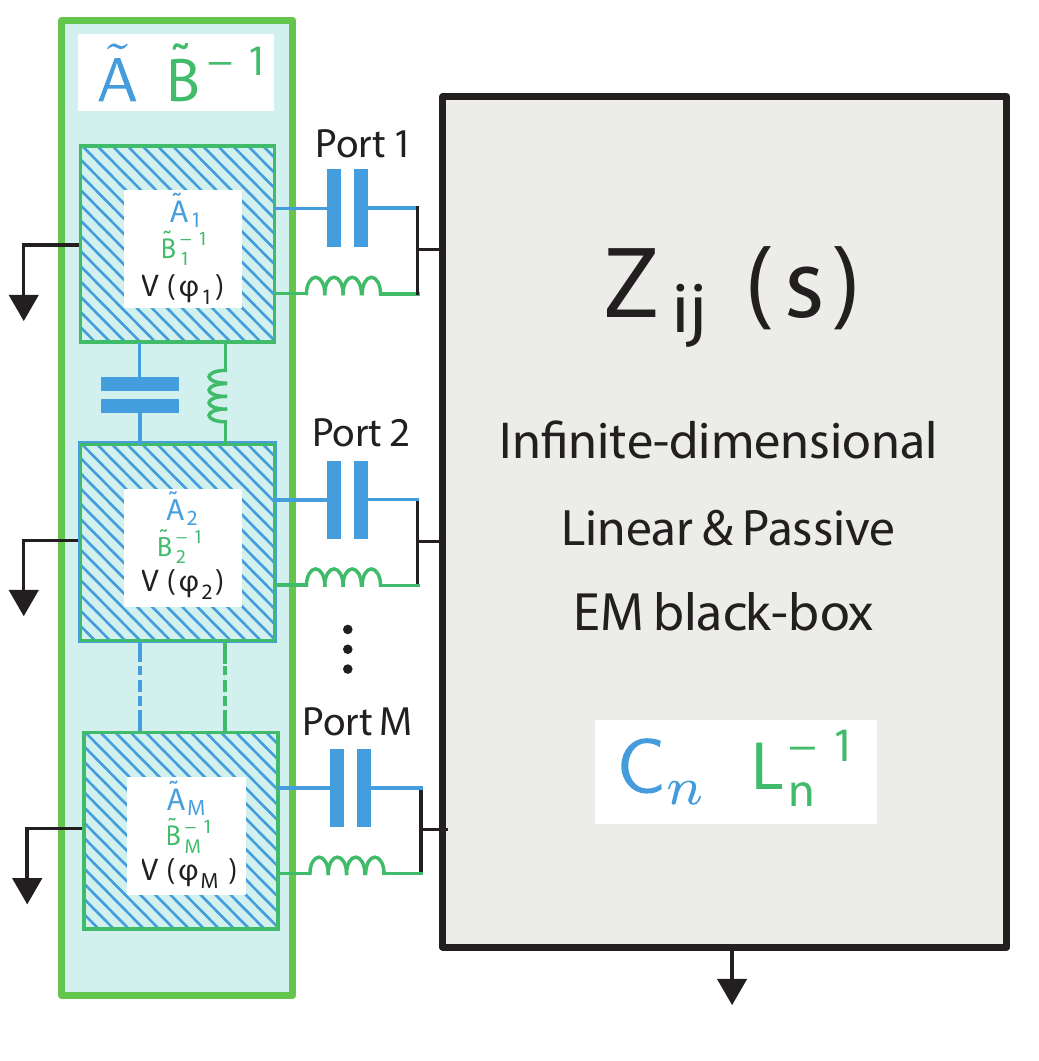}}
	\caption{\label{fig:MPort_Z_LCcoup_Networks}Infinite-dimensional multiport impedance connected to finite sized anharmonic network. A linear and passive, electromagnetic (EM) environment is modeled as a multiport impedance $Z_{ij}(s)$, fitted to a general lumped-element circuit with capacitance and inductance matrices ($\msC_n, \msL^{-1}_n$). The general infinite dimensional system is linearly coupled to anharmonic networks ($\msA_i,\msB_i^{-1},V(\bphi_i)$) with finite degrees of freedom that are also directly connected.}
\end{figure*}

We have already encountered a case of a multiport impedance in section \ref{sec:mult-netw-coupl}. Since that was a case involving a transmission line, the methods presented above were better suited for the analysis. Nonetheless, it can also be analyzed from the perspective of this section; see figure \ref{fig:MPort_Z_LCcoup_Networks} for a general multiport circuit linearly coupled to M
non-linear networks. We concentrate, as always, on capacitive coupling.

Let us consider thus a capacitance matrix of the form
\begin{equation}
\label{eq:multiportcapacitance}
\msC=
\begin{pmatrix}
\msA& -\sum_i\ba_i\bu^T_i \\ -\sum_i \bu_i\ba_i^T &  \msC_n+\sum_i\bu_i\bu_i^T
\end{pmatrix},
\end{equation}
where we assume that $\msA$ and $\msC_n$ are symmetric and positive, and that the vectors $\left\{\ba_i\right\}_{i=1}^M$ on one side and $\left\{\bu_i\right\}_{i=1}^M$ on another side are separately linearly independent. 
The notation used here is reminiscent but not completely equivalent to that in \ref{sec:mult-netw-coupl}. Namely, what were presented there as $\ba_i$ give rise to vectors here with the same notation, after padding with zeroes. Furthermore, we have chosen a different normalisation in order to unclutter formulae.

The matrix $\msC$ is a  block diagonal matrix perturbed by off-diagonal blocks each of rank $M$. This is the correct description for an $M$
port circuit, and is amenable to the inversion given in Appendix \ref{sec:multiport-impedanceapp}. The general formula presented there is however not very illuminating, and in the particular case of $\msC$ we can present the inverse in a much cleaner way, as follows:
\begin{equation}
\label{eq:inverse-multiport-c}
\msC^{-1}=
\begin{pmatrix}
\msA^{-1}+\msXi_{ij}\msA^{-1}\ba_i\ba_j^T\msA^{-1} & \msTheta_{ij} \msA^{-1}\ba_i\bu_j^T\msC_n^{-1}\\
\msGamma_{ij} \msC_n^{-1}\bu_i\ba_j^T \msA^{-1}& \msC_n^{-1}+\msLambda_{ij}\msC_n^{-1}\bu_i\bu_j^T\msC_n^{-1}
\end{pmatrix},
\end{equation}
where Einstein summation convention has been used, as it will henceforward. On demanding that this matrix indeed be the inverse we obtain four linear equations for the matrices $\msXi$, $\msTheta$, $\msGamma$ and $\msLambda$. The solution to this system of matrix equations is
\begin{align}
\label{eq:inverse-cap-coeffmatrices}
\begin{split}
\msXi=&\mu\cdot\left(\mone+\mu-\nu\cdot\mu\right)^{-1},\qquad\qquad
\msGamma=  \left(\mone+\mu-\nu\cdot\mu\right)^{-1},\\
\msLambda=& \left(\mone+\mu-\nu\cdot\mu\right)^{-1}\cdot \left(\nu-\mone\right),\quad\,
\msTheta= \mone+\mu\cdot\left(\mone+\mu-\nu\cdot\mu\right)^{-1}\cdot\left(\nu-\mone\right).
\end{split}
\end{align}
Here, the matrices $\mu$ and $\nu$ are given by
\begin{align}
\mu_{ij}=\bu_i^T\msC_n^{-1}\bu_j\quad \mathrm{and} \quad 
\nu_{ij}=\ba_i^T\msA^{-1}\ba_j.\nonumber
\end{align}
Assume that the multiport impedance presents infinite modes. A possible approach is to cutoff the number of modes in the impedance to a finite number $N$. 
Now, the issue, as pointed above for the single port case, is that the matrix $\mu$ can blow up in an $N\to\infty$ limit. That, by itself, might not be so pernicious. However, in such a situation $\msLambda$ would tend to zero, and the coupling matrix norm, defined as $\msTheta_{ij}\msGamma_{ki}\bu_j^T\msC_n^{-2}\bu_k$, could also tend to zero.

The final coupling matrix will be obtained after a canonical transformation that diagonalises the submatrix $\msD_n^{-1}=\msC_n^{-1}+\msLambda_{ij}\msC_n^{-1}\bu_i\bu_j^T\msC_n^{-1}$, and simultaneously the corresponding inductance submatrix. This can be achieved by rescaling the momenta $\bq_n$ with the square root of this matrix.  If indeed $\msD_n^{-1}$ is positive, the coupling matrix is finite.

We can follow here the steps of the analysis for the first Foster form, in which a free parameter is introduced by first dressing and then undressing the network variables with impedance variables, and in between rescaling and reordering the impedance variables. Assume that the initial network variables are collected in a vector $\bPhi_A$, while the impedance variables are $\bPhi_n$.  We shift network variables with the change of variables
\begin{equation}
\begin{pmatrix}
\bx_A\\  \bx_n
\end{pmatrix}=
\begin{pmatrix}
\mone&-\bb_i\bu_i^T\\0&\mone
\end{pmatrix}\begin{pmatrix}
\bPhi_A\\ \bPhi_n
\end{pmatrix},\nonumber
\end{equation}
where $\left\{\bb_i\right\}$ is a set of vectors to be determined later, that take the role of the $t$ parameter for the first Foster form. The capacitance matrix for the new variables reads
\begin{equation}
\msC_x=
\begin{pmatrix}
\msA& \left(\msA\bb_i-\ba_i\right)\bu_j^T\\
\bu\left(\bb_i^T\msA-\ba_i^T\right)& \msM_n
\end{pmatrix},\nonumber
\end{equation}
where
\begin{equation}
\msM_n= \msC_n+\left(\delta_{ij}-2\bb_i^T\ba_j+\bb_i^T\msA\bb_j\right)\bu_i\bu_j^T.\nonumber
\end{equation}
We change variables again, in the form
\begin{equation}
\begin{pmatrix}
\by_A\\  \by_n
\end{pmatrix}=
\begin{pmatrix}
\mone&0\\0& \msM_n^{1/2}  \end{pmatrix}
\begin{pmatrix}
\boldsymbol{\varphi}_A\\ \boldsymbol{\varphi}_n
\end{pmatrix},\nonumber
\end{equation}
leading to the capacitance matrix
\begin{equation}
\nonumber 
\msC_y=
\begin{pmatrix}
\msA& \left(\msA\bb_i-\ba_i\right)\bu_j^T\msM_n^{-1/2}\\
\msM_n^{-1/2}\bu\left(\bb_i^T\msA-\ba_i^T\right)& \mone
\end{pmatrix}.
\end{equation}
We now undo the shift of the network variables, by
\begin{equation}
\begin{pmatrix}
\bz_A\\  \bz_n
\end{pmatrix}=
\begin{pmatrix}
\mone&\bb_i\bu_i^T\msM_n^{-1/2}\\ 0&\mone
\end{pmatrix}
\begin{pmatrix}
\boldsymbol{\sigma}_A\\ \boldsymbol{\sigma}_n
\end{pmatrix},\nonumber
\end{equation}
leaving a final capacitance matrix
\begin{equation}
\msC_z=
\begin{pmatrix}
\msA& -\ba_i\bu_i^T\msM_n^{-1/2}\\ \msM_n^{-1/2}\bu_i\ba_i^T&\mone-\left(\bb_j^T\msA\bb_k-\ba_j^T\bb_k-\bb_j^T\ba_k\right) \msM_n^{-1/2}\bu_j\bu_k^T\msM_n^{-1/2}
\end{pmatrix}.\nonumber
\end{equation}
It now behoves us to invert this final capacitance matrix and demand that the inverse capacitance matrix presents no coupling between impedance modes. By construction this then entails that the corresponding submatrix is the identity matrix, and the possible coupling between impedance modes due to  inductance can be eliminated by diagonalising the corresponding inductance matrix. The condition of no coupling is seen to be achieved with the choice $\bb_j=\msA^{-1}\ba_j$. This provides us with the  matrix $\msM_n=\msC_n+\left(\delta_{ij}-\ba_i^T\msA^{-1}\ba_j\right)\bu_i\bu_j^T$. In order for the procedure to work, we require that this matrix be positive. Furthermore, the coupling matrix $\bu_i^T\msM_n^{-1}\bu_j$ should have finite components.

This presentation actually suggests a different approach. The complete succession of changes of variables is a point transformation that can be compacted to
\begin{equation}
\begin{pmatrix}
\bz_A\\ \bz_n
\end{pmatrix}=\begin{pmatrix}
\mone&0\\0& \msM_n^{1/2}  \end{pmatrix}
\begin{pmatrix}
\bPhi_A\\ \bPhi_n
\end{pmatrix}.\nonumber
\end{equation}
Thus, now consider this change of variables, with a positive operator $\msM_n$ to be determined. The corresponding capacitance matrix, starting from (\ref{eq:multiportcapacitance}), reads now
\begin{equation}
\msC_z=\begin{pmatrix}
\mone&0\\0& \msM_n^{-1/2}  \end{pmatrix} \msC\begin{pmatrix}
\mone&0\\0& \msM_n^{-1/2}  \end{pmatrix}=
\begin{pmatrix}
\msA& -\ba_i\bu_i^T\msM_n^{-1/2}\\-\msM_n^{-1/2}\bu_i\ba_i&\msM_n^{-1/2}\left(\msC_n+\bu_i\bu_i^T\right)\msM_n^{-1/2}
\end{pmatrix}.\nonumber
\end{equation}
In order for there to be no coupling amongst impedance modes in the Hamiltonian, we require that, on inverting this matrix, the corresponding submatrix be proportional to the identity operator.  Structurally, the matrix $\msC_z$ is similar to the capacitance matrix presented in (\ref{eq:multiportcapacitance}), and the inverse can be computed in a similar fashion. In fact, the condition we require for $\msM_n$ is tantamount to
\begin{equation}
\msM_n^{1/2}\left(\msC_n^{-1}+\msLambda_{ij}\msC_n^{-1}\bu_i\bu_j^T\msC_n^{-1}\right)\msM_n^{1/2}=\mone,\nonumber
\end{equation}
where the coefficients $\msLambda_{ij}$ are precisely those found earlier in (\ref{eq:inverse-cap-coeffmatrices}).
That is, $\msM_n=\msD_n^{-1}=\left(\msC_n^{-1}+\msLambda_{ij}\msC_n^{-1}\bu_i\bu_j^T\msC_n^{-1}\right)^{-1}$.
We can now see that the long process of Paladino et al.~\cite{Paladino:2003} is actually nothing else than the standard canonical analysis.

\section{Applications}
All the formal manipulations above and in the appendices below should not obscure the final objective: to provide model building tools for real devices, in which new phenomena can be uncovered, and which pave the road to more powerful quantum simulators and computers. As stated in the introduction, one of the crucial aspects of the study of multimode system quantization is to achieve faster switching times in qubits, by increasing coupling. This is usually studied in the context of spin-boson Hamiltonians, to which many of the models above can be connected. We first study generic statements about spin-boson models and convergence in transmission lines connected to qubits. Then we compute explicitly three models that have connections to existing experimental devices or proposals thereof. 
\subsection{Spin-boson models}
\label{sec:generic-behaviour}

In all the preceding results for transmission lines (TL), section \ref{sec:netw-with-transm}, the Hamiltonians have TL-Network interaction terms of four types, namely
\begin{equation}
\frac{C_{c_i}}{N_\alpha}\left(\bq_i^T \msA^{-1}_i\ba_i\right)\sum_n u_n(x_i) Q_n\,,\qquad - \frac{\bphi_i^T\bb_i}{L_{c_i}}\sum_n  u_n(x_i) \Phi_n,\nonumber
\end{equation}
or the same two, but substituting $u_n(x_i)$ with $\Delta u_n(x_i)$.

Let us now concentrate on inductive couplings. Were we to integrate out the transmission lines, their effect on the evolution of the network variables is best codified in the spectral functions
\begin{equation}
\label{eq:spectrals}
J_i^L(\omega)=\frac{\pi}{2L_c} \sum_n \frac{\left[u_n(x_i)\right]^2}{N_\alpha \omega_n} \delta(\omega-\omega_n),
\end{equation}
where $L$ stands for inductive, and correspondingly for the $\Delta u_n(x_i)$ case. Here the subindex $i$ corresponds to the relevant boundaries of the transmission lines (including possible insertion points). Compare with the last term of (\ref{eq:Ham_TL_LCcoup_2Networks3}).

The asymptotic behaviour of $u_n(x_i)$ is a consequence of the structure of the corresponding operators. Even though the underlying operators are not of the Sturm--Liouville type, see \ref{Walter_appendix}, and therefore the Sturmian theorems are not applicable, one can extract the asymptotic behaviour of $k_n$ from the spectral equations, and from here, by substitution, the asymptotic behaviour in $n$ of $u_n(x_i)$. For the cases we analyse, the eigenfunctions must have the form $\sin\left[k_n(x-x_0)\right]$, and the secular equations are generically of the form
\begin{equation}
\label{eq:genericsecular}
\tan(\xi)= \frac{ a \xi}{\xi^2-b},
\end{equation}
where $\xi$ is $k$ times some length $x_i-x_0$. The asymptotic solution of this equation (\ref{eq:genericsecular}) is
\begin{equation}
\xi_n\sim n\pi + \frac{a}{n\pi}+O\left(n^{-3}\right).\nonumber
\end{equation}
We have $u_n(x_i)\sim\sin(\xi_n)$, and thus $u_n(x_i)\sim (-1)^n a/n\pi $. Since $\omega_n=v k_n$ for some propagation speed $v$, the large $n$ behaviour of the summands is $1/n^3\sim 1/\omega^3$. We see that indeed the spectral density falls with a negative power of the frequency, and that this is the generic behaviour for all systems of transmission lines linearly coupled to lumped element networks. This model has an intrinsic ultraviolet cutoff, and there is no need for further regularisation nor renormalization.

Passing now to capacitive couplings, the analysis of Appendix \ref{sec:capac-induct-coupl} suggests that the effect of the transmission line on the evolution of the network would be encoded in a quantity proportional to $\sum_n\left[u(x_i)\right]^2\omega_n^3\delta\left(\omega-\omega_n\right)$, which, according to our analysis, would present a linear  divergence in that the large $n$ behaviour of the summands is $\sim n$, thus implying that $J(\omega)\sim\omega$ for large $\omega$.

However, let us contextualise these models. In the cases of interest to us, the network will include combinations of Josephson junctions in order to have regimes in which to operate qubits. That entails reducing the operator $\bq_i\cdot \msA^{-1}_i\ba_i$, essentially, and effectively, to a Pauli matrix multiplied by some constant. In so doing there is inevitably an energy scale, and thus a frequency, involved in the reduction process, that pertains to the network side. Comparing to the classical analysis of Appendix \ref{sec:capac-induct-coupl}, this introduces an asymmetry that curtails our formal manipulation of adding a time derivative to the Lagrangian: we do not move the derivative in $\bq_i$ to the $Q_n$s, and thus we not add a $\omega_n$ factor. Summarising, the coupling after the reduction to a qubit will be of the form
\begin{equation}
\sigma^{(i)}_x\sum_n u_n(x_i)Q_n,\nonumber
\end{equation}
and the relevant spectral function will be (up to a global constant)
\begin{equation}
J_i^{SC}(\omega)=A\sum_n \left[u_n(x_i)\right]^2\omega_n \delta\left(\omega-\omega_n\right).\nonumber
\end{equation}
Here the superindex $SC$ stands for ``spin-capacitive'' coupling. Following the analysis above for large $n$, we see that $J_i^{SC}(\omega)$ tends to zero as $1/\omega$ for large $\omega$. That is to say, in this kind of model with capacitive coupling there is a natural Drude cutoff structure in the qubit regime.

Notice that for the inductive coupling case, when the flux field can be substituted by a spin variable, this  argument is not relevant. The relevant spectral function will indeed be (\ref{eq:spectrals}), with decay $\sim\omega{^{-3}}$.

We now construct the spin-boson model in two cases. First the charge qubit coupled to a semi-infinite  transmission line, in (\ref{sec:an-example:-half}). Then the flux qubit galvanically coupled to an infinite transmission line, in \ref{sec:an-example-galvanic}.
\subsection{Charge qubit coupled to semi-infinite TL}
\label{sec:an-example:-half}
Let us consider the Lagrangian of an extension of the circuit shown in Fig. \ref{fig:TL_Ccoup_CQubit}, where the length of the line is extended $L\to\infty$, and an additional inductive coupling ($L_c$) has been introduced,
\begin{align}
L=&\frac{C_J}{2}\dot{\Phi}_J^2+U(\Phi_J)+\frac{C_c}{2}\left(\dot{\Phi}_J-\partial_t\phi(0,t)\right)^2\nonumber\\ &+\int_{\mR_+}\mathrm{d}x\,\left[\frac{c}{2}\left(\partial_t\phi\right)^2-\frac{1}{2l}\left(\partial_x\phi\right)^2\right]-\frac{1}{2L_c}\left(\Phi_J-\phi(0,t)\right)^2.\nonumber
\end{align}

Applying our techniques, and after the dust has settled, we have the Hamiltonian
\begin{align}
H=&\frac{Q_J^2}{2C_J}+\frac{1}{2L_c}\Phi_J^2+U(\Phi_J)+\int_{\mR_+}\mathrm{d}k\,\left[\frac{1}{2c}q_k^2+\frac{k^2}{2l}\phi_k^2\right]\nonumber\\ &+\frac{1}{c}\frac{C_c}{C_c+C_J}Q_J\int_{\mR_+}\mathrm{d}k\,u_k(0)q_k-\frac{1}{L_c}\Phi_J\int_{\mR_+}\mathrm{d}k\,u_k(0)\phi_k,\label{eq:examphalfhamilt}
\end{align}
where we have used the normalisation
\begin{equation}
\int_{\mR_+}\mathrm{d}x\,u_k(x)u_q(x)+ \alpha u_k(0)u_q(0)=\delta(k-q),\nonumber
\end{equation}
such that
\begin{equation}
u_k(x)= \frac{\left(k+ i \alpha k^2 - i/\beta\right) e^{ikx} + \left(k- i \alpha k^2 + i/\beta\right) e^{-ikx}}{\sqrt{2\pi\left[k^2+\left(\alpha k^2- 1/\beta\right)^2\right]}},\nonumber
\end{equation}
with the useful choices $\alpha= C_cC_J/c(C_c+C_J)$ and $\beta=L_c/l$, see \ref{subsec:infin-length-transm}. Notice that
\begin{equation}
u_k(0)=\sqrt{\frac{2k^2}{\pi(k^2+\left(\alpha k^2-1/\beta\right)^2)}}.\label{eq:uk0_LC_TL_half}
\end{equation}

Observe that, for small $k\rightarrow0$, $u_k(0)\sim\sqrt{2/\pi}\,\beta k\to0$, if $\beta$ is finite. On the other hand, if there is no inductive coupling of the transmission line to the network, $L_c\to\infty$, then $1/\beta=0$ and $u_k(0)\sim\sqrt{\pi/2}$ for $k\to0$. Looking now at the large $k$ behaviour, notice that $u_k(0)\sim\sqrt{2/\pi}/\alpha k$ if there is capacitive coupling ($C_c\neq0$). If, on the other hand, $C_c=0$, then $u_k(0)$ tends to a constant for large $k$.

In some parameter regimes of the network Hamiltonian
\begin{equation}
H_N=\frac{1}{2C_J}Q_J^2+\frac{1}{2L_c}\Phi_J^2+U(\Phi_J),\nonumber
\end{equation}
it is possible to limit the analysis to a finite dimensional subspace of energy eigenstates of the network. For definiteness assume that a two-dimensional energy eigenspace is enough to describe the most relevant phenomenology of the system, and denote an orthonormal basis of this subspace as $\left\{|+\rangle,|-\rangle\right\}$. Assume furthermore that the expectation values of $Q_J$ and $\Phi_J$ in those basis states are zero (to avoid dealing with an operator valued offset in the effective Hamiltonian). Then the effective quantum Hamiltonian is
\begin{align}
H_{\mathrm{eff}}=&\hbar\Omega \sigma_z+\int_{\mR_+}\mathrm{d}k\,\left[\frac{1}{2c}q_k^2+\frac{k^2}{2l}\phi_k^2\right]\nonumber\\
& +\frac{\alpha {\cal Q}}{C_J} \,\bn_Q\cdot\boldsymbol{\sigma}\int_{\mR_+}\mathrm{d}k\,u_k(0)q_k-\frac{{\cal F}}{\beta l}\,\bn_\Phi\cdot\boldsymbol{\sigma}\int_{\mR_+}\mathrm{d}k\,u_k(0)\phi_k,\nonumber
\end{align}
where ${\cal Q}=\left|\langle+|Q_J|-\rangle\right|$, ${\cal F}=\left|\langle+|\Phi_J|-\rangle\right|$, and $\bn_Q$ and $\bn_\Phi$ are unitary vectors in the $x-y$ plane. Write
\begin{equation}
q_k=\sqrt{\frac{\hbar\omega_k c}{2}}\left(a_k+a_k^{\dag}\right)\quad\mathrm{and}\quad \phi_k=i\sqrt{\frac{\hbar}{2\omega_k c}}\left(a_k-a_k^{\dag}\right),\nonumber
\end{equation}
where $\omega_k=v_p k=k/\sqrt{lc}$.
The relevant spectral functions are therefore
\begin{equation}
J^C(\omega)\propto \frac{\omega^3}{\left(\omega^2-\alpha\omega_\alpha^2/\beta\right)^2+\omega^2\omega_\alpha^2},\nonumber
\end{equation}
with $\omega_\alpha=v_p/\alpha$, and
\begin{equation}
J^L(\omega)
\propto  \frac{\omega\omega_{\alpha}^2}{\left(\omega^2-\alpha\omega_\alpha^2/\beta\right)^2+\omega^2\omega_\alpha^2}.\nonumber
\end{equation}
Observe that both in the presence and in the absence of inductive coupling the leading behaviour at small frequencies will be ohmic ($J(\omega)=J^C(\omega)+J^L(\omega)\sim\omega$ for small $\omega$). On the other hand, at large frequencies the leading behaviour is determined by the capacitive coupling, with $J^Q(\omega)\sim\omega^{-1}$. Finally, be aware that we have Lorentz-Drude type of spectral densities for both limits of pure capacitive ($\beta\rightarrow\infty$) and pure inductive ($\alpha\rightarrow0$) coupling~\cite{Bamba:2014}.

\subsubsection*{Quantum fluctuations of the flux and charge fields}
\begin{figure*}[h]
	\centering{\includegraphics[width=0.9\textwidth]{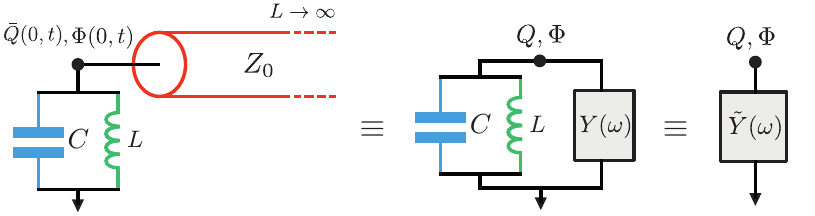}}
	\caption{\label{fig:Q_fluctuations} Semi-infinite transmission line ended with a capacitance and an inductor to ground, and equivalent models where the line is replaced by an admittance in the spirit of Devoret's Quantum fluctuations of electrical circuits~\cite{Devoret:1997}.}
\end{figure*}

Let us make use of this example to prove the connection between the methods described in this chapter and the seminal work of Devoret~\cite{Devoret:1997}. For that, we consider the circuit in Fig.~\ref{fig:Q_fluctuations}, where we have the semi-infinite transmission length ended with a capacitor and inductor (we have removed the charge qubit and renamed $C_c\rightarrow C$ and $L_c\rightarrow L$). As explained in Appendix Sec.~\ref{sec:hamilt-form} below, one can define a charge element of the complete Hilbert space
\begin{equation}
	Q(t)=\int dk \, q_k U_k=(Q(x,t),Q(0,t))
\end{equation}
where $U_k=(u_k(x), u_k(0))$ are the generalized eigenvectors of the differential operator we have used to decompose the flux field, and rewrite the above (reduced) Hamiltonian (\ref{eq:examphalfhamilt}) in terms of flux and charge fields (equivalent to $H=\int_{\mR_+} \hbar \omega_k a_k^\dag a_k$), see the general formula and further details of this subsection in the Appendix. 

It must be remarked that the true conjugate field to the flux is $\tilde{Q}(x,t)=(1+\alpha\delta(x))Q(x,t)$, i.e.  $\{\Phi(x,t),\tilde{Q}(x',t)\}=\delta(x-x')$ \cite{Malekakhlagh:2016}. However, at the end point, it can be shown that the conjugate variables are 
\begin{eqnarray}
	\{\Phi(0,t),\bar{Q}(0,t)\}=\alpha\int_{\mR_+}dk\int_{\mR_+}dk'\, u_k(0) u_{k'}(0)\{\phi_k,q_{k'}\}=1,\nonumber
\end{eqnarray}
where have defined $\bar{Q}(0,t)=\alpha Q(0,t)$ and $\Phi(0,t)=\int dk \, \phi_k u_k(0)$, and we have used the fact that  $\int_{\mR_+} dk |u_k(0)|^2=1/\alpha$.

Working in the quantized picture, we can compute averaged values for the conjugate fields at the end point given that the infinite (dressed) harmonic modes are in a thermal state, i.e. $\langle(a_k-a_k^\dag)(a_{k'}-a_{k'}^\dag)\rangle_{\rho_{\text{th}}}=-\coth(\hbar \omega_k/2 k_B T)\delta_{kk'}$ where $T$ is the temperature and $k_B$ Boltzmann's constant, 
\begin{align}
	\langle\hat{\Phi}^2(0,t=0)\rangle_{\rho_{\text{th}}}=&\frac{\hbar}{2c}\int_{\mR_+} dk\, \frac{\coth(\hbar \omega_k/2 k_B T)}{\omega_k} |u_k(0)|^2\nonumber\\
	=&\frac{\hbar Z_{LC}}{2\pi}\int_{\mR}d\omega \frac{\omega \omega_{LC}^2 Z_{LC} Z_0^{-1}}{\omega^2 Z_{LC}^2 \omega_{LC}^2 Z_0^{-2} + \left(\omega^2-\omega_{LC}^2\right)^2}\coth(\hbar \omega/2 k_B T),\nonumber
\end{align}
and 
\begin{align}
\langle\hat{\bar{Q}}^2(0,t=0)\rangle_{\rho_{\text{th}}}=\frac{\hbar}{2Z_{LC}\pi}\int_{\mR}d\omega \frac{\omega^3  Z_{LC} Z_0^{-1}}{\omega^2 Z_{LC}^2 \omega_{LC}^2 Z_0^{-2} + \left(\omega^2-\omega_{LC}^2\right)^2}\coth(\hbar \omega/2 k_B T),\nonumber
\end{align}
where $u_k(0)$ is Eq. (\ref{eq:uk0_LC_TL_half}) replacing $\alpha=C/c$ and $\beta=L/l$, $Z_{LC}=\sqrt{L/C}$, $\omega_{LC}=1/\sqrt{L C}$. Such formulas are equivalent to the corrected (3.33), and (3.34) in \cite{Devoret:1997}, with $Y(\omega)=Z_0^{-1}=\sqrt{c/l}$ following a different method\footnote{There must be a typo of $\omega_0^2$ in the numerator of (3.34) in \cite{Devoret:1997}, which is consistent with a simple dimensional analysis. There is a further $\pi$ constant mismatch in both denominators.}, readily, the interchange of the dressed admittance linear response $\tilde{Y}(\omega)$ by a continuum infinite set of harmonic oscillators. 

We recall the divergence of the charge quantum fluctuations at the end point given the infinite contribution zero-point fluctuations. It should be now clear that, on coupling the line to a network through the L and/or C lumped elements, one effectively filters the high and low frequencies, with the proper convergence of the infinite Lamb shift (or multi-qubit effective couplings).

\subsection{Flux qubit galvanically coupled to infinite TL}
\label{sec:an-example-galvanic}
\begin{figure*}[ht]
	\centering{\includegraphics[width=0.65\textwidth]{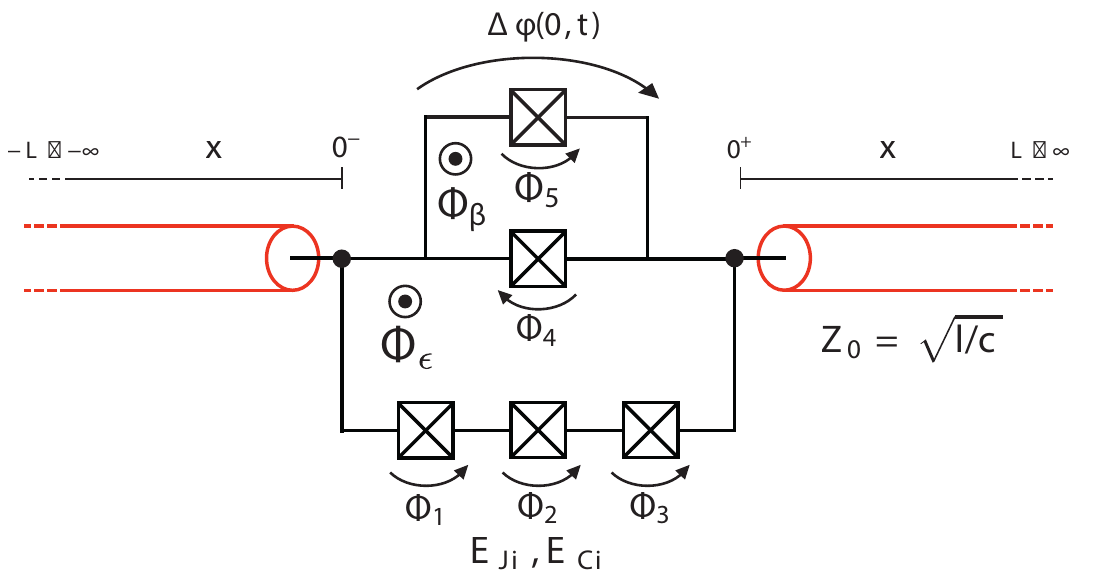}}
	\caption{\label{fig:Flux_qubit_GC} Circuit models for a capacitively-shunted flux qubit inserted in the transmission line.}
\end{figure*}

In this example, we follow the circuit layout of a flux qubit galvanically, and  tunably, coupled to a transmission line with a SQUID-loop shared between the two \cite{FornDiaz:2017}. It is represented in Fig. \ref{fig:Flux_qubit_GC}.
Clearly the variables depicted are redundant, since they fulfill  the fluxoid quantization condition on the separate loops:
\begin{equation}
\begin{split}
\varphi_{1} + \varphi_{2} + \varphi_{3} + \varphi_{4} + 2 \pi f_{\epsilon} &= 0 ,\\
\varphi_{4} + \varphi_{5} + 2 \pi f_{\beta} &= 0, \\
\varphi_{4} + \frac{\Delta \Phi \left( 0 , t \right)}{\varphi_{0}} &= 0,
\end{split}\nonumber
\end{equation}
where $f_{\epsilon}= \Phi_{\epsilon} / \Phi_q$, $f_{\beta}= \Phi_{\beta} / \Phi_q$ are the magnetic frustration in each loop. We shall use $\varphi_{1}$, $\varphi_{2}$, $\varphi_{4}$ as the independent degrees of freedom, where the fluxes are related to the phase variables through $\varphi_i=2 \pi \Phi_i/\Phi_q$. Starting from the Lagrangian that includes all these elements, with Josephson junction potentials,  linearising the terms with inductive couplings, and redefining variables in a suitable manner, we are led to an effective Lagrangian with the form of Eq. (\ref{eq:Lag_TL_LCgalvcoup_Networks_2}), in which we set $\bPhi_{\mathrm{ext}}$ to zero.

There is one crucial difference with respect to the analogous Lagrangian (\ref{eq:Lag_TL_LCgalvcoup_Networks_2}) from \ref{sec:line-galv-coupl}, namely that we take transmission lines of infinite length. We substitute the boundary conditions (\ref{eq:EVP_TL_LCgalvcoup_Networks_eq3}) by a normalizability condition. It proves convenient not to demand reality of the generalized eigenfunctions $u_k(x)$, identified by wavenumber $k$. Nonetheless, they can be selected to have $u_k^*(x)=u_{-k}(x)$. Since the flux on the lines is a real magnitude, the coefficients of its expansion
\begin{equation}
\Phi(x,t)=\int_{\mR} dk\,\Phi_k(t)u_k(x)\nonumber
\end{equation}
fulfill the relation $\Phi_k^*=\Phi_{-k}$. It is also relevant to notice that the conjugate momentum $Q_k$ is given by $c\dot{\Phi}_{-k}$ plus additional terms.

Following the same steps as in the previous section we derive the Hamiltonian
\begin{align}
H=& \frac{1}{2}\mathbf{q}^T(\msA^{-1}+ \frac{C_A^2}{\alpha c} \msA^{-1}\mathbf{a}\mathbf{a}^T \msA^{-1})\mathbf{q}+\frac{1}{2}\bphi^T\msB^{-1}\bphi+V(\bphi)\nonumber\\
&+\frac{C_A}{c} (\mathbf{q}^T\msA^{-1} \mathbf{a})\int_{\mR} dk \,~  Q_{-k}  \Delta u_k(0)  -\frac{1}{L_B} (\bphi^T\mathbf{b})\int_{\mR} dk \,~    \Phi_k  \Delta u_k(0)  \nonumber\\
&+\int_{\mR} dk \,~ \left[ \frac{|Q_k|^2}{2c}+\frac{k^2|\Phi_k|^2}{2l} \right].
\label{eq:Ham_TL_LCgalvcoup_Networks_fluxonium}
\end{align}
As expected, we have made the choices
\begin{equation}
\begin{split}
\alpha=&\frac{C_{cA}-C_A^2\mathbf{a}^T \msA^{-1}\mathbf{a}}{c}= \frac{ C}{c} \left( r_{4} + r_{5} + \frac{r_{1} r_{2} r_{3}}{ r_{1} r_{2} + r_{2} r_{3} + r_{3} r_{1} } \right),\nonumber\\
\beta=&L_{cB}/l = \frac{\varphi_{0}^{2}}{l E_{J} \left[ r_{4} + r_{5} \cos{\left( 2 \pi f_{\beta} \right)} \right]},\nonumber
\end{split}
\end{equation}
in order to eliminate the mode-mode coupling terms. Here we have introduced parameters that pertain to the experiment of reference, for later comparison.
Notice that in Eq. (\ref{eq:Ham_TL_LCgalvcoup_Networks_fluxonium})  $Q_k$ is accompanied by $\Delta u_{-k}(0)$. Now, given the relation  $\Phi_k^*=\Phi_{-k}$ and the analogous  $Q_k^*=Q_{-k}$, we define creation and annhilation operators related to the charge and flux operators 
\begin{align}
 Q_k &\equiv \sqrt{\frac{\hbar \omega_k c}{2}}(\tilde{a}_{-k} + \tilde{a}_k^{\dagger}),\nonumber\\
 \Phi_k &\equiv i\sqrt{\frac{\hbar}{2 \omega_k c}}(\tilde{a}_k - \tilde{a}_{-k}^{\dagger}),\nonumber
\end{align}
Hence the interacting part of the Hamiltonian reads
\begin{equation*}
\begin{split}
H_{\text{int}}=& \frac{C_A}{c} (\mathbf{q}^T\msA^{-1} \mathbf{a})\int_{\mR} dk \,~  Q_{-k}  \Delta u_k(0)  -\frac{1}{L_B} (\bphi^T\mathbf{b})\int_{\mR} dk \,~    \Phi_k  \Delta u_k(0)  \\
=& \frac{\gamma}{c} \left(q_{1}r_{2}+q_{2}r_{1}\right)  \int_{\mR} dk \,~ \sqrt{\frac{\hbar \omega_k c}{2}} \Delta u_k(0) (\tilde{a}_{k} + \tilde{a}_{-k}^{\dagger}) \\
&+i \frac{E_{J}}{\varphi_{0}^{2}} ~ r_{3}  \cos{\left( 2 \pi f_{\epsilon} \right) } \left( \phi_{1} + \phi_{2} \right) \int_{\mR} dk \,~    \sqrt{\frac{\hbar}{2 \omega_k c}} \Delta u_k(0)  (\tilde{a}_k - \tilde{a}_{-k}^{\dagger}),
\end{split}
\end{equation*}
again with parameters pertaining to the different elements of the system under consideration and $\gamma=r_{3}/(r_{1} r_{2} + r_{2} r_{3} + r_{3} r_{1})$.
\begin{figure*}[h]
	\centering{\includegraphics[width=0.55\textwidth]{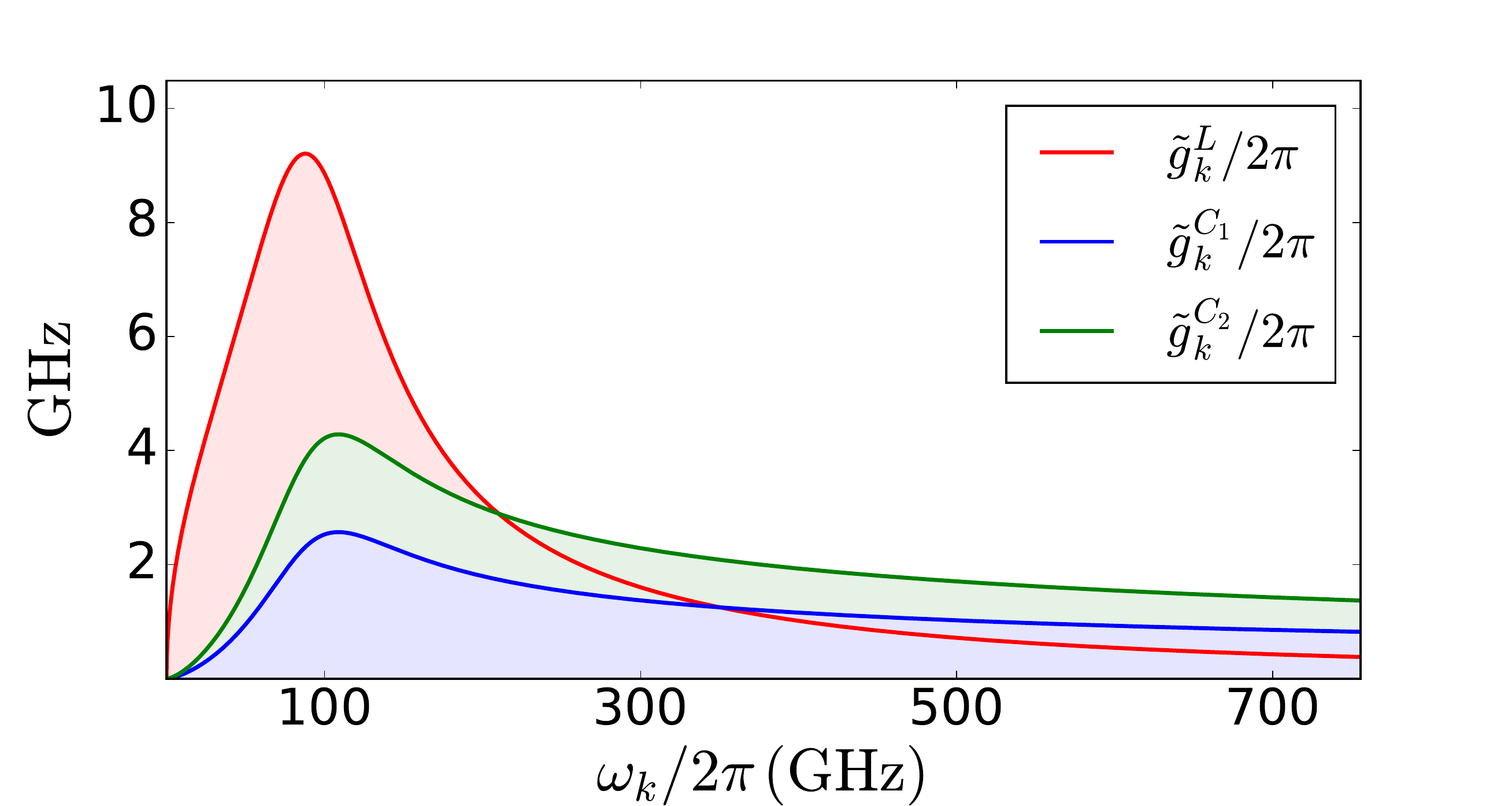}}
	\caption{\label{fig:Flux_qubit_GC_Inf_TL_gk}Capacitive and inductive coupling constants against their associated mode frequency $\omega_{k}$ for the circuit in Fig. \ref{fig:Flux_qubit_GC}. We have used the values of the experiment in \cite{FornDiaz:2017} and plotted $\tilde{g}_k=g_k/\sqrt{\beta}$ in frequency units. The inductive coupling constant behaves  as $g_{k}^{L}\propto\omega_k^{1/2}$ when $\omega_k$ tends to 0, and  as $g_{k}^{L}\propto\omega_k^{-3/2}$ for large $\omega_k$. On the other hand, the typically neglected capacitive couplings behaves as $g_{k}^{C_i}\propto\omega_k^{3/2}$ for $\omega_k\to0$ and  as $g_{k}^{C_i}\propto\omega_k^{-1/2}$ for $\omega_k\to\infty$. The natural ultraviolet cutoffs for the inductive and capacitive coupling constants are located at $\Omega_k^{(c_L)}=87.9 \,\mathrm{GHz}$ and $\Omega_k^{(c_C)}=108.8 \,\mathrm{GHz}$ respectively.}
\end{figure*}

As stated above, $u_k^*=u_{-k}$, which implies  $\Delta u_k^*(0)=\Delta u_{-k}(0)$. Let thus define $\Delta u_k(0)=|\Delta u_k(0)|\exp(i \sigma_k)$, from which $\sigma_{-k}=-\sigma_k$, and $|\Delta u_k(0)|$ is independent of the sign of $k$. We introduce a new set of canonical operators $a_k= \exp(i\sigma_k)\tilde{a}_k$ and  $a_k^{\dag}= \exp(-i\sigma_k)\tilde{a}_k^{\dag}$. The interaction part of the Hamiltonian now reads, in terms of these operators,
\begin{equation*}
\begin{split}
H_{\mathrm{int}}=& \frac{\gamma}{c} \left(q_{1}r_{2}+q_{2}r_{1}\right) \int_{\mR} dk \,~ \sqrt{\frac{\hbar \omega_k c}{2}} \left|\Delta u_k(0)\right| (a_{k} + a_{-k}^{\dagger}) \\
&+i \frac{E_{J}}{\varphi_{0}^{2}} ~ r_{3}  \cos{\left( 2 \pi f_{\epsilon} \right) } \left( \phi_{1} + \phi_{2} \right) \int_{\mR} dk \,~    \sqrt{\frac{\hbar}{2 \omega_k c}} |\Delta u_k(0)|  (a_k - a_{-k}^{\dagger})\\
=& \int_{\mR} dk \,~ g_k^{C_i}n_{i} (a_{k} + a_k^{\dagger}) +i \int_{\mR} dk \,~ g_k^{L}\varphi_{i} (a_k - a_{k}^{\dagger}),
\end{split}
\end{equation*}
where the capacitive coupling parameters are $g_k^{C_1}=r_2\gamma v_p \sqrt{\pi Z_0/R_Q} \sqrt{k}|\Delta u_k(0)|$,  $g_k^{C_2}=r_1g_k^{C_1}/r_2$, and $g_k^{L}=(E_J/\hbar) r_3 \cos(2\pi f_\epsilon)\sqrt{Z_0/\pi R_Q} |\Delta u_k(0)|/\sqrt{k}$ is the inductive one. As before, $R_Q$ is the quantum of resistance and $v_{p}$ the velocity of propagation in the line. 

The infrared and ultraviolet behaviour of these couplings are readily determined from the normalization of the $u_k$ functions. In this case, the coupling vector that fulfills the boundary conditions and normalization is
\begin{equation}
|\Delta u_k(0)|=\sqrt{\frac{2}{\pi }} \beta  \frac{|k|}{\sqrt{\beta  k^2 \left(\beta +4 \alpha  \left(\alpha  \beta  k^2-2\right)\right)+4}}.
\end{equation}

In particular, the relevant limiting behaviours of $|\Delta u_k(0)|$ are $\beta |k|/\sqrt{2\pi}$ as $k\to0$ and $1/(\alpha|k|\sqrt{2\pi})$ as $k\to \infty$. We see that the capacitive coupling  $g_k^{C_i}\sim k^{3/2}$ tends to zero faster than the inductive one $g_k^{L}\sim k^{1/2}$ as $k\to0$. Furthermore, we recover again a natural ultraviolet cutoff, in that $g_k^{C_i}\sim k^{-1/2}$ and $g_k^{L}\sim k^{-3/2}$ for large $k$, see Fig. \ref{fig:Flux_qubit_GC_Inf_TL_gk}. In other words, the inductive coupling dominates at low energies ($J(\omega)\sim J^L(\omega)$), whereas the capacitive coupling dominates at high energies ($J(\omega)\sim J^C(\omega)$).

\subsection{Charge qubits coupled to a multiport impedance}
\begin{figure*}[h!]
	\centering{\includegraphics[width=.8\textwidth]{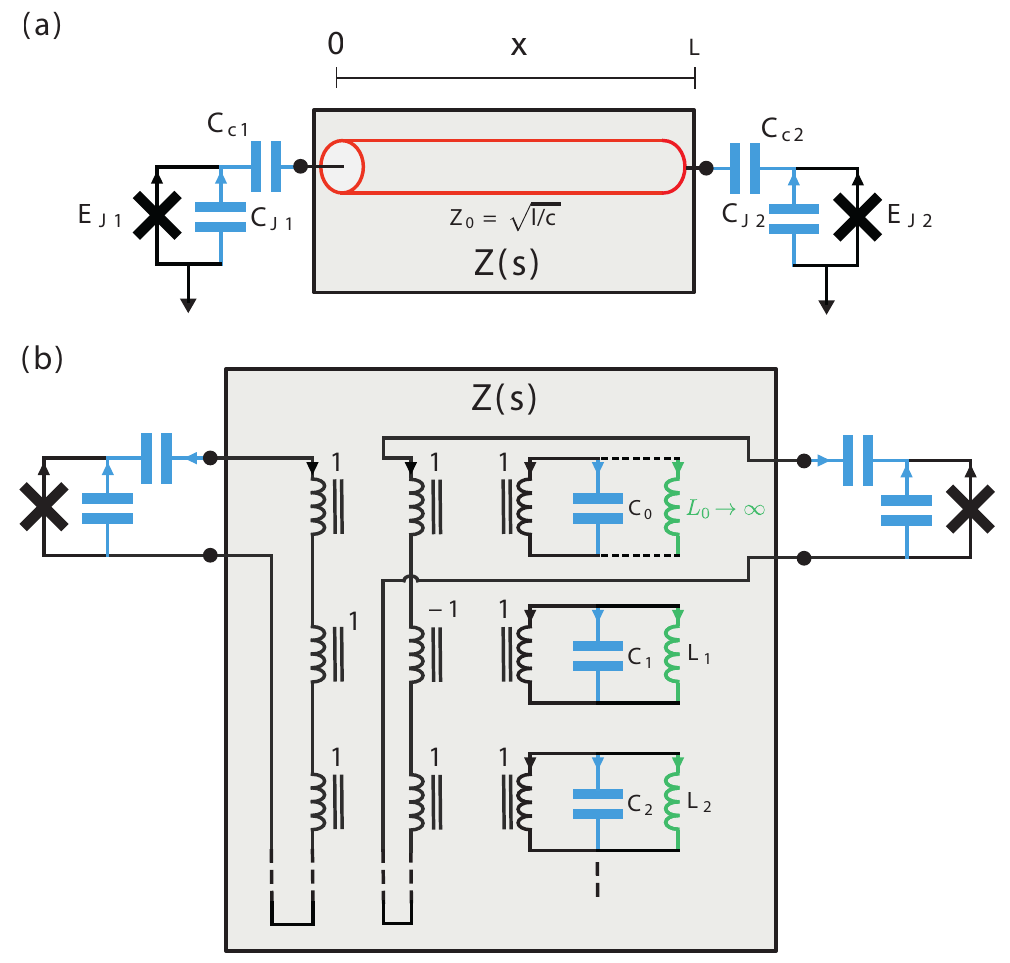}}
	\caption{\label{fig:2CQ2PortTL_v2}(a) Transmission line resonator capacitively coupled to two Josephson junctions at its ports is  modelled as a (b) 2-port impedance synthesized with a lossless multiport Foster expansion. The two port impedance $\msZ(s)$ is realized with an infinite series of stages of LC circuits coupled through a Belevitch transformer to its ports. The first stage lumped element capacitor is  $C_0 = c L$, where $c$ is the capacitance per unit length of the line and $L$ its length. The rest of the capacitors are $C_n=C_0/2$ with $n>0$. The inductors are $L_n=2Ll/n^2\pi^2$ for $n>0$. The formal expansion of $\msZ$ for the open transmission line resonator does not contain an inductor in its first stage. We introduce a virtual one in order to simplify the circuit theory analysis to derive a Hamiltonian, and then take the limit of $L_0\rightarrow\infty$ (infinite impedance of the branch) at the end. In both circuits, the arrows represent the convention used for positive current directions at each element.}
\end{figure*}

Let us finish this section with an example of a multiport impedance described by infinite degrees of freedom coupled to an anharmonic network of two degrees of freedom. In Fig. \ref{fig:2CQ2PortTL_v2}(a) we see the theoretical model of a transmission line resonator of length $L$ capacitively coupled through $C_{c\sigma}$ to two Josephson junctions of $(E_{J\sigma},C_{J\sigma})$, with $\sigma=\{1,2\}$. This circuit could be analysed as an eigenvalue problem with the theory developed in \ref{sec:mult-netw-coupl} and Appendix \ref{Walter_appendix}. However, we are going to derive the quantized Hamiltonian following the alternative presented in \ref{sec:multi-port-impedance} in order to illustrate this second method. 

To start we need to use the lumped-element equivalent circuit by which  the  transmission line, open at both ends, can be expanded. See to this point Fig. \ref{fig:2CQ2PortTL_v2}(b). Indeed, this circuit is the multiport generalization of the $1^{\mathrm{st}}$-Foster expansion used in \cite{Gely:2017} to describe the one port transmission line resonator. The classical response of the open transmission line can be encoded in a two-port impedance matrix

\begin{equation}
\msZ(s)=Z_0\begin{pmatrix}
\mathrm{coth}\left(s\sqrt{lc}L\right) &  \mathrm{csch}\left(s\sqrt{lc}L\right)\\
\mathrm{csch}\left(s\sqrt{lc}L\right) & \mathrm{coth}\left(s\sqrt{lc}L\right)
\end{pmatrix},\nonumber
\end{equation}
where $l$ and $c$ are the inductance and capacitance per unit length, $L$ is the length of the line and $Z_0=\sqrt{l/c}$ its characteristic impedance. This matrix belongs to the family of lossless positive real matrices (LPR) that can be synthesized by a passive network of inductors, capacitors and ideal transformers \cite{Newcomb:1966}, see the grey box in Fig. \ref{fig:2CQ2PortTL_v2} (b). For more details on this expansion, we refer the reader to \cite{Newcomb:1966} and Appendix \ref{sec:Mport_synthesis}. Notice that we have introducted a virtual inductance $L_0$ (whose limit is taken later as $L_0\rightarrow\infty$) to ease the network theory analysis. We can now directly apply the method developed by  Solgun and  DiVincenzo  \cite{Solgun:2015} to compute Hamiltonians of anharmonic networks coupled to a more general lossy environment described by a Brune multiport impedance. Following this reference the  equations of motion are readily derived for our set of degrees of freedom, chosen to be  the flux differences at the junctions and inductors $\bPhi=(\bPhi_J^T,\bPhi_L^T)^T=(\Phi_{J1},\Phi_{J2},\Phi_{L0},\Phi_{L1},...)^T$. The central  idea  is the elimination of the flux variables associated to the ideal transformers, since they do not store energy. The capacitance matrix of the circuit is cast into the form of (\ref{eq:multiportcapacitance}), and reads
\begin{equation}
\msC=
\begin{pmatrix}
C_{J1}+C_{c1}& 0&\sqrt{C_{c1}}\bu_1^T \\
0&C_{Jb}+C_{c2}&\sqrt{C_{c2}}\bu_2^T \\
\sqrt{C_{c1}}\bu_1& \sqrt{C_{c2}}\bu_2 &  \msC_n+\sum_i C_{ci}\bu_i\bu_i^T
\end{pmatrix},\nonumber
\end{equation}
where the submatrix is $\msC_{n}= \text{diag}(C_0, C_1, ...)$, with $C_n=C_0/2$ for $n>0$, the infinite norm coupling vectors are $\bu_1=\sqrt{C_{c1}}(1,1,1,...)^T$ and $\bu_2=\sqrt{C_{c2}}(1,-1,1,...)^T$, and, implicitly, $\ba_1=\sqrt{C_{c1}}(1,0)^T$ and $\ba_2=\sqrt{C_{c2}}(0,1)^T$. The inductance matrix is
\begin{equation}
\msL^{-1}=\begin{pmatrix}
0 & 0 & 0 \\
0 & 0 & 0\\
0 & 0 & \msL_n^{-1} 
\end{pmatrix},\nonumber
\end{equation}
with $\msL_n^{-1}=\text{diag}(L_0^{-1},L_1^{-1},...)$, and $L_n=2Ll/n^2\pi^2$ with $n\in (1,2,...)$. The Hamiltonian of the circuit can be directly derived inverting its capacitance matrix with the formula (\ref{eq:inverse-multiport-c})
\begin{equation}
H=\frac{1}{2}\bQ^{T}\msC^{-1}\bQ+\frac{1}{2}\bPhi^{T}\msL^{-1}\bPhi-\sum_{\sigma=1,2}E_{J\sigma}\cos{\varphi_{J\sigma}}.\label{eq:2CQ2PortTL_Ham}
\end{equation}
Let us consider a truncated model with $N$ inductors, and later take the limit $N\to\infty$.
The direct coupling between the two anharmonic degrees of freedom can be calculated through the upper-left submatrix, for which we need to compute the auxiliary matrices 
\begin{align}
\mu=&\begin{pmatrix}
\sum_n C_n^{-1}&\sum_n (-1)^{n+1} C_n^{-1}\\
\sum_n (-1)^{n+1} C_n^{-1}&\sum_n C_n^{-1}
\end{pmatrix}\nonumber\\
=&\begin{pmatrix}
(2N+1)C_{c1}/C_0&\pm 2\sqrt{C_{c1}C_{c2}}/C_0\\
\pm 2\sqrt{C_{c1}C_{c2}}/C_0&(2N+1) C_{c2}/C_0
\end{pmatrix}\nonumber
\end{align}
and 
\begin{equation}
\nu= -\begin{pmatrix}
C_{c1}/C_{\Sigma 1}& 0\\
0&C_{c2}/C_{\Sigma 2}
\end{pmatrix},\nonumber
\end{equation}
with $C_{\Sigma \sigma}=C_{J\sigma} + C_{c\sigma}$ for $\sigma=\{1,2\}$, and where the $+$ ($-$) sign appears when we truncate to an even (odd) number of inductors  $N$ in the circuit. The upper-left part of the inverse capacitance matrix is diagonal in the limit of infinite stages ($N\rightarrow\infty$), since  
\begin{equation}
\lim\limits_{N\rightarrow\infty}\msXi=\begin{pmatrix}
C_{\Sigma 1}/C_{J1}&0\\
0&C_{\Sigma 2}/C_{J2}
\end{pmatrix}.\nonumber
\end{equation}

\begin{figure*}[h!]
	\centering{\includegraphics[width=0.6\textwidth]{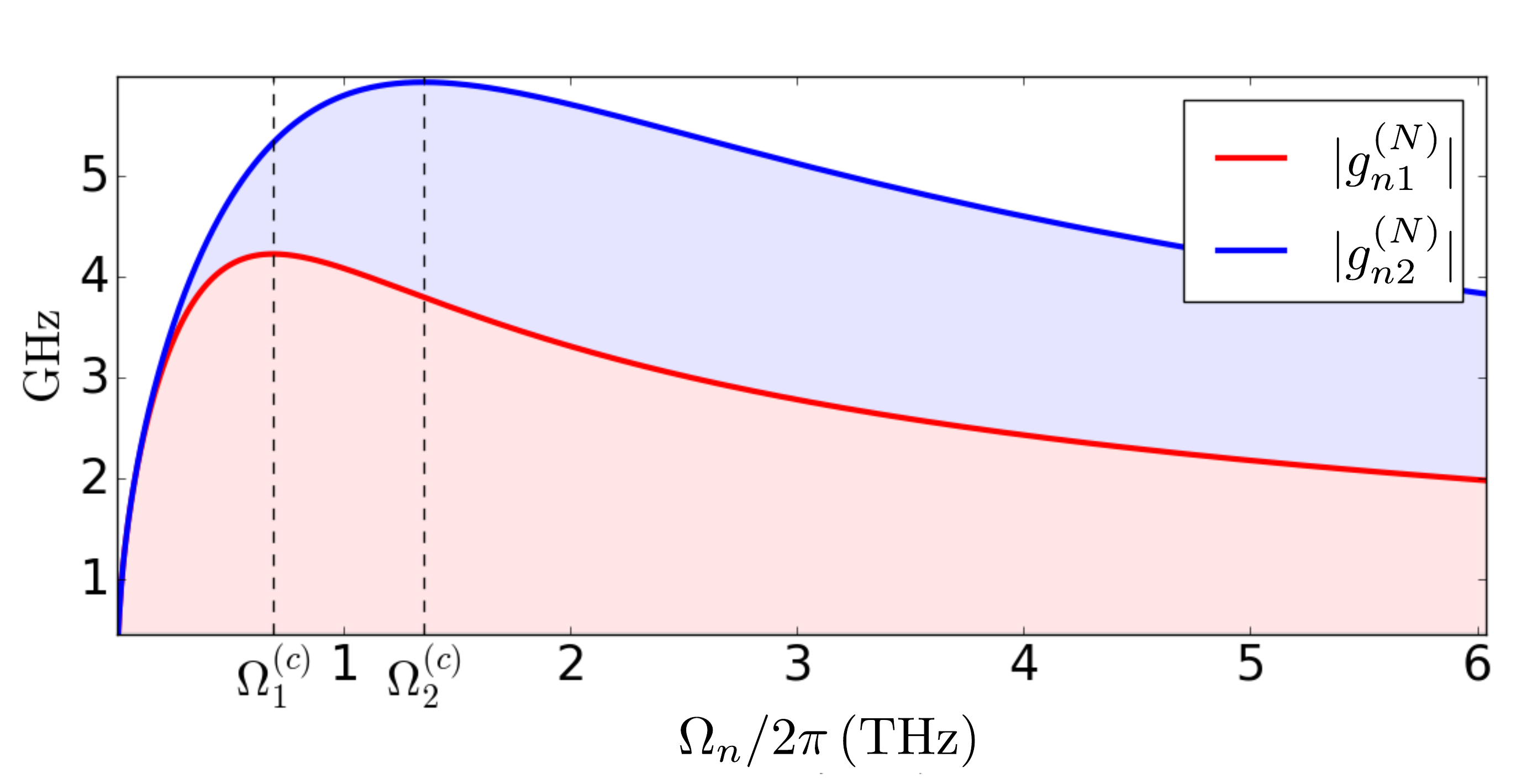}}
	\caption{\label{fig:2CQ2PortTL_gn}Capacitive coupling $g_{n\sigma}$ against its mode frequency $\Omega_{n}$ for the circuit in Fig. \ref{fig:2CQ2PortTL_v2}. We have used the same values to those of Device A in \cite{Bosman:2017} for the first junction, $C_{c1}=40.3\,\mathrm{fF}$, $C_{J1}=5.13 \,\mathrm{fF}$, while the second junction has values $C_{c2}=C_{c1}/2$ and $C_{J2}=C_{J1}/2$. The transmission line has again the parameters  $c=249\,\mathrm{pF/m}$, $l=623\,\mathrm{nH/m}$, while the length has been doubled $L=9.4\,\mathrm{mm}$, so  that the fundamental mode with frequency $\Omega_{1}= 4.26 \,\mathrm{GHz}$ enter into the more desirable experimental regime of $(4-6)\, \mathrm{GHz}$. In red (blue), $g_{n1}$ ($g_{n2}$), where the number of modes used in the numerical diagonalization has been truncated to $N=6000$: a natural cutoff of the coupling constants appear for the resonator modes at frequency $\Omega_{1}^{(c)}=685.5\,\mathrm{GHz}$ $(\Omega_{2}^{(c)}=1.35\,\mathrm{THz})$.}
\end{figure*}

This result would have also been retrieved using the continuous wave flux field expansion of \ref{sec:mult-netw-coupl}, where the infinite dimensional coupling vectors would have been purely orthogonal. Thus, we can write the inverse matrix as 
\begin{equation}
\lim\limits_{N\rightarrow\infty}\msC^{-1}\rightarrow
\begin{pmatrix}
1/C_{J1}& 0&\msTheta_{11}\frac{C_{c1}}{C_{\Sigma 1}}\bu_1^T\msC_n^{-1} \\
0&1/C_{J2}&\msTheta_{22}\frac{C_{c2}}{C_{\Sigma 2}}\bu_2^T\msC_n^{-1}\\
\msTheta_{11}\frac{C_{c1}}{C_{\Sigma 1}}\msC_n^{-1}\bu_1& \msTheta_{22}\frac{C_{c2}}{C_{\Sigma 2}}\msC_n^{-1}\bu_2 &  \msD_n^{-1}
\end{pmatrix},\nonumber
\end{equation}
where $\msTheta$ is the matrix defined in (\ref{eq:inverse-cap-coeffmatrices}). We restate now the terms in (\ref{eq:2CQ2PortTL_Ham})
\begin{equation}
H=\frac{Q_{J\sigma}^2}{2C_{J\sigma}}-E_{J\sigma}\cos{\varphi_{J\sigma}}+\msTheta_{\sigma \sigma}\frac{C_{c\sigma}}{C_{\Sigma \sigma}}Q_{J\sigma}{\bfeta}_{\sigma}^T\bQ_{L}+\frac{1}{2}\bQ_L^{T}{\msD}_n^{-1}\bQ_L+\frac{1}{2}\bPhi_L^{T}{\msL}_n^{-1}\bPhi_L,\nonumber
\end{equation}
with coupling vectors ${\bfeta}_\sigma={\msC}_n^{-1}{\bu}_\sigma$, and where we have used Einstein's summation rule for repeating greek letters. We take now safely the limit $L_0\rightarrow\infty$ and eliminate the free degree of freedom $(\Phi_{L_0},Q_{L_0})$, truncating the vectors $\tilde{\bPhi}_L=(\Phi_{L_1},\Phi_{L_2},...)^T$, $\tilde{\bQ}_L=(Q_{L_1},Q_{L_2},...)^T$ and $\tilde{\bfeta}_{\sigma}=(\eta_{\sigma 1}, \eta_{\sigma 2},...)^T$. Additionally, the matrices $\tilde{\msD}_n$ and  $\tilde{\msL}_n^{-1}$ take after ${\msD}_n$ and  ${\msL}_n^{-1}$. In fact they are the untilded versions,  pruned of their first row and column. As previously discussed, we can perform the rescaling and rotation of the harmonic variables $\bx_L=D_0^{-1/2}\msU\tilde{\msD}_n^{1/2}\tilde{\bPhi}_L$ and $\bp_L=D_0^{1/2}\msU^T\tilde{\msD}_n^{-1/2}\tilde{\bQ}_L$, where $D_0$ is a capacitance constant that mantains the units of the conjugated variables. We finally reach the normal mode structure
\begin{equation}
H=4E_{C\sigma}n_{J\sigma}^2-E_{J\sigma}\cos{\varphi_{J\sigma}}+\msTheta_{\sigma \sigma}\frac{C_{c\sigma}}{C_{\Sigma \sigma}}Q_{J\sigma}\bxi_{\sigma}^T \bp_{L}+\frac{\bp_L^2}{2 D_0}+\frac{D_0}{2}\bx_L^T \mathsf{\Omega}_n^2\bx_L,\nonumber
\end{equation}
where $\mathsf{\Omega}_n^2= D_0\msU\tilde{\msD}_n^{-1/2}\tilde{\msL}_n^{-1}\tilde{\msD}_n^{-1/2}\msU^T$ is a diagonal matrix, and ${\bxi}_{\sigma}=D_0^{-1/2} \msU^T \msD_n^{1/2} {\bfeta}_\sigma$ are the coupling vectors with finite norm even in the limit $N\rightarrow\infty$. 

We promote the conjugate variables to operators, i.e. $[\hat{x}_n,\hat{p}_n]=i\hbar$ and $[\hat{\varphi}_J,\hat{n}_J]=i$, and rewrite the harmonic sector in terms of $a_n$ and $a_n^\dagger$ in an anologous manner to previous sections to reach
\begin{equation}
H=4E_{C\sigma}\hat{n}_{J\sigma}^2-E_{J\sigma}\cos{\hat{\varphi}_{J\sigma}}+\hbar g_{n\sigma}\hat{n}_{J\sigma}(a_n+a_n^\dagger)+\hbar \Omega_{n}a_n^\dagger a_n,\nonumber
\end{equation}
where the coupling constants are redefined as $g_{n \sigma}= \msTheta_{\sigma \sigma}\frac{C_{c\sigma}}{C_{\Sigma \sigma}}\sqrt{\frac{\Omega_{n}D_0}{4\pi R_{Q}}}$ and $\Omega_{n}$ are the square root of the diagonal entries in $\mathsf{\Omega}_n^2$. In Fig. \ref{fig:2CQ2PortTL_gn} we see an example with realistic parameters where the different coupling and junction's capacitances $(C_{J\sigma}, C_{c\sigma})$ translate into different saturation frequencies for the coupling parameters. Again, both coupling parameters increase (decay) as $g_{n \sigma}\propto \sqrt{\Omega_{n}}$ ($ 1/\sqrt{\Omega_{n}}$) in the low (high) frequency limit.

Summarizing, in this chapter we have critically analysed several approaches to the quantization of lumped element networks linearly coupled to infinite dimensional systems, with a focus on the historical divergence issues in the capacitive coupling parameters. We have identified electric ($\alpha$) and magnetic ($\beta$) lengths directly related to the intrinsic soft ultraviolet cutoffs in capacitive and inductive coupling generic configurations. Both of them, can be optimally set to specific values such that the final Hamiltonian description be that of an infinite number of independent harmonic modes coupled to the finite set of degrees of freedom. We have explained the underlying mathematical structure in the study of the transmission line with point-like connections, i.e. a singular value problem for a second order differential operator, with boundary values that include the singular value itself. We provided a new proof of the associated expansion theorem. Furthermore, the methods here used to study  transmission line modes provides us with an exact expression for the Hamiltonian in both space or its reciprocal space for the fields. In doing so, one reads the $A^2$ term directly, dependent on the length parameter $\alpha$.

We have made apparent the source of issues that appear in (multiport) black-box approaches to describe general environments with infinite number of modes after truncating to $N$ harmonic modes, and subsequently taking the limit $N\to\infty$. We have again used as the central criterion that the final Hamiltonian description be that of an infinite set of independent modes, coupled to the relevant (nonharmonic) subsystem. We have shown this with a canonical transformation that is indeed determined by this criterion.

One of the main objectives of the quantization of superconducting circuits is their use for quantum information tasks. Thus, it is relevant to consider how spin-boson models arise from our presentation. In this chapter, we have shown that the spectral density for qudits capacitively/inductively coupled to transmission lines falls off with a soft power law cutoff, i.e. $g_n\sim \omega_n^{-1/2}$ when $n\rightarrow \infty$. We have also provided explicit computations for models of experimental relevance which portray this behaviour. 

Furthermore, we have given a detailed argument for the validity of standard approximations in the literature, in which the behaviour $g_n\sim \omega_n^{1/2}$ is assumed for capacitive coupling to transmission lines. In essence, such behaviour comes about when one approximates the coupling capacitor by an open connection in looking for an eigenbasis decomposition of the flux field. We have shown that this is indeed a correct assumption for the lowest frequency sector, and that it is valid for truncations to a finite number of modes if not too many are assumed present.

Looking to the future, we have presented detailed arguments at each discussed point, with the objective that this work can be a reference for initial development of new useful multi-mode models. The solid mathematical foundations in which we rely for the expansion in modes will undoubtedly be useful in other physical contexts as well, and we will be exploring further in this point. Finally, we have presented a prediction for the maximum coupling achievable with a transmission line, that has not yet been measured. Having made extensive analyses of reciprocal infinite-dimensional networks we will present in the next chapters a thorough analysis on how to include nonreciprocal elements in a systematic way to the Hamiltonian description of quantum circuits.

\chapter{Lumped Nonreciprocal Networks}
\label{chapter:chapter_4}
\thispagestyle{chapter}
\hfill\begin{minipage}{0.85\linewidth}
{\emph{Io resto interamente appagato; e mi credano certo, che se io avessi a ricominciare i miei studii, vorrei seguire il consiglio di Platone e cominciarmi dalle matematiche, le quali veggo che procedono molto scrupolosamente, né vogliono ammetter per sicuro fuor che quello che concludentemente dimostrano\\\\
(I am quite convinced; and, believe me, if I were again beginning my studies, I should follow the advice of Plato and start with mathematics, a science which proceeds very cautiously and admits nothing as established until it has been rigidly demonstrated)}}
\end{minipage}
\begin{flushright}
	\textbf{Galileo Galilei}\\
	Discorsi e dimostrazioni matematiche, intorno à due nuove scienze
\end{flushright}
\vspace*{1cm}
In electromagnetism, reciprocity is defined as the invariance of a system's linear response under interchange of sources and detectors~\cite{Caloz:2018}. Nonreciprocal (NR) elements such as gyrators~\cite{Tellegen:1948} and circulators~\cite{Carlin:1964} have been mainly used in superconducting quantum technology as noise isolators and classical information routers, i.e., out of the quantum regime, due to the size of currently available devices. Lately, there have been several proposals for building scalable on-chip NR devices based on Josephson junction-networks~\cite{Sliwa:2015,Chapman:2017,Mueller:2018}, parametric permittivity modulation~\cite{Kerckhoff:2015}, the quantum Hall effect~\cite{Viola:2014,Mahoney:2017} and mechanical resonators~\cite{Barzanjeh:2017}. Such nonreciprocal behavior presents quantum coherence properties~\cite{Mahoney:2017} and will allow novel applications in the nontrivial routing of quantum information~\cite{Lecocq:2017,Barzanjeh:2018,Metelmann:2018}. Accordingly, there is great interest in building a general framework to describe networks working fully on the quantum regime~\cite{Devoret:1997,Paladino:2003,Burkard:2004,Burkard:2005,Bourassa:2012,Nigg:2012,Solgun:2014,Solgun:2015,Mortensen:2016,Malekakhlagh:2016,Malekakhlagh:2017,Gely:2017,ParraRodriguez:2018}. 

In this chapter, we use network graph theory to derive Hamiltonians of superconducting networks that contain both nonlinear Josephson junctions and ideal lineal NR devices with frequency-independent response~\cite{Duinker:1959}. The correct treatment of such ideal devices will provide us with building blocks to describe more complex nonreciprocal linear devices~\cite{Carlin:1964} that can be treated as linear black boxes~\cite{Paladino:2003,Nigg:2012,Solgun:2014,Solgun:2015}. This theory lays the ground for the correct description of circuits in the regime where the nonreciprocal devices can be well characterized by a linear response~\cite{Sliwa:2015,Chapman:2017,Mueller:2018,Kerckhoff:2015,Viola:2014,Mahoney:2017,Barzanjeh:2017}, even if the fundamental nonreciprocal behavior is achieved by nonlinear elements~\cite{Sliwa:2015,Chapman:2017,Mueller:2018}. Outside of this regime of validity, a black-box approach is not longer useful and a microscopic description of nonreciprocal effects is imperative. We emphasize that, even though they do not exist as such in nature, ideal gyrators and circulators can be useful elements to describe efficient dynamics at certain frequency regimes. The sharp reader should not be astonished of this statement as the rest of the lumped elements, like capacitors or inductors, are also ideal and effective models. We focus on and extend the analyses of lumped-element networks of Devoret~\cite{Devoret:1997}, Burkard-Koch-DiVincenzo (BKD)~\cite{Burkard:2004}, Burkard~\cite{Burkard:2005} and Solgun and DiVincenzo~\cite{Solgun:2015}. Our extension involves, first, adding ideal gyrators and circulators described by an admittance ($\msf{Y}$) matrix to obtain quantum Hamiltonians with a countable number of flux degrees of freedom. As we will see, a bias towards a specific matrix description of NR devices (NRDs) appears useful when we want the Euler--Lagrange equations of motion to be current Kirchhoff equations in terms of flux variables. We next show how adding ideal NRDs described by impedance ($\msf{Z}$) or scattering ($\msf{S}$) matrices requires a more involved treatment, in that the system of equations must be first properly reduced. Finally, we also address canonical quantization with loop charges to treat dual circuits with $\msf{Z}$-circulators; see Ref.~\cite{Ulrich:2016} for a detailed description. We apply our theory to two useful, pedagogical and pathological circuit examples that involve the main technical issues that more complex networks could eventually present.

Our emphasis is on quantization of an electrical network, that is to say, on quantum network \emph{analysis}, and we set aside the dual problem of network \emph{synthesis}. Even so, the introduction of the techniques presented here implies that more sophisticated synthesis methods can be used for the description of quantum devices, since our analysis can be applied to a wider class of circuits than those previously considered.

\begin{figure}[h]
	\centering
	\includegraphics[width=0.7\linewidth]{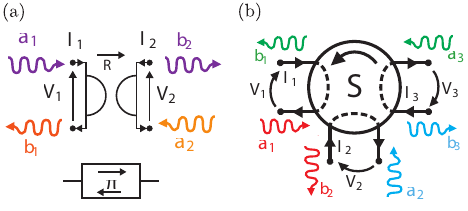}
	\caption{\label{fig:NRD_intro}(a) A 2-port gyrator: Input $a_k$ and output $b_k$ signals are related to each other through the scattering matrix $\bsb{b}=\msf{S}\bsb{a}$; with $b_2=a_{1}$ and $b_1=-a_{2}$, the element behaves as a perfect $\pi$-phase directional shifter (bottom) when impedance matched to transmission lines at ports. (b) A 3-port circulator: Input signals transform into output signals cyclically, e.g. $b_k=a_{k-1}$. Port voltages $V_k$ and currents $I_k$ can be generally related to $a_k$ and $b_k$ through Eq. (\ref{eq:consti_SG}).}
\end{figure}

Regarding the need for a more involved treatment of NR devices with immitance or scattering matrix presentations, bear in mind that, in microwave engineering, a multiport linear (black-box) device can be always described by its scattering matrix parameters $\msf{S}(\omega)$~\cite{Pozar:2009}, that relate voltages and currents at its ports $\bsb{b}=\msf{S}\bsb{a}$, with $b_k=(V_k-Z_k^* I_k)/\sqrt{\Re\{Z_k\}}$ and $a_k=(V_k+Z_k I_k)/)/\sqrt{\Re\{Z_k\}}$ being the output and input signals. The reference impedances can be chosen, for simplicity, homogeneous and real, e.g., $Z_k=R>0$. Simple properties of the scattering matrix reveal fundamental characteristics of the device. For instance, a network is reciprocal (lossless) when $\msf{S}$ is symmetric (unitary). See Fig. \ref{fig:NRD_intro} for an example of basic NR devices and their conventional symbols in electrical engineering. When ports are impedance-matched to output transmission lines (a) a 2-port (4-terminal) ideal gyrator behaves as a perfect $\pi$-phase directional shifter, i. e. $b_2=a_{1}$ and $b_1=-a_{2}$, and (b) a 3-port (6-terminal) ideal circulator achieves perfect signal circulation, e.g. $b_k=a_{k-1}$~\cite{Carlin:1964}. Other useful descriptions of multiport devices are the impedance $\msf{Z}(\omega)=R(1-\msf{S}(\omega))^{-1}(1+\msf{S}(\omega))$ and admittance  $\msf{Y}(\omega)=\msf{Z}^{-1}(\omega)$ matrices that relate port voltages and currents as $\bV=\msf{Z}\bI$ and $\bI=\msf{Y}\bV$ respectively~\cite{Pozar:2009}. Although sometimes more useful, immittance descriptions of linear devices do not always exist, and working with $\msf{S}$ can be unavoidable~\cite{Carlin:1964,Pozar:2009}. This comes about whenever the $\mathsf{S}$ matrix has $+1$ and $-1$ eigenvalues.

We organize the chapter as follows. In Sec.  \ref{sec:network-graph-theory} we present some basic aspects of network graph theory, as applicable to electrical circuits, 
with reference to the current literature on its use in  quantization. We include nonreciprocal multiport elements in the consideration. We next address, in Sec. \ref{sec:networks-with-msfy}, the construction of the Lagrangian of circuits with admittance described nonreciprocal devices and the subsequent quantization. We provide specific examples of this process. In Sec.  \ref{sec:ZS_NRD} we look into the issue of nonreciprocal devices with no admittance description. To this point we have studied circuits with flux variables. In Sec. \ref{sec:dual-quant-charge} the dual, charge variables are investigated for their use in nonreciprocal circuits.  We finish with conclusions and a perspective on future work.
\section{Network graph theory}
\label{sec:network-graph-theory}
A lumped-element electrical network is an oriented multigraph~\cite{Devoret:1997,Burkard:2004}. Each branch of the graph connects two nodes and has a direction chosen to be that of the current passing through it. A one-port element will be  assigned a branch.   The  choice of direction for the corresponding branch is arbitrary for symmetric elements. More generally,  $N$-port elements like the \emph{circulator} are represented by $N$ branches connecting $2N$ nodes pairwise~\cite{Carlin:1964}; see Fig. \ref{fig:NRD_intro}. A spanning \emph{tree} of the graph is a set of branches that connects all nodes without creating loops. The set of branches in the tree are called tree branches and all others \emph{chord} branches. Making a choice for tree and chord branches in an electrical network, we separate the currents $\bI^T=(\bI_{\mathrm{tr}}^T,\bI_{\mathrm{ch}}^T)$ and voltages $\bV^T=(\bV_{\mathrm{tr}}^T,\bV_{\mathrm{ch}}^T)$ to write Kirchhoff's equations as   
\begin{align}
\msF \bI_{\mathrm{ch}}&=-\bI_{\mathrm{tr}},\label{eq:Kirchhoff_I}\\
\msF^{T}\bV_{\mathrm{tr}}&=\bV_{\mathrm{ch}}+\dot{\mbf{\Phi}}_{\mathrm{ex}},\label{eq:Kirchhoff_V}
\end{align}
where $\msF$ is the reduced fundamental loop(/cutset) matrix describing the topology of the graph. It contains only $\{0,-1,1\}$ entries; see~\cite{Burkard:2004,Burkard:2005} for details on graph theory applied to superconducting circuits. Hence we make reference to $\mathsf{F}$ as the loop matrix. The vector of external fluxes $\mbf{\Phi}_{\mathrm{ex}}$ corresponds to the set of external fluxes threading each of the loops of the system.

The branch charge ($\bQ$)  and flux ($\mbf{\Phi}$) variables are defined from the flow variables $\bI$ and difference variables $\bV$ as $I_X(t)=\dot{Q}_X(t)$ and   $V_X(t)=\dot{\Phi}_X(t)$, where the subscript $X=C,\,L,\,J,\,G,\,T,\,R,\,Z,\,V,\,B$ denotes capacitors, inductors, Josephson junctions, nonreciprocal element branches, transformer branches, resistors, two-terminal impedances, voltage sources, and current sources, respectively. For the sake of simplicity we focus here on networks with passive and lossless elements, i.e., capacitors, inductors, Josephson junctions, nonreciprocal element branches, and transformer branches. We forward the reader to Refs.~\cite{Burkard:2004,Burkard:2005,Solgun:2014,Solgun:2015} for the inclusion of two-terminal impedances and voltage and current sources.

The constitutive equations of capacitors, inductors, and Josephson junctions are 
\begin{align}
\bQ_C&=\msC\bV_C,\label{eq:consti_C}\\
\bI_L&=\msL^{-1}\mbf{\Phi}_L,\label{eq:consti_L}\\
\mbf{\Phi}_J&=\frac{\Phi_q}{2\pi}\bvarphi_J,\label{eq:consti_J1}\\
\bI_J&=\bI_c\bsb{\mathrm{sin}}(\bvarphi_J).\label{eq:consti_J2}
\end{align}
where $\bI_c\bsb{\mathrm{sin}}(\bvarphi_J)$ is the column vector with $I_{ci}\sin(\varphi_{Ji})$ entries, $I_{c}$ the critical current of a junction and $\Phi_q$ the flux quantum. General multiport transformers (Belevitch transformers~\cite{Belevitch:1950}) have been previously added to the Burkard analysis in Ref.~\cite{Solgun:2015}. They add voltage and current constraints on the right ports in terms of its left ports and vice versa,
\begin{align}
\bI_{T}^{R}=-\msN \bI_{T}^{L}, \quad \label{eq:Belevitch_trR1}
\bV_{T}^{L}=\msN^T \bV_{T}^{R},
\end{align}
where $\msf{N}$ is the turns ratios matrix and both left and right current directions are pointing inwards. Dual transformers exist where the left-right equations (\ref{eq:Belevitch_trR1}) are inverted~\cite{Belevitch:1950,Solgun:2015}. Passing now to the focus of our study, the general constitutive equation for the ideal (frequency-independent) nonreciprocal element branches can be retrieved from the scattering matrix definition
\begin{equation}
(1-\msf{S})\bV_G(t)=(1+\msf{S})R\bI_G(t),\label{eq:consti_SG}
\end{equation}
with $R$ a constant in resistance units.

In order to carry out canonical quantization in circuits, our task will be to  simplify Kirchhoff's laws together with the constitutive relations into a set of classical Euler-Lagrange (E-L) equations, from which Hamiltonian equations can be derived through a Legendre transformation, and canonically conjugate variables can be identified. In trivial cases, as already explained, this reduces to having a kinetic matrix that is non-singular. 
\section{Networks with $\msf{Y}$ NRD}\label{sec:networks-with-msfy}
Given that Josephson junctions are nonlinear devices, E-L equations have been systematically derived in flux variables so as to have a  purely quadratic kinetic sector, e.g. Refs. ~\cite{Devoret:1997,Burkard:2004,Burkard:2005,Solgun:2014,Solgun:2015}. In particular, BKD and Burkard quantization methods are constrained, with respect to Devoret's approach, to specific topological classes of circuits to make the Hamiltonian derivation even more systematic. For instance, in BKD all the capacitors must be included in the tree, while there are no capacitor-only loops; i.e., all capacitors are tree branches, and no external impedance can appear in the tree, while Burkard quantization has dual conditions.
These assumptions about the assignment to tree and chord branches provide us with a description of the loop matrix in block matrix form, in such a way that some of the blocks are trivial. This triviality, in turn, will allow us to construct effective loop matrices by elimination of variables.

As we shall now see, those approaches can easily incorporate ideal NR elements described by the admittance matrix ($\msf{Y}$ devices) with the realistic assumption that all of their branches are independently shunted by (parasitic) capacitors. 

For instance, the BKD formalism can be extended by assuming that all ideal NR ($G$) branches are chord branches. As stated, in BKD all capacitors of the mesh have to be in the  tree branches, whereas Josephson junctions, which are always in parallel to at least one capacitor, are chosen to be chord branches. Inductors can be both in the tree ($K$) or in the chord ($L$) set. In the following, we sketch the derivation where all inductors are chord inductors. For pedagogical purposes, we derive a Burkard circuit class extension in Appendix \ref{sec_app:Burkard_extension}. Following Ref.~\cite{Solgun:2015}, we also include  Belevitch transformers in this analysis. 

The fundamental loop matrix  of a simplified BKD circuit can be written in block matrix form as
\begin{equation}
\msF=\begin{pmatrix}
\msF_{CJ}&\msF_{CL}&\msF_{CG}&\msF_{CT^{\mathrm{ch}}}\\
\msF_{T^{\mathrm{tr}}J}&\msF_{T^{\mathrm{tr}}L}&\msF_{T^{\mathrm{tr}}G}&\msF_{T^{\mathrm{tr}}T^{\mathrm{ch}}}
\end{pmatrix}.
\end{equation}
Real Josephson junctions are always in parallel to capacitors, so that $\msF_{T^{\mathrm{tr}}J}=0$. On the other hand, if all transformer left branches can be included in the tree, while transformer right branches are in the chord, then  $\msF_{T^{L}T^{R}}=\msF_{T^{\mathrm{tr}}T^{\mathrm{ch}}}=0$. We can integrate out the voltages and currents in the transformer branches~\cite{Solgun:2015} inserting  (\ref{eq:Belevitch_trR1}) into Kirchhoff's equations (\ref{eq:Kirchhoff_I}, \ref{eq:Kirchhoff_V}) and write an effective loop matrix
\begin{equation}
\msF^{\mathrm{eff}}=\begin{pmatrix}
\msF_{CJ}&\msF_{CL}^{\mathrm{eff}}&\msF_{CG}^{\mathrm{eff}}
\end{pmatrix},\label{eq:BKD_Feff}
\end{equation}
with $\msF_{CL}^{\mathrm{eff}}=\msF_{CL}+\msF_{CT^{\mathrm{ch}}}\msN \msF_{T^{\mathrm{tr}}L}$ and $\msF_{CG}^{\mathrm{eff}}=\msF_{CG}+\msF_{CT^{\mathrm{ch}}}\msN \msF_{T^{\mathrm{tr}}G}$. We insert the constitutive equations (\ref{eq:consti_L}) and the admittance version of (\ref{eq:consti_SG}), $\bI_G=\msf{Y}_G\bV_G$, into the reduced current equation to obtain a second-order equation in flux variables,
\begin{equation}
-\msC\ddot{\bPhi}_C=\bI_c\bsb{\mathrm{sin}}(\bvarphi_{C_J})+\msf{M}_0\bPhi_C+\msf{G}\dot{\bPhi}_C,
\end{equation}
where $\msf{M}_0=\msF_{CL}^{\mathrm{eff}}\msL^{-1}(\msF_{CL}^{\mathrm{eff}})^T$, $\msf{G}=\msF_{CG}^{\mathrm{eff}}\msf{Y}_G(\msF_{CG}^{\mathrm{eff}})^T$, and $\bvarphi_{C_J}=\bvarphi_{J}$ is the vector of capacitor branch phases (related to the fluxes by (\ref{eq:consti_J1})) in parallel with the junctions. $\msf{Y}_{G}$ is a skew-symmetric matrix (because it is the Cayley transform of an orthogonal matrix $\mathsf{S}$), and by construction so is $\msf{G}$ (see Appendix \ref{sec_app:Burkard_extension}). The antisymmetry associated with the first-order derivatives, together with the fact that these second-order equations have a non-singular kinetic matrix, allows us to  derive them from the Lagrangian 
\begin{equation}	L=\frac{1}{2}\left(\dot{\bPhi}_C^T\msC\dot{\bPhi}_C+\dot{\bPhi}_C^T\msf{G}{\bPhi}_C-\bPhi_C^T\msf{M}_0{\bPhi}_C\right)-U(\bvarphi_{C_J}).
\end{equation}

The conjugate charge variables are $\bsb{Q}_C=\partial{L}/\partial \dot{\bPhi}_C=\msC\dot{\bPhi}_C+\frac{1}{2}\msf{G}\bPhi_C$. Notice that conjugate charge variables are not necessarily identical to capacitor branch charge variables, which are those that appear in Eq. (\ref{eq:consti_C}). Promoting the variables to operators with canonical commutation relations $[\hat{\Phi}_{C_n},\hat{Q}_{C_m}]=i\hbar\delta_{nm}$, we derive the quantum Hamiltonian 
\begin{align}
\hat{H}=&\frac{1}{2}\left(\hat{\bsb{Q}}_C-\frac{1}{2}\msf{G}\hat{\bPhi}_C\right)^T\msC^{-1}\left(\hat{\bsb{Q}}_C-\frac{1}{2}\msf{G}\hat{\bPhi}_C\right)+\frac{1}{2}\hat{\bPhi}_C^T\msf{M}_0\hat{\bPhi}_C+U(\hat{\bvarphi}_{C_J}).\label{eq:Hamiltonian_Ydev}
\end{align}
The non linear potential is defined as $U(\hat{\bvarphi}_{C_J})=-\sum_i E_{Ji}\cos(\hat{\varphi}_{Ji})$ and the Josephson energy of each junction is $E_{Ji}=I_{ci}\Phi_q/(2\pi)$. Given the velocity-position coupling term arising from the $\msf{G}$ matrix, a form first devised in Ref.~\cite{Duinker:1959}, a diagonalization of the harmonic sector requires a symplectic transformation, that can be carried out either in the classical variables or after the canonical quantization procedure; see Appendix \ref{App_sec:symplectic}. Notice the similarity of the $\mathsf{G}$ terms to a magnetic field, and their breaking of time-reversal invariance. In the same manner as a magnetic field, these \emph{gyroscopic} terms are energy conserving. It is worth mentioning that such Hamiltonians has recently gained interest in the context of quantum information for encoding GKP states~\cite{Rymarz:2021}, and it is currently being analysed their stoquastic~\cite{Bravyi:2006} properties~\cite{Ciani:2020}. 

\subsection*{Examples}
These extended BKD and Burkard analyses can be directly applied to a huge family of circuits to derive Hamiltonians in \emph{position}-flux variables with $\msf{Y}$ NRDs. Up till now, most of the interest in  quantization of circuits has been connected with the presence of Josephson junctions. In the present analysis we combine that presence of Josephson junctions with nonreciprocal devices. We are thus motivated to keep the flux variables as the only position coordinates of a Lagrangian/Hamiltonian mechanical system. Here we demonstrate  the quantization of two circuits consisting of two Josephson junctions coupled to (i) a general 2-port nonreciprocal black box and (ii) the specific nonreciprocal impedance response of the Viola-DiVincenzo Hall effect gyrator~\cite{Viola:2014}. The first circuit is a pedagogical and useful example where the black box, in its $N$-port configuration, would represent the response of any of the given proposals in Refs.~\cite{Sliwa:2015,Chapman:2017,Mueller:2018,Kerckhoff:2015,Viola:2014,Mahoney:2017,Barzanjeh:2017} within their valid frequency range containing two gyrators. In the second circuit, we exploit a specific 2-port impedance response, which includes a gyrator, to get an easy lumped-element approximation that can be directly quantized. Extensions of these circuits with $N$-port $\msf{Y}$ circulators would also be readily treated by this formalism. We study corner cases where the circulators cannot be described by $\msf{Y}$ matrices below in Sec. \ref{sec:ZS_NRD}.

\subsubsection{NR black-box coupled to Josephson junctions}
\begin{figure}
	\centering
	\includegraphics[width=.6\linewidth]{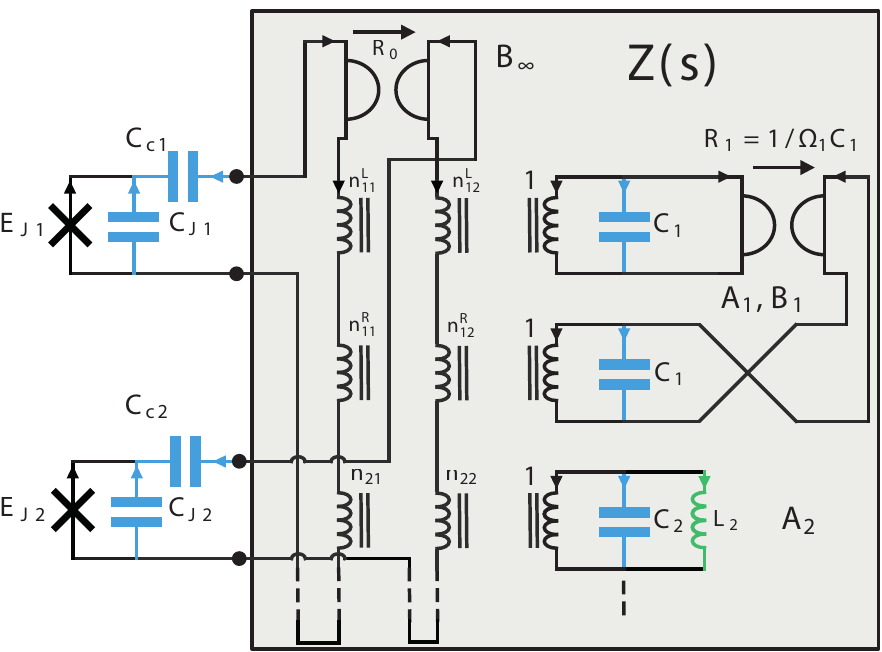}
	\caption{\label{fig:2CQ_2P_MLImpedance}Two junctions capacitively coupled to a nonreciprocal lossless impedance. Gyrator $R_0$ implements an antisymmetric pole at infinity $\msf{B}_{\infty}$. The network connected by gyrator $R_1$ yields the term $(s\msf{A}_1+\msf{B}_1)/(s^2+\Omega_1^2)$. There is a pure reciprocal stage $\msf{A}_2$. Effective tree capacitor branches are marked in red, and current directions for each branch are represented with arrows.}
\end{figure}

The first circuit consists of a 2-port nonreciprocal lossless impedance~\cite{Newcomb:1966} capacitively coupled to two charge qubits at its ports; see Fig. \ref{fig:2CQ_2P_MLImpedance}. This is a generalization of the Foster reactance-function synthesis for the 1-port reciprocal  impedance $Z(s)$, with $s=i\omega$, and a simplified version of the Brune multiport impedance expansion in Refs.~\cite{Anderson:1975,Solgun:2015}. 

A lossless multiport impedance matrix can be fraction-expanded as
\begin{equation}
\msf{Z}(s)=\msf{B}_{\infty}+s^{-1}\msf{A}_{0}+s\msf{A}_{\infty}+\sum_{k=1}^{\infty}\frac{s\msf{A}_{k}+\msf{B}_k}{s^{2}+\Omega_{k}^{2}}.\label{eq:NR_MportL_Zmat}
\end{equation}
It is easy to synthesize a lumped-element circuit that has this impedance to the desired level of accuracy; see ~\cite{Newcomb:1966}. In a lossless linear system, the $\mathsf{S}$ matrix is unitary, and therefore $\mathsf{Z}(s)=-\mathsf{Z}^\dagger(s)$ must be anti-Hermitian. If, additionally, the system is reciprocal, it must be symmetric. The only complex parameter being $s$, a lossless reciprocal impedance matrix must be odd in the variable $s$, $-\mathsf{Z}(-s)=\mathsf{Z}(s)$. Therefore, in the fraction expansion above, the $s$-odd parts correspond to reciprocal elements, while the $s$-even parts come from non-reciprocity. Thus, all  $\msf{A}$ matrices are symmetric and are implemented by reciprocal elements while $\msf{B}$ matrices are antisymmetric, and can be decomposed into networks with gyrators. $\msf{A}_0$  and $\msf{A}_\infty$ terms  correspond to the limits $L_2\rightarrow \infty$ and $C_2\rightarrow 0$, respectively, in a reciprocal stage (see Fig. \ref{fig:2CQ_2P_MLImpedance}).   $\msf{A}_\infty$ requires special treatment, but would generally be absent because of parasitic capacitors.

The general circuit implementing $\msf{Z}(s)$ contains Belevitch transformer branches~\cite{Belevitch:1950} that can be eliminated as explained above~\cite{Solgun:2015} to derive a canonical Hamiltonian. An analysis of the lossless reciprocal multiport network can be found in Ref.~\cite{ParraRodriguez:2018}. The tree and chord branch sets are divided in $\bI_{\mathrm{tr}}^T=(\bI_C^T,\bI_{T^L}^T)$ and $\bI_{\mathrm{ch}}^T=(\bI_{J}^T,\bI_L^T,\bI_{T^R}^T)$, with left (right) transformer branches being tree (chord) branches. The capacitance matrix is by construction full rank and hence invertible, \begin{equation}
\msC=\begin{pmatrix}
C_{J1}&&&&&&\\
&C_{J2}&&&&&\\
&&C_{c1}&&&&\\
&&&C_{c2}&&&\\
&&&&C_{1R}&&\\
&&&&&C_{1L}&\\
&&&&&&C_{2}\\
\end{pmatrix}.
\end{equation}
Inductive  $\msf{M}_0$ and gyration $\msf{G}$ matrices are computed using the turn ratios matrix 
\begin{equation}
\mathsf{N}=\begin{pmatrix}
n_{11}^L&0&0&n_{12}^L&0&0\\
0&n_{11}^R&0&0&n_{12}^R&0\\
0&0& n_{21}&0&0&n_{22}
\end{pmatrix}
\end{equation}
to calculate the effective loop submatrices $\msF_{CL}^{\mathrm{eff}}$, $\msF_{CG}^{\mathrm{eff}}$ in (\ref{eq:BKD_Feff}); see Appendix \ref{sec_app:NR_MportL} for an explicit form of the matrices. We recall that this analysis can be completed because the constitutive equation of the nonreciprocal elements (\ref{eq:consti_SG}) simplifies to $\bI_G=\msf{Y}_G \bV_G$, where $\bI_{G}=(I_{G0^L},I_{G0^R},I_{G1^L},I_{G1^R})^T$ and 
\begin{equation}
\msf{Y}_G=\begin{pmatrix}
\msf{Y}_{G0}&0\\0&\msf{Y}_{G1}
\end{pmatrix},
\end{equation}
with $\msf{Y}_{Gi}$ the admittance matrix for each gyrator $i\in\{0,1\}$.
\subsubsection{Hall Effect NR device}
\begin{figure}[h]
	\centering
	\includegraphics[width=.6\linewidth]{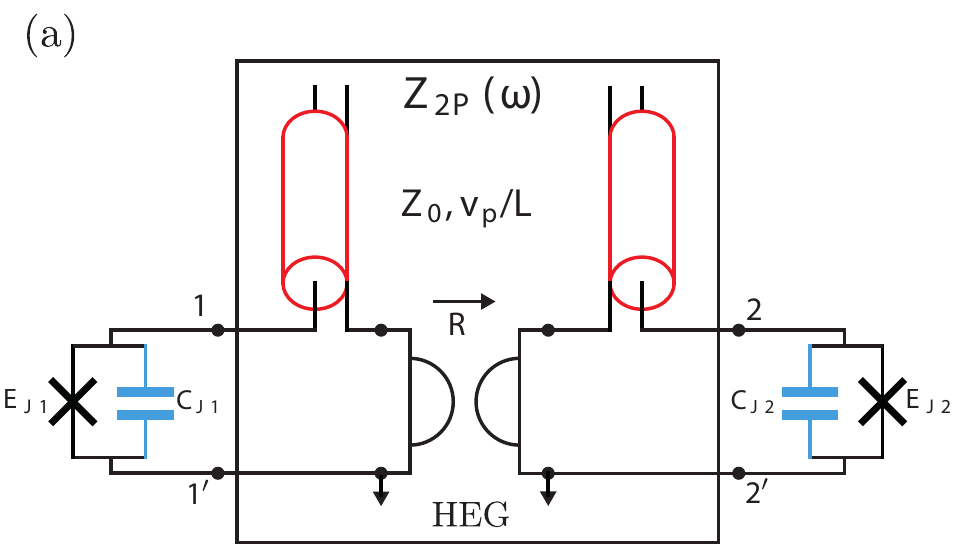}
	\vspace{.5cm}
	\includegraphics[width=.6\linewidth]{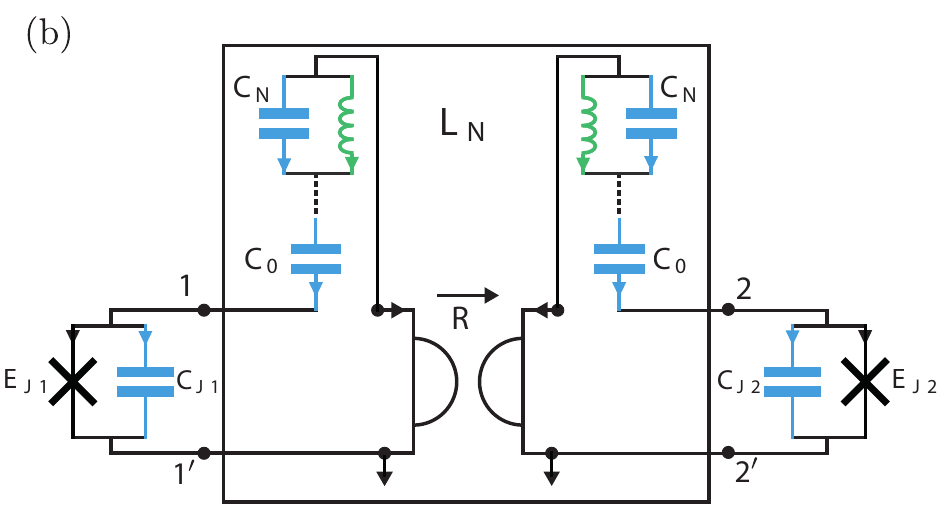}
	\caption{\label{fig:JJ_HEG_JJ} The VD Hall effect gyrator capacitively coupled to Josephson junctions. \textbf{(a) } An effective circuit of the device proposed by Viola and DiVincenzo matching the impedance response~(\ref{eqapp:Y_2P_DV}) consists of an ideal gyrator coupled to $\lambda/2$-transmission line stubs. (b) The discrete approximate circuit based on a lumped element expansion of the transmission lines~\cite{Pozar:2009} that is canonically quantized. Tree (capacitor) branches are marked in red.}
\end{figure}

The Hall effect has been proposed as instrumental in the implementation of  nonreciprocal devices. In Ref.~\cite{Viola:2014}, capacitively coupled Hall effect devices were studied by Viola and DiVincenzo in order to break time-reversal symmetry while keeping  losses  negligible. This 2-port capacitively coupled Hall bar has an impedance matrix description~\cite{Viola:2014}
\begin{equation}
\msf{Z}_{2P}(\omega)=\frac{1}{\sigma}\begin{pmatrix}
-i \cot(\omega C_L/2\sigma)&-1\\
1&-i \cot(\omega C_L/2\sigma)
\end{pmatrix},\label{eqapp:Y_2P_DV}
\end{equation}
where $\sigma$ and $C_L$ are conductance and capacitance characteristic parameters of the device, which is equivalent to that of an ideal gyrator with $R=1/\sigma$ connected in series to two $\lambda/2$-transmission line resonators of $Z_0=1/\sigma$ and $v_p/L=2\sigma/C_L$; see Fig. \ref{fig:JJ_HEG_JJ}(a). Lumped-element Foster expansions of the resonators can approximate the behavior of such a device when coupled to other lumped-element networks at its ports. This connection is achieved with lumped capacitance and inductance parameters determined by the distributed ones as $C_0=C_L/2$, $C_k=C_L/4$ and $L_k=C_L/(\sigma k \pi)^2$, for $k\in\{1,...,N\}$; see Fig. \ref{fig:JJ_HEG_JJ}(b).

We can systematically apply BKD theory and write a Lagrangian in terms of the flux branch variables of the capacitors $\bPhi_{\mathrm{tr}}^T=\bPhi_{C}^T=(\Phi_{C_{J_1}},\Phi_{C_{J_2}},\Phi_{0L},...,\Phi_{NL},\Phi_{0R},...,\Phi_{NR})$. The flux variables at the ports of the gyrators and at the tree capacitors are related by  $\bPhi_{G}=\msF_{CG}^T \bPhi_{C}$, where 
\begin{equation}
\msF_{CG}=\begin{pmatrix}
1&0\\
0&1\\
\bsb{1}_N&0\\
0&\bsb{1}_N
\end{pmatrix},
\end{equation}
with  $\bsb{1}_{N}$ an $N$-component column vector of ones. Explicitly, the three matrices describing the harmonic sector are the symmetric 
\begin{align}
\msC&=\begin{pmatrix}
C_{JL}&&&\\
&C_{JR}&&\\
&&\msC_N&\\
&&&\msC_N\\
\end{pmatrix}, \quad \mathrm{and}\quad
\msf{M}_0=\begin{pmatrix}
0&&&\\
&0&&\\
&&\msL_N^{-1}&\\
&&&\msL_N^{-1}\\
\end{pmatrix}
\end{align}
matrices, 
and the skew-symmetric nonreciprocal
\begin{equation}
\msf{G}=\frac{1}{R}\begin{pmatrix}
0&1&0&\bsb{1}_N^T\\
-1&0&-\bsb{1}_N^T&0\\
0&\bsb{1}_N&0&\bsb{1}_N\bsb{1}_N^T\\
-\bsb{1}_N&0&-\bsb{1}_N\bsb{1}_N^T&0
\end{pmatrix},
\end{equation}
where we have defined the capacitance submatrix $\msC_N=C_0\mathrm{diag}(1,1/2,...,1/2)$ and the inductance submatrix $\msL_N^{-1}=L_0^{-1}\mathrm{diag}(0,1,4,...,N^2)$,  $N$  being the number of oscillators to which we truncate the response of the resonators. Blank elements of the matrices correspond to zeros. The Hamiltonian (\ref{eq:Hamiltonian_Ydev}) can be readily computed and the canonical variables promoted to quantum operators. The diagonalization of the harmonic sector can be implemented through a symplectic transformation both before or after the quantization of variables following  Appendix \ref{App_sec:symplectic} below.
\section{Networks with $\msf{Z}$ and $\msf{S}$ NRD}
\label{sec:ZS_NRD}
The rules described above are useful to derive Hamiltonians of circuits containing ideal nonreciprocal devices characterized by a constant skew-symmetric $\msf{Y}$ matrix. However, linear systems cannot be described by admittance matrices when their $\msf{S}$ matrix has an eigenvalue $-1$. For example, ideal circulators with even (odd) number of ports,  even (odd) number of ``$-1$" entries and even (even) number of ``$1$" entries in their scattering matrix admit only $\msf{S}$-constitutive equations as in Eq. (\ref{eq:consti_SG}) (both $\msf{S}$ and $\msf{Z}$ equations)~\cite{Carlin:1964}.

\begin{figure}[h]
	\centering
	\includegraphics[width=.6\linewidth]{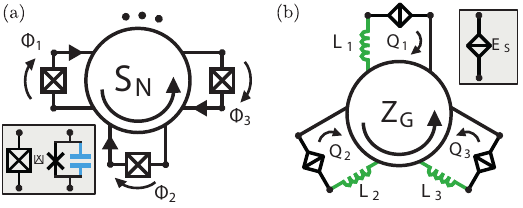}
	\caption{\label{fig:Z_S_NRCircuits} (a) $N$-port $\msf{S}$-circulator shunted by Josephson junctions. The family of $\msf{S}$ matrices of Eq. (\ref{eq:S_N_matrix}) does not have $\msf{Y}$-description, nor does it have $\msf{Z}$ for even $N$. (b) Dual circuit with a 3-port $\msf{Z}$-circulator shunted by phase-slip junctions in series with inductors.}
\end{figure}

We illustrate the problems arising when including circulators without $\msf{Y}$-descriptions with simple circuits containing 3- and 4-port circulators shunted by Josephson junctions; see Fig. \ref{fig:Z_S_NRCircuits}(a). Let us assume for concreteness that the $N$-port circulator is described by the scattering matrix
\begin{equation}
\msf{S}_{N}=(-1)^N\begin{pmatrix}
&&&1\\
1&&&\\
&\ddots&&\\
&&1&
\end{pmatrix},\label{eq:S_N_matrix}
\end{equation} 
blank elements being zero. This family of circulators cannot be assigned a $\msf{Y}$-matrix, nor do they have a $\msf{Z}$-description for even $N$. We depart from BKD and Burkard rules and choose as tree branches the circulator branches,  $\bI_{\mathrm{tr}}=\bI_G$, and capacitors and Josephson junction branches as chord branches $\bI_{\mathrm{ch}}^T=(\bI_J^T,\bI_C^T)$. Kirchhoff's laws can be simply written as $-\bI_G=\bI_C+\bI_J$ and $\bV_G=\bV_C=\bV_J=\dot{\bPhi}$, choosing $\msf{F}_{GC}=\msf{F}_{GJ}=\mathbbm{1}$. Without loss of generality and in the interest of clarity let us assume that all Josephson junctions have homogeneous capacitances $C_{i}=C$.
Introducing Kirchhoff's and constitutive equations for capacitors ($\bI_C=C \ddot{\boldsymbol{\Phi}}$) and junctions ($\bI_J=\nabla_{\bPhi}U\left(\boldsymbol{\Phi}\right)$) into (\ref{eq:consti_SG}) results in
\begin{align}
-R\left(\mathbbm{1}+\mathsf{S}\right)(C \ddot{\boldsymbol{\Phi}}+\nabla_{\bPhi}U(\boldsymbol{\Phi}))=\left(\mathbbm{1}-\mathsf{S}\right)\dot{\bPhi},\label{eq:S_circuit_eom}
\end{align}
with $\nabla_{\bPhi}U\left(\boldsymbol{\Phi}\right)=\left(U_1'(\Phi_1),U_2'(\Phi_2),...\right)^T\,$. Let $\msf{P}=\bsb{v}_{-1}\bsb{v}_{-1}^T$  be the projector onto the eigenspace of $\msf{S}$ such that $\msf{P}\msf{S}=-\msf{P}$, as it is the case for the family of matrices (\ref{eq:S_N_matrix}). Equation (\ref{eq:S_circuit_eom}) implies $\msf{P}\dot{\bPhi}=0$; i.e., there is a frozen combination of fluxes, which corresponds to a degenerate kinetic matrix that makes the Legendre transformation impossible to perform.  A simple solution is to change coordinates to single out the frozen variable from the dynamical ones, and remove it through a projection of Eqs. (\ref{eq:S_circuit_eom}) into $\msf{Q}=\mathbbm{1}-\msf{P}$. Integrating the frozen variable, we can express $\bPhi=\alpha\bsb{v}_{-1}+\sum_{n=1}^{N-1}\bsb{w}_n f_n$, where $\{\bsb{w}_n\}$ is a real basis expanding the projector $\msf{Q}$, $\alpha$ an initial-value flux constant, and $\{f_n\}$ the reduced set of degrees of freedom.   For the four-port case we have  the following systems of equations, $\bsb{v}_{-1}^T\dot{\bPhi}=0$ and 
\begin{align}
C\ddot{f}_1&=-\partial \tilde{U}_\alpha(\boldsymbol{f})/\partial f_1\label{eq:S_circuit_f2_eom}\\
C\ddot{f}_2&=R^{-1}\dot{f}_3-\partial \tilde{U}_\alpha(\boldsymbol{f})/\partial f_2\label{eq:S_circuit_f3_eom}\\
C\ddot{f}_3&=-R^{-1}\dot{f}_2-\partial \tilde{U}_\alpha(\boldsymbol{f})/\partial f_3\label{eq:S_circuit_f4_eom}
\end{align}
with the definition $\tilde{U}_\alpha(\boldsymbol{f})=U(\mathbf{\Phi}(\alpha,\bsb{f}))$ and $\boldsymbol{f}=\left(f_1,f_2,f_3\right)$. A similar system of equations can be derived for the three-port case except for (\ref{eq:S_circuit_f2_eom}), associated with eigenvalue $\lambda=1$ and only appearing in the four-port case; see Appendix \ref{sec:upperc-vari-circ} for the general $N$-port solution. Finally, the quantized Hamiltonian with fully dynamical variables is  
\begin{align}
\hat{H}&=\frac{1}{2C}\left({\hat{\bsb{Q}}}-\tfrac{1}{2}\msf{G}_{\msf{Q}}\hat{\bsb{f}}\right)^T\left(\hat{\bsb{Q}}-\tfrac{1}{2}\msf{G}_{\msf{Q}}\hat{\bsb{f}}\right)+\tilde{U}_\alpha(\hat{\boldsymbol{f}})\nonumber
\end{align}
with $\bsb{Q}=\partial L/\partial \dot{\bsb{f}}$ the conjugated charge variables, and the skew-symmetric matrix reads
\begin{equation}
\msf{G}_{\msf{Q}}=\frac{1}{R}\begin{pmatrix}
0&0&0\\	0&0&1\\	0&-1&0
\end{pmatrix}.\nonumber
\end{equation}
Had $-1$ not been  an eigenvalue of $\msf{S}$, all initial variables would have been dynamical. Generally, there is a coordinate transformation for any ideal circulator such that $\msf{G}$ is block diagonal, with  $2\!\!\times\!\!2$ blocks, and, possibly, one zero in the diagonal associated with  eigenvalue $+1$ (see Appendix \ref{sec:upperc-vari-circ}).

\section{Dual quantization in charge variables}\label{sec:dual-quant-charge}
The procedures  explained above are useful to derive Lagrangians with flux variables as \emph{positions} in a mechanical system. Equivalent descriptions of linear systems are possible with charge-position variables, with E-L voltage equations, or with a mixed combination of both flux and charge variables. Indeed,  fluxes have been used as position variables in the context of superconducting qubits because the Josephson junction has a nonlinear current-voltage constitutive equation (\ref{eq:consti_J2}). Thus,  the Lagrangian of a circuit with these elements and $\msf{Z}$ circulators in charge variables results in nonlinear kinetic terms. Although possible, dealing with such terms is usually more cumbersome. 

In recent years, the \emph{phase-slip} (PS) junction ~\cite{Mooij:2006,Arutyunov:2008}, a nonlinear low-dissipative element in charge variable, has been implemented in superconducting technology~\cite{Astafiev:2012,Peltonen:2013,Belkin:2015}. This element has a  constitutive equation  dual  to that of the Josephson junction; i.e., its voltage drop is  $V_{P}(t)=V_c\sin(\pi Q_{P}/e)$, and it is usually represented as in Fig. \ref{fig:Z_S_NRCircuits}(b) in green. Quantization of circuits with PS junctions and ideal $\msf{Z}$-NR elements in charge variables can be implemented directly, using the constitutive equation $\bV_G=\msf{Z}_G \bI_G$. For example, the circuit in Fig. \ref{fig:Z_S_NRCircuits}(b) with a $\msf{Z}_G$-circulator, the dual circuit of Fig. \ref{fig:Z_S_NRCircuits}(a),  has the dual Lagrangian interaction term $L_{G}=(1/2)\dot{\bsb{Q}}\msf{Z}_G\bsb{Q}$ and the quantum Hamiltonian
\begin{equation}
\hat{H}=\frac{1}{2}\left(\hat{\bPhi}-\tfrac{1}{2}\msf{Z}_G\hat{\bsb{Q}}\right)^T\msL^{-1}\left(\hat{\bPhi}-\tfrac{1}{2}\msf{Z}_G\hat{\bsb{Q}}\right)\nonumber+U(\hat{\bsb{Q}}),\label{eq:Hamiltonian_PS_Zdev}
\end{equation}
where $\msL$ is the diagonal inductance matrix and $U(\hat{\bsb{Q}})=-\sum_i E_{Si}\cos(\pi \hat{Q}_i/e)$. We forward the reader to Ref.~\cite{Ulrich:2016} for a systematic quantization method of circuits with loop charges~\cite{YurkeDenker:1984}.

In summary, we have presented in this chapter a general framework to quantize canonically superconducting circuits with Josephson junctions and ideal linear nonreciprocal devices. We have introduced systematic rules for quantizing classes of circuits with ideal admittance-described nonreciprocal devices in flux variables. In such a scheme we have derived the Hamiltonian of Josephson junctions capacitively coupled to both a general linear nonreciprocal 2-port black box and the Viola-DiVincenzo gyrator at its ports. These two examples show the crucial elements that we address in the general construction, and will be of interest in their own right in forthcoming experimental devices. We have given an explicit method to quantize $N$-port ideal $\msf{Z}$ and $\msf{S}$ circulators shunted by Josephson junctions in flux variables, by careful elimination of frozen variables. Finally, we discussed an extension of these procedures to quantize circuits in terms of charge variables, a dual method of special importance when dealing with circuits containing nonreciprocal elements and phase-slip junctions. On the quest to quantize generic superconducting circuits we need further work to add distributed elements, e.g., infinite transmission lines, to the analysis. That is the task left for the next chapter \ref{chapter:chapter_5}.
\chapter{Distributed and Lumped Nonreciprocal Networks}
\label{chapter:chapter_5}
\thispagestyle{chapter}
\hfill\begin{minipage}{0.85\linewidth}
{\emph{In speaking of the energy of the field, however, I wish to be understood literally. All energy is the same as mechanical energy, whether it exists in the form of motion or in that of elasticity, or in any other form. The energy in electromagnetic phenomena is mechanical energy
}}
\end{minipage}
\begin{flushright}
	\textbf{James Clerk Maxwell}\\
	A Dynamical Theory of the Electromagnetic Field
\end{flushright}
\vspace*{1cm}
Up to now, we have considered circuits with either distributed elements or nonreciprocal elements. It is thus natural to ask the question of how can those two be brought together in a complete Hamiltonian theory. In this chapter we propose a systematic procedure to quantize canonically Hamiltonians of light-matter models of transmission lines point-wise coupled through generic linear ideal circulator systems. This theory combines and generalizes all the the above chapters on canonical quantization in circuit QED with point-like coupling and nonreciprocal devices. Up to now, we have mainly used the node-flux variable description and briefly introduced the loop-charge dual variable choice for quantizing circuits. We introduce now a description in terms of \emph{both} flux and charge variables \cite{Jeltsema:2009,Ulrich:2016} that allows us to complete the separation of variables program to fullness. We make essential use of electromagnetic duality to show that the apparent redundancy is not a obstacle but an advantage, when one intends to construct Hamiltonians of transmission lines and nonreciprocal ideal devices. Furthermore, when coupled to other circuit lumped elements this construction mantains good ultraviolet behaviour  \cite{Malekakhlagh:2017,ParraRodriguez:2018}, as we exemplify at the end with a circuit containing a Josephson junction coupled through a transmission line to an ideal circulator. Finally we show progress on the quantization of frequency dependent nonreciprocal black-boxes coupled to transmission lines and nonlinear lumped elements. The main problem is reduced to the analysis of series/parallel coupling configurations of transmission lines, capacitors, inductors and ideal gyrators/circulators. Some particular cases are found to be trivial extensions of the ideal nonreciprocal coupling. A road towards the final Hamiltonian without nondynamical variables or free-particle dynamics is proposed. This theory enhances the quantum engineering toolbox to analyse and design complex superconducting circuits based on nonreciprocal elements. 

\section{Transmission lines in the doubled space}
\begin{figure}[b!]
	\centering
	\includegraphics[width=0.59\linewidth]{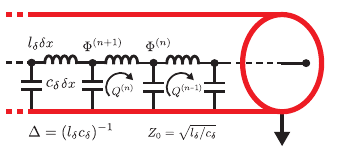}
	\caption{Differential model of a transmission line with primary parameters, i.e. inductance ($l_\delta$) and capacitance ($c_\delta$) per unit length, or secondary parameters, i.e. phase velocity ($\Delta$) and characteristic impedance ($Z_0$). Node-flux $\Phi_i$ and loop-charge $Q_i$ variables are a redundant set of degrees of freedom.}
	\label{fig:TL_ds}
\end{figure}

Transmission lines (TLs) are physical media that confine electromagnetic fields, effectively reducing the number of relevant dimensions to one \cite{Pozar:2009}. In particular, Maxwell's equations for transversal-electromagnetic modes supported inside of $N$ TLs simplify to distributed Kirchhoff's equations, commonly known as {\it telegrapher's equations} \cite{Pozar:2009}, which in the continuum limit are expressed in terms of flux $\bPhi(x_n,t)=\bPhi^{(n)}(t)$ and charge fields $\bQ(x_n,t)=\bQ^{(n)}(t)$ as 
\begin{align}
	\mcd\ddot{\bPhi}(x,t)=\dot{\bQ}'(x,t),\quad
	\mld\ddot{\bQ}(x,t)=\dot{\bPhi}'(x,t),\label{eq:telegraphers}
\end{align}
where $\mcd$, $\mld$ are macroscopic capacitance and inductance per-unit-length diagonal matrices describing the $N$ lines, see Fig. (\ref{fig:TL_ds}) for an example of one TL. In what follows the time and space variables will be implicit. We have defined the flux and charge real field vectors as  ${\bPhi}=({\Phi}_1, {\Phi}_2, ..., {\Phi}_N)^T$, and ${\bQ}=({Q}_1, {Q}_2, ..., {Q}_N)^T$ respectively. Recall that (\ref{eq:telegraphers}) are nothing but the continuous limit ($\delta x\to0$) of Eqs. (\ref{eq:LC_EOMs_doubled_space}) of chapter~\ref{chapter:chapter_1}.
For the sake of clarity, we will work through the rest of the chapter with rescaled flux and charge fields,  $\sqrt{\mcd}\bPhi\rightarrow\bPhi$ and $\bQ/\sqrt{\mcd}\rightarrow\bQ$. Then Eqs. (\ref{eq:telegraphers}) are the
Euler-Lagrange (E-L) equations for the {\it telegrapher's} Lagrangian \cite{Jeltsema:2009}
\begin{equation}
	L_{\mathrm{TG}}= \int_{\mcl{I}}dx\,\frac{1}{2}\left[\dot{\bPhi}^T\dot{\bPhi} +\dot{\bQ}^T\msDelta^{-1}\dot{\bQ}-\dot{\bQ}^T\bPhi'-\dot{\bPhi}^T\bQ'\right]\label{eq:TL_Lag_ds},
\end{equation}
where $\msDelta=\msf{c}_\delta^{-1/2}\msf{l}_\delta^{-1}\msf{c}_\delta^{-1/2}$ is the  velocity matrix, and $\mcl{I}$ is the interval of the TL, either finite or the half-line. Notice that $\bPhi$ and $\bQ$ are \emph{not} conjugate variables. Both of them are coordinates in the configuration space.

Electromagnetic duality suggests however that this description is redundant, and that one can work with either just flux or just charge fields, the counterparts appearing, now yes, as the conjugate momenta in a Hamiltonian. Under rearrangement,  Eqs. (\ref{eq:telegraphers}) become the wave equations
\begin{align}
	\ddot{\bPhi}(x,t)-\msDelta\bPhi''(x,t)&=0,\label{eq:wave_Phi}\\
	\ddot{\bQ}(x,t)-\msDelta\bQ''(x,t)	&=0,\label{eq:wave_Q}
\end{align}
as derived from the more-commonly-used Lagrangians in the superconducting quantum technologies community \cite{YurkeDenker:1984,Chakravarty:1986,Yurke:1987,Werner:1991,Devoret:1997,Paladino:2003,Blais:2004,Houck:2008,Bourassa:2009,Clerk:2010,Koch:2010,Filipp:2011,Bergenfeldt:2012,Bourassa:2012,Peropadre:2013,Sundaresan:2015,Malekakhlagh:2016,Mortensen:2016,Roy:2016,Vool:2017,Roy:2018,ParraRodriguez:2018,Malekakhlagh:2017} 
\begin{align}
	L_{\bPhi}=&\int_{\mcl{I}}dx\,\frac{1}{2}\left[\dot{\bPhi}^T\dot{\bPhi}-(\bPhi')^T\msDelta\bPhi'\right],\label{eq:TL_Lag_Phi}\\
	L_{\bQ}=&\int_{\mcl{I}}dx\,\frac{1}{2}\left[\dot{\bQ}^T\msDelta^{-1}\dot{\bQ}-(\bQ')^T\bQ'\right].\label{eq:TL_Lag_Q}
\end{align}
Although the earliest works \cite{YurkeDenker:1984,Chakravarty:1986,Yurke:1987,Werner:1991} used the charge field Lagrangian, problems arose in the addition of nonlinear elements. For instance, Yurke and Denker \cite{YurkeDenker:1984} did not derive a Hamiltonian,  and Yurke  \cite{Yurke:1987} suggested that canonical quantization be completed with Dirac's procedure \cite{Dirac:1950,Dirac:1959}. Notably, Werner and Drummond \cite{Werner:1991} derived the first canonical Hamiltonian in a mix of flux variables for discrete elements, and charge fields for transmission lines. 

However, these efforts did not have continuity, and the flux-field Lagrangian became the standard tool for  canonical quantization of superconducting circuits \cite{Devoret:1997,Paladino:2003,Blais:2004,Houck:2008,Bourassa:2009,Clerk:2010,Koch:2010,Filipp:2011,Bourassa:2012,Bergenfeldt:2012,Peropadre:2013,Nigg:2012,Solgun:2014,Sundaresan:2015,Solgun:2015,Malekakhlagh:2016,Mortensen:2016,Roy:2016,Vool:2017,Malekakhlagh:2017,Roy:2018,ParraRodriguez:2018}. Some issues arose in the precise derivation of the Hamiltonian \cite{Blais:2004,Houck:2008,Filipp:2011,Kockum:2014,Sundaresan:2015,Bourassa:2009,Peropadre:2013,Roy:2016,Vool:2017,Roy:2018}, regarding the presence or otherwise \cite{Paladino:2003,Bourassa:2009,Bergenfeldt:2012,Snyman:2015,Mortensen:2016,Gely:2017,Malekakhlagh:2017,Wiegand:2020} of divergences in Lamb-shifts and  effective couplings appeared in the literature. Correct solutions to these points have also been presented  \cite{Paladino:2003,Bourassa:2009,Bergenfeldt:2012,Gely:2017,Malekakhlagh:2017,Nigg:2012,Solgun:2014,Solgun:2015}. These divergences were actually artificial and avoidable artifacts, and the role of the flux-field Lagrangian as a predictive tool was restored, see chapter \ref{chapter:chapter_3}. However, to date the flux-field Lagrangian approach has proven unable to handle the quantization of general nonreciprocal systems.

\subsection{Reduced space operators}
Let us now review the standard separation of variables analysis (i.e. the normal mode construction for TLs) for the successful quantization of  Lagrangian (\ref{eq:TL_Lag_Phi}) in a form  relevant for our extension. The analysis of (\ref{eq:TL_Lag_Q}) is completely analogous. In essence, one looks for an expansion of the fields in terms of an orthonormal basis for the relevant Hilbert space, with the coefficients of the expansion becoming the new dynamical variables to be quantized. Such a basis is determined by  a self-adjoint differential operator ${\cal L}$ acting on a domain in the Hilbert space of multicomponent functions of position with as many components as lines, namely $N$. For clarity, we will denote the multicomponent functions of position as $\bU(x)$ while we keep the notation $\bPhi(x,t)$ for the dynamical fields. The boundary conditions of the problem at hand are part of the definition of the domain, such as, as we assume in this section, open-boundary conditions  ($\msDelta\bPhi'(0,t)\equiv\msDelta\bPhi_0'=0$) at one end of the lines. In the case of interest, the action of the  operator is
\begin{equation}
	\mL \bU=-\msDelta \bU''\,.
\end{equation}

This operator is positive, and we denote its (generalized) eigenvalues as $\omega^2$, choosing $\omega$ to be nonnegative real. As the equation is multicomponent, degeneracies might arise, depending on the symmetry of the circuit, and we use a discrete degeneracy index $\lambda$ to denote degenerate eigenstates, ${\cal L}\bU_{\omega\lambda}=\omega^2\bU_{\omega\lambda}$, orthonormalised with respect to the natural inner product. Since the differential expression and the boundary conditions of the operator ${\cal L}$ are real, the operator is real, and a real basis can always be chosen.

For definiteness we complete the description  with the flux-field Lagrangian, as the analysis for the charge one is completely analogous, see Appendix \ref{appendix_d}. We expand the flux fields in the eigenbasis,  $\bPhi=\int_{\mR_+}d\Omega\,F_{\omel}(t)\bU_{\omel}(x)$ and substitute in  the Lagrangian  to obtain $L_{\bPhi}=\frac{1}{2} \int d\Omega \left(\dot{F}_\omel^2-\omega^2F_\omel^2\right)$
where $\Omega=(\omega,\lambda)$, and  $\int d\Omega\equiv\sum_{\lambda}\int_{\mR_+} d\omega$. Now the set $F_\omel(t)$ are the new dynamical variables. There is no obstacle to carrying out the Legendre transformation, resulting in the Hamiltonian $H_{\bPhi}=\frac{1}{2}\int d\Omega \left(\Pi_\omel^2+\omega^2F_\omel^2\right)$, identical for both the flux-field and charge presentations,  where the canonical momenta are $\Pi=\partial L_{\bPhi}/\partial \dot{F}$. The process of canonical quantization is now straightforward, by promoting the conjugate pairs of variables to quantum operators, and using annihilation and creation forms one recovers the standard expression
\begin{eqnarray}
	H_{\bPhi}=\sum_{\lambda=1}^{N} \int d\omega\,\hbar \omega\,{a}_\omel^\dag {a}_\omel\equiv H_{\bQ},\label{eq:TL_LU_Ham}
\end{eqnarray}
where we have discarded the infinite constant term  $\int d\Omega\frac{1}{2}$, and made  the degeneracy index sum explicit again. 

\subsection{Double space operator}
Let us now apply the same procedure to the  telegrapher's Lagrangian (\ref{eq:TL_Lag_ds}). Crucially, this involves a \emph{doubled}
space, with $\bW$ the new multicomponent functions, now with $2N$ components. For later convenience we arrange $\bW$ as a doublet of two $N$ component functions, $\bU$ and $\bV$, corresponding respectively to fluxes and charges.  We know however that only $N$ components can be physical. Considering now the operator defined by ${\cal L}\bW=-\msDelta\bW''$ on the elements of its domain. The compact notation signifies that $\msDelta$ applies to both $\bU$ and $\bV$ components. The boundary conditions, for simplicity, will again be open boundary conditions at one end of the lines, $\msDelta\bU'_0=\bV_0=0$.

This new operator is again positive and real, and we shall use the corresponding real eigenbasis $\bW_\ome$, (generalised) orthonormal with respect to the inner product
\begin{eqnarray}
	\langle\bW_1,\bW_2\rangle=\int_{\mcl{I}} dx\, \left[\bU_1^T  \bU_2+\bV^T_1\msDelta^{-1}\bV_2\right]\,.\label{eq:Inner_product_ds}
\end{eqnarray}
The degeneracy index, which we will denote by $\epsilon,\epsilon'$, now runs up to $2N$.

\subsection*{Telegrapher's symmetry}
We  use  the electromagnetic duality symmetry  given by the exchange of electric and magnetic fields to show that the redundant description is actually identical to the previous one. As is well known, that duality is not local when expressed in terms of the fundamental potential vector field, and we should not expect that here it be a mere rotation of flux and charge fields.  In fact,  we introduce the {\it telegrapher's} operator $\mT$, acting on the elements of the domain of ${\cal L}$ as
\begin{equation}
	\mT\bW=-i\begin{pmatrix}
		\bV'\\\msDelta\bU'
	\end{pmatrix}.\label{eq:TL_T_op_def}
\end{equation}

This is a discrete symmetry that commutes with the fundamental operator ${\cal L}$.  Furthermore,  $\mT^2={\cal L}$, and its spectrum is given by $\pm i\omega$. However, it is purely imaginary and, therefore, its action on the real basis $\bW_\ome$ cannot be merely a sign. Alternatively, notice that it interchanges non trivially flux and charge components. Thus, on each degeneracy eigenspace it will act as $\mT\bW_\ome=\omega \mt_{\epsilon\epsilon'}\bW_{\omega\epsilon'}$. The matrix $\mt$  must be therefore imaginary and idempotent. We conclude that, by  rearranging the real basis in each degeneracy subspace, it can always be written as $\mt=-\sigma_y\otimes\mone_N$, with $\mone_N$ the identity matrix on $N$ dimensional vectors, see Appendix \ref{appendix_d}.

\subsection{Lagrangian and Hamiltonian in double space}
We now have all the necessary tools to complete the analysis in the doubled case. Expand the flux and charge doublet in the orthonormal basis $\bW_\ome$,
\begin{equation}
	\label{eq:Phi_Q_expansion}
	\begin{pmatrix}
		\bPhi\\ \bQ 
	\end{pmatrix}=\int d\Omega\,X_\ome\bW_\ome=\int d\Omega\,X_{\ome}\begin{pmatrix}
		\bU\\\bV
	\end{pmatrix}_{\ome}\,,
\end{equation}
and rewrite the Lagrangian (\ref{eq:TL_Lag_ds})  as 
\begin{eqnarray}
	L=\frac{1}{2}\int d\Omega\, \dot{X}_{\ome}(\dot{X}_{\ome}-i\omega\mt_{\epsilon'\epsilon}X_{\omega\epsilon'}).\label{eq:Lag_TL_modes_ds}
\end{eqnarray}
Now the dynamical variables are $X_\ome$, as expected. Notice that the matrix $\mt$ that implements the telegrapher's symmetry appears explicitly in the Lagrangian. It now behoves us to prove that, albeit different at a first glance from $L_{\bPhi}$ above, equation (\ref{eq:Lag_TL_modes_ds}) produces equivalent dynamics for the flux and charge variables.

On writing $\mt$ as $-\sigma_y\otimes\mone_N$, it is convenient to rearrange the degeneracy indices accordingly, with $u/v$ being the indices for the action of the Pauli matrix, while the $N$ dimensional factor corresponds to an index $\lambda,\lambda'$. In other words, the single degeneracy index $\epsilon$, ranging from 1 to $2N$, is replaced by a doublet $\alpha\lambda$, with $\alpha$ taking values $u$ and $v$, and $\lambda$ from 1 to $N$. Let us denote $X_{\omega(u\lambda)}$ by $F_{\omega\lambda}$ and $X_{\omega(v\lambda)}$ by $G_\omel$.
The Lagrangian is rewritten as 
\begin{eqnarray}
	L=\frac{1}{2}\int d\Omega\,\left[ \dot{F}_{\omel}^2+\dot{G}_{\omel}^2+\omega\left(\dot{G}_\omel  F_{\omel}-\dot{F}_\omel G_\omel\right)\right],\nonumber
\end{eqnarray}
where now the implicit sum runs over $\int d\Omega\equiv\sum_{\lambda=1}^{N}\int d\omega$.
Notice two salient facts: the kinetic term is nondegenerate, and this Lagrangian is amenable to Legendre transform to provide us with a Hamiltonian, first, and, second, the term in parenthesis is of magnetic nature. Its antisymmetry is directly inherited from the antisymmetry of $\mt$, hence from the telegrapher's duality symmetry. In fact, the Hamiltonian reads
\begin{eqnarray}
	H=\frac{1}{2}\int d\Omega\,\left[\left(\Pi_\omel+\frac{\omega}{2}G_{\omel}\right)^2+\left(P_\omel-\frac{\omega}{2}F_{\omel}\right)^2\right],\nonumber
\end{eqnarray}
with  two pairs of conjugated variables with Poisson brackets $\{F_{\omel},\Pi_{\omega'\lambda'},\}=\{G_{\omel},P_{\omega'\lambda'},\}=\delta_{\omega\omega'}\delta_{\lambda\lambda'}$. The identification of the structure as being magnetic reveals to us the dynamical content of this Hamiltonian. Namely, under the canonical transformation $\tilde{F}_\omel=\frac{1}{2}F_\omel-\frac{1}{\omega}P_\omel$, $\tilde{\Pi}_\omel=\Pi_\omel+\frac{\omega}{2}G_\omel$, $\tilde{G}_\omel=\frac{1}{2}G_\omel-\frac{1}{\omega}\Pi_\omel$ and $\tilde{P}_\omel=P_\omel+\frac{\omega}{2}F_\omel$, 
the Hamiltonian reaches its final form
\begin{align}
	H&=\frac{1}{2}\int d\Omega\,\left(\tilde{\Pi}_\omel^2+\omega\tilde{F}_{\omel}^2\right)\equiv_{\mathrm{q.}}\sum_{\lambda}^{N}\int d\omega\, \hbar\omega\, a_{\omel}^\dag a_\omel\,.\label{eq:H_quantized_ds_modes}
\end{align}
Although at first sight in (\ref{eq:Lag_TL_modes_ds}) there are $2N$ modes in each degeneracy space, our treatment reveals that half of those are \emph{nondynamical} ($\tilde{G}_\omel$ and $\tilde{P}_\omel$), in that they have no evolution under the physical Hamiltonian (\ref{eq:H_quantized_ds_modes}). Once this result has been achieved, quantization follows in the same manner as before. 

In summary, to this point we have shown that the telegrapher's Lagrangian (\ref{eq:TL_Lag_ds}) is amenable to quantization, and that it controls the correct number of degrees of freedom. The most standard way to quantize circuits, nicely summarized in Devoret's \cite{Devoret:1997}, begins by writing a set of equations of motion (EOMs), and continues by finding a Lagrangian whose E-L equations correspond to. Notice that if we were to start from the telegrapher's equations we would first need to reduce them to the wave equation in the standard (reduced) approach and then understand those as the E-L equations of the Lagrangians presented as $L_{\bPhi}$ and $L_{\bQ}$ in (\ref{eq:TL_Lag_Phi}) and (\ref{eq:TL_Lag_Q}). We are following the same route: we start directly from the telegrapher's equations and understand them as E-L for $L_{\mathrm{TG}}$. The redundancy of the telegrapher's equations is not made to disappear in  the EOMs but rather in the Hamiltonian. This we have achieved by introducing a systematic analysis, based on the telegrapher's duality operator. It is actually not necessary to use the doubled space for the description of the systems above. We shall however use this systematic procedure to handle the quantization of a system for which it will prove crucial, with inclusion of ideal nonreciprocal elements.

\section{Connection to ideal circulators}
\begin{figure}[h]
	\centering
	\includegraphics[width=0.33\linewidth]{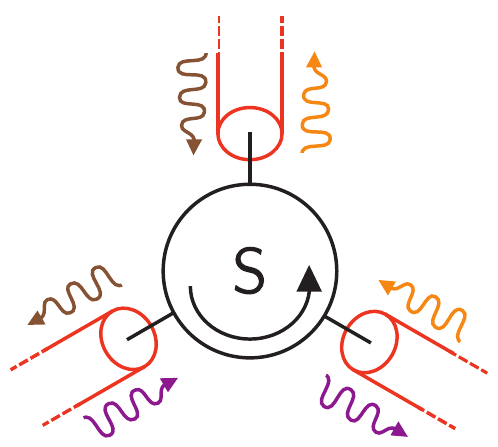}
	\caption{Transmission lines coupled to an ideal circulator. We assume a common ground plane for the lines and circulator ports.}
	\label{fig:S_3ports}
\end{figure}

Let us briefly recap the fundamentals of linear lossless nonreciprocal devices \cite{Duinker:1959,Tellegen:1948,Carlin:1964} from previous chapter \ref{chapter:chapter_4} , and explain their inclusion in Lagrangian models with transmission lines, e.g. a 3-port circulator connected directly to three TLs as in Fig. (\ref{fig:S_3ports}).
A linear ideal NR system can be generically described by a unitary, non-symmetric (frequency-independent) scattering matrix $\mS$. This matrix relates the amplitude and phase of input and output signals $\bsb{b}^{\mathrm{out}}=\msf{S}\bsb{b}^{\mathrm{in}}$, or voltages and currents at the ports
\begin{equation}
	(1-\msf{S})\dot{\bPhi}_0=R(1+\msf{S})\dot{\bQ}_0,\label{eq:NR_constitutive}
\end{equation}
where $R>0$ is a reference resistance, and $\mS$ is a constant matrix. We assume here and in the rest of the chapter that all frequency-dependence of $\mS(\omega)$ could have been extracted in a network of capacitors and inductors  \cite{Carlin:1964,Newcomb:1966}. Such more involved networks require more careful treatment of the double-space variables as we will show below, and we restrict here the analysis to ideal NR devices. The constitutive Eq. (\ref{eq:NR_constitutive}) can be simplified to admittance $\dot{\bQ}_0=\bar{\mY}\dot{\bPhi}_0$ and impedance $\bar{\mZ}\dot{\bQ}_0=\dot{\bPhi}_0$ equations when neither $+1$ nor $-1$ are eigenvalues of $\mS$ \cite{Carlin:1964}, where $\bar{\mY}$ and $\bar{\mZ}$ are skew-symmetric real matrices. We consider here nonreciprocal systems with an immittance description, see Appendix \ref{appendix_d} for the discussion of degenerate cases.
\subsection{Obstacles in the reduced description}
As we shall now show, there is a fundamental obstacle in treating  systems of transmission lines coupled with  ideal nonreciprocal elements just using flux variables. In order to make the issue clearer, consider first transmission lines capacitively coupled to other degrees of freedom, using flux variables.  The conservation of charge at the coupling leads to $\mathsf{\Delta}\boldsymbol{\Phi}_0'= \mA\left(\ddot{\boldsymbol{\Phi}}_0-\ddot{\boldsymbol{\Phi}}_L\right)$, where $\mA=\mathsf{c}_\delta^{-1/2}\mathsf{C}\mathsf{c}_\delta^{-1/2}$ is a (rescaled) capacitance coupling matrix and $\boldsymbol{\Phi}_L$ denotes other, discrete, degrees of freedom. Clearly, unless there is perfect cancellation of $\ddot{\boldsymbol{\Phi}}_0$ and  $\ddot{\boldsymbol{\Phi}}_L$ at all times, no normal mode analysis is feasible with standard Sturm--Liouville boundary conditions for the wave equation part, as made explicit in chapter \ref{chapter:chapter_3}.  In this type of situation an analysis in terms of the (reduced) system, relying on just flux variables, is indeed possible by extending the type of boundary conditions, operator, and inner product under consideration.

Let us now consider transmission lines coupled in one end through an ideal (frequency independent) nonreciprocal element with an admittance presentation,  and their description with flux variables. The Lagrangian is
\begin{equation}
	\label{eq:simpleYflux}
	L=L_{\boldsymbol{\Phi}}+\frac{1}{2}\dot{\boldsymbol{\Phi}}_0^T\mathsf{Y}\boldsymbol{\Phi}_0\,,
\end{equation}
with $\mathsf{Y}$ the skew-symmetric  rescaled admittance matrix, $\mathsf{Y}=\mathsf{c}_\delta^{-1/2}\bar{\mathsf{Y}}\mathsf{c}_\delta^{-1/2}$. The Euler--Lagrange equations derived from this Lagrangian are given by the wave equations (\ref{eq:wave_Phi}) and the conservation of charge at the end
\begin{eqnarray}
	\label{eq:ELsimpleYflux}
	\mathsf{\Delta}\boldsymbol{\Phi}'_0&=& \mathsf{Y}\dot{\boldsymbol{\Phi}}_0\,.
\end{eqnarray}
This is a well posed problem, and explicit solutions can be found for particular configurations. For instance, a system with two lines, identity velocity matrix and $\mathsf{Y}=i\sigma^y$ rescaled admittance has the general solution
\begin{equation}
	\label{eq:example}
	\boldsymbol{\Phi}\to
	\begin{pmatrix}
		f(t-x)+g(t+x)\\ g(t-x)-f(t+x)
	\end{pmatrix}
	\,.
\end{equation}
By inspection we notice that it is dynamically equivalent to a wave equation in one dimension. The issue now is to develop a systematic procedure to construct canonical variables and a Hamiltonian amenable to canonical quantization. In order to do so, we  look for normal modes and the corresponding expansion. It immediately becomes clear that no Sturm--Liouville boundary conditions are available to us in this flux presentation, since the charge conservation equation $\mathsf{\Delta}\boldsymbol{\Phi}'_0= \mathsf{Y}\dot{\boldsymbol{\Phi}}_0$ involves the time derivative, which would lead, under separation of variables, to a frequency dependent boundary condition for the normal form, $\mathsf{\Delta}\boldsymbol{U}'=-i\omega \mathsf{Y}\boldsymbol{U}$. The technique used in chapter \ref{chapter:chapter_3} to address a similar issue in capacitive coupling, namely the consideration of a radically different kind of self-adjoint operator, is now not available to us, because of the skew-symmetry of the admittance, as opposed to the symmetry of the capacitance matrix $\mA$ presented above. That symmetry suggested an extension of the Hilbert space, which, of course, includes the definition of an inner product which involved this symmetric matrix. No such avenue is opened to us by the admittance matrix.

There is an alternative argument that relies on the mathematical literature and the breaking of time reversal invariance. Consider that the system under description does indeed break time reversal symmetry, in such a way that this breaking is localized on a boundary. This is the case for the Lagrangian \eqref{eq:simpleYflux}. This implies that the charge conservation law at the boundary will, under separation of variables, present a linear term in frequency. On the other hand, the eigenvalues of the relevant second order differential operator, were it to exist, depend on frequency as $\omega^2$. The mathematical literature for eigenvalue problems in which the boundary condition involves the eigenvalue, following the seminal work of Walter \cite{Walter:1973} and Fulton  \cite{Fulton:1977}, provides us with self-adjoint operators if the boundary condition depends \emph{linearly} on the eigenvalue. If time reversal symmetry is broken on the boundary, however, the boundary condition depends on the square root of the eigenvalue. Therefore the construction we relied on for capacitive coupling (in the flux description) is no longer available.

This set of arguments demonstrates that a description only in terms of fluxes is inadequate to provide us with the systematic canonical quantization we desire. This does not mean, at all, that canonical quantization cannot be achieved. For instance, direct inspection of Eq. \eqref{eq:example} shows that the second component can be understood as conjugate to the first one. Indeed, in that simple situation a Sturm--Liouville separation of variables is accessible by describing one line in fluxes and the second line in charges, and the procedure provides us with the desired result. This comes about because charge conservation reads in this case, in terms of components,
\begin{eqnarray}
	\delta_1\Phi_1'(0,t)&=& \mathsf{Y}_{12}Q'_2(0,t)\,,\label{eq:firsttwist}\\
	\mathsf{Y}_{21}\dot{\Phi}_1(0,t)&=&\dot{Q}_2(0,t)\,.\label{eq:sectwist}  
\end{eqnarray}
Redefining space-time units so as to have $\delta_1=1$ and $\mathsf{Y}_{12}=-\mathsf{Y}_{21}=1$, in correspondence to example above, Eq. \eqref{eq:example}, we  define a  differential operator $\mathcal{L}$ acting on the doublet $\bW=
\begin{pmatrix}
	U&V
\end{pmatrix}^T$ as $\mL \bW=-\bW''$, with boundary conditions $U(0)=-V(0)$ and $U'(0)=V'(0)$. It is self-adjoint when using the standard inner product, and expansion on the corresponding basis leads directly to diagonalized quantization.

In generalizing this idea to a larger number of transmission lines, the assignment of flux or charge character to individual transmission lines is not straightforward: it would involve identiying the canonical form of the skew-symmetric matrix $\mY$ and changing variables accordingly. If  couplings with  capacitive and inductive aspects are also taken into account, no direct assignment of flux or charge character to individual transmission lines is productive either.

\subsection{Solution in the doubled space}
Having looked into the obstacles associated with the description in terms only of fluxes, we present a general solution based on the doubled space description. Let us consider an $N$-port nonreciprocal element connected to $N$ semi-infinite lines as in Fig. (\ref{fig:S_3ports}). The Lagrangian of the system in the doubled space can be written as 
\begin{equation}
	L = L_{\mathrm{TG}}+\frac{1}{2}\left\{
	\begin{aligned}
		&\dot{\bQ}_0^T\bPhi_0+ \dot{\bPhi}_0^T\msf{Y}\bPhi_0,\quad\\
		&\dot{\bPhi}_0^T\bQ_0+ \dot{\bQ}_0^T\msf{Z}\bQ_0,\quad
	\end{aligned}
	\right. \label{eq:TL_YZ_Lag_ds}
\end{equation}
working with rescaled flux and charge fields, and matrices $\mY$ and $\mZ=\sqrt{\mcd}\bar{\mZ}\sqrt{\mcd}$. The constitutive equation (\ref{eq:NR_constitutive}) written in admittance form is the Euler-Lagrange boundary condition equation in the corresponding (\ref{eq:TL_YZ_Lag_ds}). Since the boundary condition involves both fluxes and charges, it seems natural to address the issue also with the doubled approach, with  necessary adaptations for this condition. We concentrate here on the admittance case, as the impedance one is analogous. The relevant operator ${\cal L}$ acts on a doublet $\bW$ of $N$ component functions, $\bU$ and $\bV$, belonging to a domain restricted by the conditions
\begin{equation}
	\label{eq:Domain_L_Y}
	\bV_0 = \mY \bU_0,\qquad \mathrm{and}\qquad\msDelta\bU'_0=\mY \bV'_0\,,
\end{equation}
and, as before, 
\begin{equation}
	\mL\bW=-\begin{pmatrix}
		\msDelta\bU''\\\msDelta\bV''
	\end{pmatrix}.
\end{equation}
The corresponding inner product is again (\ref{eq:Inner_product_ds}), and the telegrapher's duality symmetry, defined as before, but now on the current domain, mantains its crucial properties, all the way to its representation by means of $\mt$ and its properties.

Since the boundary  terms cancel on expanding the Lagrangian \eqref{eq:TL_YZ_Lag_ds} in an eigenbasis $\bW_\ome$ of ${\cal L}$, \begin{equation}
	\dot{\bQ}_0^T\bPhi_0+ \dot{\bPhi}_0^T\msf{Y}\bPhi_0=	\int d\Omega d\Omega' \dot{X}_{\ome}\bU_{\ome}^T(0)(\mY^T+\mY)\bU_{\omep}(0)X_{\omep}=0,\nonumber
\end{equation}
due to $\mY$ being skew-symmetric, the same steps as before lead us to a properly quantized Hamiltonian, of the form of Eq. \eqref{eq:H_quantized_ds_modes}, and, as before, there are  $N$ degrees of freedom per frequency mode. Again, even if we have introduced an apparent redundancy in the description of the system, there is a clear and systematic identification of the dynamical variables, and the number of these is the correct one.

In summary, we have presented a systematic quantization procedure for ideal nonreciprocal devices coupled to transmission lines, by using the doubled presentation of the telegrapher's equation and discarding the uncoupled nondynamical sector.

\section{Nonlinear networks}
\begin{figure}[]
	\centering
	\includegraphics[width=.6\linewidth]{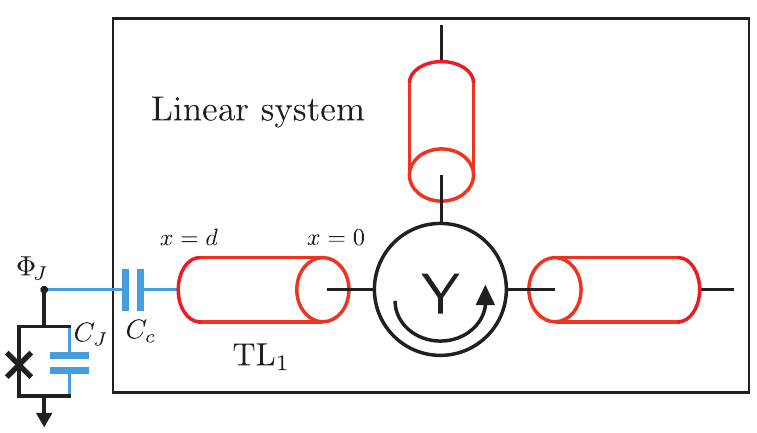}
	\caption{Josephson junction connected to transmission lines resonators and a $\mY$-circulator. The linear sector of the Hamiltonian is described by an infinite set of harmonic oscillators.}
	\label{fig:FL_NL_network}
\end{figure}

In order to show the power of our approach, we now expand the construction presented in previous chapter \ref{chapter:chapter_3} to describe also networks of ideal circulators connected to finite-length transmission lines with capacitive connections to Josephson junctions as in Fig. \ref{fig:FL_NL_network}. Particular as this example might be, it still captures all the essential difficulties of much more general situations. More concretely, similar constructions to what we are now going to present would apply to capacitive/inductive insertions in the line or connection to more general meshes like in the chapters above.

In addition to the boundary conditions (\ref{eq:Domain_L_Y}) at the circulator, the analysis of the circuit in Fig. \ref{fig:FL_NL_network} requires imposing current conservation at the $x=d$ endpoint of the first transmission line, $\mathrm{TL}_1$. This extra condition is not of Sturm--Liouville type, however, and, as shown in chapter \ref{chapter:chapter_3}, it is necessary to consider an operator that does not act merely on the multicomponent Hilbert space of the lines, but rather on the direct sum of that Hilbert space and a boundary finite dimension Hilbert space, such that its elements are of the form $\mW=\left(\bW,w\right)$. In the case at hand $w$ is a real number. In this expanded space, with the corresponding inner product, ${\cal L}$ acts on its domain, defined by the boundary conditions (\ref{eq:Domain_L_Y}) and $(\msDelta\bU'_d)_\perp=(\bV_d)_\perp=0$ together with the restriction $w=\alpha(\bsb{n}\cdot \bU_d)$, as $\mL\mW=(-\msDelta\bW'', -(\bsb{n}\cdot\msDelta\bU'_d))$, where $\bsb{n}$ is an $N$ dimensional vector and $\perp$ denotes orthogonal to $\bsb{n}$. $\alpha$ is a free parameter which can be optimally set to the value $C_s/c=C_c C_J/[c(C_c + C_J)]$, in order that the Hamiltonian will not have TL mode-mode couplings ($A^2$ diamagnetic term), see chapter \ref{chapter:chapter_3} and Appendix \ref{appendix_d} for further details. For simplicity, we have assumed open boundary conditions in the endpoints of the other lines ($x=d$).

Following the same procedure as before, namely expansion in the eigenbasis of this operator to obtain a Lagrangian in new dynamical variables, for which Legendre transformation is well defined, and the identification of the dynamical sector of proper dimension (equivalent elimination of nondynamical variables to previous section), we derive the quantum Hamiltonian 
\begin{align}
	H=\frac{Q_J^2}{2C_J}-E_J\cos(\varphi_J)+\sum_{n,\lambda}^{\infty,N} \hbar\omega_n\, a_{n\lambda}^\dag a_{n\lambda} +\xi Q_J\sum_{n,\lambda}\left[r_{n\lambda} a_{n\lambda}+r^*_{n\lambda} a_{n\lambda}^\dagger\right],\nonumber
\end{align}
where $r_{n\lambda}=\sqrt{\frac{\hbar\omega_n}{2}}  (u_{n u\lambda}+i u_{n v\lambda})$, with $u_{n\epsilon}=\bsb{n}\cdot(\bU_{n\epsilon})_d$  the eigenfunctions at the coupling point $\Phi_1(x, t)=\sum_{n,\epsilon}X_{n\epsilon}(t)(\bsb{n}\cdot\bU_{n\epsilon}(x))$, and $\xi=\frac{C_c}{(C_c+C_J)\sqrt{c_\delta}}$. As suggested above,  the computation of Lamb-shifts for this Hamiltonian, which will be proportional to $\chi\propto \sum_{n,\lambda}|r_n|^2/\omega_n=\hbar/(2\alpha)<\infty$, proves to be convergent, since, as previously shown in previous chapters \ref{chapter:chapter_2} and \ref{chapter:chapter_3} for reciprocal networks, $u_{n\epsilon}\propto 1/n$ when $\omega_n\rightarrow\infty$, see Appendix \ref{appendix_d}. 

To our knowledge, this is the first Hamiltonian description combining transmission lines, nonreciprocal elements and Josephson junctions in an exact manner. In contrast to the procedures described in chapter \ref{chapter:chapter_4} for pure lumped-element networks with nonreciprocal elements, the methods here take advantage of the inherent properties of the transmission lines, and allow us to construct a unique eigenvalue problem (for a self-adjoint operator) to not just quantize but further diagonalize the linear sector even in the presence of the nonreciprocal elements. Furthermore, one may safely take the infinite-length limit of the lines (continuum spectrum) and still find meaningful Hamiltonians predicting divergence-free Lamb shifts or effective couplings. 
\section{Towards a generic linear boundary condition}
\begin{figure}[h]
	\centering
	\includegraphics[width=.7\linewidth]{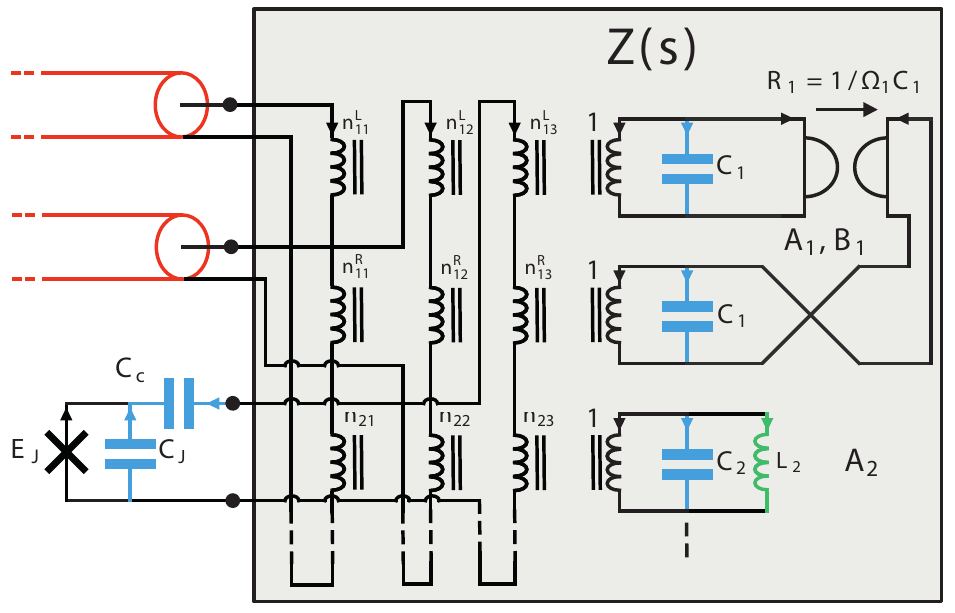}
	\caption{Josephson junction connected to a transmission line through a 3-port nonreciprocal impedance $\mZ(\omega)$ black-box.}
	\label{fig:1TL_1JJcCoupled_2port_Z}
\end{figure}

We are going to finish this chapter with a first approach to the analysis of the generic circuit presented in the introduction containing transmission lines, a linear nonreciprocal (impedance) black-box and a Josephson junction, see Fig. \ref{fig:1TL_1JJcCoupled_2port_Z}. This is a minimal generalization of  Fig. \ref{fig:2CQ_2P_MLImpedance} which contains the fundamental issues for quantizing electric circuits, i.e. two continuum infinite-dimensional systems (semi-infinite transmission lines with homogeneous $c_\delta$ and $l_\delta$), a truncated discrete infinite-dimensional nonreciprocal linear system (the black-box) and a nonlinear degree of freedom (the junction).
For the sake of concreteness and simplicity, let us consider a 3-port impedance that can be decomposed in one nonreciprocal and one reciprocal pole; 
\begin{equation}
	\msZ(s)=\sum_k \frac{s \mA_k +\mB_k}{s^2 +\Omega_k^2}= \frac{s \mA_1 +\mB_1}{s^2 +\Omega_1^2}+ \frac{s \mA_2}{s^2 +\Omega_2^2},
\end{equation}
with $s=i\omega$. As we have seen in previous chapters, the Belevitch transformer imposes two sets of constraints between voltages and currents on the right and left of its ports (\ref{eq:Belevitch_trR1}) that can be eliminated to write the (Euler-Lagrange) current equations
\begin{equation}
	\begin{pmatrix}
	\dot{\bQ}_0\\\frac{-\partial U(\Phi_J)}{\partial \Phi_J}
	\end{pmatrix}=\tilde{\msA}\begin{pmatrix}
	\ddot{\bPhi}_0\\\ddot{\Phi}_J
\end{pmatrix}+\tilde{\mY}\begin{pmatrix}
\dot{\bPhi}_0\\\dot{\Phi}_J
\end{pmatrix}+\tilde{\msB}^{-1}\begin{pmatrix}
\bPhi_0\\\Phi_J
\end{pmatrix},\label{eq:E_L_2TL_J_NRBox}
\end{equation}
where we have defined the matrices
\begin{align}
	\tilde{\msA}&=\frac{1}{c_\delta}(\msLambda \mN)^{-1}\msC(\mN^T\msLambda ^T)^{-1}+\begin{pmatrix}
		0&0&0\\0&0&0\\0&0&C_J/c_\delta
	\end{pmatrix},\nonumber\\
	\tilde{\mY}&=\frac{1}{c_\delta}(\msLambda \mN)^{-1}\bar{\mY}(\mN^T\msLambda ^T)^{-1},\nonumber\\
	\tilde{\msB}^{-1}&=\frac{1}{c_\delta}(\msLambda \mN)^{-1}\msL^{-1}(\mN^T\msLambda ^T)^{-1},\nonumber
\end{align}
and the voltages on the left of the transformer are related to the transmission lines and junction drops by $\msLambda\bV_L=(\dot{\bPhi}_0^T, \dot{\Phi}_J)^T$. Realize that in order to perform the full elimination of the internal degrees of freedom of the box, we have assumed the matrix $(\msLambda\msN)$ to be invertible. More general cases, e.g. with more discrete stages, will require the introduction of additional degrees of freedom. The Lagrangian for the above coupled equations (\ref{eq:E_L_2TL_J_NRBox}) is 
\begin{align}
	L=L_{\text{TG}}+\frac{1}{2}\left(\dot{\bQ}_0^T\bPhi_0 +\dot{\bPsi}^T\tilde{\msA}\dot{\bPsi}+ \bPsi^T\tilde{\msB}^{-1}\bPsi-\dot{\bPsi}^T\tilde{\mY}\bPsi\right)-U(\Phi_J),
\end{align}
with $\bPsi=(\bPhi_0^T, \Phi_J)^T$. Notice that the tilde matrices contain the constraints of the multiport transformer. The difficulty to quantize this system resides in the involved boundary condition for the transmission lines. We section the analysis required to derive an exact Hamiltonian for this circuit by studying the generalized parallel (this case) and series coupling types of boundary conditions. 
\subsection{Parallel configuration}
\label{Sec:parallel}
\begin{figure}[h]
	\centering
	\includegraphics[width=0.43\linewidth]{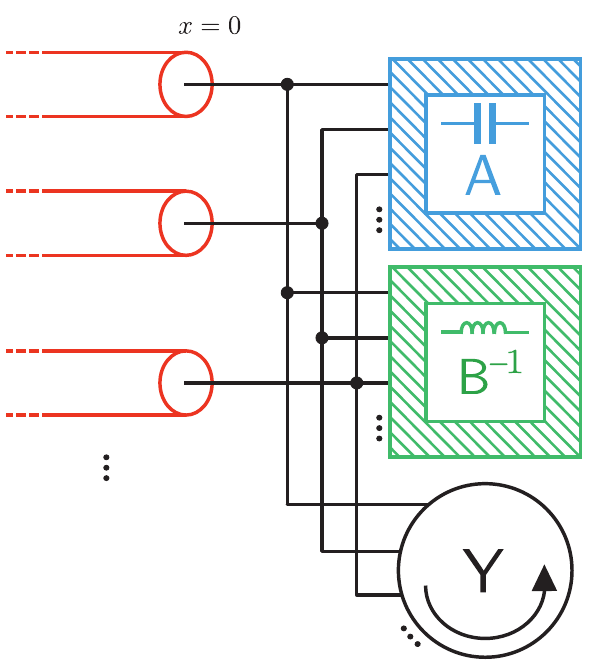}
	\caption{Transmission lines coupled in a parallel configuration to lumped networks of capacitors $(\mA)$, inductors $(\mB)$ and admittance-described circulator $\mY$.}
	\label{fig:Parallel_config}
\end{figure}

In the previous subsection we have considered only two transmission lines for the sake of clarity. Now, we consider the more general problem of $N$ semi-infinite transmission lines (half lines) connected in a parallel configuration to lumped networks of capacitors and inductors, described by the symmetric (and rescaled) matrices $\mA=\mcd^{-1/2}\msf{C}\mcd^{-1/2}$, and $\mB^{-1}=\mcd^{-1/2}\msf{L}^{-1}\mcd^{-1/2}$ respectively, see Fig. (\ref{fig:Parallel_config}). As previously advanced, an ideal nonreciprocal element described by an admittance matrix $\mY$ enters more naturally in this description and we refer to Appendix \ref{appendix_d} for degenerate cases when it does not exist, as well as for cases where the capacitance or inductive matrices are singular.
A Lagrangian describing the circuit in Fig. (\ref{fig:Parallel_config}) reads 
\begin{align}
	L &= L_{\mathrm{TG}}+\frac{1}{2}\left(\dot{\bQ}_0^T\bPhi_0+\dot{\bPhi}_0^T\mA\dot{\bPhi}_0-{\bPhi}_0^T\mB^{-1}{\bPhi}_0+\dot{\bPhi}_0^T\msf{Y}\bPhi_0\right) \label{eq:Lag_TL_ABY_ds}.
\end{align}
The lumped element terms are minimally described by flux variables when they share the same nodes. The Euler-Lagrangian equation at the boundary becomes
\begin{align}
	\dot{\bQ}_0&=\mA\ddot{\bPhi}_0+\mB^{-1}{\bPhi}_0+\msf{Y}\dot{\bPhi}_0,\label{eq:PC_BC_current_law}
\end{align}
which corresponds to the current flowing out of the line being equal to the sum of currents for all lumped element components at the common nodes. It must be remarked that the networks do not contain internal free nodes, and thus any voltage or current can always be written in terms of all the flux variables at the boundary of the lines. As in previous chapter \ref{chapter:chapter_3}, the capacitive (inductive) coupling boundary condition requires the use of an eigenbasis in an enlarged Hilbert space for flux (charge) field descriptions. As we are using here the doubled description in flux and charge fields, we will find an eigenbasis of $\mcl{H}=(L^2(\mR_+)\otimes\mathbbm{C}^{2N})\oplus\mathbbm{C}^{2N}$. The natural extension of the domain of the differential operator introduced above for ideal nonreciprocal b.c. is 
\begin{align}
	\mcl{D}(\mL)=&\left\{\left(\bsb{W},\bsb{w}\right),\bsb{W}=\begin{pmatrix}
		\bU\\\bV
	\end{pmatrix}(x)\in AC^1(\mR_+)\otimes\mathbbm{C}^{2n},\right.\nonumber\\
	&\left.\bsb{w}=\begin{pmatrix}
		\mA \bU\\\bV-\mA \bV' - \mY \bU
	\end{pmatrix}_0\in\mathbbm{C}^{2N}\right\},\label{eq:Domain_Luv_TL_ABY}
\end{align}
where the action of the new operator on its elements $\mcl{W}\in\mD(\mL)$ is
\begin{align}
	\mL \mcl{W} &=-\left(\msDelta\bsb{W}'',\tilde{\bsb{w}}=\begin{pmatrix}
		\msDelta\bU'-\mB^{-1}\bU-\mY \bV'\\\mB^{-1}\bV'
	\end{pmatrix}_0\right).	\label{eq:L_operator_TL_ABY}
\end{align}

This operator is self-adjoint with respect to the enhanced inner product
\begin{align}
	\langle \mcl{W}_1,\mcl{W}_2\rangle&=\int_{\mR_+}dx \,\bsb{W}_1^\dag \msSigma \bsb{W}_2 +\bsb{w}_1^\dag\msGamma\bsb{w}_2,\label{eq:innerproduct_ABY}
\end{align}
with $\msSigma=\text{diag}(\mathbbm{1},\msDelta^{-1})$ and $\msGamma=\text{diag}(\mA^{-1},\mB)$, see Appendix \ref{appendix_d}. The solution of the eigenvalue problem $\mL\mcl{W}=\omega^2\mcl{W}$, gives us the required basis $\mW_{\ome}$ for quantizing the full system. Again, we jointly expand the flux and charge fields as in (\ref{eq:Phi_Q_expansion}). It must be noted, that given that $\mY=0$ the operator and its domain disentangle and $\mL=\mLU\oplus\mLV$. Given the reality condition of the flux and charge fields and the reality of the self-adjoint operator $\mL$, we are free to choose a real basis $\mW_{\ome}=\mW_{\ome}^*$ and real coordinates $X_{\ome}\in\mR$. We define the action of the new duality operator $\mT$ on the basis of $\mL$ now as
\begin{align}
	\mT\mcl{W}_{\ome}&=-i\left(\begin{pmatrix}
		\bV'\\ \msDelta\bU'
	\end{pmatrix},\begin{pmatrix}
		\mA \bV'\\\msDelta\bU'-\mA\msDelta\bU''-\mY\bV'
	\end{pmatrix}_0\right)_{\ome},\label{eq:T_op_TL_ABY}
\end{align}
and by linearity extended to the whole Hilbert space. Again, a pure imaginary representation of this operator can be found for an orthonormal basis in two blocks of $X_{\ome}=(F_{\omega\lambda },G_{\omega\lambda})^T$ such that $\mt_{\epsilon'\epsilon}=\sigma_y\otimes\mone_{N}$, where $\epsilon,\epsilon'\leq 2N$ and $\lambda\leq N$, i.e. $\mR^{2N}\equiv\mR^{N}\oplus\mR^{N}$. The reason for this imaginary representation, as pointed out in previous subsection (), is the commutativity $[\mT,\mL]=0$, which allows for simultaneous diagonalization, together with the reality of the basis of $\mL$ and the coefficients of $\mT$ being purely imaginary. The Lagrangian (\ref{eq:Lag_TL_ABY_ds})  in modes can be written as 
\begin{align}
	L=&\,\frac{1}{2}\int d\Omega d\Omega'\, \left[\dot{X}_{\ome}(\dot{X}_{\ome}-i\omega\mt_{\epsilon'\epsilon}X_{\omega\epsilon'})\right.\nonumber\\
	&-(\dot{X}_{\ome}-i\omega\mt_{\epsilon'\epsilon}X_{\omega\epsilon'}) \mk^{\omega\omega'}_{\epsilon\gamma}(\dot{X}_{\omega'\gamma}-i\omega\mt_{\gamma'\gamma}X_{\omega'\gamma'})\left.\right],\label{eq:Lag_TL_ABY_modes_ds}
\end{align}
where 
\begin{align}
	\mk^{\omega\omega'}_{\epsilon\gamma}&=(\bV-\mA \bV' - \mY \bU)_{\omega\epsilon0}^T\mB(\bV-\mA \bV' - \mY \bU)_{\omega'\gamma0}\nonumber\\
	&=\frac{1}{(\omega\omega')^2}(\bV'_{\omega\epsilon0})^T\mB^{-1}\bV'_{\omega'\gamma0}
\end{align}
is the symmetric (hermitian for complex bases) kernel $(\mk^{\omega\omega'}_{\epsilon\gamma})^T=\mk^{\omega'\omega}_{\gamma\epsilon}=\mk^{\omega\omega'}_{\epsilon\gamma}$ of the projector ($\mK^2=\mK$) with action $\mK[\mW_{\ome}]=\sum_{\gamma}\int d\omega'\, k^{\omega\omega'}_{\epsilon\gamma}\mW_{\omega'\gamma}$, and we have used the notation $\bW_{\omega\epsilon0}\equiv\bW_{\omega\epsilon}(0)$, see Appendix \ref{appendix_d} for the full derivation of the parallel configuration Lagrangian. Interestingly, $\mK$ is only present when there is a network of inductors $\mB^{-1}\neq0$ and we use as basis the eigenspace of the operator (\ref{eq:L_operator_TL_ABY}). Otherwise, it must be appreciated that the kinetic matrix in the double-coordinate space is singular due to the projector. Thus, the general reduction of degrees of freedom with inductors (and capacitors) and nonreciprocal devices in a parallel coupling configuration is much more involved and it remains as an open problem. In the following, we are going to present two completed particular cases, and a counting argument for the expected result in the general Lagrangian (\ref{eq:Lag_TL_ABY_modes_ds}).
\subsubsection{Circuit without inductors ($\msB^{-1}=0$)}
Let us begin with the simplest case of all. The Lagrangian (\ref{eq:Lag_TL_ABY_modes_ds}) without inductors $L|_{\msB^{-1}=0}$ does not contain the projector, and can be reduced to the case of (\ref{eq:Lag_TL_modes_ds}) using as basis the spectral decomposition of the simplified operator 
\begin{align}
	\mcl{D}(\mL|_{\msB^{-1}=0})=&\,\left\{\left(\bsb{W},\bsb{w}\right),\bsb{W}(x)\in AC^1(\mcl{I})\otimes\mathbbm{C}^{2N},\,\bsb{w}=
	\mA \bU_0,\right.\nonumber\\
	&\left.\bV_0=\mA \bV'_0 + \mY \bU_0\right\},\label{eq:Domain_Luv_TL_AY}\\
	\mL|_{\msB^{-1}=0} \mcl{W} =&\left(-\msDelta\bsb{W}'',\tilde{\bsb{w}}=\begin{pmatrix}
	-\msDelta\bU'+\mY \bV'
	\end{pmatrix}_0\right),	\label{eq:L_operator_TL_AY}
\end{align}
to the Hamiltonian (\ref{eq:H_quantized_ds_modes}) where the reduction of modes and its full diagonalization is straightforward. Be aware that the inner product (\ref{eq:innerproduct_ABY}) must be reduced accordingly, in such a way that the matrix $\msB$ does not appear, i.e. $\msGamma=\msA^{-1}$. It must be appreciated here that, because we are using this specific complete basis (and inner product) in the Lagrangian expansion, the kinetic term is full rank and permits an easy Legendre transformation. 

\subsubsection{Circuit without nonreciprocal element ($\mY=0$)}
Let us now turn to the case when the nonreciprocal element is missing. This case can be directly treated with the reduced basis expansion in the spirit of previous chapter~\ref{chapter:chapter_3} to directly get Hamiltonian (\ref{eq:TL_LU_Ham}), but we analyse it here with the redundant basis for a better illustration of the difficulties found in the analysis of the full case. 

As we show in Appendix \ref{appendix_d}, it is generically possible to pick an orthonormal real basis $\mW_{\ome}$ of the Hilbert space when there are only capacitors and inductors for which the kernel is
\begin{align}
	\mk^{\omega\omega'}_{\epsilon\gamma}=\begin{pmatrix}
		0&0\\
		0&\bar{\mk}^{\omega\omega'}_{\lambda\lambda'}
	\end{pmatrix},\label{eq:k_mat_AB}
\end{align}
while preserving the structure of the matrix $\mt=-\sigma_y\otimes\mone_{N}$. Up to a total derivative term the Lagrangian (\ref{eq:Lag_TL_ABY_modes_ds}) is recast into 
\begin{equation}
	L|_{\mY=0}=\frac{1}{2}\int d\Omega d\Omega'\,\left[ \dot{F}_{\omel}^2+\dot{g}_{\omel}^2+\omega\left(\dot{g}_\omel  F_{\omel}-\dot{F}_\omel g_\omel\right)-F_{\omel}(\omega\bar{\mk}_{\lambda\lambda'}^{\omega\omega'} \omega')F_{\omega\lambda'}\right],\label{eq:Lag_TL_AB_ds}
\end{equation}
where $g_{\omel}=\sum_{\lambda'}\int d\omega'\, (\delta_{\omega\omega'}\delta_{\lambda\lambda'}-\bar{\mk}_{\lambda\lambda'}^{\omega\omega'})G_{\omega'\lambda'}$ must be understood as the projection of the $G_{\omel}$ variables in the orthogonal subspace $\mK^\perp=1-\mK$, thus eliminating the zeroes in the kinetic term. 

This allows us to formally perform a Legendre transformation $\Pi_{\omel}=\partial L|_{\mY=0}/\partial\dot{F}_\omel$ and $p_{\omel}=\partial L|_{\mY=0}/\partial\dot{g}_\omel$ and get the Hamiltonian
\begin{align}
	H|_{\mY=0}=&\,\frac{1}{2}\int d\Omega  d\Omega'\left[\left(\Pi_{\omel}+\frac{\omega}{2}g_\omel\right)^2+F_{\omel}\omega\bar{\mk}_{\lambda\lambda'}^{\omega\omega'} \omega'F_{\omega\lambda'}\right.\nonumber\\
	&\left.+\left(p_{\omel}-\frac{\omega}{2}F_\omel\right)(\delta_{\omega\omega'}\delta_{\lambda\lambda'}-\bar{\mk}_{\lambda\lambda'}^{\omega\omega'})\left(p_{\omelp}-\frac{\omega'}{2}F_{\omelp}\right)\right]\nonumber
\end{align}
Although we have partially removed some nondynamical variables from the Lagrangian, further steps must be done to reveal the independent unconstrained set of conjugate variables. For the sake of clarity, let us make explicitly two consecutive canonical transformations that will unfold the result. First, a shift of momenta with the opposite conjugated variables 
\begin{align}
	\tilde{F}_\omel&=F_\omel, \quad\tilde{\Pi}_\omel=\Pi_\omel+\frac{\omega}{2}g_\omel,\nonumber\\
	\tilde{g}_\omel&=g_\omel,\quad \tilde{p}_\omel=\int d\Omega'(\delta_{\omega\omega'}\delta_{\lambda\lambda'}-\bar{\mk}_{\lambda\lambda'}^{\omega\omega'})\left(p_\omelp+\frac{\omega'}{2}F_\omelp\right),\nonumber
\end{align}
where we have made explicit the fact that $\mK^\perp[p_\omel]=p_\omel$ to get  
\begin{align}
	H|_{\mY=0}&=\frac{1}{2}\int d\Omega\left[\tilde{\Pi}_{\omel}^2+\left(\tilde{p}_{\omel}-\omega\tilde{F}_\omel\right)^2\right]\nonumber\\
	&\equiv\frac{1}{2}\int d\Omega\left[\bar{\Pi}_{\omel}^2+\omega^2\bar{F}_\omel^2\right]=_{\mathrm{quant.}}\sum_{\lambda}^{N}\int d\omega\, \hbar\omega\, \hat{a}_{\omel}^\dag\hat{a}_\omel.\label{eq:H_ABY_quant}
\end{align}
The second shift,  $\bar{F}_\omel=\tilde{F}_\omel-\omega\tilde{p}_\omel$, and $\bar{g}_\omel=\mK^\perp[\tilde{g}_\omel-\omega\tilde{\Pi}_\omel]$ with trivial change of momenta, has been performed in the second line and we have skipped the final canonical quantization procedure of the infinite set of harmonic oscillators. As expected, we end up with the same amount of $N$ dynamical and harmonic degrees of freedom per frequency mode because the nondynamical pairs $(\bar{g}_{\omel},\bar{p}_{\omel})$ have disappeared.

\subsubsection{Hamiltonian for a general circuit}
Having discussed the above two particular cases, we turn back now to the Lagrangian (\ref{eq:Lag_TL_ABY_modes_ds}). It is indeed possible to derive a formal Hamiltonian from this Lagrangian even with a singular kinetic term. Up to a total derivative, the Lagrangian can be written as, 
\begin{align}
	L=&\int d\Omega \left[\dot{X}_{\ome}^0(\dot{X}_{\ome}^0- \msLambda_{\epsilon\epsilon'}^\omega X_{\omega\epsilon'}^0)-X_{\omega\epsilon}^1\msLambda_{\epsilon\epsilon'}^\omega \dot{X}_{\omega\epsilon'}^1\right]\nonumber\\
	&+\int d\Omega d\Omega'\left[X_{\omega\epsilon'}\msLambda_{\epsilon'\epsilon}^\omega \mk_{\epsilon\gamma}^{\omega\omega'}\msLambda_{\gamma\gamma'}^{\omega'} \dot{X}_{\omega'\gamma'}\right],
\end{align}
where we have defined the \emph{anti-symmetric} matrix $\msLambda_{\epsilon\epsilon'}^\omega=i\omega\mt_{\epsilon'\epsilon}$, and the reduced (projected) coordinates $X_{\omega\epsilon}^0=\int d\Omega'(\delta_{\omega\omega'}\delta_{\epsilon\epsilon'}-\mk_{\epsilon\epsilon'}^{\omega\omega'})X_{\omep}$ and $X_{\omega\epsilon}^1=\int d\Omega'\mk_{\epsilon\epsilon'}^{\omega\omega'}X_{\omep}$. This Lagrangian is a generalization of previous (\ref{eq:Lag_TL_AB_ds}) where a block-matrix representation of $\mK$ operator like (\ref{eq:k_mat_AB}) does not exist while preserving $\mt_{\epsilon\epsilon'}=-\sigma_y\otimes\mone_N$, see the related Appendix. Although there is no standard kinetic term for $X_{\ome}^1$, the Lagrangian belongs to the class of first-order descriptions of quadratic systems in the spirit of Faddeev-Jackiw~\cite{Faddeev:1988}. Let us derive the Hamiltonian here by writing first a Routhian, i.e. performing a partial Legendre transformation to the $X_{\ome}^0$ variables $\Pi_{\ome}^0=\partial L/\partial \dot{X}_{\ome}^0$, 
\begin{align}
	R=\int d\Omega( \Pi_{\ome}^0\dot{X}_{\ome}^0)-L,
\end{align}
and converting back to the \emph{first-order} Lagrangian 
\begin{align}
		L=&\int d\Omega (\Pi_{\ome}^0\dot{X}_{\ome}^0-X_{\omega\epsilon}^1\msLambda_{\epsilon\epsilon'}^\omega \dot{X}_{\omega\epsilon'}^1)-H,\nonumber
\end{align}
from which one can read, \emph{à la} Fadeev-Jackiw, the Hamiltonian
\begin{align}
	H=&\int d\Omega (\Pi_{\ome}^0- \tfrac{1}{2}\int d\Omega'\msTheta_{\epsilon\epsilon'}^{\omega\omega'} X_{\omega'\epsilon'}^0)(\Pi_{\omega\epsilon}^0- \tfrac{1}{2}\int d\Omega''\msTheta_{\epsilon\gamma}^{\omega\omega''} X_{\omega''\gamma}^0)\nonumber\\
	&+\int d\Omega d\Omega' \left[X_{\omega\epsilon'}\msLambda_{\epsilon'\epsilon}^\omega \mk_{\epsilon\gamma}^{\omega\omega'}\msLambda_{\gamma\gamma'}^{\omega'} X_{\omega'\gamma'}\right],\label{eq:H_ABY_firstorder}
\end{align}
where we have defined the again anti-symmetric 
\begin{equation}
	\msTheta_{\epsilon\epsilon'}^{\omega\omega'}=\int d\Omega''(\delta_{\omega\omega''}\delta_{\epsilon\epsilon''}-\mk_{\epsilon\epsilon''}^{\omega\omega''})\msLambda_{\epsilon''\gamma}^{\omega''}(\delta_{\omega''\omega'}\delta_{\gamma\epsilon'}-\mk_{\gamma\epsilon'}^{\omega''\omega'}).\nonumber
\end{equation}
As expected, $(X_{\ome}^0,\Pi_{\omep}^0)$ are pairs of conjugate variables with a canonical Poisson bracket $\{X_{\ome}^0,\Pi_{\omep}^0\}=\delta_{\omega\omega'}\delta_{\epsilon\epsilon'}$, whereas $X_{\ome}^1$ encode noncanonical conjugated pairs through the nontrivial Poisson bracket $\{X_{\ome}^1,X_{\omep}^1\}=(\msXi_{\epsilon\epsilon'}^{\omega\omega'})^{-1}$, with the definition of the projected matrix $\msXi_{\epsilon\epsilon'}^{\omega\omega'}=\int d\Omega''(\mk_{\epsilon\epsilon''}^{\omega\omega''})\msLambda_{\epsilon''\gamma}^{\omega''}(\mk_{\gamma\epsilon'}^{\omega''\omega'})$. Here, we have assumed the existence of the inverse of the matrix $\msXi_{\epsilon\epsilon'}^{\omega\omega'}$, which could happen only in the case of even number of transmission lines. A completely general procedure would require projecting out an odd space with trivial dynamics, and taking the inverse in the nontrivial subspace. 

As in previous cases this Hamiltonian contains at most $N$ harmonic oscillators per frequency subspace as we are going to prove in the following dimensionality analysis. However, one must appreciate the extra difficulty in finding an existent, although not yet found, symplectic transformation that explicitly shows this statement, due to the presence of the $X_{\ome}^1$ variables.  Given the quadratic nature of the Hamiltonian, we can write its matrix acting on the phase space Hilbert space $\tilde{\mcl{H}}=\mcl{H}_0\oplus\mcl{H}_0\oplus\mcl{H}_1$ in a compact form
\begin{equation}
	\msf{h}=\begin{pmatrix}
	\msSigma_+\msSigma_- -\msTheta^2/4&\msTheta/2&\msSigma_+\msXi\\
	\msTheta^T/2&\mone&0\\
	\msXi^T\msSigma_-&0&\msXi^T\msXi
	\end{pmatrix},
\end{equation}
where we have defined as operators $\msSigma_-=\msSigma_+^T=\mk \msLambda(1-\mk)$,  $\msTheta=(1-\mk)\msLambda(1-\mk)$, and  $\msXi=\mk \msLambda\mk$, such that the full Hamiltonian (\ref{eq:H_ABY_firstorder}) is $H=\bra{\bz}\msf{h}\ket{\bz}$ with 
\begin{equation}
	\ket{\bz}=\begin{pmatrix}
	\ket{\bx_0}\\\ket{\bpi_0}\\\ket{\bx_1}
	\end{pmatrix}.\nonumber
\end{equation}
Here we have defined the Hilbert subspaces as $\mcl{H}_0=(1-\mk)\mcl{H}$, $\mcl{H}_1=\mk\mcl{H}$, and the (doubled) generalized-coordinates subspace  $\mcl{H}=L^2[\mR_+]\otimes\mR^{2N}$. Recall that $\msLambda$ is an anti-symmetric operator and thus $-\msTheta^2\geq0$. On the other hand, the symplectic form written in a compact notation reads
\begin{equation}
	\msf{\msOmega}=\begin{pmatrix}
	0&\mathbbm{1}&0\\-\mathbbm{1}&0&0\\0&0&\tilde{\msOmega}
	\end{pmatrix},
\end{equation}
where $\tilde{\msOmega}=\msXi^{-1}$. Given the assumption of nonsingularity of the  symplectic form for the $\ket{\bx_1}$ variables, the kernel of the Hamiltonian matrix can be obtained from solving the equation $\msf{h}\ket{\bz}=0$. Its solution entails the relations $\ket{\bpi_0}=\frac{\msTheta}{2}\ket{\bx_0}$ and $\ket{\bx_1}=\msXi^{-1}\msSigma_-\ket{\bx_0}$, i.e. the kernel is isomorphic to $\mcl{H}_0$. Meaning that, after taking out the trivial component, the reduced phase space is isomorphic to $\mcl{H}_0\oplus\mcl{H}_1=\mcl{H}=L^2[\mR_+]\otimes\mR^{2N}$. In other words, this proves that the non-trivial phase space has the same dimension as the non-doubled description in terms of only flux variables and thus the maximum number of harmonic oscillators per frequency is $N$. 

We remark that further work will be required to find the symplectic transformation that brings the Hamiltonian (\ref{eq:H_ABY_firstorder}) form into its diagonal and non-redundant shape. In doing so, special attention will have to be given for the cases where the symplectic form $\tilde{\Omega}$ is singular (as for example with a circuit containing an odd number of transmission lines connected), which will undoubtedly require the elimination of free-particle dynamics by means of the application of a generalized Williamson's theorem~\cite{Williamson:1936}.

\subsection{Series configuration}
\label{Sec:series}
\begin{figure}[h!]
	\centering
	\includegraphics[width=0.55\linewidth]{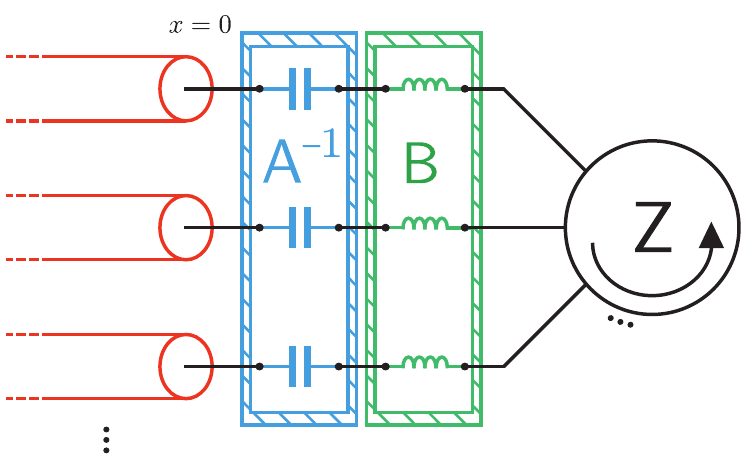}
	\caption{Transmission lines coupled in a series configuration to lumped networks of capacitors $(\mA)$, inductors $(\mB)$ and impedance-described circulator $(\mZ)$.}
	\label{fig:Series_config}
\end{figure}

We end up this section with a brief comment on the electromagnetic duality symmetry in the context of electric circuits. The flux-charge duality becomes entangled with the series-parallel configuration duality \cite{Ulrich:2016}. Similarly to previous section, consider now the problem of $N$ semi-infinite transmission lines connected at the boundary (without loss of generality, $x=0$) in a series configuration with three networks of capacitors (${\mA}^{-1}=\mld^{-1/2}\msf{C}^{-1}\mld^{-1/2}$), inductors (${\mB}=\mld^{-1/2}\msf{L}\mld^{-1/2}$), and nonreciprocal ideal devices ($\mZ=\mld^{-1/2}\bar{\msf{Z}}\mld^{-1/2}$), see Fig. (\ref{fig:Series_config}), where the rescaling of the variables has been reciprocal to the parallel configuration.
Notice that the natural description for the nonreciprocal elements is an impedance matrix, such that analogously to Eq. (\ref{eq:Lag_TL_ABY_ds}) we have the Lagrangian 
\begin{align}
	L =&\, \int_{\mathbbm{R}^+}dx\,\frac{1}{2}\left[\dot{\bQ}^2 +\dot{\bPhi}^T\msDelta^{-1}\dot{\bPhi}+\bQ^T\dot{\bPhi}'+\bPhi^T\dot{\bQ}'\right]\nonumber\\
	&+\frac{1}{2}\left(\dot{\bPhi}_0^T\bQ_0+\dot{\bQ}_0^T{\mB}\dot{\bQ}_0-{\bQ}_0^T{\mA}^{-1}{\bQ}_0+\dot{\bQ}_0^T\msf{Z}\bQ_0\right) \label{eq:TL_ABZ_Lag_Phi_Q2}.
\end{align}
and now the voltage equation at the boundary
\begin{equation}
	\dot{\bPhi}_0={\mB}\ddot{\bQ}_0+{\mA}^{-1}{\bQ}_0+\mZ\dot{\bQ}_0.\label{eq:SC_BC_current_law}
\end{equation}
Realize that this Lagrangian is equivalent to (\ref{eq:Lag_TL_ABY_ds}) by interchanging $\bPhi\rightleftarrows\bQ$, and thus a dual operator to (\ref{eq:L_operator_TL_ABY}) can be defined, and an analogous analysis can be performed. 

In summary, in this chapter we have put forward a consistent quantization procedure for  superconducting circuits modeled in a redundant flux-charge description with a corresponding double-space basis. The apparent redundancy is eliminated by making use of a duality symmetry. Such a doubled basis becomes mandatory for the correct identification and quantization of the dynamical  degrees of freedom in circuits with ideal nonreciprocal devices, e.g. circulators connected to transmission lines. We have applied the theory to a circuit with a Josephson junction connected through a transmission line to a circulator, deriving a Hamiltonian in which there is no mode-mode coupling among the dressed normal modes, and which is free of artificial divergence issues. Finally, we have made an extension of the redundant doubled-basis idea to treat more complex circuits with generic linear lossless lumped boundary conditions made of capacitors, inductors and ideal circulators. We have explicitly shown the reduction of redundant nondynamical for particular cases without inductors or without circulators. We have also made a counting argument for the generic boundary condition with the three types of elements proving that the expected maximum number of harmonic oscillators per frequency will be recovered. 

Further forthcoming work is required to find the symplectic transformation that makes this statement explicit. As a result, the generic basis to expand the flux and charge fields in transmission lines to couple to other lumped-element nonlinear networks in a divergence-free manner will be found.


\chapter{Conclusions and Outlook}
\label{chap_conclusions}
\thispagestyle{chapter}
\hfill\begin{minipage}{0.85\linewidth}
{\emph{Ezina ekinez egina\\\\
		(Through hard work, the impossible becomes possible)
}}
\end{minipage}
\begin{flushright}
	{Basque proverb}
\end{flushright}
\vspace*{1cm}
In this Thesis, we have put forward analytical tools to quantize canonically superconducting circuits in the context of Kirchhoff's laws, explicitly showing that the divergence-free nature originated in the macroscopic classical theory. In doing so, we have answered positively the main question posed in chapter~\ref{chapter:chapter_1}, which we rewrite again here: 

{\it Is it possible to find systematically a convergent quantum theory of superconducting chips from lumped and distributed Kirchhoff's equations?} Yes, indeed.  

In short, we have studied the different divergence issues appearing in an illustrative and minimal circuit QED set-up containing a multi-mode transmission line resonator capacitively coupled to a Josephson junction. We have extended this analysis to a catalogue of multiple infinite dimensional systems linearly coupled to finite non-harmonic degrees of freedom. Moreover, we have introduced the ideal nonreciprocal elements in an exact manner into the effective Hamiltonian descriptions of lumped element circuits. Finally, we have introduced the more generic doubled-space description for deriving exact Hamiltonians transmission lines coupled through nonreciprocal linear systems to nonharmonic degrees of freedom. More specifically: 

In chapter \ref{chapter:chapter_2}, we have analysed a multi-mode quantum Rabi model of circuit QED from a macroscopic lumped element equivalent circuit. We have explicitly shown the convergence of the Lamb shift in the absence of any high frequency cutoff, arising from a natural renormalization of the Hamiltonian parameters with increasing number of harmonic modes. We have also studied the implications of a finite junction capacitance, which introduces a natural electrical length cutoff in the coupling to high frequency modes. We have shown that when constructing a quantum Rabi model from capacitively coupled lumped element circuits, it is crucial to include the natural renormalization from an exact Legendre transformation to get correct Hamiltonian parameters from the macroscopic values of the circuit elements. Furthermore, we have shown a connection between Hamiltonian models with truncated number of modes and low-energy approximations of the infinite dimensional model of the transmission line resonator. We point out the usefulness of this approach in the context of ultra-strong coupling regime experiments where taking into account multi-mode effects is compulsory. 

In chapter \ref{chapter:chapter_3}, we have critically analysed a number of approaches to the quantization of superconducting circuits with an infinite dimensional environment, mainly transmission lines and generic multiport immittance black-boxes, with particular interest on the issue of divergences in the Lamb-shifts or effective (adiabatic) couplings predicted by capacitive coupling constants. With respect to the transmission lines, we have made use of solid mathematical constructions, i.e. singular value problems for second order differential operators, with boundary values that include the singular value itself, to correctly describe capacitive (and inductive) divergence-free coupling parameters to nonlinear (lumped-element) networks. In doing so, we have identified the fundamental electric and inductive lengths defining the cutoffs of such parameters. They have been optimally selected with the criterium that the final Hamiltonian have no TL mode-mode couplings. Interestingly, the coupling parameter for either pure capacitive or inductive length is of Lorentz-Drude type, with a soft decay $g_n\sim \omega_n^{-1/2}$. In transforming back to fields in the Hamiltonian description, one may directly read the $A^2$ diamagnetic term, dependent on the respective length parameter. An analogous analysis has been performed with multiport linear reciprocal black-boxes with an infinite set of modes. There, a lumped-element model has been truncated to $N$ number of modes before taking the infinite limit. On the criterion that the final Hamiltonian description be that of an infinite set of independent harmonic modes coupled to a finite set of variables, we have shown the convergence of the infinite dimensional limit within the canonical transformation. We have performed this analysis on a catalogue of linear coupling configurations and proved the exact same behaviour. Furthermore, we have shown the connection between the methods here described, and other ways of computing the quantum fluctuations of the flux and charge fields of the semi-infinite transmission line ended in an LC resonator, which is equivalent to the RLC circuit. Looking into the future, it would be interesting to experimentally check the prediction for the maximum coupling achievable with a transmission line, which can be improved with a coupling capacitor model of finite length. 

From the thorough analyses of chapters \ref{chapter:chapter_2} and \ref{chapter:chapter_3}, with the formulation of multi-mode quantum Rabi models in the context of Kirchhoff's equations presenting a natural decoupling of light and matter, one should be able to achieve the similar divergence-free Hamiltonian models in other (nonrelativistic) quantum mechanical set-ups, e.g. atoms coupled to waveguide modes (of cavities) in the dipole approximation where an effective atom length, e.g. Bohr radius, is phenomenologically invoked\footnote{A first, but unfinished work in this direction may be found in Appendix G in \cite{Malekakhlagh:2016} with an effective quantization of the hydrogen atom in a one dimensional cavity in the dipole approximation.}.

In chapter \ref{chapter:chapter_4}, we have described a procedure to add ideal nonreciprocal elements to the exact Hamiltonian descriptions of lumped element networks in a generalization of the standard techniques based on network graph theory and the choice of flux variables as degrees of freedom. We have exemplified this technique with two circuits. Firstly, a two-port Viola-DiVincenzo gyrator connected to Josephson junctions. Secondly, we have quantized the equivalent lumped-element circuit for a generic two-port nonreciprocal impedance. We have also discussed a technical issue regarding the introduction of ideal nonreciprocal elements in a flux variable description which lack of admittance description, and we proved that the problem simplifies to remove extra constraints, with a reduction of independent variables. Finally, we have discussed the dual quantization method in terms of loop-charges which could be particularly useful in future superconducting technologies based on phase-slip junctions and nonreciprocal elements. In the same direction as later chapter \ref{chapter:chapter_5}, one can look for generalizations of Hamiltonian descriptions based on a doubled or mixed configuration space starting from a redundant Lagrangian description in terms of node fluxes and loop charges. In this way,  the lumped-element multiport admittance black-box can be trivially quantized in dual manner to the impedance black-box treated in this chapter. 

In chapter \ref{chapter:chapter_5}, we have presented a more general canonical quantization technique for circuits described in terms of a redundant flux-charge description in the configuration space. Instead of eliminating the redundancy in the Euler-Lagrange equations of motion (Lagrangian), we do so in the phase space (Hamiltonian) by making use of a duality symmetry. This double-space basis becomes the most efficient starting point to derive exact Hamiltonians of networks containing arbitrary number of transmission lines pointwise coupled by ideal nonreciprocal elements due to the time-reversal symmetry breaking term mixing configuration-space fields in a nontrivial way. A differential Sturm-Liouville operator complete basis is naturally found in the doubled space which corresponds to the normal mode basis that exactly diagonalizes the Hamiltonian. In a generalization of previous chapter \ref{chapter:chapter_3}, we extend the technique to describe point (capacitive/inductive) linear connections to the transmission line and we find the exact Hamiltonian of a circuit containing a Josephson junction capacitively coupled to a transmission line connected to two others by a circulator. Naturally, the divergence-free properties of previous chapters are preserved as we enhance the Hilbert space on which the functions we use to develop the flux and charge fields live, and the TL mode-mode couplings can be also exactly eliminated in the Hamiltonian by the same optimal criterion of chapter \ref{chapter:chapter_3}. Finally, we have performed for the first time an analysis on how to extend the theory to quantize circuits containing transmission lines coupled by frequency-dependent nonreciprocal devices. The problem reduces to the analysis of circuits with general boundary conditions with capacitors, inductors and nonreciprocal ideal elements. Working in the double-space basis, we have demonstrated the elimination of redundant nondynamical variables for particular cases without inductors or without circulators at the boundary, where the latter case is also amenable by the reduced basis. We have given a proof for the nontrivial dimension of the phase space for the Hamiltonian of the generic linear coupling boundary. However, future work will be required to find the symplectic transformation that brings the full Hamiltonian into its diagonal basis.

On the whole, this Thesis expands the theory to find canonical Hamiltonian models for circuits based on Kirchhoff's laws, a matter of special relevance in deriving models for superconducting quantum technologies. We expect that the results presented here will help analyze, design and synthesize new superconducting circuits taking into account without restraint, the complex infinite-dimensional nature of coupled light-matter systems, as well as new devices that effectively break time-reversal symmetry. Intrinsically, these circuits have the potential to both unveil unsolved mysteries of the universe, and give a technological leap to humankind. In addition, we surmise that the theory of self-adjoint operators on which many results of this Thesis rest will be of interest to the mathematical community, in particular, the use of a doubled Hilbert space to circumvent second-order boundary value problems with the square-root of eigenvalues in the boundary condition. 

{\raggedleft\vfill{
	\begin{flushright}
		\emph{Comment me suis-je échappé?\\
			Avec difficulté\\
			Comment ai-je planifié ce moment?\\
			Avec plaisir}\\
		\vspace*{0.5cm}
		\textbf{Alexandre Dumas} \\
		{Le Comte de Monte-Cristo}
	\end{flushright}
	}}

\cleardoublepage

\addcontentsline{toc}{chapter}{APPENDICES}
 
\appendix
\renewcommand{\chaptermark}[1]{\markboth{\textit{Appendix \thechapter.  #1}}{}}

\chapter{Further Details on the Convergence of the QRM in cQED}
\label{appendix_a}
\section{Derivation of the circuit Hamiltonian}
The input impedance of a shorted transmission line, at a distance $\lambda_0/4$ from the short (see Ref.~\cite{Pozar:2009}) is given by
\begin{equation}
Z(\omega) = i Z_0 \tan\bigg(\frac{\pi}{2}\frac{\omega}{\omega_0}\bigg)\ ,
\end{equation}
where $Z_0$ is the characteristic impedance of the waveguide, $\omega_0/2\pi$ is the resonance frequency and $\lambda_0$ the wavelength of the fundamental mode of the quarter wave resonator when the AA is replaced by an open termination. The partial fraction expansion of the tangent
\begin{equation}
\tan(z) = \sum_{m=0}^\infty \frac{-2z}{z^2-(m+\frac{1}{2})^2\pi^2}
\end{equation}
leads to an expression for the resonators imput impedance which is equal to that of an infinite number of parallel LC resonators. Each of them corresponds to a resonance mode
\begin{align}
Z(\omega) &= i Z_0 \tan\bigg(\frac{\pi}{2}\frac{\omega}{\omega_0}\bigg) = \sum_{m=0}^\infty \frac{1}{iC_0\omega+\frac{1}{iL_m\omega}}\ ,\nonumber\\
C_0 &= \frac{\pi}{4\omega_0 Z_0}\ ,\quad L_m = \frac{1}{(2m+1)^2}\frac{4Z_0}{\pi \omega_0}\ .\nonumber
\end{align}

\begin{figure}[]
	\centering
	\includegraphics[width=0.6\textwidth]{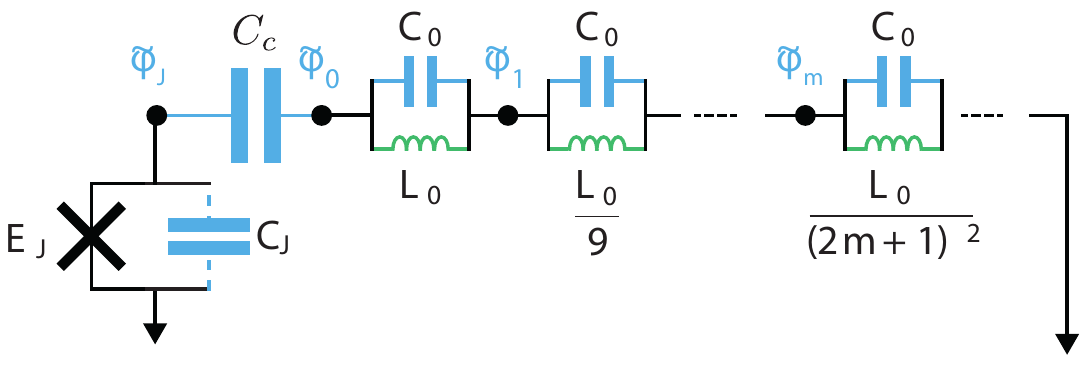}
	\caption{Circuit quantized in this section. In accordance with Ref.~\cite{Devoret:1997}, the degrees of freedom of the circuit are chosen to be the fluxes (indicated in blue) at the nodes of the circuit.}
	\label{fig:circuit_notations}
\end{figure}

We truncate the system to the first $M$ resonators and use the tools of circuit quantization to obtain the corresponding Hamiltonian. Following the methodology given in Refs.~\cite{Devoret:1997} and \cite{Girvin:2014}, we start by defining a set of nodes of the circuit and their corresponding fluxes. We define the flux $\tilde{\phi}$ from the voltage $v$ of that node to ground as
\begin{equation}
\tilde{\phi}(t) = \int_{-\infty}^t v(t')dt'\ .
\end{equation}
As described in Fig. \ref{fig:circuit_notations}, the node corresponding to the superconducting island of the AA is denoted by the subscript $J$, and we number from $0$ to $M-1$ the nodes corresponding to the fluxes from the $m$-th LC oscillator to the coupling capacitor. The Lagrangian of the system is given by
\begin{equation}
\begin{split}
L&= C_J\frac{\dot{\tilde{\phi}}_J^2}{2} + C_c\frac{(\dot{\tilde{\phi}}_J-\dot{\tilde{\phi}}_0)^2}{2} + \sum_{m=0}^{m<M-1} C_0\frac{(\dot{\tilde{\phi}}_m-\dot{\tilde{\phi}}_{m+1})^2}{2} + C_0\frac{(\dot{\tilde{\phi}}_{M-1})^2}{2} \\
&+E_J\cos\bigg(2\pi\frac{\tilde{\phi}_J}{\Phi_q}\bigg) - \sum_{m=0}^{m<M-1} (2m+1)^2\frac{(\tilde{\phi}_m-\tilde{\phi}_{m+1})^2}{2L_0} - (2M-1)^2\frac{(\tilde{\phi}_{M-1})^2}{2L_0}\ ,
\end{split}\nonumber
\end{equation}
where $\Phi_q = h/2e$ corresponds to the flux quantum. We now make the change of variables $\phi_m = \tilde{\phi}_{m} - \tilde{\phi}_{m+1}$ for $0\le m<M-1$, leaving the remaining two variables unchanged $\phi_{M-1} = \tilde{\phi}_{M-1}$ and $\phi_J = \tilde{\phi}_J$. The Lagrangian then reads
\begin{equation}
\begin{split}
L&= C_J\frac{\dot{\phi}_J^2}{2} + C_c\frac{\big(\dot{\phi}_J-\sum_{m=0}^{m<M}\dot{\phi}_m\big)^2}{2} + \sum_{m=0}^{m<M} C_0\frac{(\dot{\phi}_m)^2}{2}\\
&+E_J\cos\bigg(2\pi\frac{\phi_J}{\Phi_q}\bigg) - \sum_{m=0}^{m<M} (2m+1)^2\frac{(\phi_m)^2}{2L_0}\ .
\end{split}\nonumber
\end{equation}
Now the variables $\dot{\phi}_m$ correspond directly to the voltage difference across the capacitance of the $m$-th LC oscillator. With the objective of writing a Hamiltonian, it is useful to express the capacitive part of the Lagrangian in matrix notation
\begin{equation}
L= \frac{1}{2}\boldsymbol{\dot{\phi}}^T \msC \boldsymbol{\dot{\phi}} +E_J\cos\bigg(2\pi\frac{\phi_J}{\Phi_q}\bigg) - \sum_{m=0}^{m<M} (2m+1)^2\frac{(\phi_m)^2}{2L_0}\ ,\nonumber
\end{equation}
with the definition of the flux vector $\bphi=(\phi_J, \phi_0,\phi_1,...,\phi_{M-1})^T$, and the capacitance matrix
\begin{equation}
\msC = 
\begin{pmatrix}
C_J+C_c     & -C_c    &  -C_c     & -C_c      & \cdots & -C_c \\
-C_c      &  C_0+C_c     & C_c & C_c &  \cdots & C_c  \\
-C_c      &  C_c     & C_0+C_c & C_c &  & \\
-C_c      &  C_c     & C_c & C_0+C_c &  & \\
\vdots & \vdots &  &        &     \ddots &\\
-C_c      &  C_c     &  &  &  & C_0+C_c\\
\end{pmatrix}\ .\nonumber
\end{equation}
The canonical momenta (dimensionally charges) are equal to
\begin{equation}
q_i = \frac{\partial L}{\partial \dot{\phi}_i} = \msC_{ij}\dot{\phi}_i
\end{equation}
using Einstein summation convention for repeated indices. The Hamiltonian $H = q_i\dot{\phi}_i -L$ is then given by
\begin{equation}
H = \frac{1}{2}\boldsymbol{q}^T\msC^{-1}\boldsymbol{q}  -E_J\cos\left(2\pi\frac{\phi_J}{\Phi_q}\right) + \sum_{m=0}^{m<M} (2m+1)^2\frac{(\phi_m)^2}{2L_0}\ ,
\end{equation}
with the vector of charges $\bsb{q}=(q_J, q_0,q_1,...,q_{M-1})^T$, and the inverse of the capacitance matrix is
\begin{equation}
\begin{split}
\msC^{-1} &= \frac{1}{C_0(MC_cC_J+C_0(C_c+C_J))}\\
&\times\begin{pmatrix}
C_0^2+MC_0C_c    & C_0C_c     &  C_0C_c       & \cdots  \\
C_0C_c    &  C_0(C_c+C_J)+(M-1)C_cC_J    & -C_JC_c  &   \cdots   \\
C_0C_c    &  -C_JC_c     &\ddots &   \\
\vdots & \vdots &  &           \ddots \\
\end{pmatrix}\ .
\end{split}\nonumber
\end{equation}
It is easy to check this result in a very general way by veryfing that $\msC\msC^{-1} = \msC^{-1}\msC = \mone$. We now quantize the canonical variables $q_i\rightarrow\hat{q_i}$, $\phi_i\rightarrow\hat{\phi_i}$, postulating the commutation relation $[\hat{\phi}_i,\hat{q}_j] = i\hbar\delta_{ij}$. This results in the Hamiltonian
\begin{equation}
\hat{H}^{(M)} = \hat{H}^{(M)}_{\text{AA}} + \hat{H}^{(M)}_{\text{cav}} + \hat{H}^{(M)}_{\text{int}}\ .
\end{equation}
The AA Hamiltonian is defined as
\begin{equation}
\hat{H}^{(M)}_{\text{AA}} = \frac{1}{2C^{(M)}_{\text{AA}}}\hat{q}_J^2 -E_J\cos\bigg(2\pi\frac{\phi_J}{\Phi_q}\bigg)\ ,
\end{equation}
where the atoms capacitance is given by
\begin{equation}
C^{(M)}_{\text{AA}} = \frac{C_0(MC_cC_J+C_0(C_c+C_J))}{C_0^2+MC_0C_c}\ .
\end{equation}
Usually, the charge is expressed in number of Cooper pairs $\hat{q}_J = 2e\hat{n}_J$ and the charging energy is given by $E_C^{(M)} = e^2/2C^{(M)}_{\text{AA}}$, resulting in the Hamiltonian
\begin{equation}
\hat{H}^{(M)}_{\text{AA}} = 4E_C^{(M)}\hat{n}_J^2 - E_J\cos\left(2\pi\frac{\hat{\phi}_J}{\Phi_q}\right)\ .
\end{equation}
In chapter~\ref{chapter:chapter_2} we introduced the superconducting phase difference accross the junction as $\hat{\varphi}_J = 2\pi\frac{\hat{\phi}_J}{\Phi_q}$. The cavity Hamiltonian is equal to
\begin{equation}
\hat{H}^{(M)}_{\text{cav}} = \sum_{m =0}^{m<M}\frac{1}{2C^{(M)}_0}\hat{q}_m^2  + \sum_{m=0}^{m<M} (2m+1)^2\frac{(\phi_m)^2}{2L_0}\ ,
\end{equation}
where the effective capacitance of each oscillator is given by
\begin{equation}
C_0^{(M)} = \frac{C_0(MC_cC_J+C_0(C_c+C_J)}{(M-1)C_cC_J+C_0(C_c+C_J)}\ .
\end{equation}
We define the creation and annihilation operators
\begin{equation}
\hat{\phi}_m = \frac{-i}{\sqrt{2m+1}}\sqrt{\frac{\hbar}{2}\sqrt{\frac{L_0}{C_0^{(M)}}}}(\hat{a}_m-\hat{a}_m^\dagger)\ ,
\end{equation}
\begin{equation}
\hat{q}_m = \sqrt{2m+1}\sqrt{\frac{\hbar}{2}\sqrt\frac{C_0^{(M)}}{L_0}}(\hat{a}_m+\hat{a}_m^\dagger)\ ,
\end{equation}
reducing the cavity Hamiltonian to 
\begin{equation}
\hat{H}^{(M)}_{\text{cav}} = \sum_{m =0}^{m<M}\hbar\omega_m^{(M)}\hat{a}^\dagger\hat{a}\ ,
\end{equation}
\begin{equation}
\omega_m^{(M)} = \frac{2m+1}{\sqrt{L_0C_0^{(M)}}}\ ,
\end{equation}
where we have dropped the constant energy contributions $\hbar\omega_m^{(M)}/2$. The quantum voltage of each mode is
\begin{equation}
\hat{V}_m^{(M)} = \frac{\hat{q}_m}{C_0^{(M)}} = V_{\text{zpf},m}^{(M)}(\hat{a}_m+\hat{a}_m^\dagger) ,
\end{equation}
defining the zero point fluctuations of the $m$-th mode by $V_{\text{zpf},m}^{(M)} = \sqrt{2m+1}\sqrt{\frac{\hbar\omega^{(M)}_0}{2C^{(M)}_0}}$.\\
The interaction term $\hat{H}^{(M)}_{\text{int}}$ is given by
\begin{equation}
\hat{H}^{(M)}_{\text{int}} = \sum_{m=0}^{m<M}\sum_{m'=m+1}^{m'<M}G^{(M)}_{m,m'}(a_m+a_m^\dagger)(a_{m'}+a_{m'}^\dagger)
+ \sum_{m=0}^{m<M}\hbar\bar{g}_{m}^{(M)}\hat{n}_J(a_m+a_m^\dagger)\ ,
\end{equation}
where $G_{m,m'}^{(M)}$ quantifies the coupling between the $m$-th and $m'$-th modes of the resonator through the presence of the capacitances introduced by the AA
\begin{equation}
G_{m,m'}^{(M)} = -\frac{C_0C_cC_J}{MC_cC_J+C_0(C_c+C_J)} \frac{(C_0^{(M)})^2}{C_0^2}V_{\text{zpf},m}^{(M)}V_{\text{zpf},m'}^{(M)}\ ,
\end{equation}
and $\hbar\bar{g}_{m}^{(M)}= \beta^{(M)}V^{(M)}_{\text{zpf},m} 2e$ quantifies the coupling between the $m$-th mode of the resonator and the AA. It is weighted by the capacitance ratio
\begin{equation}
\beta^{(M)} = \frac{C_0C_c}{MC_cC_J+C_0(C_c+C_J)} \frac{C_0^{(M)}}{C_0}\ .
\end{equation}
We can also write the Hamiltonian in the basis of eigenstates of the AA Hamiltonian. Defining the eigenstates $\{\ket{i}^{(M)}\}$ and eigenvalues $\epsilon_i^{(M)}$ by $\hat{H}^{(M)}_{\text{AA}}\ket{i}^{(M)} = \hbar\epsilon_i^{(M)}\ket{i}^{(M)}$ and making the transformation $\hat{H}^{(M)} \rightarrow \sum_i\ket{i}^{(M)}\bra{i}^{(M)}\hat{H}^{(M)}\sum_j\ket{j}^{(M)}\bra{j}^{(M)}$ we obtain the final form of the Hamiltonian

\begin{align}
\begin{split}
\hat{H}^{(M)} &= \sum_{m=0}^{m<M}\hbar\omega^{(M)}_m\hat{a}_m^\dagger \hat{a}_m + \sum_i\hbar\epsilon_i^{(M)} \ket{i}^{(M)}\bra{i}^{(M)} \\
& + \sum_{i,j}\sum_{m=0}^{m<M}\hbar g^{(M)}_{m,i,j}\ket{i}^{(M)}\bra{j}^{(M)}(\hat{a}_m+\hat{a}_m^\dagger)
\\
& +\sum_{m=0}^{m<M}\sum_{m'=m+1}^{m<M}G^{(M)}_{m,m'}(\hat{a}_m+\hat{a}_m^\dagger)(a_{m'}+a_{m'}^\dagger)
\end{split}
\label{eq:circuit_Hamiltonian}
\end{align}
where the coupling $g_{m,i,j}$ is given by:
\begin{equation}
\hbar g_{m,i,j} = V^{(M)}_{\text{zpf},m}2e\beta^{(M)}\bra{i}^{(M)}\hat{n}_J\ket{j}^{(M)}\ .
\end{equation}

If $C_J=0$, this Hamiltonian reduces to the one given in chapter~\ref{chapter:chapter_2}. If not we can make use of a Bogoliubov transformation to express it in terms of the eigenmodes of the resonator as described in the Sec.~\ref{sec:Bogo}. This would allow us to recover the form of the Hamiltonian given in chapter~\ref{chapter:chapter_2}. Alternatively the above hamiltonian can be diagonalized as it is to obtain an energy spectrum.

\section{Dispersive shift of coupled LC oscillators}
\begin{figure}[]
	\centering
	\includegraphics[width=0.25\textwidth]{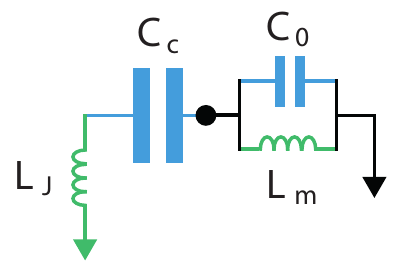}
	\caption{Circuit of a linearized AA (series LC oscillator) coupled to a single mode $m$ of the resonator. Studying this circuit provides a good approximation of the Lamb shift in the Transmon regime $E_J\gg E_C$. The coupling point is shown as a blue dot.}
	\label{fig:circuit_shift}
\end{figure}

In this section, we derive the Lamb shift of a linearized AA (a series LC oscillator) dispersively coupled to a single resonator mode (a parallel LC oscillator) for the case $C_J=0$ as shown in Fig. \ref{fig:circuit_shift}. The atom is linearized by discarding the purely non-linear part of the Josephson junction, leaving an inductor $L_J = \hbar^2/4e^2E_J$ \cite{Nigg:2012}. We find that this shift gives a good approximation of the Lamb shift of high modes in the Transmon regime $E_J\gg E_C^{(0)}$. We denote by $L_J$ the inductance of the linearized AA and by $L_m$ and $C_0$ the inductance and capacitance of a coupled parallel LC oscillator representing a bare resonator mode. The dispersive approximation assumes 
\begin{equation}
\omega_m\gg\omega_a\ ,
\label{eq:first_ass}
\end{equation} 
where $\omega_a$ is the resonance frequency of the bare linearized atom $\omega_a = 1/\sqrt{L_J C_c}$ and $\omega_m$ is the resonance frequency of the bare mode resonator $\omega_m = 1/\sqrt{L_m C}$. This condition is assumed to be met due to a small mode inductance
\begin{equation}
L_J\gg L_m
\label{eq:second_ass}
\end{equation} 
as is the case for high frequency modes $m\gg1$. Resonance is reached when the input impedance of the parallel LC oscillator is equal to minus that of the series LC oscillator, which is equivalent to a boundary condition of matching voltage and current at the coupling point shown in Fig. \ref{fig:circuit_shift}. This condition reads
\begin{equation}
\frac{1}{iC_c\omega}+iL_J\omega = - \frac{1}{iC_0\omega+\frac{1}{iL_m\omega}}\ .
\end{equation}
Introducing the bare resonance frequencies corresponding to both resonators shunted to ground at the coupling point, this equation can be rewritten
\begin{equation}
\omega^4-\omega^2\bigg(\omega_m^2+\omega_a^2 + \frac{L_m}{L_J}\omega_m^2\bigg)+\omega_m^2\omega_a^2=0\ .
\end{equation}
This equation has two positive solutions
\begin{equation}
\omega_{\pm} = \frac{\omega_m}{\sqrt{2}}\sqrt{1+\eta\bigg(1+\frac{C_0}{C_c}\bigg)\pm\sqrt{\bigg(1+\eta\bigg(1+\frac{C_0}{C_c}\bigg)\bigg)^2-4\frac{C_0}{C_c}\eta}}\ ,
\end{equation}
where we introduced the quantity $\eta = L_m/L_J$. In the assumption of Eq. (\ref{eq:second_ass}), we obtain to first order in $\eta$ the resonance frequency
\begin{equation}
\omega_-\simeq\omega_a-\frac{\omega_a}{2}\frac{L_m}{L_J}\ ,
\label{eq:classical_lamb_shift_1}
\end{equation}
which yields the value of this shift $\bar{\chi}_m = -\frac{\omega_a}{2}\frac{L_m}{L_J}$. If we introduce the Josephson energy through $L_J = \hbar^2/4e^2E_J$, the atomic frequency $\hbar\omega_a = \sqrt{8E_JE_C^{(0)}}$ and the coupling $\hbar \gamma_m = 2e\sqrt{\frac{\hbar\omega_m}{2C_0}}\bigg(\frac{E_J}{32E_C^{(0)}}\bigg)^\frac{1}{4}$, this shift can be written in the language of the Rabi Hamiltonian following
\begin{equation}
\bar{\chi}_m = -2\frac{\big(\omega_a\big)^2\big(\gamma_m\big)^2}{\omega_m^3}\ .
\end{equation}

Extrapolating this formula for the case of a non-linearized atom in the Transmon regime by making the approximatinos $\omega_a^{(0)}\simeq\omega_a$ and $g_m^{(0)} \simeq \gamma_m$ we obtain the formula for the shift presented in  chapter~\ref{chapter:chapter_2}
\begin{equation}
\chi_m \simeq -2\frac{\big(\omega_a^{(0)}\big)^2\big(g_m^{(0)}\big)^2}{\omega_m^3}\ .
\end{equation}

\section{Numerical methods}
In order to perform numerical calculations, we first diagonalize the AA Hamiltonian $\hat{H}^{(M)}_{\text{AA}}$ (also known as Cooper pair box Hamiltonian) in the charge basis $\{\ket{n_J}\}_{n_J=-N_{\text{max}}, .., +N_{\text{max}}}$ where $\ket{n_J}$ is an eigenstate of $\hat{n}_J$. In this basis the Josephson junction term is given by (see Ref.~\cite{Schuster:2007})
\begin{equation}
\cos (\hat{\varphi}_J) = \frac{1}{2}\sum_{N=-\infty}^{+\infty} \ket{N}\bra{N+1} + \ket{N+1}\bra{N}\ .
\end{equation}
The basis is truncated to a certain number of Cooper pairs $\pm N_{\text{max}}$. We found that using more than $N_{\text{max}} = 20$ has little impact on simulation results for our set of example parameters. After diagonalization of $H_{\text{AA}}$ we can inject the values for $\epsilon_i^{(M)}$ and $\bra{i}^{(M)}\hat{n}_J\ket{j}^{(M)}$ into the Hamiltonian $\hat{H}^{(M)}$ which we in turn diagonalize. Numerical calculations are performed using the Python library QuTIP \cite{Johansson:2013}.\\ 

\begin{figure}[]
	\centering
	\includegraphics[width=0.45\textwidth]{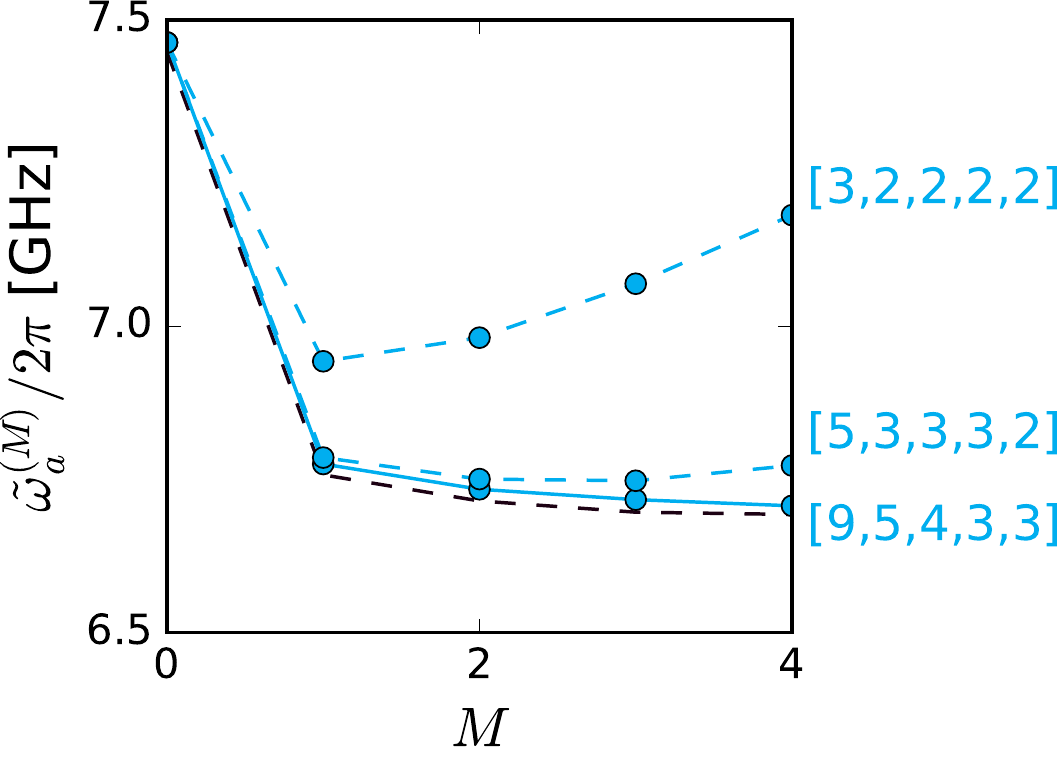}
	\caption{The blue dots correspond to calculations of the dressed first AA transition frequency as we increase the number of modes included in the circuit model. Associated to each line is an array of integers, the first number of the array corresponds to the number of AA levels included in the model, and the following numbers correspond to the number of photon levels included, ordered with increasing mode frequency. Including less photon levels in the modes and in the AA leads to divergences. The dashed black line is the result of applying black box quantization to the $M$ mode lumped element equivalent circuit of the system.}
	\label{fig:numerics}
\end{figure}

What must ensue is a careful choice of the size of the Hilbert space, namely the number of photon levels $n_m$ for the mode $m$ as well as the number of AA levels $n_a$. Note that the size of the Hilbert space scales as $2^{n_a}\prod_{m=0}^{m<M} n_{m}$. We find that a high number of photon levels are needed for convergence. This is particularly true for the modes which are the closest (in frequency) to $\omega^{(0)}_{ge}$. This is illustrated in Fig. \ref{fig:numerics} and explains the difficulty of providing a good estimate for the effective Lamb shift through a simple application of perturbation theory. \\

\section{Bogoliubov transformation} \label{sec:Bogo}
In the case $C_J \ne 0$, one way to recover the form of the Hamiltonian presented in chapter~\ref{chapter:chapter_2} is through a Bogoliubov transformation as described in this section. In the Hamiltonian given by Eq. (\ref{eq:circuit_Hamiltonian}), the energy of the bare resonator modes and the mode-mode coupling term correspond to the Hamiltonian of $M$ coupled harmonic oscillators
\begin{equation}
\hat{H}'= \sum_{m=0}^{m<M}\hbar\omega^{(M)}_m\hat{a}_m^\dagger \hat{a}_m + \sum_{m=0}^{m<M}\sum_{m'=m+1}^{m<M}G^{(M)}_{m,m'}(\hat{a}_m+\hat{a}_m^\dagger)(a_{m'}+a_{m'}^\dagger)\ ,
\label{eq:Hprime}
\end{equation}
which can be diagonalized through a Bogoliubov transformation even for $M$ on the order of thousands \cite{Javanainen:1996}. We start by writting Eq. (\ref{eq:Hprime}) as follows
\begin{equation}
\hat{H}'= \sum_{m,m'=0}^{m,m'<M}[\msf{\Theta}_{m,m'}(\hat{a}_m a_{m'} + \hat{a}_m^\dagger a_{m'}^\dagger) + \msf{\Xi}_{m,m'}(\hat{a}_m^\dagger a_{m'}+\hat{a}_m^\dagger a_{m'}^\dagger)]\ ,
\end{equation}
or, in matrix notation
\begin{equation}
\hat{H}'= \hat{\ba} ^T
\msf{h}'
\hat{\ba}\ ,
\end{equation}
where $\hat{\ba}$ is a vector of the annihilation and creation operators
\begin{equation}
\hat{\ba}^T = [\hat{a}_0, \hat{a}_1, ..., \hat{a}_{M-1},\hat{a}_0^\dagger,\hat{a}_1^\dagger,...,\hat{a}_{M-1}^\dagger]\ ,
\end{equation}
and $\msf{h}'$ is the matrix
\begin{equation}
\msf{h}'= \begin{bmatrix}
\msf{\Theta}     & \msf{\Xi}     \\
\msf{\Xi}    &   \msf{\Theta}     \\
\end{bmatrix}\ .
\end{equation}
In this case $\msf{\Theta}_{m,m'} = \msf{\Xi}_{m,m'} = G_{m,m'}^{(M)}/2$ if $m \ne m'$ and $\msf{\Theta}_{m,m} = 0$, $ \msf{\Xi}_{m,m} = \hbar \omega_m^{(M)} /2$ otherwise. We now find a matrix that maps $\hat{\ba}$ to a new set of creation and annihilation operators $\hat{\bb}$,
\begin{equation}
\hat{\bb}^T = [\hat{b}_0, \hat{b}_1, ..., \hat{b}_{M-1},\hat{b}_0^\dagger,\hat{b}_1^\dagger,...,\hat{b}_{M-1}^\dagger] ,
\end{equation} 
which diagonalize $\hat{H}'$ whilst maintaining the expected commutation relations $[\hat{b}_m,\hat{b}_{m'}^\dagger] = \delta_{m,m'}$. Following the methodology described in Ref. \cite{Javanainen:1996}, we introduce the matrix
\begin{equation}
\msJ= 
\begin{bmatrix}
0     & \mone     \\
-\mone     &   0    \\
\end{bmatrix}\ ,
\end{equation}
where $\mone$ is an $M\times M$ identity matrix. Diagonalizing the matrix $\msf{h}'\msJ$ yields eigenvalues that come in pairs such that if $\mu$ is an eigenvalue, then $-\mu$ is too. We order the eigenvalues and eigenstates such that the negative eigenvalues come first, in order of increasing absolute value, and the corresponding positive eigenvalues next, in the same order. We use the following notation for these eigenvalues
\begin{equation}
[-\mu_0,-\mu_1, ..., -\mu_{M-1}, \mu_0,\mu_1, ..., \mu_{M-1}]\ .
\end{equation}
We then construct a matrix $\msf{F}$ with the eigenvectors as columns and normalize them such that the $\msf{F}$ is sympletic: $\msf{F}^T\msJ\msf{F}=\msJ$. To do so, we normalize each eigenvector $v_m$ such that $\sum_{i=0}^{i<2M}(v_m)_i^2 = 1$ and flip the sign of certain eigenvectors such that the first coeffecient of $v_m$ (with eigenvalue $-\mu_m$) has the same sign as the $M$-th coefficient of $v_{m+M}$ (with eigenvalue $\mu_m$). The matrix thus constructed should be of the form
\begin{equation}
\msf{F}= 
\begin{bmatrix}
\msA     & \msB     \\
\msB     &   \msA     \\
\end{bmatrix}\ .
\end{equation}
\begin{figure}[]
	\centering
	\includegraphics[width=0.45\textwidth]{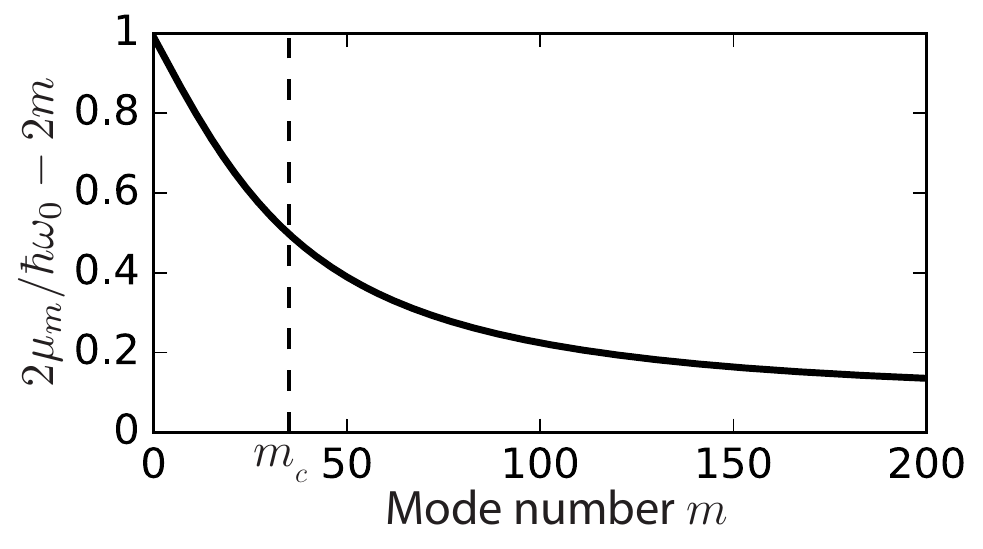}
	
	\caption{Mode frequency as a function of the number of modes in the case $C_J=5$ fF (other circuit parameters fixed in the corresponding chapter). The critical mode $m_C$ marks a transition from the regime where the resonator acts as a $\lambda/4$ resonator, with frequencies $2\mu_m\simeq (2m+1)\omega_0$ to a regime where the resonator becomes a $\lambda/2$ resonator, with eigenfrequencies $2\mu_m\rightarrow 2m\omega_0$.}
	\label{fig:fn_vs_m}
\end{figure}

By defining the vector of annihilation and creation operators $\hat{\bb}$ as 
\begin{equation}
\hat{\ba} = \begin{bmatrix}
\msA     & -\msB     \\
- \msB     &   \msA     \\
\end{bmatrix}\hat{\bb}\ , 
\end{equation}
we have defined a basis which diagonalizes $\hat{H}'$
\begin{equation}
\hat{H}'= \sum_{m=0}^{m<M}2\mu ^{(M)}_m\hat{b}_m^\dagger \hat{b}_m\ ,
\end{equation}
the new eigenenergies in fact being given by twice the positive eigenvalues of the previously diagonalized matrix. In this basis, the atom-mode interaction term becomes
\begin{equation}
\hat{H}^{(M)}_{\text{int}} = \sum_{i,j}\sum_{m=0}^{m<M}\bigg[\sum_{m'=0}^{m'<M}\hbar g^{(M)}_{m',i,j}(\msA-\msB)_{m',m}\bigg]\ket{i}^{(M)}\bra{j}^{(M)}(\hat{b}_m+\hat{b}_m^\dagger)\ ,
\end{equation}
and we recover the extended Rabi Hamiltonian structure by defining the coupling as
\begin{equation}
g^{(M)}_{m,i,j} = \sum_{m'=0}^{m'<M}\hbar g^{(M)}_{m',i,j}(\msA-\msB)_{m',m}\ .
\end{equation}

\begin{figure}[h]
	\centering
	\includegraphics[width=0.25\textwidth]{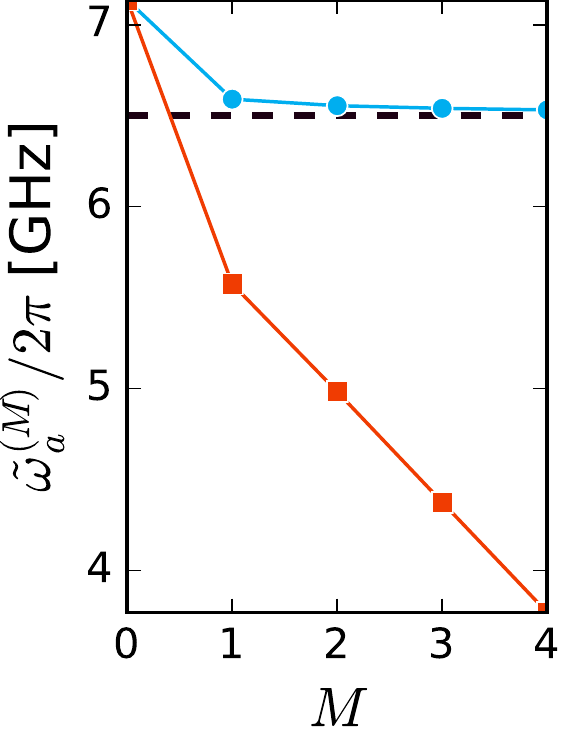}
	
	\caption{Convergence of the spectrum in the case $C_J=5$ fF (other circuit parameters fixed in chapter~\ref{chapter:chapter_2}). We plot the $\ket{g}\rightarrow\ket{e}$ transition frequency of the atom dispersively shifted by $M$ resonator modes as a function of $M$. Dots (squares) correspond to an exact diagonalization of the circuit (non-)renormalized multi-mode Rabi model. The frequency obtained by black-box quantization (dashed line) provides a point of reference corresponding to the case when all modes are included for the linearized system, and quartic anharmonicities are added for a truncated 3-mode model.}
	\label{fig:spectrum}
\end{figure}

This coupling strength was plotted in chapter~\ref{chapter:chapter_2}. In Fig.~\ref{fig:fn_vs_m}, we plot the frequencies of the newly defined eigenmodes of the resonator. As expected, these transition from the eigenfrequencies of a $\lambda/4$ resonator to those of a $\lambda/2$. In Fig. \ref{fig:spectrum}, we show the same plot as in Fig. 3 of the corresponding chapter but for $C_J=5$ fF, the result of diagonalizing the Hamiltonian derived above.

\section{Dressing of the atomic charging energy}

The Hamiltonian $H^{(M+1)}$ with $M+1$ bosonic modes coupled to the Josephson junction (Eq. (1) of the chapter~\ref{chapter:chapter_2}) lives in the Hilbert space $H^{(M+1)}=H_{n_J}\otimes_{m \leq M}H_m$

\begin{align}
H^{(M+1)}&= \hbar\omega_M a_M^{\dagger}a_M+\hbar\bar{g}_M(a_M+a_M^{\dagger})\hat{n}_J + H^{(M)}\nonumber\\
&=\hbar\omega_M \left(a_M^{\dagger}+\frac{\bar{g}_M}{\omega_M}\hat{n}_J\right)\left(a_M+\frac{\bar{g}_M}{\omega_M}\hat{n}_J\right) + H^{(M)}-\hbar\frac{\bar{g}_M^2}{\omega_M}\hat{n}_J^2\nonumber\\
&= \hbar\omega_M b_M^\dagger b_M + \tilde{H}^{(M)},
\end{align}

where we have defined the bosonic operators 
\begin{align}
b_M &= a_M + \frac{\bar{g}_M}{\omega_M}\hat{n}_J,\nonumber\\
b_M^\dagger &= a_M^\dagger + \frac{\bar{g}_M}{\omega_M}\hat{n}_J,\nonumber
\end{align}

and the Hamiltonian
\begin{equation}
\tilde{H}^{(M)} = H^{(M)}-\hbar\frac{\bar{g}_M^2}{\omega_M}\hat{n}_J^2
\end{equation}

We look for an effective Hamiltonian which approximates the low energy part of $H^{(M+1)}$. The pair $b_M$, $b_M^\dagger$ are canonically conjugate, $\left[b_M,b_M^\dagger\right]=1$. Thus, $b_M^\dagger b_M$ is a number operator. If $\hbar\omega_M$ is much larger than the characteristic energy of the low energy sector of $\tilde{H}^{(M)}$, the low energy sector of $H^{(M+1)}$ will be well approximated by setting $b_M^\dagger b_M$ to zero. That is, by studying the restriction of $H^{(M+1)}$ to the vacuum subspace of $b_M$, namely 
\begin{align}
\mathcal{S}^{(M+1)}=\{\Ket{\Psi}/ \,\,\,b_M\Ket{\Psi}=0\}.
\end{align}
In order that the separation of scales that has been assumed indeed holds, it is also imperative that $\tilde{H}^{(M)}$ acting on $\mathcal{S}^{(M+1)}$ results in states neighbouring $\mathcal{S}^{(M+1)}$. That is to say, that the commutator $\left[b_M,\tilde{H}^{(M)}\right]$ acting on $\mathcal{S}^{(M+1)}$ be small. In the case at hand, 

\begin{align}
\left[b_M,\tilde{H}^{(M)}\right]&=\frac{\bar{g}_M}{\omega_M}\left[\hat{n}_J,H^{(M)}\right]\nonumber\\
&=i\frac{E_J\bar{g}_M}{\omega_M}\sin(\varphi_J),
\end{align}
so if $E_J, \bar{g}_M\ll \omega_M$ then we can say that the commutator above is small, and that $\tilde{H}^{(M)}|_{\mathcal{S}^{(M+1)}}$ will provide a good effective Hamiltonian for $H^{(M+1)}$. Notice that these conditions are increasingly better fulfilled with growing mode number $M$ for the model in the above chapter~\ref{chapter:chapter_2}. We now construct explicitly the effective Hamiltonian $\tilde{H}^{(M)}|_{\mathcal{S}^{(M+1)}}$. The subspace $\mathcal{S}^{(M+1)}$ can be expanded in the following basis
\begin{equation}
\Ket{\alpha_{n_J}}^{(M+1)} = \Ket{n_J}\Ket{\beta}^{(M)}\Ket{z_M = -\gamma_M n_J},
\end{equation}
where vectors $\Ket{\beta}^{(M)}$ form a basis of the truncated subspace $\otimes_{m< M}H_m$, $\Ket{z_{M}}$ is a coherent state for the $(M+1)$-th mode and $\gamma_M = \bar{g}_M/\omega_M$. The original bosonic $a_M$ and Cooper-Pair number $\hat{n}_J$ operators act on this basis as
\begin{align}
a_M\Ket{\alpha_{n_J}}^{(M+1)}&=-\gamma_M n_J\Ket{\alpha_{n_J}}^{(M+1)},\nonumber\\
\hat{n}_J\Ket{\alpha_{n_J}}^{(M+1)}&=n_J\Ket{\alpha_{n_J}}^{(M+1)}.\nonumber
\end{align}
Thus, the matrix elements of $H^{(M+1)}|_{\mathcal{S}^{(M+1)}}$ are
\begin{align}
\Bra{\alpha_{n_J}}^{(M+1)}H^{(M+1)}\ket{\alpha_{M_J}}^{(M+1)}&= \left<-\gamma_N n_J|-\gamma_N M_J\right>\bra{\alpha_{n_J}}^{(M)}\tilde{H}^{(M)}\ket{\alpha_{M_J}}^{(M)}\nonumber\\
&=e^{-\gamma_M^2 \left(n_J-M_J\right)^2/2}\bra{\alpha_{n_J}}^{(M)}\tilde{H}^{(M)}\ket{\alpha_{M_J}}^{(M)}\nonumber\\
&\approx\bra{\alpha_{n_J}}^{(M)}\tilde{H}^{(M)}\ket{\alpha_{M_J}}^{(M)},
\end{align}
where the last line gives us a further approximation, valid if $\gamma_M\ll1$ and the low energy states of $\tilde{H}^{(M)}$ have small dispersion for $\hat{n}_J$. If these indeed hold, $\tilde{H}^{(M)}$ itself is a good effective Hamiltonian for $H^{(M+1)}$. We can iterate this procedure down to a mode $L$ for which the above conditions still holds. For $M\rightarrow\infty$, an effective Hamiltonian with $L$ modes is then given by 

\begin{equation}
H=\sum_{m=0}^{m<L}\hbar\omega_{m}^{(L)}a_{m}^{\dagger}a_{m}+ \hbar\bar{g}_{m}^{(L)}\left(a_{m}+a_{m}^{\dagger}\right)\hat{N}_{J}+4\tilde{E}_{C}^{(L)}\hat{N}_{J}^{2}-E_{J}\,\mathrm{cos}(\hat{\varphi}_J),
\end{equation}
\begin{equation}
\tilde{E}_C^{(L)} = \lim_{M\rightarrow\infty}E_C^{(M)} - \sum_{m\geq L}^{M}\hbar(\bar{g}^{(M)}_{m})^{2}/4\omega_{m} .
\end{equation}

\chapter{Further Details on Reciprocal Distributed Networks}
\label{appendix_b}

\section{Main mathematical results}
\label{Walter_appendix}
The main tools used in chapter~\ref{chapter:chapter_3} have been (i) mapping the transmission line Lagrangians from a field presentation to a mode description, that takes into account the coupling at \emph{points} (through lumped elements) with other circuit elements, and (ii) manipulations of vectors $\bu$ and matrices $\bu\bu^T$ as in the finite vector case even for full mode expansions.

Both these aspects can be justified in a Hilbert space context (even though more general presentations could be possible) by addressing there the interesting problem of boundary conditions that incorporate the singular value, as we presently see.

In this Appendix we first consider finite transmission lines and then half-line transmission lines. With lesser detail we signal two other configurations, and then we give a general description of the recipe we have applied.

\subsection{Finite length transmission lines}
\label{sec:finite-length-transm}

Consider the following singular value problem:
\begin{align}
- u''(x)  &= k^2 u(x)\,\quad x\in(0,L),\label{eq:appwalt1}\\
u(L)&=0,\\
\frac{1}{\beta}u(0)- u'(0)&=  \alpha k^2 u(0).\label{eq:appwalt3}
\end{align}
$\alpha$ and $\beta$ are constants with dimension of length.

The presence of the singular value $k^2$ in the boundary condition at $0$ means that this is not a Sturm--Liouville problem, and the usual oscillation and expansion theorems are therefore not applicable. Nonetheless, it is easy to establish a secular equation, which the singular values must fulfill, and identify the corresponding functions. In the example at hand, the functions are
\begin{equation}
\nonumber
u_n(x)=N_n\sin\left[k_n(L-x)\right], 
\end{equation}
for singular value $k_n$, where the normalization factor $N_n$ will be fixed later, and the secular equation reads
\begin{equation}
\label{eq:appsecular}
\frac{\alpha}{L}\left(kL\right)^2\sin(kL) - (kL) \cos(k L) - \frac{L}{\beta} \sin(k L)=0.
\end{equation}

Using this secular equation it is easy to establish that the functions $u_n(x)$ are orthogonal with respect to the inner product
\begin{equation}
\label{eq:appinnerprod}
\langle u,v\rangle = \int_0^L \mathrm{d}x\,\bar{u}(x) v(x) +\alpha \bar{u}(0) v(0),
\end{equation}
with positive $\alpha$.

\subsubsection{Expansion theorem}
\label{sec:appexpansion}
We now need to establish an expansion theorem, that guarantees that any function $f$ defined in the interval $(0,L)$ can be written as a superposition $f=\sum_n f_n u_n$ with coefficients $f_n$, and in such a way that the value at the endpoint is also recovered. This will be achieved by identifying problem (\ref{eq:appwalt1}-\ref{eq:appwalt3}) with an eigenvalue problem for a self-adjoint operator, following the idea presented in \cite{Walter:1973}. We present here an alternative proof of that expansion theorem.

Thus, consider the Hilbert space ${\cal H}=L^2\left[(0,L)\right]\oplus \mathbb{C}_\alpha$, with elements $U=(u,a)$, where $u\in L^2\left[(0,L)\right]$ and $a$ is a complex number. The definition as a direct sum entails the inner product
\begin{equation}
\label{eq:biginnerprod}
\langle U,V\rangle = \int_0^L \mathrm{d}x\,\bar{u}(x) v(x) +\alpha \bar{a} b,
\end{equation}
for elements $U=(u,a)$ and $V=(v,b)$. Again, $\alpha$ is a positive length.

Let us now define an operator $A$, of inverse length squared dimension,  with domain
\begin{equation}
\nonumber
D(A)= \left\{U=(u,a)\in {\cal H}|u,u'~\mathrm{are}~ AC[0,L]\,,\,u(L)=0\,,\,a=\lim_{x\to 0}u(x)\right\}
\end{equation}
where $AC[0,L]$ denotes absolutely continuous in the interval, 
and acting on elements of its domain as
\begin{equation}
\nonumber
AU=(-u'',-u'(0)/\alpha+u(0)/\alpha\beta).
\end{equation}
$\beta$ is another positive length.

It is easy to check that this is a symmetric operator, i.e., $\langle U,AV\rangle=\langle AU,V\rangle$ for all $U,V\in D(A)\subset {\cal H}$. It is also easy to check that it is a monotone (accretive, dissipative) operator: for all $U\in D(A)$
\begin{equation}
\label{eq:monotone}
\langle U,AU\rangle= \int_0^L \mathrm{d}x\,|u'|^2+ \frac{1}{\beta}|u(0)|^2\geq 0.
\end{equation}

We now prove that $A$ is maximal monotone. That is, we prove that the range of $1+L^2A$ is the whole Hilbert space ${\cal H}$. In other words that for all $F\in {\cal H}$ there exists $U\in D(A)$ such that $U+L^2AU=F$. We insert the length squared $L^2$ for dimensional reasons, but we need not use precisely the length of the interval; multiplying this length by any number provides us with the same result. The problem now consists in showing the existence of solutions $u$, such that both $u$ and $u'$ are absolutely continuous, for the problem
\begin{align}
\label{eq:appexistence}
-L^2 u''+u&= f,\nonumber\\
u(L)&=0,\\
\left(1+\frac{L^2}{\alpha\beta}\right)u(0)- \frac{L^2}{\alpha}u'(0)&= a, \nonumber                                                                     
\end{align}
where $F=(f,a)$, with $f\in L^2[(0,l)]$ and $a$ a complex number. Make the change of variable
\begin{equation}
\nonumber
u(x)=v(x)+ \frac{a\sinh\left(1-x/L\right)}{\sinh(1)\left(1+L^2/\alpha\beta\right)+\cosh(1) L/\alpha}.
\end{equation}
We now have to prove the existence of a solution $v$, such that $v,v' \in AC[0,L]$, for the following problem with inhomogeneous term $f\in L^2[(0,l)]$:
\begin{align}
\nonumber
 -L^2 v''+v&= f,\nonumber\\
 v(L)&= 0,\nonumber\\
\left(1+\frac{L^2}{\alpha\beta}\right)v(0)- \frac{L^2}{\alpha}v'(0)&= 0.\nonumber
\end{align}
This existence has been well established from the Sturm--Liouville case (see for instance \cite{Reed:1975}), whence the existence (and uniqueness) of solutions for problem (\ref{eq:appexistence}) is obtained, and thus maximal monotony of $A$. It follows that it is self-adjoint (see, for instance, proposition 7.6 in \cite{Brezis:2010}), and the spectral theorem of self-adjoint operators provides us with the expansion theorem we desired. Namely, the operator $A$ has a discrete spectrum $\{k^2_n\}_{n=0}^\infty$, solutions of (\ref{eq:appsecular}), $k_n$ being inverse lengths, with eigenvectors $U_n=\left(u_n(x),u_n(0)\right)$ where
\begin{equation}
\nonumber
u_n(x)= N_n \sin\left[k_n(L-x)\right],
\end{equation}
and the normalisation can be chosen as
\begin{equation}
\nonumber
N_n^2=\left[\frac{L}{2} +\left(\frac{\alpha}{2}+ \frac{1}{2\beta k_n^2}\right)\sin^2(k_nL)\right]^{-1}
\end{equation}
for the eigenvectors $\left\{U_n\right\}_{n=0}^\infty$ to form an orthonormal basis, $\left\langle U_n,U_m\right\rangle=\delta_{nm}$. Notice that we have chosen a real basis, and we will use this fact in the formulae that follow in this subsection \ref{sec:finite-length-transm}.
An element $F=(f,a)\in {\cal H}$ admits an expansion $F=\sum_{k=0}^\infty \langle U_k,F\rangle U_k$. Consider now an element $f\in L^2[(0,L)]$, with continuous representative $\tilde{f}$,  and extend it to ${\cal H}$ as $F=(f,\tilde{f}(0))$. We thus obtain the expansion we desired:
\begin{equation}
\nonumber
f(x)= \sum_{n=0}^\infty\langle U_n,F\rangle u_n(x) = \sum_{n=0}^\infty u_n(x)\left[\alpha u_n(0)\tilde{f}(0)+\int_0^L \mathrm{d}\xi\,u_n(\xi) f(\xi)\right].
\end{equation}

\subsubsection{Secondary inner product}
\label{sec:second-inner-prod}
We shall now prove (with the normalisation here used) equation (\ref{eq:TL_LCcoup_Network_ortho_2}). In what follows, the first step is integration by parts, the second makes use of the fact that $-u''_m(x)=k_m^2 u_m(x)$, the third relates the computation to the orthogonality $\langle U_n,U_m\rangle =\delta_{nm}$, and the last one introduces the boundary condition of (\ref{eq:appwalt1}):
\begin{align}
\nonumber
\int_0^L\mathrm{d}x\,u_n'(x)u_m'(x)&= -\int_0^L\mathrm{d}x\,u_n(x)u_m''(x)-u_n(0)u_m'(0)\nonumber\\
&= k_m^2 \int_0^L\mathrm{d}x\,u_n(x)u_m(x) -u_n(0)u_m'(0)\nonumber\\
&= k_m^2\left[\langle U_n,U_m\rangle -\alpha u_n(0)u_m(0)\right]-u_n(0)u_m'(0)\nonumber\\
&= k_m^2 \delta_{nm} -u_n(0)\left[\alpha k_m^2 u_m(0)+u_m'(0)\right]\nonumber\\
&= k_m^2 \delta_{nm} -\frac{1}{\beta}u_n(0)u_m(0).
\end{align}
Alternatively, let us consider the quantity
\begin{equation}
\label{eq:betainner}
\left\langle U_n,F\right\rangle_{1/\beta}=\int_0^L\mathrm{d}x\,f'(x)u'_n(x) + \frac{1}{\beta}a u_n(0),
\end{equation}
for $F=(f,a)\in{\cal H}$. It is in fact well defined by integration by parts, and it thus follows that
\begin{equation}
\label{eq:betaalphaequiv}
\left\langle U_n,F\right\rangle_{1/\beta}= k_n^2\left\langle U_n,F\right\rangle,
\end{equation}
where the right hand side is the inner product we have used above, namely (\ref{eq:biginnerprod}).

We shall now extend this operation to an inner product in ${\cal H}$. As we have shown above, \ref{sec:appexpansion}, an element $F\in{\cal H}$ can be expanded as $F=\sum_nF_nU_n$, and the inner product reads $\langle F,G\rangle=\sum_n\bar{F}_nG_n$. Then we define $\langle F,G\rangle_{1/\beta}$ as
\begin{equation}
\label{eq:betainnergen}
\left\langle F,G\right\rangle_{1/\beta}=\sum_{n=0}^\infty k_n^2 \bar{F}_nG_n.
\end{equation}
Notice that this is an inner product: it is not degenerate because $k_n^2>0$ for all $n$ (v. Eq. (\ref{eq:monotone})). In fact, this is the inner product associated with the natural quadratic form induced by $A$, as presented in Eq. (\ref{eq:monotone}).

We immediately obtain an orthonormal basis $V_n=U_n/k_n$ with respect to this inner product, and we are led to the corresponding expansion theorem,
\begin{equation}
\label{eq:betaexpansion}
F=\sum_{n=0}^\infty \langle V_n,F\rangle_{1/\beta} V_n.
\end{equation}
\subsubsection{Sum rules}
\label{sec:appsumrules}

The expansion in ${\cal H}$ indicated above provides us with sum rules that prove very useful in our analysis of circuits. First, consider the special element of ${\cal H}$ given by $U=(0,1)$. Since it admits an expansion, it is the case that
\begin{equation}
\label{eq:firstsumrule}
1= \alpha\sum_{n=0}^\infty u_n^2(0).
\end{equation}
Notice that, asymptotically,
\begin{equation}
\label{eq:asymptotics}
k_n\sim \frac{n\pi}{L}+ \frac{L}{\alpha n \pi}+O(n^{-2}),
\end{equation}
which ensures convergence, since $u_n(0)\sim (-1)^n 2/\alpha n\pi$.

The sum rule (\ref{eq:firstsumrule})  is explicitly proven as follows:
\begin{align}
\label{eq:ruleproof}
(0,1)&= \sum_{n=0}^\infty \langle U_n,(0,1)\rangle U_n= \sum_{n=0}^\infty \alpha u_n(0) U_n\nonumber\\
&= \left(\sum_{n=0}^\infty \alpha u_n(0) u_n(x), \alpha\sum_{n=0}^\infty u_n^2(0)\right)
\end{align}

In chapter~\ref{chapter:chapter_3} we have used a different notation, namely $\left|\bu\right|^2=N_\alpha/\alpha c$. The vector $\bu$ is related to the sequence $\{u_n(0)\}$ by an overall normalization factor $\sqrt{N_\alpha /c}$.

Let us now obtain another sum rule by expanding $(0,1)\in{\cal H}$ in the basis $V_n$, orthonormal with respect to $\langle \bullet,\bullet\rangle_{1/\beta}$. Clearly $\left\langle V_n, (0,1)\right\rangle_{1/\beta}=v_n(0)/\beta=u_n(0)/\beta k_n$, and it follows that
\begin{equation}
\label{eq:secondsumrule}
1=\frac{1}{\beta}\sum_{n=0}^\infty \frac{u_n^2(0)}{k_n^2}.
\end{equation}
These sum rules can be understood from a more general point of view in the doubled space in further Appendix~\ref{appendix_d}, with multiple lines connected through general lumped-element networks. 
\subsection{Infinite length transmission lines}
\label{subsec:infin-length-transm}
Let us now consider that the interval is only semibounded, i.e. $(0,\infty)$. As we shall see, the corresponding operator $A$ does not have a discrete spectrum, but is nonetheless self-adjoint, and an expansion theorem, in the form of an integral transform, does hold. Thus, we now examine the problem
\begin{align}
\label{eq:appwaltinf}
- u''(x)&= k^2 u(x),\,\quad x\in(0,\infty),\\
\frac{1}{\beta}u(0)- u'(0)&=  \alpha k^2 u(0),
\end{align}
where, again,
$\alpha$ and $\beta$ are positive constants with dimension of length, and we require (square) normalizability of $u$. Clearly there are no strong solutions to this problem. Setting aside functional details, it can nonetheless be checked that the following $u_k(x)$ functions are generalised orthonormal with respect to the inner product
\begin{equation}
\label{eq:appinnerprodinfty}
\langle u,v\rangle = \int_{\mR_+} \mathrm{d}x\,\bar{u}(x) v(x) +\alpha \bar{u}(0) v(0)\,:
\end{equation}
\begin{equation}
\label{eq:genorthornor}
u_k(x)= \frac{\left(k+ i \alpha k^2 - i/\beta\right) e^{ikx} + \left(k- i \alpha k^2 + i/\beta\right) e^{-ikx}}{\sqrt{2\pi\left[k^2+\left(\alpha k^2- 1/\beta\right)^2\right]}}. 
\end{equation}
Observe that we have again chosen real $u_k(x)$, and furthermore that $u_{-k}(x)=-u_k(x)$, which allows us to restrict ourselves to positive $k$.
They are generalised orthonormal in the sense that
\begin{equation}
\label{eq:orhtoinfty}
\langle u_k,u_q\rangle= \delta(k-q).
\end{equation}
In order to prove this statement directly it is convenient to use the distributional identity
\begin{equation}
\label{eq:distribeq}
\int_{\mR_+}\mathrm{d}x\, e^{i(k-q) x}= \pi \delta(k-q) + i \mathrm{P}\frac{1}{k-q},
\end{equation}
where $\mathrm{P}$ denotes principal part.

\subsubsection{Expansion theorem}
\label{sec:expansion-theorem-cont}

We shall follow the scheme presented in \ref{sec:appexpansion} to prove a corresponding expansion theorem.
We shall make use of the well known fact that, in physics terms, the free Hamiltonian in the half-line is well defined and self-adjoint once we fix at the origin the condition
\begin{equation}
\label{eq:bchamhalf}
\cos(\theta)u(0)+\sin(\theta) l u'(0)=0,
\end{equation}
with $\theta$ and angle and $l$ the unit length. Notice that for the non-relativistic free particle there is no natural length. We are free to choose any length, and in fact nothing crucial in our proof will depend on the choice adopted. Thus, we select $l=\sqrt{\alpha\beta}$.
Let us denote the free Hamiltonian in the half-line with this choice of length and condition (\ref{eq:bchamhalf}) $H_\theta$, and its domain by $D(H_\theta)\subset L^2\left[(0,\infty)\right]$. In fact, we shall also make the choice
\begin{equation}
\label{eq:thetachoice}
\tan(\theta)= -\frac{1}{2}\sqrt{\frac{\beta}{\alpha}},
\end{equation}
for later convenience.

Let us now introduce an operator $A$ acting on $D(A)\subset {\cal H}= L^2\left[(0,\infty)\right]\oplus \mathbbm{C}_\alpha$, given by
\begin{equation}
\label{eq:domainf}
D(A)=\left\{U=(u,a)\in{\cal H}|u\in H^2[(0,\infty)],\,a=\lim_{x\to0}u(x)\right\}.
\end{equation}
$H^2[(0,\infty)]$ is the Sobolev space $H^2[(0,\infty)]=W^{2,2}[(0,\infty)]$ (for details, see for instance \cite{Brezis:2010} or \cite{Lions:2011}).
The inner product in ${\cal H}$ is of course
\begin{equation}
\label{eq:innerinf}
\left\langle U,V\right\rangle=\left\langle(u,a),(v,b)\right\rangle=\int_{\mR_+}\mathrm{d}x\,\bar{u}(x)v(x)+ \alpha\bar{a}b.
\end{equation}

The operator $A$ acts on its domain as
\begin{equation}
\label{eq:actionAinf}
AU=(-u'',-u'(0)/\alpha+u(0)/\alpha\beta).
\end{equation}
Both $u(0)$ and $u'(0)$ are understood as limits.
It is again easy to prove that it is a symmetric operator, and that it is monotone. In order to prove that is maximal monotone we examine the invertibility of $1+ \alpha\beta A$ in ${\cal H}$. This requires us studying the problem
\begin{align}
\label{eq:appexistenceinf}
-\alpha\beta u''+u&= f,\nonumber\\
2u(0)- \beta u'(0)&= a,
\end{align}
for $f\in L^2[(0,\infty)]$, $a\in\mathbbm{C}$, and $u$ square summable. Make the change of variable
\begin{equation}
\label{eq:chofvarinf}
u(x)=\frac{a}{2+\sqrt{\beta/\alpha}} e^{-x/\sqrt{\alpha\beta}}+v(x).
\end{equation}
We now have to study the problem
\begin{align}
\label{eq:vexistenceinf}
-\alpha\beta v''+v&= f,\nonumber\\
2v(0)- \beta v'(0)&= 0,
\end{align}
and the boundary condition at the origin is $v(0)+\tan(\theta)\sqrt{\alpha\beta}v'(0)=0$, with the choice (\ref{eq:thetachoice}). As $H_\theta$ is selfadjoint and positive, the existence (and uniqueness) of the solution of this problem is guaranteed, whence the existence (and uniqueness) of the solution of (\ref{eq:appexistenceinf}) for all $F\in{\cal H}$ follows. As a consequence, there is a unique self-adjoint extension of $A$ as defined above, and we obtain the expansion theorem we desire: for all $F\in{\cal H}$ we have $F=\int_{\mR_+}\mathrm{d}k\,\langle U_k,F\rangle U_k$, using physics notation, where $U_k=(u_k(x),u_k(0))$, with $u_k(x)$ defined in (\ref{eq:genorthornor}). Restricting ourselves to $f\in L^2[(0,\infty)]$, we obtain
\begin{align}
\label{eq:infexpansion}
f_k&= \int_{\mR_+}\mathrm{d}x\, u_k(x) f(x) + \alpha u_k(0) f(0)\nonumber\\
\mathrm{and}\quad f(x)&= \int_{\mR_+}\mathrm{d}k\, f_k u_k(x),
\end{align}
with the caveats more usual for Fourier transforms.

\subsubsection{Secondary inner product}
\label{sec:second-inner-prod-inf}

We now proceed to construct a second expansion by considering a new inner product. As in the finite interval case, it is defined from the natural quadratic form induced by $A$,
\begin{equation}
\label{eq:quadinf}
\langle U,U\rangle_{1/\beta}=\left\langle U,AU\right\rangle=\int_{\mR_+}\mathrm{d}x\, |u'(x)|^2+ \frac{1}{\beta} |u(0)|^2.
\end{equation}
The extension of this inner product to ${\cal H}$ can be presented via polarization identities, via a Parseval identity, or alternatively by the spectral theorem, namely
\begin{equation}
\label{eq:secondinnerinf}
\left\langle F,G\right\rangle = \int_{\mR_+}\frac{\mathrm{d}k}{k^2}\bar{F}_kG_k,
\end{equation}
where
\begin{equation}
\label{eq:fkexp}
F_k=\left\langle U_k,F\right\rangle=\int_{\mR_+}\mathrm{d}x\,u_k(x)f(x)+ \alpha u_k(0) a
\end{equation}
for $F=(f,a)\in{\cal H}$, and analogously for $G$. The kernel of the integration is $1/k^2$, thus seemingly producing a non-integrable singularity at the origin. A more careful analysis implies analyzing $u_k(x)u_k(y)/k^2$, which is in fact regular at the origin $k=0$. It is therefore feasible to establish a (generalised) orthonormal basis with respect to this inner product, $\left\langle V_k,V_q\right\rangle_{1/\beta}=\delta(k-q)$, by $V_k=U_k/k$. A new expansion theorem follows, in the form
\begin{equation}
\label{eq:secondexpinf}
F=\int_{\mR_+}\mathrm{d}k\,\left\langle V_k,F\right\rangle_{1/\beta}V_k.
\end{equation}
\subsubsection{Sum rules}
\label{sec:infsumrules}

Let us now consider sum rules, analogous to (\ref{eq:firstsumrule}). In particular, let us expand $F=(0,1)$ in the form  $F=\int_{\mR_+}\mathrm{d}k\,\langle U_k,F\rangle U_k$ . Clearly, $\langle U_k,F\rangle= \alpha u_k(0)$. Thus the expansion reads, for the second component of $F$,
\begin{align}
\label{eq:contsumrule}
1&= \int_{\mR_+}\mathrm{d}k\, \alpha u_k^2(0)= \frac{\alpha}{\pi} \int_o^\infty\mathrm{d}k\,\frac{2 k^2}{k^2+\left(\alpha k^2-1/\beta\right)^2}\nonumber\\
&= \frac{1}{\pi}\int_{-\infty}^{+\infty}\mathrm{d}q\,\frac{q^2}{\left(q^2-x^2\right)^2+q^2},
\end{align}
where $x=\sqrt{\alpha/\beta}$. Remember that we are using a real basis $u_k(x)$. This last integral can be computed explicitly by residues, and it is actually independent of the real variable $x$, thus providing us with an independent check of the expansion and the sum rule derived thereof, namely
\begin{equation}
\label{eq:contsumrule2}
\int_{\mR_+}\mathrm{d}k\,u_k^2(0)= \frac{1}{\alpha}.
\end{equation}
The sum rule analogous to (\ref{eq:secondsumrule}) follows from the expansion (\ref{eq:secondexpinf}), and  reads
\begin{equation}
\label{eq:contsecondsumrule}
\beta=\int_{\mR_+}\mathrm{d}k\,\frac{u_k^2(0)}{k^2}=\frac{\alpha}{\pi}\int_{\mR_+}\frac{\mathrm{d}q}{q^2+\left(q^2-\alpha/\beta\right)^2}.
\end{equation}
Again, this integral can be explicitly computed, and the sum rule is checked. Indeed, let
\begin{equation}
g(q,x)=\frac{1}{q^2+(q^2-x^2)},\nonumber
\end{equation}
with which we compute
\begin{align}
\label{eq:explicitint}
I_g(x)&=\frac{1}{\pi} \int_{\mR_+}\frac{\mathrm{d}q}{q^2+\left(q^2-x^2\right)^2}\nonumber\\
&= 2i\left[\lim_{q\to i(1+\sqrt{1-4x^2})/2}\left(q-\frac{i}{2}(1+\sqrt{1-4x^2})\right)g(q,x)\right.\nonumber\\
&\quad\left.+\lim_{q\to i(1-\sqrt{1-4x^2})/2}\left(q-\frac{i}{2}(1-\sqrt{1-4x^2})\right)g(q,x)\right]=\frac{1}{x^2}.
\end{align}

\subsection{Other configurations}
\label{sec:other-configurations}

Let us now consider other configurations. We shall not give all the details, which can be filled out following the scheme in the previous subsections.

\subsubsection{Galvanic coupling}
\label{sec:galvanic-coupling}

Let the Hilbert space be $L^2[(-L,0)]\oplus L^2[(0,L)]\oplus\mathbbm{C}_\alpha$, with elements $F=(f_-,f_+,a)$. The inner product is
\begin{equation}
\label{eq:galvanicinner}
\left\langle (f_-,f_+,a),(g_-,g_+,b)\right\rangle=\int_{-L}^0\mathrm{d}x\,\bar{f}_- g_-+\int_0^L\mathrm{d}x\,\bar{f}_+ g_++ \alpha \bar{a}b.
\end{equation}

We define an operator $A$ with domain
\begin{align}
\label{eq:galvanicformal}
D(A)=&\left\{U=(u_-,u_+,\Delta)|u_-,u_-'\in AC[(-L,0)],\,u_+,u_+'\in AC[(0,L)],\right.\nonumber\\ 
&\left. u_+(L)=u_-(-L)=0,\,u_-'(0)=u_+'(0),\,\Delta = u_+(0)-u_-(0)\right\},
\end{align}
acting on its domain as $AU=(-u_-'',-u_+'',- u'(0)/\alpha+\Delta/\alpha\beta)$, with $\Delta=u_+(0)-u_-(0)$.
It is clear that it is a  symmetric operator, and it is positive since
\begin{equation}
\label{eq:uaugalv}
\left\langle U,AU\right\rangle=\int_{-L}^0\mathrm{d}x\,|u_-'|^2+\int_{0}^L\mathrm{d}x\,|u_+'|^2+\frac{1}{\beta}\left|\Delta\right|^2.
\end{equation}

The eigenvalue and eigenvector equation, $AU=k^2U$, give us the following boundary value problem:
\begin{align}
-u''_-&=k^2 u_-,\qquad -u''_+=k^2 u_+,\nonumber\\
u'_-(0)&=u'_+(0),\qquad u_+(L)=u_-(-L)=0,\label{eq:eqgalvanichilb}\\
-u'_-(0)&= \left(\alpha k^2 -\frac{1}{\beta}\right)\left[u_+(0)-u_-(0)\right].\nonumber
\end{align}
The secular equation reads
\begin{equation}
\label{eq:seculargalvanic}
k\cos(kL)=2\left(\alpha k^2- \frac{1}{\beta}\right)\sin(kL).
\end{equation}
Notice that in fact this operator is definite positive, and $0$ is not an eigenvalue.

It will be no surprise at this point that we define a secondary inner product as the natural quadratic form induced by $A$, with reference to (\ref{eq:uaugalv}), thus justifying Eq. (\ref{eq:TL_TL_LCgalvcoup_Networks_ortho_2}).

We can also obtain sum rules, such as
\begin{equation}
\label{eq:galvanic-sum-rule}
\frac{1}{\alpha}=\sum_{n=0}^\infty \frac{\left|\Delta_n\right|^2}{\langle U_n,U_n\rangle}.
\end{equation}

\subsubsection{Point insertion}
\label{sec:point-insertion}
Again, consider the Hilbert space ${\cal H}=L^2[(0,L)]\oplus \mathbbm{C}\oplus L^2[(L,\infty)]$, and an operator $A$ acting on $U=(u_-,u_+,a=u_+(L)=u_-(L))\in D(A)\subset{\cal H}$ as $(-u_-'',-u_+'',u(L)/\alpha\beta -(u_+'(L)-u_-'(L))/\alpha) $, where we demand that $u_+(L)=u_-(L)$, $u(0)=0$ and the values at the endpoints are understood as limits . The eigenvector and eigenvalue equation reads
\begin{align}
-u''_-&=k^2 u_-,\qquad-u''_+=k^2 u_+,\nonumber\\
u_-(L)&=u_+(L),\qquad u_-(0)=0,\nonumber\\
u_-'(L)&= u_+'(L)+\left(\alpha k^2 -\frac{1}{\beta}\right) u(L).\nonumber
\end{align}

\subsection{The recipe}
\label{sec:recipe}

Let us summarize in a general way the point of view presented in this section. We are considering Hilbert spaces of the form ${\cal H}=\left(\oplus_{k=1}^{N}L^2[I_k]\right)\oplus\left(\oplus_{j=1}^M \mathbbm{C}^{n_j}_{\alpha_j}\right)$. There are $N$ intervals $I_K$ and $M$ relevant boundaries. In fact we are considering  second order differential operators acting on single functions, whence for all $M$ relevant boundaries we have $n_j=1$. The number of relevant boundaries will depend on the system we model. The largest possible number corresponds to sequences of $N$ finite length transmission lines such that all endpoints are relevant boundaries, thus $M=2N$. We are considering the standard inner product in each interval, and the second order differential operation ${\cal L}:u\to -u''$ in each. More general second order Sturm--Liouville differential operators, ${\cal L}:u\to -(p u')'+q u'$ ($p>0$) can also be considered, and, by proper modifications of the weight, other second order differential operators. 

In this manner, the elements $U\in{\cal H}$ are of the form $U=\left(u_1,\ldots,u_N,a_1,\ldots,a_m\right)$, $V=\left(v_1,\ldots,v_N,b_1,\ldots,b_m\right)$. If required, we will explicitly denote the element a number component belongs to, as in $a_j^U$, $a_j^V$.  The inner product is determined by the direct sum structure as
\begin{equation}
\nonumber 
\langle U,V\rangle=\sum_{k=1}^N\int_{I_k}\mathrm{d}x\,\bar{u}_k(x)v_k(x)+\sum_{j=1}^M \alpha_j\bar{a}_jb_j.
\end{equation}
We shall consider operators $A$ that will act on the function components as Sturm--Liouville operators, ${\cal L}_k$ acting on the $k$-th component. Their domains will be determined by the finiteness or otherwise of the intervals. For finite intervals, typically we shall require $u_k$ and $u'_k$ to be $AC[I_k]$, absolutely continuous in the corresponding interval.

An endpoint is a \emph{relevant} boundary, i.e. there is a number component associated to it, if the domain of $A$ is not restricted by a boundary condition on $u$ at that endpoint. The definition of the domain of $A$ includes a condition on the number components ($a_j$) at relevant boundaries. A boundary can be associated to just one interval or it can be associated to two intervals, if it is a common endpoint. If a relevant boundary, with index $j$, is associated to just one interval, $I_k$, the corresponding number component $a_j$ will be determined by the (free) value of the limit of $u_k$ when tending to that boundary. If, however, the relevant boundary with index $j$ is associated, as their common endpoint $x_k$, to two intervals $I_k$ and $I_{k+1}$, there will be a condition for $U$ to belong to the domain in terms of the continuity of either $u$ or its derivative, i.e., either $\lim_{x\to x_k^-} u_k(x)=\lim_{x\to x_k^+} u_{k+1}(x)$ or similarly for the derivatives, and the number component $a_j$ will be determined by the common value or by the jump in the functions. Finally, $AU=\left({\cal L}_iu_1,\ldots,{\cal L}_Nu_N,d_1/\alpha_1 +a_1/\alpha_1\beta_1,\ldots, d_M/\alpha_M+a_M/\alpha_M\beta_M\right)$, where $d_j$ is a linear combination of the limits of derivatives of the functions at the $j$-th relevant boundary. Again, we will explicity denote $d_j^U$, $d_j^V$, etc., if required. The eigenvalue and eigenvector equation is thus
\begin{align}
\nonumber 
{\cal L}_iu_i&=k^2 u_i,\nonumber\\
d_j+\frac{1}{\beta_j}a_j&=\alpha_j k^2 a_j,\nonumber
\end{align}
together with some boundary conditions. For our procedure to be well defined, the boundary terms arising from
\begin{equation}
\nonumber 
\int_{I_k}\mathrm{d}x\,\left[\bar{v}_k(x) {\cal L}_ku_k(x)-\overline{\left({\cal L}_kv_k(x)\right)}u_k(x)\right]
\end{equation}
must be cancelled by
\begin{equation}
\nonumber 
\sum_{j=1}^M \left(\bar{a}_j^vd_j^u-\bar{d}_j^v a_j^u\right).
\end{equation}
This ensures symmetry. Furthermore, we require that  $\left\langle U,AU\right\rangle$ be lower bounded. By shifting some ${\cal L}_k$ by a constant (alternatively, by demanding that they all be positive) the demand is positivity. This can generally be achieved if the boundary terms
\begin{equation}
\nonumber 
\sum_{k=1}^N\left[-p_k\bar{u}_ku_k'\right]_{\partial I_k}
\end{equation}
are cancelled by $\sum_{j=1}^M\bar{a}_jd_j$ on the domain of $A$.

Once symmetry and positivity are ensured, it remains to be examined whether indeed we are led to a self-adjoint operator, whence an expansion theorem would follow. We have addressed this issue in the examples by studying the existence (and uniqueness) of solutions to $\left(1+L^2A\right)U=F$ for $F\in{\cal H}$. This amounts to studying the system of equations
\begin{equation}
\nonumber 
u_k+L^2{\cal L}_k u_k = f_k
\end{equation}
for $k=1,\ldots,N$ together with conditions
\begin{equation}
\nonumber 
a_j^U+\frac{L^2}{\alpha_j}\left(d_j^U+ \frac{a_j^U}{\beta_j}\right)=a_j^F
\end{equation}
for $j=1,\ldots,M$, and the required boundary conditions for elements $U\in D(A)$. Let $j$ be the index of  the only boundary relevant to the $k$-th element (for definiteness; extensions are straightforward); this entails the idea that the other endpoint of $I_k$ has associated homogeneous boundary conditions. We construct $u_k^0$, a solution of the problem
\begin{align}
\nonumber 
u_k+L^2{\cal L}_ku_k&=0,\nonumber\\
a_j^U+\frac{L^2}{\alpha_j}\left(d_j^U+ \frac{a_j^U}{\beta_j}\right)&=a_j^F,\nonumber\\
\mathrm{other~B.C.}&&\nonumber
\end{align}
Notice that the relevant bounday condition involves $u_k$ linearly. Once this has been achieved, one makes the change of variables $u_k=u_k^0+v_k$, and we are led to the study of $v_k+L^2 {\cal L}_kv_k=f_k$ with homogeneous boundary conditions. For the systems we have considered, this has a unique solution, and the existence and uniqueness of solutions to   $\left(1+L^2A\right)U=F$  has been thus established. This, in turn, establishes that $F\in{\cal H}$ can be expanded as $F=\sum_{n=0}^\infty\langle U_n,F\rangle U_n$, with $U_n$ orthonormal eigenvectors of $A$.

\subsubsection{Side results}
\label{sec:side-results}

As a side product of the process we obtain sum rules, generically by expanding special elements of the Hilbert space, of the form $u_k=0$ and one of the $a_j$ set to one while the others are zero. This gives us
\begin{equation}
\label{eq:genericsumrule}
\frac{1}{\alpha_j}=\sum_{n=0}^\infty\left|a_j^{U_n}\right|^2
\end{equation}
for normalised $U_n$ eigenvectors.

Another side result is what we have termed the secondary inner product. We have relied in our proofs on the positivity of $\left\langle U,AU\right\rangle$ for $U\in D(A)$ (in fact based on physical reasons: it has to be associated to the harmonic approximation for small oscillations), and this provides us, by extension to the whole Hilbert space ${\cal H}$, with a positive definite quadratic form. From the expansion theorem, $F=\sum_n\langle U_n,F\rangle U_n$, denoting the coefficients of $F$ (resp. $G$) in the orthonormal basis $U_n$ with eigenvalues $k_n^2$ (orthonormal with respect to the initial product $\langle\cdot,\cdot\rangle$) as $F_n$ (resp. $G_n$), the new inner product  $\langle\cdot,\cdot\rangle_{1/\beta}$ is given as
\begin{equation}
\nonumber 
\langle F,G\rangle_{1/\beta}= \sum_{n=0}^\infty\frac{1}{k_n^2} \bar{F}_nG_n.
\end{equation}

\subsection{Alternative mathematical approaches}
\label{sec:altern-math-appr}

\subsubsection{Trace operator}
\label{sec:trace-operator}

We have restricted ourselves to the Hilbert space setting, due to the later application to quantization. Nonetheless, a number of questions regarding these expansions can also be analysed in terms of Sobolev spaces, for which the concept of \emph{trace} (in the sense of trace of an element $u\in W^{1,p}(\Omega)$ which is understood as the ``boundary function'' $u|_{\partial\Omega}$, see \cite{Lions:2011} ) appears. That context is natural in order to study geometrically the transformation from Lagrangian to Hamiltonian in cases such as those we consider (see \cite{Barbero:2015,Barbero:2017}  for such a viewpoint).

\subsubsection{Delta distribution}
\label{sec:delta-function}

An alternative approach, still in the Hilbert space context, is to consider the Hilbert space $L^2\left[[0,L);\mu\right]$, where the measure presents a point mass in the initial point. Even more, the idea can be extended to the case of measures with additional point masses, both in the interior and at the endopoints. This is, for instance, the concrete presentation that appears in \cite{Walter:1973}.  It is also related to the computations of \cite{Malekakhlagh:2016,Malekakhlagh:2017}.

Undoubtedly this is a feasible route; we have preferred to set it aside to avoid problems in moving to the Hamiltonian formalism. It should be noted that constructing precisely the functional analytic details need not be trivial at all, though.

\section{Capacitive to inductive coupling}
\label{sec:capac-induct-coupl}
A common presentation of spin-boson Hamiltonians is related to the seminal work of Caldeira and Leggett \cite{CaldeiraLeggett:1981,CaldeiraLeggett:1983,Leggett:1984}. There a system, with coordinate $q$, is coupled to a bath of linear oscillators, with coordinates $x_n$, and the coupling of interest is of the form $q\sum_n c_n x_n$. Caldeira and Leggett thoroughly analyse other possibilities, in particular those of the form $\dot{q}\sum_n c_n x_n$ (or equivalently $-q\sum_n c_n\dot{x}_n$), and show with an example how to relate both forms. In fact, as they also point out, this is achieved through a canonical transformation, and there is no point transformation that can reproduce it. For completeness we present this canonical presentation here, and then we study  the mapping from a capacitive $\dot{q}\sum_n c_n \dot{x}_n$ to the inductive form $q\sum_n c_n x_n$, which we will see is a point transformation.

Thus, let us first consider the Lagrangian 
\begin{equation}
\nonumber 
L=\frac{m}{2}\dot{q}^2-V(q)+\frac{1}{2}\dot{\bx}^T\msM\dot{\bx}-\frac{1}{2}\bx^T\msLambda^2\bx-{q}\bc^T\dot{\bx},
\end{equation}
where there is a single system variable $q$, and a bath of harmonic oscillators, with position variables $\bx$, are coupled to the system via a coupling vector $\bc$ in the interaction term $-{q}\bc^T\dot{\bx}$. Let the corresponding canonical momenta be $p$ and $\bp$. The Hamiltonian is
\begin{equation}
\nonumber 
H=\frac{p^2}{2m}+\frac{1}{2}\bp^T\msM^{-1}\bp+q\bc^T\msM^{-1}\bp+\frac{q^2}{2}\bc^T\msM^{-1}\bc+V(q)+\frac{1}{2}\bx^T\msLambda^2\bx.
\end{equation}
With the canonical transformation $\bpi=-\bx$, $\bxi=\bp$, and going back to the Lagrangian, one obtains
\begin{equation}
\nonumber 
\tilde{L}=\frac{m}{2}\dot{q}^2-V(q)-\frac{q^2}{2}\bc^T\msM^{-1}\bc+\frac{1}{2}\dot{\bxi}^T\msLambda^{-2}\dot{\bxi}-\frac{1}{2}\bxi^T\msM^{-1}\bxi-q\bc^T\msM^{-1}\bxi.
\end{equation}
The coupling indeed is now of inductive form, as expected. Notice however that the variable $\xi$ has dimensions of momenta, and that has to be taken into account to determine the spectral density, for instance. On computing explicitly, the spectral density reads, formally,
\begin{equation}
\nonumber 
J(\omega)=\bc^T\msM^{-1}\msLambda\msO^T\delta\left(\omega-\msOmega\right)\msOmega^{-1}\msO\msLambda\msM^{-1}\bc.
\end{equation}
Here $\msO$ is the orthogonal matrix diagonalizing $\msLambda\msM^{-1}\msLambda$, namely
\begin{equation}
\nonumber 
\msO\msLambda\msM^{-1}\msLambda\msO^T=\msOmega^2,
\end{equation}
with diagonal $\msOmega$. The definition we use for the spectral density $J(\omega)$ is best expressed in terms of the classical equations of motion for the variable $q$. The source term for  $q$ due to the dynamics of $\bxi$ is reexpressed, after solving the classical equations of motion for $\bxi$, in the form $\int_0^t\mathrm{d}\tau\int\mathrm{d}\omega J(\omega)\sin\omega(t-\tau)\,q(\tau)$, which provides the definition of $J(\omega)$. 

Dimensionally, $\msLambda^2$ is mass times frequency squared, $\msM$ is mass, $\msO$ adimensional, so the terms bracketed by the $\bc$ vectors (setting aside the delta) are frequency over mass.

This example shows that the identification of the spectral density is coupling dependent (as would be obvious from dimensional analysis). We now study capacitive coupling in this regard.  In particular, 
we will compute this process explicitly for the following general example:
\begin{equation}
\label{eq:CLcapacitiveLagr}
L=\frac{m}{2}\dot{q}^2-V(q)+\sum_n \left[\frac{m_n}{2}\dot{x}_n^2-\frac{m_n\omega_n^2}{2}{x}_n^2\right]-\dot{q}\sum_n c_n \dot{x}_n.
\end{equation}
We shall use a more compact notation that also provides a slight generalisation, namely
\begin{equation}
\nonumber 
L=\frac{m}{2}\dot{q}^2-V(q)+\frac{1}{2}\dot{\bx}^T\msM\dot{\bx}-\frac{1}{2}\bx^T\msLambda^2\bx-\dot{q}\bc^T\dot{\bx}.
\end{equation}

In this case the  canonical transformations can be reduced to a point transformation applied to the initial Lagrangian, namely
\begin{equation}
\label{eq:contacttrans}
\bx=\bx_2+q\,\msM^{-1}\bc,
\end{equation}
and the new Lagrangian reads
\begin{align}
\nonumber 
\tilde{L}=&\, \frac{1}{2}\left(m+\bc^T\msM^{-1}\bc\right)\dot{q}^2-V(q)-\frac{1}{2}\bc^T\msM^{-1}\msLambda^2\msM^{-1}\bc\,q^2\nonumber\\
& +\frac{1}{2}\dot{\bx}_2^T\msM\dot{\bx}_2-\frac{1}{2}\bx_2^T\msLambda^2\bx_2-q\bc^T\msM^{-1}\msLambda^2\bx_2.\nonumber
\end{align}
For the specific case of (\ref{eq:CLcapacitiveLagr}) we have
\begin{align}
\nonumber 
\tilde{L}=&\,\frac12\left(m+\sum\frac{c_n^2}{m_n}\right)\dot{q}^2- V(q)-q^2\sum\frac{c_n^2\omega^2_n}{2m_n}\nonumber\\
& +\sum\left(\frac{m_n}{2}\dot{\xi}_n^2-\frac{m_n\omega^2_n}{2}\xi_n^2\right)- q\sum c_n\omega_n^2\xi_n,\nonumber
\end{align}
with the new variables $\xi_n=x_n-c_n q/m_n$, 
and one obtains a spectral density
\begin{equation}
\nonumber 
J(\omega)=\sum\frac{c_n^2\omega^3_n}{m_n} \delta(\omega-\omega_n).
\end{equation}

The transformation (\ref{eq:contacttrans})  is implemented  in the quantum case by a unitary transformation. That is,
\begin{equation}
\nonumber 
\hat{U}=\exp\left[-i\hat{q}\bc^T\msM^{-1}\hat{\bp}/\hbar\right]=\exp\left[-\frac{i}{\hbar}\sum_n\frac{c_n}{m_n}qp_n\right].
\end{equation}

\section{Hamiltonian formalism with transmission lines}
\label{sec:hamilt-form}

\subsection{The continuum Hamiltonian}
\label{sec:cont-hamilt}
In order to compare with \cite{Malekakhlagh:2016,Malekakhlagh:2017,Bamba:2014}, it might be convenient to write the continuum version of some of the Hamiltonians presented here. We shall carry out this task, fully explicitly, in the case of Hamiltonian (\ref{eq:Ham_LCcoup_Network2}), using the notation of subsection \ref{subsubsec:chap2_mixed_linear_coupling}. To obtain this expression, notice that we have started from a Lagrangian with a continuum part, namely (\ref{eq:Lag_TL_LCcoup_Network}). Next we have expressed the flux field $\Phi(x,t)$ as the infinite sum $\Phi(x,t)=\sum_n\Phi_n(t)u_n(x)$, where $u_n(x)$ are the function component of $U_n=(u_n(x),u_n(0))\in {\cal H}=L^2[(0,L)\oplus \mathbbm{C}_\alpha$. These basis vectors are orthogonal according to
\begin{equation}
\nonumber 
\left\langle U_n,U_m\right\rangle= c\left(\int_0^L\mathrm{d}x\,u_n(x)u_m(x)+ \alpha\, u_n(0) u_m(0)\right)=N_\alpha \delta_{nm}.
\end{equation}
The dimension of $N_\alpha$, namely capacity, has been chosen  so that the function elements $u(x)$ are adimensional. Furthermore $\alpha$ has dimension of length, as we stated previously.  Since the dimension of the flux field is voltage times time, we see that the dimension of $\Phi_n$ is $\left[\Phi\right]/[u_n]= V\cdot T$, with $V$ standing in for voltage, and $T$ for time. Later, we have $C$ for capacity.

The variables $Q_n=\partial L/\partial\dot\Phi_n$ have dimension $\left[Q_n\right]=[L]\cdot T/[\Phi_n]= C\cdot V^2\cdot T/ \left(V\cdot T\right)=C V$, i.e., charge.
In order to obtain a Hamiltonian with continuum component, we construct a function of time  that takes values in ${\cal H}$, $Q(t)$, as
\begin{equation}
\nonumber 
Q(t)=\sum_{n=0}^\infty Q_n(t) U_n=\left(Q(x,t),Q(0,t)\right).
\end{equation}
Observe that
\begin{equation}
\nonumber 
\left\langle Q(t),Q(t)\right\rangle=\sum_{n=0}^\infty Q_n^2(t) N_\alpha=c\left[ \int_0^L\mathrm{d}x\, Q^2(x,t)+\alpha\, Q^2(0,t)\right].\nonumber
\end{equation}
We can thus substitute in the Hamiltonian (\ref{eq:Ham_LCcoup_Network2}) to obtain
\begin{align}
\nonumber 
H=&\, \frac{1}{2}\bq^T(\msA^{-1}+ \frac{C_c^2}{\alpha c} \msA^{-1}\ba\ba^T \msA^{-1})\bq+\frac{1}{2}\bphi^T\msB^{-1}\bphi+V(\bphi)\nonumber\\
&+\frac{c\alpha}{2N^2_\alpha} Q^2(0,t)+ \frac{C_c}{N_{\alpha}} (\bq^T\msA^{-1} \ba)Q(0,t)+ \frac{1}{2L_g}\Phi^2(0,t)-\frac{1}{L_g} (\bphi^T\bb)\Phi(0,t)\nonumber\\ 
&+\int_0^L\mathrm{d}x\,\left[\frac{c}{2N^2_\alpha}Q^2(x,t)+\frac{1}{2l}\left(\partial_x\Phi(x,t)\right)^2\right].\nonumber
\end{align}
Here we have made use of Eq. (\ref{eq:TL_LCcoup_Network_beta_fix})  to substitute $\beta=L_g/l$. Bear in mind that a definite choice for $\alpha$ has been made, $\alpha=C_c(1-C_c\ba^T\msA^{-1}\ba)/c$. On first sight it might look as if this expression for the Hamiltonian had a major flaw, namely that it explicitly depends on the arbitrary constant $N_\alpha $ we have introduced. In fact, this constant fixes the unit of charge we use, and since only the combination $Q(x,t)/N_\alpha$ appears, there is no free parameter in the Hamiltonian.
\subsection{Canonical variables}
\label{sec:canonical-variables}
In all our analysis we have constructed Lagrangian functions which are quadratic in the derivatives, whence the Hamiltonian is derived in the standard way. As is well known, Hamiltonian dynamics is not fully determined by the Hamiltonian. Additionally the Poisson bracket is necessary to produce the relevant vector field. Starting from the Lagrangian (in general cases) this Poisson bracket is fully determined, and, in most cases, it is taken for granted, since the standard procedure introduces the canonical momenta. 

We have actually followed this route, in that the Poisson bracket has been systematically taken to be
\begin{equation}
\nonumber 
\left\{F,G\right\}=\sum_{i=1}^{N_{\mathrm{disc}}}\left(\frac{\partial F}{\partial \phi_i}\frac{\partial G}{\partial q_i}-\frac{\partial F}{\partial q_i}\frac{\partial G}{\partial \phi_i}\right)
+\sum_{n=0}^\infty \left(\frac{\partial F}{\partial \Phi_n}\frac{\partial G}{\partial Q_n}-\frac{\partial F}{\partial Q_n}\frac{\partial G}{\partial \Phi_n}\right),
\end{equation}
for $F$ and $G$ functions of the canonical variables, $F\left(\bq,\bQ;\bphi,\bPhi\right)$, and similarly for $G$. 

In keeping with the notation of chapter~\ref{chapter:chapter_3}, we denote with $\bphi$ the variables associated with lumped element networks, and with $\bPhi$ those associated to transmission lines. $\bq$ and $\bQ$ are the corresponding canonical moments.

Consider now one transmission line (i.e. one interval of the real line, $I$), with flux field $\Phi(x,t)$ and charge field $Q(x,t)$. They are not canonically conjugate in general:
\begin{align}
\nonumber 
\left\{\Phi(x,t),Q(x',t)\right\}=&\sum_{n,m}\left\{\Phi_n(t),Q_m(t)\right\}u_n(x)u_m(x')\\
=&\sum_{n} u_n(x)u_n(x').\nonumber
\end{align}
When the expansion functions form a real orthonormal basis with respect to the inner product $\langle f,g\rangle=\int_I\mathrm{d}x\bar{f}(x)g(x)$, the expansion theorem can be reexpressed as the equality $\sum_n u_n(x)u_n(x')=\delta(x-x')$, thus proving $\Phi(x,t)$ and $Q(x,t)$ are canonically conjugate in such a case.

Let us assume that $u_n(x)$ are the function components of an orthonormal basis $U_n$ with respect to an inner product of the form we consider, $\left\langle (u,a),(v,b)\right\rangle=\int_I\bar{u}v+ \alpha\bar{a}b$. Then, \emph{formally}, we obtain
\begin{equation}
\label{eq:deltas}
\delta(x-x')=\left[1+\alpha\delta(x')\right]\sum_n u_n(x)u_n(x').
\end{equation}
Let us prove this statement. The expansion theorem, in the form presented in \ref{Walter_appendix}, informs us that $f(x)=\sum f_n u_n(x)$, where $f_n=\langle U_n,F\rangle$, and $F=(f(x),f(0))k$ (setting $0$ as the relevant boundary), i.e., $f_n=\int_I f u_n +\alpha f(0)u_n(0)$. Hence
\begin{align}
\nonumber 
f(x)&= \alpha f(0)\sum_n u_n(0)u_n(x)+\int_I\mathrm{d}x'\,\left[\sum_n u_n(x)u_n(x')\right] f(x')\nonumber,\\
&=\int_I\mathrm{d}x'\,\left[1+\alpha\delta(x')\right]\left[\sum_n u_n(x)u_n(x')\right] f(x'),\nonumber
\end{align}
thus concluding (\ref{eq:deltas}). Now define, formally,
\begin{equation}
\nonumber 
\tilde{Q}(x,t)=\left[1+\alpha\delta(x)\right] Q(x,t).
\end{equation}
By explicit computation one obtains (formally!)
\begin{equation}
\nonumber 
\left\{\Phi(x,t),\tilde{Q}(x',t)\right\}= \left[1+\alpha\delta(x')\right]\sum_n u_n(x)u_n(x')=\delta(x-x').
\end{equation}
Prima facie, the variables $\Phi(x,t)$ and $\tilde{Q}(x,t)$ are canonically conjugate, and they could be used for (field) quantization. This is in fact (although the way to this point is very different) what was proposed in \cite{Malekakhlagh:2016}. Note however that one cannot make sense of quantities such as $\tilde{Q}^2$ without some regularisation and prescription, since  $\tilde{Q}$ is only defined in a distributional way. In reference \cite{Malekakhlagh:2016} this problem is avoided since in fact they can refer back to what we have denoted as untilded charge field.

For reference, let us write the Poisson bracket for $F$ and $G$ functionals of the flux and the charge field:
\begin{align}
\nonumber 
\left\{F,G\right\}=&\,\int_I\mathrm{d}x\,\left[\frac{\delta F}{\delta\Phi(x)}\frac{\delta G}{\delta Q(x)}-\frac{\delta F}{\delta Q(x)}\frac{\delta G}{\delta \Phi(x)}\right]\\
&-\alpha\int_I\mathrm{d}x\,\left(\sum_n u_n(x)u_n(0)\right)\left[\frac{\delta F}{\delta\Phi(x)}\frac{\delta G}{\delta Q(0)}-\frac{\delta F}{\delta Q(0)}\frac{\delta G}{\delta \Phi(x)}\right].\nonumber
\end{align}

\section{Inversion of infinite matrices}
\label{sec:infinite-matrix-inversion}

\subsection{Single port impedance}
\label{sec:single-port-imped}

A crucial aspect of our analysis, in its different forms, is the inversion of infinite dimensional capacitance matrices, presented in block-matrix format. In the single port case we mostly analyze, in which we couple a transmission line or a more general single port non-dissipative, passive, linear impedance with infinite modes to a network with a finite number of degrees of freedom, the coupling submatrix is of rank one. This is made explicit in our presentation in that we write the capacitance coupling block-matrix as $-C_c\ba\bu^T$ (Eq. (\ref{eq:C_LCcoup_Network})), $-C_A\ba\Delta\bu^T$ (Eq. (\ref{eq:C_TL_LCgalvcoup_Networks})), $-C_c\bu\bv^T$ (Eq. (\ref{eq:C_TL_LCcoup_TL})).  In a manner reminiscent of the Sherman--Morrison formula, we find that for invertible $\msA_1$ and $\msA_2$ and rank one $\msD$ the following generally holds:
\begin{align}
\label{eq:sherman-morrison}
&\begin{pmatrix}
\msA_1&\msD\\ \msD^\dag&\msA_2
\end{pmatrix}^{-1}=
\begin{pmatrix}
\msA_1^{-1}&0\\0&\msA_2^{-1} \end{pmatrix}\\
&+\frac{1}{1-\mathrm{Tr}\left(\msD\msA_2^{-1}\msD^\dag\msA_1^{-1}\right)}
\begin{pmatrix} \msA_1^{-1}\msD\msA_2^{-1}\msD^\dag\msA_1^{-1}&
-\msA_1^{-1}\msD\msA_2^{-1}\\  -\msA_2^{-1}\msD^\dag\msA_1^{-1}&  \msA_2^{-1}\msD^\dag\msA_1^{-1}\msD\msA_2^{-1}
\end{pmatrix}.\nonumber
\end{align}
This formula can be checked directly, making use that for a rank one operator $\msD:{\cal V}_1\to{\cal V}_2$, and if the adjoint and the traces exist,
\begin{equation}
\nonumber 
\msD\msA\msD^\dag\msB\msD= \mathrm{Tr}\left[\msD\msA\msD^\dag\msB\right]\msD\qquad \mathrm{and}\qquad
\msD^\dag\msB\msD\msA\msD^\dag= \mathrm{Tr}\left[\msD\msA\msD^\dag\msB\right]\msD^\dag,\nonumber
\end{equation}
where $\msA:{\cal V}_1\to{\cal V}_1$ and $\msB:{\cal V}_2\to{\cal V}_2$.

The condition for (\ref{eq:sherman-morrison}) to hold is that the trace $\mathrm{Tr}\left(\msD\msA_2^{-1}\msD^\dag\msA_1^{-1}\right)$ exist and be different from one.
\subsection{Multiport impedance}
\label{sec:multiport-impedanceapp}
The analogous inversion for the multiport case is necessarily more involved. We have presented an explicit case in section \ref{sec:mult-netw-coupl}, and we look at the more general situation in \ref{sec:multi-port-impedance}. The essential result, as in the Woodbury--Sherman--Morrison case, is that perturbing an invertible operator with an operator of finite rank should produce, for the new inverse, again a perturbation of the same rank.  The (necessarily formal) inversion formula we require goes as follows. Let
\begin{equation}
\nonumber 
\msC=
\begin{pmatrix}
\msA&\msD\\ \msD^T&\msC_1
\end{pmatrix}
\end{equation}
be a real matrix, with $\msA$ and $\msC_1$ invertible and $\msD$ of finite rank. Its inverse if it, together with the correct subelements, exists, is given by
\begin{align}
\nonumber 
\msC^{-1}&=
\begin{pmatrix}
\msA^{-1}&0\\0&\msC_1^{-1}
\end{pmatrix}+\begin{pmatrix}
	\msf{E}&\msf{I}\\\msf{R}&\msf{W}
\end{pmatrix},\nonumber\\
\msf{E}&=\msA^{-1}\msD\msC_1^{-1}\left(\mathbbm{1}-\msD^T\msA^{-1}\msD\msC_1^{-1}\right)^{-1}\msD^T\msA^{-1},\nonumber\\
\msf{I}&=-\msA^{-1}\msD\msC_1^{-1}\left(\mathbbm{1}-\msD^T\msA^{-1}\msD\msC_1^{-1}\right)^{-1},\nonumber\\
\msf{R}&=-\msC_1^{-1}\msD^T\msA^{-1}\left(\mathbbm{1}-\msD\msC_1^{-1}\msD^T\msA^{-1}\right)^{-1},\nonumber\\
\msf{W}&=\msC_1^{-1}\msD^T\msA^{-1}\left(\mathbbm{1}-\msD\msC_1^{-1}\msD^T\msA^{-1}\right)^{-1}\msD\msC_1^{-1}.\nonumber
\end{align}
The proof of this formula is by direct substitution. Both in the infinite and in the finite dimension case, and as we analyze for multiport impedances in \ref{sec:multi-port-impedance}, the invertibility of the capacitance matrix is conditioned on the existence of the inverse $\left(\mathbbm{1}-\msD\msC_1^{-1}\msD^T\msA^{-1}\right)^{-1}$. In the infinite dimension case there can be further subtleties. This formula reduces to (\ref{eq:sherman-morrison}) when the rank of $D$ is one. To see this one has to realize that $\left(\mathbbm{1}-\msD\msC_1^{-1}\msD^T\msA^{-1}\right)^{-1}$ is led by $\msC_1^{-1}$, and analogously $\left(\mathbbm{1}-\msD\msC_1^{-1}\msD^T\msA^{-1}\right)^{-1}$ only appears preceded by $\msA^{-1}$. This allows the reduction to the quantity $\left[1-\mathrm{Tr}\left(\msD\msC_1^{-1}\msD^\dag\msA^{-1}\right)\right]^{-1}$.

\section{Zeros and infinities for  $1^{\mathrm{st}}$-Foster form impedance}
\label{App:1st_Foster}
We observe that formally the coupling vector $\be_n$ in the capacitance matrix (\ref{eq:Paladino_Cmat}) has infinite norm when the impedance has to be represented with an infinite set of stages. This by itself is not an issue.  However, in the computation of the inverse capacitance matrix the crucial quantity $\be_n^T \msC_n^{-1} \be_n$ does appear, and whenever this tends to infinity the inverse is poorly defined, if not altogether nonsensical.

Let us be explicit, and apply to the capacitance matrix
\begin{equation}
\msC_{x}=\begin{pmatrix}
a & b\be_n^T\\
b\be_n & \msM_{n}
\end{pmatrix}.\nonumber 
\end{equation}
with  $\msM_{n}=\msC_n+d \be_n\be_n^T$, the formal inversion formula of Eq. (\ref{eq:sherman-morrison}). It is indeed applicable, since the coupling matrix is indeed rank one. Notice now that the formal expression $\mathrm{Tr}\left(\msD\msA_2^{-1}\msD^{\dag}\msA_1^{-1}\right)$ becomes $(b^2/a)\be_n^T\msM_n^{-1}\be_n$. It is thus incumbent on us to compute the inverse $\msM_n^{-1}$. It is clearly of the form $\msC_n^{-1}+\gamma\msC_n^{-1}\be_n\be_n^T\msC_n$, where the coefficient $\gamma$ must obey the equation
\begin{equation}
\nonumber 
\gamma+d+d \gamma \be_n^T \msC_n^{-1} \be_n=0,
\end{equation}
i.e., formally,
\begin{equation}
\label{eq:gamma-value}
\gamma=\frac{-d}{1+d \be_n^T \msC_n^{-1} \be_n}.
\end{equation}
Thus, we see that
\begin{equation}
\nonumber 
\frac{b^2}{a}\be_n^T\msM_n^{-1}\be_n= \frac{b^2}{a}\be_n^T \msC_n^{-1} \be_n \left(1- \frac{d \be_n^T \msC_n^{-1} \be_n}{1+d \be_n^T \msC_n^{-1} \be_n}\right)=\frac{b^2}{a}\frac{\be_n^T \msC_n^{-1} \be_n}{1+d \be_n^T \msC_n^{-1} \be_n}.
\end{equation}
This quantity can therefore tend to a finite limit in the pathological situation we consider, namely $b^2/ad$. Notice however that $\gamma$ from Eq. (\ref{eq:gamma-value}) tends to zero. Carrying out with the analysis, the final result is that in the limit in which $\be_n^T \msC_n^{-1} \be_n\rightarrow\infty$ the inverse capacitance matrix presents no coupling whatsoever between the impedance and the network variables.

Reconsider now the quantity $b^2/ad$. As stated in the previous Appendix \ref{sec:infinite-matrix-inversion}, it has to be different from one for the inversion to be possible, in the pathological situation we consider. Looking back to the original parameters of section \ref{sec:1st-foster-expansion}, in particular the matrix (\ref{eq:Paladino_Cmat}), we have that it reads $C_B/C_\Sigma=C_B/(C_A+C_B)$. If the capacity $C_A$, external to the one port impedance, were 0, the capacitance matrix would not be invertible in the case $\be_n^T \msC_n^{-1} \be_n\to\infty$. Notice that, in the same limit and for those parameters, we have
\begin{equation}
\label{eq:1st_Foster_sum_rule_eMe}
\be_n^T\msM_n^{-1}\be_n\to \frac{1}{C_B}.
\end{equation}
The result that there is an overcounting of degrees of freedom if $C_A=0$ and $\be_n^T \msC_n^{-1} \be_n\to\infty$  was implicit in the divergence of the charge energy found in the Hamiltonian in chapter \ref{chapter:chapter_2} for the specific case of a transmission line resonator coupled to a charge qubit, when the capacitance of the Josephson junction $C_J$ was taken to zero. Furthermore, we have shown the relation (\ref{eq:1st_Foster_sum_rule_eMe}) which will be useful in the  derivation of the corresponding Hamiltonian.

\section{$2^{\mathrm{nd}}$-Foster form admittance quantization}
\label{App:2nd_Foster}
In this section, we derive a Hamiltonian of an anharmonic flux variable coupled to a lossless admittance $Y(s)$ decomposed in the $2^{\mathrm{nd}}$-Foster form. This expansion has been widely used to describe the effect of a general, lossy, environment seen by a harmonic oscillator \cite{Devoret:1997}, by taking a continuous limit, which we do not carry out here. We study the differences and similarities to the previous section describing the circuit analysed by Paladino et al. \cite{Paladino:2003}. 

\begin{figure*}[]
	\centering{\includegraphics[width=0.5\textwidth]{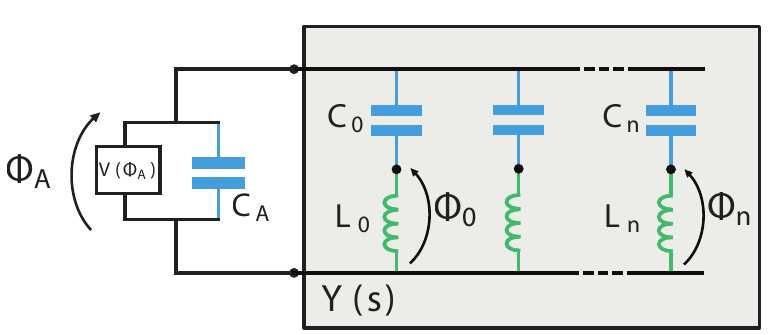}}
	\caption{\label{fig:2nd_Foster_form}$2^{\mathrm{nd}}$-Foster form admittance coupled to anharmonic variable. One port admittance $Y(s)$ modelled by  parallel $L_n,\,C_n$ oscillators is coupled to a flux variable non-anharmonic potential shunted by a capacitance. With the choice of flux variables in the inductors, we derive a Hamiltonian with capacitive coupling.}
\end{figure*}

First of all and contrary to the first Foster form, it must be noticed that the expansion of the admitance in this circuit allows only the description of electromagnetic environments with poles at frequency $s=0$, i.e. $\lim\limits_{s \rightarrow 0}Z(s)=\infty$. Secondly, this configuration has internal variables already separated from the network variables it is connected to, see Fig. \ref{fig:2nd_Foster_form}. Choosing as the internal degrees the flux diferences in the inductors (capacitors) we can derive a Hamiltonian with capacitive (inductive) coupling to the external variables. Here, we perform the analysis with the more cumbersome capacitive coupling in contrast with the inductive coupling done in \cite{Devoret:1997}, in order to compare it with the previous calculation of the above section in  Appendix \ref{App:1st_Foster}. 
The Lagrangian of the system at hand can be directly written as
\begin{align}
L&= \frac{C_A}{2} \dot{\Phi}_A^2+\sum_{n}\left[\frac{C_n}{2} \left(\dot{\Phi}_A-\dot{\Phi}_{n}\right)^2-\frac{1}{2L_{n}}\Phi_{n}^2\right]-V(\Phi_A)\nonumber\\
&=\frac{1}{2}\dot{\bPhi}^T \msC\dot{\bPhi}-\frac{1}{2}\bPhi^T \msL^{-1}\bPhi - V(\Phi_A),\nonumber
\end{align}
where $\bPhi^T=(\Phi_A,\bPhi_{n}^T)=(\Phi_A, \Phi_0, \Phi_1, ...)$, the capacitance matrix reads
\begin{equation}
\msC=\begin{pmatrix}
C_{\Sigma} & -\bc_n^T\\
-\bc_n & \msC_{n}
\end{pmatrix},\nonumber
\end{equation}
where we defined $C_{\Sigma}=C_A+\sum_{n}C_n$, $\msC_n=\mathrm{diag}(C_1, C_2...)$, the variable $\Phi_A$ couples to $\bPhi_{n}$ through the vector $\bc_n=(C_1, C_2,...)^T$, and the inductance matrix is 
\begin{equation}
\msL^{-1}=\begin{pmatrix}
0 & 0\\
0 & \msL_{n}^{-1}
\end{pmatrix},\nonumber
\end{equation}
with $\msL_n^{-1}=\mathrm{diag}(L_1^{-1}, L_2^{-1}...)$. In contrast with the analysis of the $1^{\mathrm{st}}$-Foster expansion, we can directly invert the capacitance matrix
\begin{equation}
\msC^{-1}=\begin{pmatrix}
C_A^{-1} & C_A^{-1}\mathbf{1}_n^T\\
C_A^{-1}\mathbf{1}_n & \msC_{n}^{-1}+C_A^{-1} \mathbf{1}_n \mathbf{1}_n^T
\end{pmatrix},\nonumber
\end{equation}
where again $\mathbf{1}_n=(1,1,...)^T$ is the vector with the dimension of the Hilbert space describing the admitance to derive a Hamiltonian,
\begin{equation}
H=\frac{1}{2}\bq^T\msC^{-1}\bq+\frac{1}{2}\bphi^T\msL^{-1}\bphi+V(\Phi_A).\nonumber
\end{equation}
Here the conjugate variables to the fluxes are the charges $\bq=\partial L/\partial \bphi$. To simplify the analysis, we make a first rescaling of the variables $\bp=(q_A,\bp_n^T)^T=\left(q_A,\left[C_0^{1/2}\msC_n^{-1/2}\bq_n\right]^T\right)^T$, with its corresponding change in fluxes $\bpsi=(\Phi_A,\bpsi_n^T)^T=\left(\Phi_A,\left[C_0^{-1/2}\msC_n^{1/2}\bphi_n\right]^T\right)^T$ such that the Hamiltonian transforms into
\begin{equation}
H_I=\frac{1}{2}\bp^T\msC_I^{-1}\bp+\frac{1}{2}\bpsi^T\msL_I^{-1}\bpsi+V(\psi_A),\nonumber
\end{equation}
with 
\begin{equation}
\msC_I^{-1}=\begin{pmatrix}
C_A^{-1} & C_A^{-1}\be_n^T\\
C_A^{-1}\be_n & \msM_n^{-1}
\end{pmatrix},\nonumber
\end{equation}
where the coupling vector $\be_n=C_0^{-1/2}\msC_n^{1/2}\mathbf{1}_n$ and $\msM_n^{-1}=C_0^{-1}\mathbbm{1}+C_A^{-1} \be_n \be_n^T$. On the other hand, we have a nondiagonal inductance matrix 
\begin{equation}
\msL_I^{-1}=\begin{pmatrix}
0 & 0\\
0 & (\msL_n^{I})^{-1}
\end{pmatrix},\nonumber
\end{equation}
with the submatrix $(\msL_n^{I})^{-1}=C_0 \msC_n^{-1/2}\msL_{n}^{-1}\msC_n^{-1/2}$. We can diagonalize together the capacitance $\msM_n^{-1}$ and inductance $(\msL_n^{I})^{-1}$ submatrices with a rescaling and unitary transformation of the charges $\brho=(q_A,\brho_n^T)^T=\left(q_A,\left[M_0^{1/2} \msU\msM_n^{-1/2}\bp_n\right]^T\right)^T$ and their conjugate fluxes $\boldsymbol{\chi}=(\Phi_A,\boldsymbol{\chi}_n^T)^T$, where $M_0$ is a finite constant with units of capacitance,
\begin{equation}
H_{II}=\frac{1}{2}\brho^T\msC_{II}^{-1}\brho+\frac{1}{2}\boldsymbol{\chi}^T\msL_{II}^{-1}\boldsymbol{\chi}+V(\Phi_A),\nonumber
\end{equation}
with the final capacitance and 
\begin{align}
\msC_{II}^{-1}&=\begin{pmatrix}
C_A^{-1} & C_A^{-1}\bff_n^T\\
C_A^{-1}\bff_n & M_0^{-1} \mathbbm{1}
\end{pmatrix},\qquad 
\msL_{II}^{-1}=\begin{pmatrix}
0 & 0\\
0 & (\msL_n^{II})^{-1}
\end{pmatrix},\nonumber
\end{align}
with the coupling vectors are $\bff_n=M_0^{-1/2}\msM_n^{1/2}\msU^T\be_n$, and the diagonal inductance submatrix is  $(\msL_n^{II})^{-1}=C_0^{-1}\msU^{T}\msM_n^{1/2}(\msL_{n}^{I})^{-1}\msM_n^{1/2}\msU$. We can rewrite the Hamiltonian as 
\begin{equation}
H=\frac{q_A^{2}}{2C_A}+V(\Phi_A)+\frac{1}{C_A}q_A\sum_{n}f_{n}\rho_{n}+\sum_{n}\frac{1}{2}\left[\frac{\rho_{n}^2}{M_0}+M_0 \Omega_{n}^{2}\chi_{n}^{2}\right],\nonumber
\end{equation}
where we have defined the frequencies $\Omega_n=1/\sqrt{M_0 L_n^{II}}$. Analogously to the analysis of the $1^{\mathrm{st}}$-Foster form, the coupling vector to the variables describing the admitance has finite norm even when the number of harmonic variables tends to infinity, i.e. 
\begin{equation}
\lim\limits_{\left|\be_n\right|^2\rightarrow\infty}\left|\bff_n\right|^2=\lim\limits_{\left|\be_n\right|^2\rightarrow\infty}M_0^{-1}\be_n^T \msM_n \be_n=C_A/M_0.\nonumber
\end{equation}

\section{Lossless transmission line impedance expansion}
\label{sec:Mport_synthesis}
A 2-port lossless transmission resonator can be characterized by its inductance $l$ and capacitance $c$ per unit length, and its total length $L$. Its impedance matrix 
\begin{equation}
\msZ(s)=Z_0\begin{pmatrix}
\mathrm{coth}\left(s\sqrt{lc}L\right) &  \mathrm{csch}\left(s\sqrt{lc}L\right)\\
\mathrm{csch}\left(s\sqrt{lc}L\right) & \mathrm{coth}\left(s\sqrt{lc}L\right)
\end{pmatrix},\nonumber
\end{equation}
is lossless and positive real (LPR) \cite{Newcomb:1966} in Laplace space. This is a generalization of the Foster reactance-function synthesis for the one-port circuit, and a simplified version of the Brune multiport impedance expansion used by Solgun and DiVincenzo \cite{Solgun:2015}. We can fraction-expand the formulae of the hyperbolic functions
\begin{align}
\coth\left(s\right)&=\frac{1}{s}+\sum_{n=1}^{\infty}\frac{2s}{s^{2}+n^{2}\pi^{2}},\qquad
\mathrm{csch}\left(s\right)=\frac{1}{s}+\sum_{n=1}^{\infty}\frac{2s(-1)^{n}}{s^{2}+n^{2}\pi^{2}},\nonumber
\end{align}
and find the decomposition of the impedance 
\begin{equation}
\msZ(s)=s^{-1}\msA_{0}+\sum_{n}^{\infty}\frac{s\msA_{n}}{s^{2}+\Omega_{n}^{2}},\nonumber
\end{equation}
where  we have defined the matrices
\begin{align}
s^{-1}\msA_{0}&=\frac{1}{cL}\begin{pmatrix}
1&1\\
1&1
\end{pmatrix},\label{eq:Mport_synthesis_A0}\\
\msA_{n}&=\frac{2}{cL}\begin{pmatrix}
1&(-1)^{n}\\
(-1)^{n}&1
\end{pmatrix},\label{eq:Mport_synthesis_An}
\end{align}
and the frequencies $\Omega_{n}^2=\frac{n^{2}\pi^{2}}{lcL^{2}}$. Such an expansion is a consequence of $\msZ=-\msZ^{\dagger}$ and the PR property. Following Sec. (7) in \cite{Newcomb:1966}, it is easy to synthesize a lumped-element circuit that has this impedance to the desired level of accuracy. The matrix (\ref{eq:Mport_synthesis_A0}), which corresponds to the pole at $s=0$, can be decomposed into
\begin{equation}
s^{-1}\msA_{0}= \begin{pmatrix}
1\\
1
\end{pmatrix}\left[\frac{1}{scL}
\right] \begin{pmatrix}
1&1
\end{pmatrix},\label{eq:Mport_synthesis_A0decomp}
\end{equation}
while the matrices (\ref{eq:Mport_synthesis_An}) with poles at the frequencies $\Omega_n$ can be expanded in 
\begin{equation}
\frac{s\msA_{n}}{s^{2}+\Omega_{n}^{2}}= \begin{pmatrix}
1\\
(-1)^{n}
\end{pmatrix}\left[\frac{2s/cL}{s^{2}+\Omega_{n}^{2}}
\right] \begin{pmatrix}
1&(-1)^{n}
\end{pmatrix}.\label{eq:Mport_synthesis_Andecomp}
\end{equation}

The first term in (\ref{eq:Mport_synthesis_A0decomp}) is implemented with a Belevitch transformer \cite{Belevitch:1950} of turn-ratios  $\msT_{0}=\left[1\,\,1\right]$ and a capacitor of capacitance $C_{0}=cL$. Each term in (\ref{eq:Mport_synthesis_Andecomp}) is synthesized via transformers $\msT_{n}=\left[1\,\,(-1)^{n}\right]$ and a capacitor of capacitance $C_{n}=cL/2$ shunted by an inductor of inductance $L_{n}=2lL/n^2 \pi^2$. Connecting all the stages we finally arrive to the circuit equivalent sketched in Fig. \ref{fig:2CQ2PortTL_v2}.

\chapter{Further Details on Nonreciprocal Lumped Networks}
\label{appendix_c}

\section{Extended Burkard analysis}
\label{sec_app:Burkard_extension}
We extend Burkard \cite{Burkard:2005} and Solgun-DiVincenzo \cite{Solgun:2015} analyses to include ideal multiport NR $\mY$-devices under the assumption that each branch of a NRD is shunted by a capacitor in the circuit independently. Relaxing the requirements of the BKD analysis, we allow nonreciprocal branches to appear both in the tree and in the chord set. We divide the tree and chord currents and voltages for the different components of the circuit in the following way:
\begin{align}
\bI_{\mathrm{tr}}^T&=\left(\bI_{J}^T,\bI_{L}^T,\bI_{G^{\mathrm{tr}}}^T,\bI_{T^{\mathrm{tr}}}^T\right),\qquad\,\,\,\,\,\,
\bI_{\mathrm{ch}}^T=\left(\bI_{C_{J}}^T,\bI_{C}^T,\bI_{G^{\mathrm{ch}}}^T,\bI_{T^{\mathrm{ch}}}^T\right),\nonumber\\
\bV_{\mathrm{tr}}^T&=\left(\bV_{J}^T,\bV_{L}^T,\bV_{G^{\mathrm{tr}}}^T,\bV_{T^{\mathrm{tr}}}^T\right),\quad
\bV_{\mathrm{ch}}^T=\left(\bV_{C_{J}}^T,\bV_{C}^T,\bV_{G^{\mathrm{ch}}}^T,\bV_{T^{\mathrm{ch}}}^T\right),\nonumber
\end{align}
where we have added gyrator branches to  both branch sets. We can write Kirchhoff's current laws without external fluxes for simplicity, 
\begin{align}
\msF \bI_{\mathrm{ch}}=&-\bI_{\mathrm{tr}},\nonumber\\
\msF^{T}\bV_{\mathrm{tr}}=&\bV_{\mathrm{ch}},\nonumber 
\end{align}
making use of the fundamental loop matrix $\msF$; see Refs. \cite{Burkard:2004,Burkard:2005} for a detailed analysis of graph theory applied to superconducting circuits, that can be partitioned as 
\begin{equation}
\msF=\begin{pmatrix}
\mathbbm{1}&\msF_{JC}&\msF_{JG^{\mathrm{ch}}}&\msF_{JT^{\mathrm{ch}}}\\
0&\msF_{LC}&\msF_{LG^{\mathrm{ch}}}&\msF_{LT^{\mathrm{ch}}}&\\
0&\msF_{G^{\mathrm{tr}}C}&\msF_{G^{\mathrm{tr}}G^{\mathrm{ch}}}&\msF_{G^{\mathrm{tr}}T^{\mathrm{ch}}}&\\
0&\msF_{T^{\mathrm{tr}}C}&\msF_{T^{\mathrm{tr}}G^{\mathrm{ch}}}&\msF_{T^{\mathrm{tr}}T^{\mathrm{ch}}}
\end{pmatrix}.
\end{equation}
We eliminate ideal transformer branches $\bI_{T}^T=(\bI_{T^{\mathrm{tr}}}^T,\bI_{T^{\mathrm{ch}}})^T$   \cite{Solgun:2015}, which do not store energy and are not degrees of freedom of the system, by making use of Kirchhoff's current law for tree transformer branches and the current constraint equation of the transformer (\ref{eq:Belevitch_trR1}),
\begin{align}
\bI_{T^{\mathrm{tr}}}=&-(\msF_{T^{\mathrm{tr}}C}\bI_{C}+\msF_{T^{\mathrm{tr}}G^{\mathrm{ch}}}\bI_{G^{\mathrm{ch}}}),\nonumber\\
\bI_{T^{\mathrm{ch}}}=&-\msN \bI_{T^{\mathrm{tr}}},\nonumber
\end{align}
with $\msN$ the turns ratios matrix. Here we have assumed that transformer tree (left) branches are not shunted by transformer chord (right) branches, i.e., $\msF_{T^{\mathrm{tr}}T^{\mathrm{ch}}}=0$ \cite{Belevitch:1950,Solgun:2015}. We can thus express the current in the right branches of Belevitch transformer as
\begin{equation}
\bI_{T^{\mathrm{ch}}}=\msN(\msF_{T^{\mathrm{tr}}C}\bI_{C}+\msF_{T^{\mathrm{tr}}G^{\mathrm{ch}}}\bI_{G^{\mathrm{ch}}}).\nonumber
\end{equation}
We write tree Josephson, inductor, and NR tree branch currents as a function of only capacitor and NR chord branch currents,
\begin{align}
-\bI_{J}=&\bI_{C_{J}}+\msF_{JC}\bI_{C}+\msF_{JG^{\mathrm{ch}}}\bI_{G^{\mathrm{ch}}}+\msF_{JT^{\mathrm{ch}}}\bI_{T^{\mathrm{ch}}}\nonumber\\
=&\bI_{C_{J}}+\msF_{JC}^{\mathrm{eff}}\bI_{C}+\msF_{JG^{\mathrm{ch}}}^{\mathrm{eff}}\bI_{G^{\mathrm{ch}}}\label{eq:NR_Burkard_IJ_KKeff}\\
-\bI_{L}=&\msF_{LC}^{\mathrm{eff}}\bI_{C}+\msF_{LG^{\mathrm{ch}}}^{\mathrm{eff}}\bI_{G^{\mathrm{ch}}},\label{eq:NR_Burkard_IL_KKeff}\\
-\bI_{G^{\mathrm{tr}}}=&\msF_{G^{\mathrm{tr}}C}^{\mathrm{eff}}\bI_{C}+\msF_{G^{\mathrm{tr}}G^{\mathrm{ch}}}^{\mathrm{eff}}\bI_{G^{\mathrm{ch}}}.\label{eq:NR_Burkard_IGtr_KKeff}
\end{align}
Here, we have defined effective loop submatrices \cite{Solgun:2015}
\begin{align}
\msF_{XC}^{\mathrm{eff}}=&\msF_{XC}+\msF_{XT^{\mathrm{ch}}}\msN\msF_{T^{\mathrm{tr}}C},\nonumber\\
\msF_{XG^{\mathrm{ch}}}^{\mathrm{eff}}=&\msF_{XG^{\mathrm{ch}}}+\msF_{XT^{\mathrm{ch}}}\msN\msF_{T^{\mathrm{tr}}G^{\mathrm{ch}}},\nonumber
\end{align}
with $X=\{J,L,G^{\mathrm{tr}}\}$, that have real entries instead of the usual ternary set $\{-1,1,0\}$ for branches with currents in the same or opposite direction, or out of the loop, respectively. 

Using Kirchhoff's current law and the capacitor constitutive equation, we write the inductors in terms of the junction and inductor voltages,
\begin{align}
\bI_{C_J}&=\dot{\bQ}_{C_J}=\msC_J\dot{\bV}_J,\label{eq:NR_MportL_IC1}\\
\bI_C&=\msC\left((\msF_{JC}^{\mathrm{eff}})^T\dot{\mbf{V}}_J+(\msF_{LC}^{\mathrm{eff}})^T\dot{\mbf{V}}_L+(\msF_{G^{\mathrm{tr}}C}^{\mathrm{eff}})^T\dot{\mbf{V}}_{G^{\mathrm{tr}}}\right).\label{eq:NR_MportL_IC2}
\end{align}
We rewrite again current-voltage constitutive relations for inductors and junctions, Eqs. (\ref{eq:consti_L}-\ref{eq:consti_J2}) in chapter~\ref{chapter:chapter_4}, for the symmetric elements,
\begin{align}
\bI_J=&\bI_c \bsb{\mathrm{sin}}(2\pi\bPhi_J/\Phi_q)=-\nabla_{\bPhi_J}U(\bPhi_J),\label{eq:NR_Burkard_IJ_const1}\\
\bI_L=&\msL^{-1}\bPhi_L,\label{eq:NR_Burkard_IL_consti}
\end{align}
while that for the $\mY$-NR branches, Eq. (\ref{eq:consti_SG}) in chapter~\ref{chapter:chapter_4} can be decomposed into  
\begin{equation}
\begin{pmatrix}
\bI_{G^{\mathrm{tr}}}\\	\bI_{G^{\mathrm{ch}}}
\end{pmatrix}=\begin{pmatrix}
\mY_{G^{\mathrm{tr}}G^{\mathrm{tr}}}&\mY_{G^{\mathrm{tr}}G^{\mathrm{ch}}}\\	\mY_{G^{\mathrm{ch}}G^{\mathrm{tr}}}&\mY_{G^{\mathrm{ch}}G^{\mathrm{ch}}}
\end{pmatrix} \begin{pmatrix}
\bV_{G^{\mathrm{tr}}}\\	\bV_{G^{\mathrm{ch}}}
\end{pmatrix}.\label{eq:NR_Burkard_IG_consti}
\end{equation}

Introducing Kirchhoff's voltage law in the current-voltage relation for chord NR branches we derive
\begin{align}
\bI_{G^{\mathrm{ch}}}=&(\mY_{G^{\mathrm{ch}}G^{\mathrm{tr}}}+\mY_{G^{\mathrm{ch}}G^{\mathrm{ch}}}(\msF_{G^{\mathrm{tr}}G^{\mathrm{ch}}}^{\mathrm{eff}})^T)\bV_{G^{\mathrm{tr}}}\label{eq:NR_Burkard_IGch_KKeff}\\
&+\mY_{G^{\mathrm{ch}}G^{\mathrm{ch}}}\left[(\msF_{LG^{\mathrm{ch}}}^{\mathrm{eff}})^T \bV_L+(\msF_{JG^{\mathrm{ch}}}^{\mathrm{eff}})^T \bV_J\right].\nonumber
\end{align}
Substituting Eqs. (\ref{eq:NR_Burkard_IJ_const1}-\ref{eq:NR_Burkard_IGch_KKeff}) in (\ref{eq:NR_Burkard_IJ_KKeff}, \ref{eq:NR_Burkard_IL_KKeff} and \ref{eq:NR_Burkard_IGtr_KKeff}) we have the equations of motion of the circuit that can be derived from the Lagrangian,
\begin{align}
L=\frac{1}{2}\dot{\bPhi}^T\mcl{C}\dot{\bPhi}-\frac{1}{2}\bPhi^T\msM_0 \bPhi+\frac{1}{2}\dot{\bPhi}^T\msG\bPhi-U(\bPhi_J),\nonumber
\end{align}
with $\bPhi^T=(\bPhi_J^T,\bPhi_L^T,\bPhi_{G^{\mathrm{tr}}}^T)$. The symmetric capacitive and inductive matrices read
\begin{align}
\mcl{C}&=\begin{pmatrix}
\msC_J & 0&0\\0&0&0\\0&0&0
\end{pmatrix}+\mcl{F}_{C}^{\mathrm{eff}}\msC(\mcl{F}_{C}^{\mathrm{eff}})^T,\quad \msf{M_0}=\mcl{I}_L\msL^{-1}\mcl{I}_L^T,\nonumber
\end{align}
where we defined
\begin{align}
\mcl{F}_{X}^{\mathrm{eff}}=&\begin{pmatrix}
\msF_{JX}^{\mathrm{eff}}\\	\msF_{LX}^{\mathrm{eff}}\\\msF_{G^{\mathrm{tr}}X}^{\mathrm{eff}}
\end{pmatrix},\quad
\mcl{I}_L=\begin{pmatrix}
0\\
\mathbbm{1}_{L}\\
0\\
\end{pmatrix},\nonumber
\end{align}
and $X=\{C,G^{\mathrm{ch}}\}$. The skew-symmetric matrix is
\begin{align}
\msG=&\,\mcl{I}_{G^{\mathrm{tr}}}\mY_{G^{\mathrm{tr}}G^{\mathrm{tr}}}\mcl{I}_{G^{\mathrm{tr}}}^T+\mcl{F}_{G^{\mathrm{ch}}}^{\mathrm{eff}}\mY_{G^{\mathrm{ch}}G^{\mathrm{ch}}}(\mcl{F}_{G^{\mathrm{ch}}}^{\mathrm{eff}})^T\label{eq:NR_Burkard_Gmat}\\
&+\mcl{F}_{G^{\mathrm{ch}}}^{\mathrm{eff}}\mY_{G^{\mathrm{ch}}G^{\mathrm{tr}}}\mcl{I}_{G^{\mathrm{tr}}}+\mcl{I}_{G^{\mathrm{tr}}}^T\mY_{G^{\mathrm{tr}}G^{\mathrm{ch}}}(\mcl{F}_{G^{\mathrm{ch}}}^{\mathrm{eff}})^T,\nonumber
\end{align}
with the identity vector
\begin{equation}
\mcl{I}_{G^{\mathrm{tr}}}=\begin{pmatrix}
0\\
0\\
\mathbbm{1}_{G^{\mathrm{tr}}}
\end{pmatrix}.\nonumber
\end{equation}
Explicitly,
\begin{equation}
\msG=\begin{pmatrix}
\msG_{JJ}&\msG_{JL}&\msG_{JG^{\mathrm{tr}}}\\
-\msG_{JL}^T&\msG_{LL}&\msG_{LG^{\mathrm{tr}}}\\
-\msG_{JG^{\mathrm{tr}}}^T&-\msG_{LG^{\mathrm{tr}}}^T&\msG_{G^{\mathrm{tr}}G^{\mathrm{tr}}}
\end{pmatrix},\nonumber
\end{equation}
where all the submatrices are defined as
\begin{align}
\msG_{JJ}=&\msF_{JG^{\mathrm{ch}}}^{\mathrm{eff}}\mY_{G^{\mathrm{ch}}G^{\mathrm{ch}}}(\msF_{JG^{\mathrm{ch}}}^{\mathrm{eff}})^T\nonumber,\\
\msG_{JL}=&\msF_{JG^{\mathrm{ch}}}^{\mathrm{eff}}\mY_{G^{\mathrm{ch}}G^{\mathrm{ch}}}(\msF_{LG^{\mathrm{ch}}}^{\mathrm{eff}})^T\nonumber,\\
\msG_{LL}=&\msF_{LG^{\mathrm{ch}}}^{\mathrm{eff}}\mY_{G^{\mathrm{ch}}G^{\mathrm{ch}}}(\msF_{LG^{\mathrm{ch}}}^{\mathrm{eff}})^T\nonumber,\\
\msG_{JG^{\mathrm{tr}}}=&\msF_{JG^{\mathrm{ch}}}^{\mathrm{eff}}(\mY_{G^{\mathrm{ch}}G^{\mathrm{tr}}}+\mY_{G^{\mathrm{ch}}G^{\mathrm{ch}}}(\msF_{G^{\mathrm{tr}}G^{\mathrm{ch}}}^{\mathrm{eff}})^T)\nonumber,\\	\msG_{LG^{\mathrm{tr}}}=&\msF_{LG^{\mathrm{ch}}}^{\mathrm{eff}}(\mY_{G^{\mathrm{ch}}G^{\mathrm{tr}}}+\mY_{G^{\mathrm{ch}}G^{\mathrm{ch}}}(\msF_{G^{\mathrm{tr}}G^{\mathrm{ch}}}^{\mathrm{eff}})^T)\nonumber,\\
\msG_{G^{\mathrm{tr}}G^{\mathrm{tr}}}=&\msF_{G^{\mathrm{tr}}G^{\mathrm{ch}}}^{\mathrm{eff}}\mY_{G^{\mathrm{ch}}G^{\mathrm{ch}}}(\msF_{G^{\mathrm{tr}}G^{\mathrm{ch}}}^{\mathrm{eff}})^T+\mY_{G^{\mathrm{tr}}G^{\mathrm{tr}}}\nonumber\\
&+\msF_{G^{\mathrm{tr}}G^{\mathrm{ch}}}^{\mathrm{eff}}\mY_{G^{\mathrm{ch}}G^{\mathrm{tr}}}+\mY_{G^{\mathrm{tr}}G^{\mathrm{ch}}}(\msF_{G^{\mathrm{tr}}G^{\mathrm{ch}}}^{\mathrm{eff}})^T.\nonumber
\end{align}
The Hamiltonian of this system is
\begin{equation}
H=\frac{1}{2}\left(\bQ-\frac{1}{2}\msG\bPhi\right)^T\mcl{C}^{-1}\left(\bQ-\frac{1}{2}\msG\bPhi\right)+\frac{1}{2}\bPhi^T\msM_0 \bPhi+U(\bPhi_J)\label{eqapp:H_BurkardG},
\end{equation}
where $\bQ=\partial L/\partial \bPhi$ are the conjugate charges to the flux variables. Canonical quantization follows promoting the variables to operators with commutation relations $[\Phi_i,Q_j]=i\hbar$. 
\section{NR multiport impedance coupled to Josephson junctions}
\label{sec_app:NR_MportL}
We explicitly compute matrices of Hamiltonian (\ref{eq:Hamiltonian_Ydev}) for circuit in Fig. \ref{fig:2CQ_2P_MLImpedance}, both in the chapter~\ref{chapter:chapter_4}, of a nonreciprocal 2-port lossless impedance \cite{Newcomb:1966} capacitively coupled to Josephson junctions. 

The tree and chord branch sets are divided in $\bI_{\mathrm{tr}}^T=(\bI_C^T,\bI_{T^L}^T)$ and $\bI_{\mathrm{ch}}^T=(\bI_{J}^T,\bI_L^T,\bI_{T^R}^T)$, with left (right) transformer branches being tree (chord) branches. A general turns ratios matrix for the Belevitch transformer is
\begin{equation}
\mathsf{N}=\begin{pmatrix}
n_{11}^L&0&0&n_{12}^L&0&0\\
0&n_{11}^R&0&0&n_{12}^R&0\\
0&0& n_{21}&0&0&n_{22}
\end{pmatrix}.
\end{equation}
We will calculate with it the effective loop matrix (\ref{eq:BKD_Feff}) and get Hamiltonian (\ref{eq:Hamiltonian_Ydev}) in chapter~\ref{chapter:chapter_4}. The capacitance matrix $\msC=\mathrm{diag}(C_{J1},
C_{J2}, C_{c1}, C_{c2}, C_{1R}, C_{1L}, C_{2})$ is full rank. The inductive $\msM_0$ matrix can be computed with the loop submatrix
\begin{equation}
\msF_{CL}^{\mathrm{eff}}=\msF_{CL}=
\begin{pmatrix}
\mymathbb{0}_{M}&0\\
0&1 \\
\end{pmatrix}
\end{equation}
where $M=J+g+G1+L$. $\{J,g,G1,L\}$ are, respectively, the number of (i) Josephson junctions (2), (ii) coupling capacitors (2), (iii) gyrator-shunted capacitors (2), and (iv) inductors ($L$). $\mymathbb{0}_{M}$ represents a zero square matrix of $M$ dimension. The skew-symmetric gyration matrix $\msG$ can be computed using the effective loop submatrix,
\begin{align}
\msF_{CG}^{\mathrm{eff}}=\msF_{CG}+\msF_{CT^{R}}\msN \msF_{T^{L}G}
=
\begin{pmatrix}
1 & 0 & 0 & 0 \\
0 & 1 & 0 & 0 \\
1 & 0 & 0 & 0 \\
0 & 1 & 0 & 0 \\
{n_{11}^L} & {n_{12}^L} & 1 & 0 \\
{n_{11}^R} & {n_{12}^R} & 0 & 1 \\
{n_{21}} & {n_{22}} & 0 & 0 \\
\end{pmatrix}\nonumber
\end{align} 
which is calculated through the turn ratios matrix $\msf{N}$ and the submatrices
\begin{align}
\msF_{CG}=&
\begin{pmatrix}
1 & 0 & 0 & 0 \\
0 & 1 & 0 & 0 \\
1 & 0 & 0 & 0 \\
0 & 1 & 0 & 0 \\
0 & 0 & 1 & 0 \\
0 & 0 & 0 & 1 \\
0 & 0 & 0 & 0 \\
\end{pmatrix},\quad
\msF_{CT^{R}}=\begin{pmatrix}
0 & 0 & 0 \\
0 & 0 & 0 \\
0 & 0 & 0 \\
0 & 0 & 0 \\
1 & 0 & 0 \\
0 & 1 & 0 \\
0 & 0 & 1 \\
\end{pmatrix},\quad
\msF_{T^{L}G}=
\begin{pmatrix}
1 & 0 & 0 & 0 \\
1 & 0 & 0 & 0 \\
1 & 0 & 0 & 0 \\
0 & 1 & 0 & 0 \\
0 & 1 & 0 & 0 \\
0 & 1 & 0 & 0 \\
\end{pmatrix}.\nonumber
\end{align}

This analysis can be performed because the constitutive equation of the nonreciprocal elements (\ref{eq:consti_SG}) in chapter~\ref{chapter:chapter_4} could be simplified to $\bI_G=\mY_G \bV_G$, where $\bI_{G}=(I_{G0^L},I_{G0^R},I_{G1^L},I_{G1^R})^T$ and 
\begin{equation}
\mY_G=\begin{pmatrix}
\mY_{G0}&0\\0&\mY_{G1}
\end{pmatrix},
\end{equation}
with
\begin{equation}
\mY_{Gi}=\frac{1}{R_i}\begin{pmatrix}
0&1\\-1&0
\end{pmatrix},
\end{equation}
the admittance matrix for each gyrator $i\in\{0,1\}$. The final gyration matrix is
\begin{equation}
\msG=\msF_{CG}^{\mathrm{eff}}\mY_{G}(\msF_{CG}^{\mathrm{eff}})^T.
\end{equation}

\section{Symplectic diagonalization}
\label{App_sec:symplectic}
We discuss now the procedure to diagonalize the quadratic sector of Hamiltonian (\ref{eq:Hamiltonian_Ydev}) from a more general perspective with respect to Appendix~\ref{appendix_a}. We can perform a canonical change of variables $\bQ_C=\msC^{1/2}\msf{O}^T\bq$,  $\bPhi_C=\msC^{-1/2}\msf{O}^T\bff$ such that we diagonalize the pure capacitive and inductive sectors of the Hamiltonian,
\begin{align}
H=\frac{1}{2}(\bq^T,\bff^T)\begin{pmatrix}
\mathbbm{1}&\msf{\Gamma}\\
\msf{\Gamma}^T&\msf{\Omega}^2
\end{pmatrix}\begin{pmatrix}
\bq\\
\bff
\end{pmatrix}+U(\bff),\label{eq:Hamiltonian_NRD_2}
\end{align}
with the definitions $\msf{\Gamma}=-\frac{1}{2}\msf{O}\msC^{-1/2}\msG\msC^{-1/2}\msf{O}^T$ and  $\msf{\Omega}^2=\msf{O}\msC^{-1/2}\msL^{-1}\msC^{-1/2}\msf{O}^T-\msf{\Gamma}^2$ a diagonal matrix. The conjugate variables ($\bq,\bff$) are canonical in that $\{q_i,f_j\}=\delta_{ij}$. The presence of the antisymmetric matrix $\msf{\Gamma}$ in the harmonic part of the Hamiltonian leads to new normal frequencies that are greater or equal to those without it. In order to carry out canonical quantization of this Hamiltonian it is convenient to proceed with the symplectic diagonalization of the harmonic part. Consider thus the matrix
\begin{equation}
\label{eq:hamharmonicmatrix}
\msf{H}_h=\begin{pmatrix}
\mathbbm{1}&\msf{\Gamma}\\
\msf{\Gamma}^T&\msf{\Omega}^2
\end{pmatrix}\,.
\end{equation}
Since this matrix is symmetric and definite positive, the corresponding theorem of Williamson \cite{Laub:1974} holds that it can be brought to the canonical form $\msf{D}=\mathrm{diag}\left(\msf{\Lambda},\msf{\Lambda}\right)$, with $\msf{\Lambda}$ a definite positive diagonal matrix, by a symplectic transformation $\msf{S}$. That is, $\msf{S}^T\msf{H}_h\msf{S}=\msf{D}$ with symplectic matrix $\msf{S}$. The determination of the symplectic eigenvalues and of the canonical symplectic transformation can be achieved by considering the matrix $\msf{H}_h\msf{J}$, with
\begin{equation}
\label{eq:jmatrix}
\msf{J}=\begin{pmatrix}
0&\mathbbm{1}\\
-\mathbbm{1}&0
\end{pmatrix}.
\end{equation}
Its eigenvalues form conjugate pure imaginary pairs, $\pm i \lambda_j$, where the positive numbers $\lambda_j$ are the diagonal elements of $\msf{\Lambda}$.  Choose an eigenvector $\mathbf{v}_j$ corresponding to $i\lambda_j$. Its complex conjugate, $\mathbf{v}_j^*$, is an eigenvector with $-i\lambda_j$ eigenvalue.  Organize the column eigenvectors in a matrix $\msf{F}=\begin{pmatrix}\mathbf{v}_1&\mathbf{v}_2&\cdots&\mathbf{v}_N&\mathbf{v}_1^*&\cdots&\mathbf{v}_N^*
\end{pmatrix}
$. Normalize the vectors by the condition $\msf{F}\msf{F}^\dag=\msf{H}_h$. Define a matrix function  $\msf{S}_{\msf{V}}=\left(\msf{F}^\dag\right)^{-1}\msf{V}\msf{D}^{1/2}$ acting on unitaries $\msf{V}$. It is clearly the case that, for all unitaries $\msf{V}$ and phase choices for the eigenvectors $\mathbf{v}_j$, $\msf{S}_{\msf{V}}^\dag \msf{H}_h\msf{S}_{\msf{V}}=\msf{D}$, since $\msf{F}^{-1}\msf{H}_h\left(\msf{F}^\dag\right)^{-1}=\mathbbm{1}$. The unitary $\msf{V}$ is determined by the requirement that it provide us with a symplectic matrix, $\msf{S}^T\msf{J}\msf{S}=\msf{J}$. {\it Inter alia}, this means that $\msf{S}$ is real. In fact, the choice
\begin{equation}
\label{eq:vchoice}
\msf{V}=\frac{1}{\sqrt{2}}
\begin{pmatrix}
1&i\\ 1&-i
\end{pmatrix}
\end{equation}
achieves this objective. This can be readily checked by noticing that
\begin{equation}
\label{eq:vtimesf}
\msf{V}^{\dag}\msf{F}^\dag=\frac{1}{\sqrt{2}}
\begin{pmatrix}
\mathbf{v}_1^\dag+\mathbf{v}_1^T\\
\vdots\\
\mathbf{v}_N^\dag+\mathbf{v}_N^T\\
-i\left( \mathbf{v}_1^\dag-\mathbf{v}_1^T\right)\\
\vdots   
\end{pmatrix}
\end{equation}
is explicitly real in this case, 
so $\msf{S}^{-1}=\msf{D}^{-1/2} \msf{V}^{\dag}\msf{F}^\dag$ is seen to be real. Furthermore, this choice also determines $\msf{S}$ as symplectic.

In the new variables, $\left(\boldsymbol{\xi}^T\ \boldsymbol{\pi}^T\right)=\mathsf{S}^T\left(\mathbf{q}^T\ \mathbf{f}^T\right)$, the quadratic part of the Hamiltonian is diagonal. They can now be canonically quantized, in the form $\xi_n=(a_n+a_n^\dag)/\sqrt{2}$, $\pi_n=-i(a_n-a_n^\dag)/\sqrt{2}$.    
\section{Reduction of variables in circuits without $\mY$  ideal NR devices}
\label{sec:upperc-vari-circ}
We formalize and generalize the problem of the quantization of circuits in flux variables with linear NR devices that are only described by a constitutive equation through $\msf{S}$.  Further below, we apply this method to the derivation of the circuits in Fig. \ref{fig:Z_S_NRCircuits}(a) in chapter~\ref{chapter:chapter_4}.

We start from the equation of motion (\ref{eq:S_circuit_eom}), that we rewrite as  
\begin{align}
\label{eqapp:S_circuit_eom}
\left(\mathbbm{1}+\mathsf{S}\right)(\msC \ddot{\boldsymbol{\Phi}}+\nabla_{\boldsymbol{\Phi}} U(\boldsymbol{\Phi}))=-R^{-1}\left(\mathbbm{1}-\mathsf{S}\right)\dot{\bPhi},
\end{align}
with $\nabla_{\boldsymbol{\Phi}}U(\boldsymbol{\Phi})=\left(U_1'(\Phi_1),U_2'(\Phi_2),...\right)^T\,$, and $U_i'(\Phi_i)=E_{Ji}\sin(\Phi_i)$. $\msC$ is a non-degenerate capacitance matrix. 
An ideal $N$-port circulator can always be described by a scattering matrix 
\begin{equation}
\msf{S}=\begin{pmatrix}
&&&s_N\\
s_1&&&\\
&\ddots&&\\
&&s_{N-1}&
\end{pmatrix},\label{eqapp:S_matrix}
\end{equation}
where each non-zero element can only be $s_k=\pm1$. By a correct choice of terminals, it can be proven that there are only two canonical types of ideal $N$-port circulators: those with values ($s_k=1$) in all their entries, and others with all ($s_k=1$), except for one ($s_j=-1$); see Ref. \cite{Carlin:1964} for further details.

The eigenvalue equation of the scattering matrix can be retrieved noticing that $\msf{S}^N=\prod_k s_k \mathbbm{1}$,
\begin{equation}
\lambda^N=\prod_k s_k =\pm 1.
\end{equation}
The eigenvalues of the scattering matrix lie on the unit circle, $e^{i\epsilon\pi/N} e^{2 i\pi n/N}$ with $n\in\{0,N-1\}$, and $\epsilon$ either 0 or 1. The eigenvalue $\lambda=-1$ appears with multiplicity one for $N$ even ($N$ odd)  with $\prod s_k=1$ ($\prod s_k=-1$). On the other hand, the eigenvalue $\lambda=1$ is present also with multiplicity one for $N$ both even and odd  when $\prod s_k=+1$.  All other eigenvalues come in pairs of complex conjugate values ($\lambda_k$ and $\lambda_k^*$).

Let us assume that $\msf{S}$ presents eigenvalue $-1$. We define the projector $\msf{P}=\bv_{-1}\bv_{-1}^T$ such that $\msf{S}\msf{P}=-\msf{P}=\msf{P}\msf{S}$, where $\bv_{-1}$ is the normalized eigenvector corresponding to the eigenvalue $-1$. We complete the identity with the projector $\mQ=\mathbbm{1}-\msf{P}$, which also commutes with $\msf{S}$; $[\msf{S},\mQ]=[\msf{S},\msf{P}]=0$. It is trivial to prove that $\msf{P}$ is real and that thus so it is $\mQ$. If $-1$ is an eigenvalue, it always has multiplicity 1. Then, given that $\msf{S}=\msf{S}^*$,
\begin{align}
\left(\msf{S}\bv_{-1}\right)^*=&-\bv_{-1}^*=\msf{S}\bv_{-1}^*,\\
\msf{S}\bv_{-1}=&-\bv_{-1}.
\end{align}
The above two equations can only be true if $\bv_{-1}=\bv_{-1}^*$. 
Then, applying $\msf{P}$ to Eq. (\ref{eqapp:S_circuit_eom}), we have
\begin{equation}
\msf{P}\dot{\bPhi}=0.
\end{equation}
This equation can be integrated, so that the flux variable vector is expressed as
\begin{equation}
\bPhi=\msf{P}\bPhi+\mQ\bPhi=\alpha \bv_{-1}+ \bPsi,\label{appeq:phi_exp}
\end{equation}
where we defined $\bPsi=\mQ\bPhi$, and $\alpha$ is an initial-value constant in flux units. Inserting the above expression in the equation of motion and applying $\mQ$ on the left, we have
\begin{align}
\mQ\left(\mathbbm{1}+\mathsf{S}\right)\mQ(\msC_{\mathsf{Q}} \ddot{\boldsymbol{\Psi}}+\mQ\nabla_{\boldsymbol{\Psi}}\tilde{U}_\alpha(\boldsymbol{\Psi}))=\nonumber
-R^{-1}\mQ\left(\mathbbm{1}-\mathsf{S}\right)\mQ\dot{\bPsi},\nonumber\\
\end{align}
with $\msC_{\mQ}=\mQ\msC\mQ$  a new symmetric reduced capacitance matrix, and $\tilde{U}_\alpha(\boldsymbol{\Psi})=U(\mQ\boldsymbol{\Phi}+\alpha\bv_{-1})$ the new potential. The differential nabla operator on the original flux variables becomes  $\nabla_{\bPhi}=\mQ\nabla_{\bPsi}+\bv_{-1}\partial_{\alpha}$. In this new $N-1$ dimensional space, the remnant of $\mathsf{Q}\left(\mathbbm{1}+\mathsf{S}\right)\mathsf{Q}$  is invertible. Formally, we derive in this reduced space the Euler-Lagrange equation 
\begin{align}
\mathsf{C}_{\mQ} \ddot{\boldsymbol{\Psi}}+\mQ\nabla_{\boldsymbol{\Psi}}\tilde{U}_\alpha(\boldsymbol{\Psi})=
-\msG_{\mQ}\dot{\bPsi},\label{appeq:S_circuit_EL_formal}
\end{align}
with $\msG_{\mQ}=R^{-1}(\mQ\left(\mathbbm{1}+\mathsf{S}\right)\mQ)^{-1}(\mQ\left(\mathbbm{1}-\mathsf{S}\right)\mQ)$, again understood in the reduced space. There,  $\mathsf{G}_{\mathsf{Q}}$ is the Cayley transform of an orthogonal matrix, and thus a skew-symmetric matrix. 

Let us illustrate the procedure with the choice of a specific decomposition of the real projector $\mQ$. Consider   $\bv_k$ and its complex conjugate $\bv_k^*$ to be orthogonal vectors in the subspace complementary to $\msf{P}$. It is  then easy to prove that real $\mathrm{Re}\{\bv_k\}=(\bv_k+\bv_k^*)/2$ and imaginary parts $\mathrm{Im}\{\bv_k\}=-i(\bv_k-\bv_k^*)/2$ are orthogonal vectors, again orthogonal to the $\msf{P}$ eigenspace. This assumption will hold if the vector $\bv_k$ is an eigenvector of $\mathsf{S}$ with complex eigenvalue. If the eigenvalue $\lambda=1$ is present, its associated eigenvector is also real; the proof is completely analogous to the above for the eigenvector $\bv_{-1}$. Normalizing all vectors, we can write
\begin{align}
\mQ&=\bv_{1}\bv_{1}^T+\sum_k \bx_k \bx_k^T + \by_k \by_k^T=\sum_{n=1}^{N-1} \bw_n \bw_n^T,\label{appeq:Z_S_Q_w}
\end{align}
with $\bx_k=\mathrm{Re}\{\bv_k\}/||\mathrm{Re}\{\bv_k\}||$ and $\by_k=\mathrm{Im}\{\bv_k\}/||\mathrm{Im}\{\bv_k\}||$,  $k$ running through all the vectors coming in complex conjugate pairs. In general, let us denote by  $\bw_n$ those real orthonormal vectors spanning the orthogonal space.

Using this nomenclature and Eq. (\ref{appeq:phi_exp}) we write 
\begin{align}
\bPhi&=\alpha \bv_{-1}+\sum_n f_n \bw_n=\msM\begin{pmatrix}
\alpha\\
\bff
\end{pmatrix},\label{appeq:Z_S_phi_M_f}
\end{align}
with $f_n=\bw_n^T\bPhi$, and $\msM=[\bv_{-1},\bw_1,\bw_2,...]$ an orthogonal matrix, i.e., $\msM\msM^T=\mathbbm{1}$. The nabla operator can be rewritten as 
\begin{equation}
\nabla_{\bPhi}=(\msM^{-1})^T\begin{pmatrix}
\frac{\partial}{\partial \alpha}\\
\frac{\partial}{\partial f_1}\\
\vdots
\end{pmatrix}=\msM\begin{pmatrix}
\frac{\partial}{\partial \alpha}\\
\nabla_{\bff}
\end{pmatrix}.\label{appeq:Z_S_nabla_phi_f}
\end{equation}
Finally, inserting the above decompositions  (\ref{appeq:Z_S_phi_M_f},\ref{appeq:Z_S_nabla_phi_f},\ref{appeq:Z_S_Q_w}) in Eq. (\ref{appeq:S_circuit_EL_formal}), we  rewrite the equation of motion
\begin{align}
\sum_{n,m,l} \bw_n (\mathbbm{1}+\msf{S})_{nm}\left[ (\msC)_{ml}\ddot{f}_l+\partial_{f_m}\tilde{U}_{\alpha}(\bff)\right]=-R^{-1}\sum_{n,l}  \bw_n(\mathbbm{1}-\msf{S})_{nl} \dot{f}_l,
\end{align}
with $(\msf{A})_{rt}=\bw_r^T\msf{A}\bw_t$, together with $\dot{\alpha}=0$. Multiplying from the left with the real row vectors $\{\bw_n^T\}$,  and inverting the first matrix on the left-hand side, we arrive at an explicit form of Eq. (\ref{appeq:S_circuit_EL_formal})
\begin{equation}
(\msC)_{ml}\ddot{f}_l+\partial_{f_m}\tilde{U}_{\alpha}(\bff)
=-(\msG_{\mQ})_{ml}\dot{f}_l,\nonumber\\
\end{equation}
where we have defined $(\msG_{\mQ})_{ml}=R^{-1}(\mathbbm{1}+\msf{S})_{mn}^{-1}(\mathbbm{1}-\msf{S})_{nl}$ and we have used Einstein's notation of summation over repeated indices. Here, we can identify $\bPsi\equiv(0,\bff)$ in Eq.~(\ref{appeq:S_circuit_EL_formal}). Furthermore, the matrix $\mathsf{C}_{\mathsf{Q}}$ has as matrix elements in this basis precisely $ (\msC)_{ml}$. The Lagrangian without constraints and full-rank kinetic matrix with such equations of motion  
\begin{align}
L=\frac{1}{2}\left(\dot{\bff}^T\msC_{\mQ}\dot{\bff}+\dot{\bff}^T \msG_{\mQ}\bff\right)-\tilde{U}_{\alpha}(\bff),\nonumber
\end{align}
with $\bff=\left(f_1,f_2,...\right)$. The quantized Hamiltonian is  
\begin{align}
\hat{H}=\frac{1}{2}\left({\hat{\bQ}}-\frac{1}{2}\msG_{\mQ}\hat{\bff}\right)^T\msC_{\mQ}^{-1}\left(\hat{\bQ}-\frac{1}{2}\msG_{\mQ}\hat{\bff}\right)+\tilde{U}_\alpha(\hat{\bff}),\nonumber
\end{align}
again with  $\bQ=\partial L/\partial \dot{\bff}$ the conjugated charge variables, which are promoted to operators. 

\subsection*{Examples}
Let us now use this general theory to quantize the specific cases illustrated in chapter~\ref{chapter:chapter_4}. The scattering matrix of Eq. (15) introduced in the circuits in Fig.~\ref{fig:Z_S_NRCircuits}(a) in the chapter~\ref{chapter:chapter_4},
\begin{equation}
\msf{S}_{N}=(-1)^N\begin{pmatrix}
&&&1\\
1&&&\\
&\ddots&&\\
&&1&
\end{pmatrix},\label{eqapp:S_N_matrix}
\end{equation} 
has $-1$ eigenvalues for all $N$ and $+1$ eigenvalues for even-$N$ numbers of ports. Notice that in the analysis of the equations of motion above we have not made use of the canonical form of $\mathsf{S}$ matrices mentioned earlier, and indeed this example does not and needs not conform to that canonical presentation.

\subsubsection{3-port case}
The eigenvalues and eigenvectors for $N=3$ are  $\blambda_3=(-1,\lambda_3,\lambda_3^*)$ and $\msf{V}_3=[\bv_{-1},\bv_3,\bv_3^*]^T$, respectively, with $\lambda_3=e^{2\pi i/3}$ and $\bv_3=(e^{2\pi i/3},e^{-2\pi i/3},1)/\sqrt{3}$.  The eigenvalue $\lambda=-1$ of $\msf{S}_N$, present in this family of matrices, is associated with the constraint $\bv_{-1}^T\dot{\bPhi}=0$ where $\bv_{-1}=(1,1,1)/\sqrt{3}$ is the normalized eigenvector.

We can apply the theory described above to compute the projectors 
\begin{align}
\msf{P}=&\bv_{-1}\bv_{-1}^T=\frac{1}{3}\begin{pmatrix}
1&1&1\\
1&1&1\\
1&1&1
\end{pmatrix},\\
\mQ=&\mathbbm{1}-\msf{P}=\frac{1}{3}\begin{pmatrix}
2 & -1 & -1 \\
-1 &2 & -1 \\
-1 & -1 & 2
\end{pmatrix}.
\end{align}
The reduced capacitance matrix is 
\begin{equation}
\msC_{\mQ}=\frac{1}{2}\left(
\begin{array}{cc}
\frac{1}{3} \left(C_1+C_2+4 C_3\right) & \frac{C_2-C_1}{ \sqrt{3}} \\
\frac{C_2-C_1}{ \sqrt{3}} &  \left(C_1+C_2\right) \\
\end{array}
\right),
\end{equation}
while the gyration matrix is
\begin{equation}
\msG_{\mQ}=\frac{1}{R\sqrt{3}}\begin{pmatrix}
0&1\\
-1&0
\end{pmatrix}.
\end{equation}
Finally, the potential function $\tilde{U}_\alpha=U(\msM(\alpha,\bff^T)^T)$, with $U(\bPhi)=-\sum_{i=1}^{3}E_{Ji}\cos(\Phi_i)$ and 
\begin{equation}
\msM=\begin{pmatrix}
\frac{1}{\sqrt{3}} & \frac{1}{\sqrt{3}} & \frac{1}{\sqrt{3}} \\
-\frac{1}{\sqrt{6}} & -\frac{1}{\sqrt{6}} & \sqrt{\frac{2}{3}} \\
\frac{1}{\sqrt{2}} & -\frac{1}{\sqrt{2}} & 0 
\end{pmatrix}.
\end{equation}
\subsubsection{4-port case}
The eigenvalues and eigenvectors for $N=4$ are  $\blambda_4=(-1,1,\lambda_4,\lambda_4^*)$  and $\msf{V}_4=[\bv_{-1},\bv_{1},\bv_4,\bv_4^*]^T$, respectively, with $\lambda_4=i$ and $\bv_{4}=(-i,-1,i,1)/2$.  The eigenvalue $\lambda=-1$ of $\msf{S}_N$, present in this family of matrices, is associated with the constraint $\bv_{-1}^T\dot{\bPhi}=0$ where $\bv_{-1}=(-1,1,-1,1)/2$ is the normalized eigenvector.

The inhomogeneous capacitance matrix is 
\begin{equation}
\msC_{\mQ}=\left(
\begin{array}{ccc}
\frac{C_1+C_2+C_3+C_4}{4} & \frac{C_4-C_2}{2 \sqrt{2}} & \frac{C_3-C_1}{2 \sqrt{2}} \\
\frac{C_4-C_2}{2 \sqrt{2}} & \frac{C_2+C_4}{2}  & 0 \\
\frac{C_3-C_1}{2 \sqrt{2}} & 0 & \frac{C_1+C_3}{2}  \\
\end{array}
\right),
\end{equation}
that reduces to $\msC_{\mQ}=C\mathbbm{1}$ for $C_i=C$. On the other hand, the gyration matrix has now a zero column and row corresponding to the eigenvalue $\lambda=+1$,
\begin{equation}
\msG_{\mQ}=\frac{1}{R}\begin{pmatrix}
0&0&0\\	0&0&1\\	0&-1&0
\end{pmatrix}.
\end{equation}
Given the complex-conjugate pairwise nature of the eigenvalues and eigenvectors, the gyration matrix can always be written in a basis with $2\!\!\times\!\!2$ blocks, except for the  row and the column of zeros corresponding to the $+1$ eigenvalue. Finally, we have the potential function $\tilde{U}_\alpha=U(\msM(\alpha,\bff^T)^T)$, with $U(\bPhi)=-\sum_{i=1}^{3}E_{Ji}\cos(\Phi_i)$ and 
\begin{equation}
\msM=\begin{pmatrix}
-\frac{1}{2} & \frac{1}{2} & -\frac{1}{2} & \frac{1}{2} \\
\frac{1}{2} & \frac{1}{2} & \frac{1}{2} & \frac{1}{2} \\
0 & -\frac{1}{\sqrt{2}} & 0 & \frac{1}{\sqrt{2}} \\
-\frac{1}{\sqrt{2}} & 0 & \frac{1}{\sqrt{2}} & 0
\end{pmatrix}.
\end{equation}

\chapter{Further Details on Distributed and Nonreciprocal Networks}
\label{appendix_d}
\section{Reduced space operator}
We formally define the differential operators for the half-line in the reduced space  by their domain
\begin{align}
\mD(\mL)=&\{\bsb{E},\,\bsb{E} \in \mathbbm{C}^N\otimes AC^1(\mcl{I}),\, \msf{D}_{\bE}\bsb{E}_0+\msF_{\bE} \bsb{E}'_0=0 \},\label{eq:TL_LU_VE_def}
\end{align}
with $\bE$ either $\bU$ or $\bV$, $\msf{D}_{\bU}=\msF_{\bV}=0$, $\msF_{\bU}=\msDelta$, and $\msD_{\bV}=\mathbbm{1}$, and by  their action on elements of the domain $\bE\in\mD(\mL)$, $\mL \bE=-\msDelta \bE''$.

We associate the notation $\bU$ with fluxes and $\bV$ with charges throughout. We use for convenience two different inner products, $\langle \bU_1,\bU_2\rangle=\int_{\mcl{I}}dx\, \bU_{1}(x)\bU_2(x)$ for the flux and $\langle \bV_1,\bV_2\rangle=\int_{\mcl{I}}dx\, \bV_{1}(x)\msDelta^{-1}\bV_2(x)$ for the charge presentation, with which one can construct orthonormal bases for the different spaces.  

We denote with $\bE_{\omega\lambda}$ the generalized eigenvectors of the operator $\mL$, where $\omega\in\mR_+$ is a continuous parameter and $\lambda$ is a discrete degeneracy index bounded to the number of lines $N$, that solve the Sturm-Liouville eigenvalue problems $\mL \bE_{\omel}=\omega^2 \bE_{\omel}$, with positive eigenvalue $\omega^2$. The reality of the coefficients of the differential expressions and the boundary conditions translates into the reality of the operator, $\mL=\mL^*$, and thus a real basis can always be chosen.

\section{Double space operator for ideal nonreciprocal elements}
\label{App_sec:Self_adjoint_op}
The domain of the self-adjoint operator proposed for the mode decomposition of the admittance-described NR element connected to $N$ semi-infinite lines is
\begin{align}
\mcl{D}(\mL)=&\left\{\bW(x)\in AC^1(\mcl{I})\otimes\mathbbm{C}^{2N},\,\bV_0 = \mY \bU_0,\,\msDelta\bU'_0=\mY \bV'_0\right\}\,.
\end{align}

In what follows we systematically understand $\bW$ to be a doublet of $N$ component functions,
$\bW^T=(\bU^T\,\bV^T)$. The operator acts as $\mL \bW =-\msDelta\bW''$. It can be easily checked that under the inner product 
\begin{align}
\langle\bW_1,\bW_2\rangle_{\mcl{H}_2}=\int_{\mcl{I}}dx \,\bW_1^\dag \msSigma  \bW_2\label{eq:inner_product_lines_only}
\end{align}
with $\msSigma=\text{diag}(\mathbbm{1},\msDelta^{-1})$, the operator is self-adjoint, as follows. For $\bW_2$ to be in $\mD(\mL^\dag)$, there needs to exist a $\bW_3$ such that $\langle \mL\bW_1,\bW_2\rangle-\langle\bW_1,\bW_3\rangle=0$ holds $\forall\bW_1\in\mD(\mL)$, in which case one defines $\bW_3=\mL^\dag\bW_2$. Now, by integration by parts  
\begin{align}
&\langle \mL\bW_1,\bW_2\rangle-\langle\bW_1,-\msDelta\bW_2''\rangle=\nonumber\\
&\bV_1'(0)^\dag\left[\mY^T\bU_2(0)+\bV_2(0)\right]-\bU_1^\dag(0)\left[\msDelta \bU_2'(0)+\mY^T\bV_2'(0)\right]\,.
\end{align}
Since $\bV'(0)$ and $\bU(0)$ are not determined by the boundary conditions, only their relations to other quantities, in order for the RHS to be zero for all $\bW\in{\cal D}({\cal L})$ the terms in square brackets have to be zero. We thus see that the domains of ${\cal L}$ and ${\cal L}^\dag$ match. The requirement of absolutely continuous derivative is necessary for the applicability of integration by parts.
\subsection{Positivity}
The above defined operator (in this case defined in the interval $\mcl{I}=\mR^+$) is a monotone (accretive) operator 
\begin{align}
\langle \bW, \mL\bW\rangle=&\int_{\mR^+} dx\,(\bW^\dag)'\msSigma\msDelta\bW'> 0
\end{align}
$\forall \,\bW\in \mcl{D}(\mL)$ as $\msSigma\msDelta$, is a real positive symmetric matrix.

\subsection{$\mT$ Symmetry}
Let us define a transformation $\mT$ on the domain of $\mL$ by 
\begin{align}
\mT\bW=&-i\begin{pmatrix}
\bV'\\ \msDelta\bU'
\end{pmatrix}\nonumber\,.
\end{align}
Since $W\in\mD(\mL)$ is absolutely continuous, this is well defined. Now, the crucial property of this transformation is that applied to eigenvectors that obey $\mL\bW=\omega^2\bW$ it is the case that $\mT\bW$ belongs to the domain of $\mL$ and is again an eigenvector of $\mL$ with the same eigenvalue. Thus, extended by linearity, the two of them commute \[\mL\left[\mT\bW\right]=\mT\left[\mL\bW\right]\,.\]
It follows that, unless $\mT$ acts trivially on the $\omega^2$ eigenspace, that eigenvalue is degenerate. Thus, generically, the presence of this symmetry give rise to an at least two-fold degeneracy of the eigenvalues. This transformation implements electromagnetic duality in the present context.

\subsection{Orthonormal eigenbasis}
Let us find now an orthonormal basis for the spectral decomposition of the differential operator. The general solution for the system of ordinary differential equations $\mL\bW=\omega^2\bW$ is 
\begin{align}
\bW_{\ome}(x)=\mcl{N}\left[\cos\left(\omega x\msDelta^{-\frac{1}{2}}\right)\begin{pmatrix}
\msDelta^{-\frac{1}{2}}\be_c\\ \msDelta^{\frac{1}{2}}\br_c
\end{pmatrix}+ \sin\left(\omega x\msDelta^{-\frac{1}{2}}\right)\begin{pmatrix}
\be_s\\ \br_s
\end{pmatrix}\right].\notag
\end{align}
We have introduced an additional normalization parameter $\mcl{N}$ for later convenience. Introducing the general solution in the  boundary conditions  one can fix  $2N$ constants of the general solution as 
\begin{align}
\br_c=&\tilde{\mY}\be_c,\notag\\
\be_s=&\tilde{\mY}\br_s,\notag
\end{align}
where  we have rescaled the admittance matrices with the velocity matrix, i.e. $\tilde{\mY}=\msDelta^{-\frac{1}{2}}\mY\msDelta^{-\frac{1}{2}}$. In general it might be the case that   $\br_c$, $\be_r$, and the free $\br_s\equiv \br$ and $\be_c\equiv \be$  depend on frequency. We now demand (generalized) orthonormality, 
\begin{align}
\langle\bW_{\ome},\bW_{\omep}\rangle=\delta_{\omega\omega'}\delta_{\epsilon\epsilon'}.
\end{align}

Inserting the general solution with the restrictions set by the boundary conditions, and making use of the following identities 
\begin{align}
\int_{\mR^+}dx\,\cos(\omega x\msDelta^{-\frac{1}{2}})\cos(\omega' x\msDelta^{-\frac{1}{2}})&=\frac{\pi}{2}\delta_{\omega\omega'}\msDelta^{\frac{1}{2}},\notag\\
\int_{\mR^+}dx\,\sin(\omega x\msDelta^{-\frac{1}{2}})\sin(\omega' x\msDelta^{-\frac{1}{2}})&=\frac{\pi}{2}\delta_{\omega\omega'}\msDelta^{\frac{1}{2}},\label{eq:identities_int_cossine}\\
\int_{\mR^+}dx\,\cos(\omega x\msDelta^{-\frac{1}{2}})\sin(\omega' x\msDelta^{-\frac{1}{2}})&=\frac{1}{2}\msDelta^{\frac{1}{2}}\mcl{P}\left(\frac{1}{\omega+\omega'}-\frac{1}{\omega-\omega'}\right),\notag
\end{align}
we rewrite the orthonormality conditions as
\begin{equation}
\label{eq:morthono}
\frac{\pi}{2}\mcl{N}_\ome^2\begin{pmatrix}
\be\\\br
\end{pmatrix}_\ome^T \mcM\begin{pmatrix}
\be\\\br
\end{pmatrix}_{\omega\epsilon'}=\delta_{\epsilon\epsilon'}\,,
\end{equation}
where 
\begin{align}
\mcM=&\begin{pmatrix}
\mM&0\\0&\mM
\end{pmatrix}=\mcM^T,\\
\mM=& \msDelta^{-\frac{1}{2}}+\mY^T\msDelta^{\frac{1}{2}}\mY,
\end{align}
whence we see that $\mM$ is symmetric. In this way we have rewritten the orthonormality conditions as an algebraic problem, namely that of finding vectors that are orthogonal with respect to the quadratic form determined by $\mcM$. This is tantamount to its diagonalization problem, clearly. Because of its structure, we can assert that  its   eigenvalues $m_\lambda$, are at least doubly degenerate. They are independent of $\omega$, in contrast to more general possibilities. Furthermore we do know that it is diagonalizable, which allows us to assert that the degeneracy index $\epsilon$ runs from 1 to $2N$ for all eigenvalues $\omega$ of $\mL$. Each eigenvalue $m_\lambda$ of $\mcM$ is  associated with a pair of orthogonal eigenvectors $(\be^T, 0)_{\lambda}^T$ and $(0, \be^T)_{\lambda}^T$, also independent of $\omega$. These two sets allow us additionally  to separate the degeneracy index $\epsilon$ running from $1\leq\epsilon\leq 2N$ into a $\{u,\,v\}$ index and a $1\leq\lambda\leq N$ such that we write the orthonormal basis 
\begin{align}
\bW_{\omega (u\lambda)}&=\sqrt{\frac{2}{\pi m_\lambda}}\cos\left(\omega x\msDelta^{-\frac{1}{2}}\right)\begin{pmatrix}
\msDelta^{-\frac{1}{2}}\be\\ \msDelta^{\frac{1}{2}}\tilde{\mY}\be
\end{pmatrix}_{\lambda},\notag\\
\bW_{\omega (v\lambda)}&=\sqrt{\frac{2}{\pi m_\lambda}} \sin\left(\omega x\msDelta^{-\frac{1}{2}}\right)\begin{pmatrix}
\tilde{\mY}\be\\ \be
\end{pmatrix}_{\lambda}.\notag
\end{align}

\subsubsection*{Telegrapher's matrix representation}
It follows from the details of the computation above that in this basis the matrix representation of the $\mT$ operator  has the explicit block shape
\begin{align}
\langle\bW_{\ome},\mT\bW_{\omep}\rangle&=\delta_{\omega\omega'}\omega\mt_{\epsilon'\epsilon}=\delta_{\omega\omega'}\omega\begin{pmatrix}
0&\mt_{uv}\\\mt_{vu}&0
\end{pmatrix}^T\nonumber\\
&=\delta_{\omega\omega'}\omega(\sigma_y\otimes\mone_{N})=\delta_{\omega\omega'}\omega\begin{pmatrix}
0&-i\mone_{N}\\i\mone_{N}&0
\end{pmatrix}.\notag
\end{align}
\section{$\msf{S}$ matrix degenerate case}
\label{AppSec:Degenerate_case_Smat}
Here, we briefly discuss how the analysis should be updated when  admittance or impedance matrices do not exist for the ideal nonreciprocal element. The constitutive equation for the nonreciprocal system is
\begin{equation}
(1-\mS)\dot{\bPhi}_0=R^{-1}(1+\mS)\dot{\bQ}_0.
\end{equation}

Let us consider the case where we want an admittance description as for the impedance case we have the dual problem. As discussed in \cite{ParraRodriguez:2019}, we may project on the space of eigenvalue $-1$ to find the constraint 
\begin{equation}
\mP_1\dot{\bPhi}_0=0,
\end{equation}
from where one can reduce the number of free coordinates to $N-1$. An admittance equation can be written as 
\begin{equation}
\tilde{\bQ}_0=\tilde{\mY}\tilde{\bPhi}_0,
\end{equation}
where the matrix $\tilde{\mY}=R^{-1}(\mQ_1(1+\mS)\mQ_1)^{-1}\mQ_1(1-\mS)\mQ_1$, and the flux $\tilde{\bPhi}_0$ and charge $\tilde{\bQ}_0$ vectors have $N-1$ entries. The domain of the operator to treat this case is updated to be 
\begin{align}
\mcl{D}(\mL)=&\left\{\bW(x)\in AC^1(\mcl{I})\otimes\mathbbm{C}^{2N},\,\mQ_1\bV_0 = \tilde{\mY}\mQ_1 \bU_0,\,\mQ_1\msDelta\bU'_0=\tilde{\mY} \mQ_1\bV'_0\right\}.\nonumber
\end{align}

\section{Analysis of the example}
\label{AppSec:example_3TL_Y_JJ}
Here we show the full computation of the Hamiltonian for the circuit in Fig. 3 of the main text. To do that, we need to enlarge the differential operator \cite{ParraRodriguez:2018} to describe networks of ideal circulators connected to transmission lines which may have capacitive connections to nonlinear networks. The domain of the new operator with an additional boundary for the first line is
\begin{align}
\mcl{D}(\mL)=&\left\{(\bW(x),w),w=\alpha (\bsb{n}\cdot \bU_d)\in \mR,\bV_0 = \mY \bU_0,\, (\msDelta\bU'_d)_\perp=0=(\bV_d)_\perp\right\},\nonumber
\end{align}
where $\bsb{n}$ is a vector projecting on the first transmission line function, i.e. $\bsb{n}=
\begin{pmatrix}
1&0&0
\end{pmatrix}^T$, and $\perp$ refers to its orthogonal. $\alpha$ is a free parameter which will be optimally set the value $\alpha_s=C_s/c_\delta=C_c C_J/[c_\delta(C_c + C_J)]$, such that the Hamiltonian will not have mode-mode couplings, see \cite{ParraRodriguez:2018}. For simplicity, we assumed open boundary conditions (current equals to zero) in the other lines. A suitable inner product is determined by the parameter $\alpha$, and for elements $\mcl{W}\in\mD(\mL)$
\begin{equation}
\langle \mcl{W}_1,\mcl{W}_2\rangle=\int_{\mcl{I}} dx\, \bW_1 \msSigma \bW_2+\frac{w_1 w_2}{\alpha},\nonumber
\end{equation}
where $\msSigma$ is defined as previously in (\ref{eq:inner_product_lines_only}). The action of the operator on its elements now reads $\mL\mW=(-\msDelta\bW'', -(\bsb{n}\cdot\msDelta\bU'_d))$. It is easy to check that $\mL$ is self-adjoint with this inner product, which means that its eigenvectors form a basis. The Lagrangian of the system is written as
\begin{align}
L=L_{\text{TG}}+L_{\mY}+\frac{\alpha_c}{2}\left(\bsb{n}\cdot\dot{\bPhi}_d-\frac{\dot{\Phi}_J}{\sqrt{\alpha_\Sigma}}\right)^2+\frac{\alpha_J}{2\alpha_\Sigma}\dot{\Phi}_J^2+E_J\cos(\varphi_J),\nonumber
\end{align}
where we have rescaled the flux and phase variables such that $\alpha_i = C_i/c_\delta$, $C_\Sigma=(C_J+C_c)$, $\varphi_J=2\pi\Phi_J/(\Phi_q \sqrt{C_\Sigma})$, and $\Phi_q$ is the flux quantum. We must solve the eigenvalue problem  $\mL\mW_{n\epsilon}=\omega_n^2\mW_{n\epsilon}$  to expand the fields in the eigenbasis of the differential operator and rewrite the Lagrangian as 
\begin{align}
L=&\,\frac{1}{2}\sum_{n,\epsilon} \dot{X}_{n\epsilon}(\dot{X}_{n\epsilon}-\omega_n\msJ_{\epsilon\epsilon'}X_{n\epsilon'})+\frac{\alpha_-}{2}(\dot{X}_{n\epsilon} u_{n \epsilon})^2-\frac{\alpha_c}{\sqrt{\alpha_\Sigma}} (\dot{X}_{n\epsilon} u_{n\epsilon})\dot{\Phi}_J\nonumber\\
&+\frac{\dot{\Phi}_J^2}{2}+E_J\cos(\varphi_J)\nonumber
\end{align}
where $\alpha_-=\alpha_c-\alpha$,  and  $u_{n\epsilon}=\bsb{n}\cdot(\bU_{n\epsilon})_d$. We picked the basis for which $\mt=-i\sigma_y=\msJ^T$, and is normalized as $\langle \mW_{n\epsilon},\mW_{m\epsilon'}\rangle=\delta_{nm}\delta_{\epsilon\epsilon'}$. We can write it in a more compact notation
\begin{align}
L=\frac{1}{2}\dot{\bar{\bX}}^T\mC \left(\dot{\bar{\bX}}-\msG\bar{\bX}\right)+E_J\cos(\varphi_J),\nonumber
\end{align}
where $\bar{\bX}=(\Phi_J,\bX^T)=(\Phi_J, F_{11}, ...,F_{1N}, G_{11},...,G_{1N},...,F_{21},...,G_{21}, ...)^T$, and the matrices are defined as 
\begin{align}
\mC&=\begin{pmatrix}
1&-\frac{\alpha_c }{\sqrt{\alpha_\Sigma}}\bu^T\\-\frac{\alpha_c }{\sqrt{\alpha_\Sigma}}\bu & \mone + \alpha_- \bu\bu^T
\end{pmatrix},\quad
\msG=\begin{pmatrix}
0&0&0&\\0& \omega_1 \msJ&0&\\0&0& \omega_2 \msJ&\\&&&\ddots
\end{pmatrix},\nonumber
\end{align}
with $\bu=(u_{11}, u_{12}, ..., u_{1(2N)}, u_{21},....)^T$. We perform a Legendre transformation 
$\bar{\bPi}=\partial L/\partial \dot{\bar{\bX}}=(Q_J,\bPi^T)=(Q_J, \pi_{11},..., \pi_{1N}, p_{11},...,\pi_{1N},...,\pi_{21},...,p_{21},...)^T$, by formally inverting the kinetic matrix $\mC$. We fix $\alpha=\alpha_s$, such that the mode-mode coupling disappears, see Eqs. (16-18) 
in \cite{ParraRodriguez:2018}
\begin{equation}
\left.\mC^{-1}\right|_{\alpha=\alpha_s}=\begin{pmatrix}
\frac{\alpha_\Sigma}{\alpha_J}&\frac{\alpha_c}{\sqrt{\alpha_\Sigma}} \bu^T\\\frac{\alpha_c}{\sqrt{\alpha_\Sigma}} \bu& \mone
\end{pmatrix},\nonumber
\end{equation}
where the infinite-length coupling vector has finite norm $\left|\bu\right|^2=1/\alpha=1/\alpha_s$ \cite{Walter:1973,ParraRodriguez:2018}. We derive the Hamiltonian 
\begin{align}
H=&\,\frac{1}{2}\left(\bar{\bPi}+\frac{1}{2}\msG\bar{\bX}\right)^T\mC^{-1} \left(\bar{\bPi}+\frac{1}{2}\msG\bar{\bX}\right)+E_J\cos(\varphi_J)\nonumber\\
=&\,\frac{1}{2}\sum_{n,\epsilon}\left( \Pi_{n\epsilon}+\frac{1}{2}\omega_n\msJ_{\epsilon\epsilon'}X_{n\epsilon'}\right)^2+\gamma Q_J u_{n\epsilon}\left( \Pi_{n\epsilon}+\frac{1}{2}\omega_n\msJ_{\epsilon\epsilon'}X_{n\epsilon'}\right)\nonumber\\
&+\frac{\alpha_\Sigma}{2\alpha_J}Q_J^2-E_J\cos(\varphi_J),\nonumber
\end{align}
where $\gamma=\alpha_c/\sqrt{\alpha_\Sigma}$. One can apply now the canonical transformation that eliminates the $N$ nondynamical coordinates per frecuency
\begin{align}
\tilde{F}_{n\lambda}&=\frac{1}{2}F_{n\lambda}-\frac{1}{\omega_n}P_{n\lambda},\qquad \tilde{\pi}_{n\lambda}=\pi_{n\lambda}+\frac{\omega_n}{2}G_{n\lambda},\nonumber\\
\tilde{G}_{n\lambda}&=\frac{1}{2}G_{n\lambda}-\frac{1}{\omega_n}\pi_{n\lambda},\qquad
\tilde{p}_{n\lambda}=p_{n\lambda}+\frac{\omega_n}{2}F_{n\lambda},\nonumber 
\end{align}
and rescale back the Josephson conjugate variables $Q_J\rightarrow Q_J/\sqrt{C_\Sigma}$ and $\varphi_J\rightarrow \varphi_J\sqrt{C_\Sigma}$, to derive the Hamiltonian 
\begin{align}
H=\frac{1}{2}\sum_{n,\epsilon}\left( \tilde{\pi}_{n\lambda}^2+\omega_n^2\tilde{F}_{n\lambda}^2\right)+\xi Q_J \left(u_{n u \lambda}\tilde{\pi}_{n\lambda}-u_{n v \lambda}\omega_n\tilde{F}_{n\lambda}\right)+\frac{Q_J}{2C_J}-E_J\cos(\varphi_J),\nonumber
\end{align}
where $\xi=\gamma/\sqrt{C_\Sigma}=\alpha_c/(\alpha_\Sigma \sqrt{c_\delta})$. We promote the conjugate variables to quantized operators and define the annihilation and creation pair as $\tilde{\pi}_{n\lambda}=\sqrt{\hbar\omega_n/2}(a_{n\lambda}+a_{n\lambda}^\dag)$, and $\tilde{F}_{n\lambda}=i\sqrt{\hbar/(2\omega_n)}(a_{n\lambda}-a_{n\lambda}^\dag)$. Finally,
\begin{align}
H=\frac{Q_J^2}{2C_J}-E_J\cos(\varphi_J)+\sum_{n,\lambda}^{\infty,N} \hbar\omega_{n}\, a_{n\lambda}^\dag a_{n\lambda} +\xi Q_J\sum_{n,\lambda}\left[r_{n\lambda}^* a_{n\lambda}+r_{n\lambda} a_{n\lambda}^\dagger\right]\nonumber
\end{align}
where $r_{n\lambda}=\sqrt{\frac{\hbar\omega_n}{2}}  (u_{n u\lambda}+i u_{n v\lambda})$. This coupling parameter allows the computation of convergent Lamb-shifts (and effective multi-partite couplings) as shown by   \cite{Malekakhlagh:2017,Gely:2017,ParraRodriguez:2018} that are proportional to 
\begin{equation}
\chi\propto \sum_{n,\lambda}|r_n|^2/\omega_n=\hbar/(2\alpha)<\infty, \nonumber
\end{equation}
given that the eigenfunctions will decay as $u_{n\epsilon}\propto 1/n$ when $\omega_n\rightarrow\infty$.
\section{Double space operator for the full case}
Let us prove now how one may extend the basis to analyse transmission lines connected by generic linear boundary conditions. Be $\mA$ and $\mB$ full-rank real symmetric matrices and $\mY$ a skew-symmetric matrix. The domain of the self-adjoint operator proposed for the mode decomposition of the parallel configuration circuit is
\begin{align}
\mcl{D}(\mL)=&\left\{\left(\bsb{W},\bsb{w}\right), \bsb{W}\in AC^1(\mcl{I})\otimes\mathbbm{C}^{2N},\bsb{w}=\begin{pmatrix}
\mA \bU\\\bV-\mA \bV' - \mY \bU
\end{pmatrix}_0\in\mathbbm{C}^{2N}\right\},\nonumber
\end{align}
where the operator on elements of its domain acts as 
\begin{align}
\mL \mcl{W} &=\left(-\msDelta\bsb{W}'',\tilde{\bsb{w}}=-\begin{pmatrix}
\msDelta\bU'-\mB^{-1}\bU-\mY \bV'\\\mB^{-1}\bV'
\end{pmatrix}_0\right).\label{eq:L_op_appendix_full}
\end{align}
It can be easily checked that under the inner product 
\begin{align}
\langle \mcl{W}_1,\mcl{W}_2\rangle&=\int_{\mcl{I}}dx \,\bsb{W}_1^\dag \msSigma \bsb{W}_2 +\bsb{w}_1^\dag\msGamma\bsb{w}_2
\end{align}
with $\msSigma=\text{diag}(\mone,\msDelta^{-1})$ and $\msGamma=\text{diag}(\mA^{-1},\mB)$ this operator is self-adjoint. We sketch the proof here for the sake of completeness which follows closely that in the first section of this Appendix for the ideal nonreciprocal boundary condition. For $\mW_1$ to be in $\mD(\mL^\dag)$, there needs to exist a $\mW_3$ such that \begin{align}
&\langle \mW_1,\mL\mW_2\rangle-\langle\mW_3,\mW_2\rangle=0.
\end{align}
holds $\forall\mW_2\in\mD(\mL)$, in which case one defines $\mW_3=\mL^\dag\mW_1$. Now, by integration by parts  
\begin{align}
\langle \mW_1,\mL\mW_2\rangle-\langle\mW_3,\mW_2\rangle&=\left[\bsb{W}_1^\dag \msSigma\msDelta\bsb{W}_2' -(\bsb{W}_1^\dag)' \msDelta\msSigma\bsb{W}_2\right]_0&\nonumber\\
&-\bsb{w}_1^\dag\msGamma \begin{pmatrix}
\msDelta\bU_2'-\mB^{-1}\bU_2-\mY \bV_2'\\\mB^{-1}\bV_2'
\end{pmatrix}_0&\nonumber\\
&-\bsb{w}_3^\dag \msGamma\begin{pmatrix}
\mA \bU_2\\\bV_2-\mA \bV_2' - \mY \bU_2
\end{pmatrix}_0,&
\end{align}
where $\bsb{w}_1^\dag=\begin{pmatrix}
\bsb{a}^\dag&\bsb{b}^\dag
\end{pmatrix}$, and $\bsb{w}_3^\dag=\begin{pmatrix}
\bsb{c}^\dag&\bsb{d}^\dag 
\end{pmatrix}$. The four equations for the action and domain of the adjoint operator are explicitly
\begin{align}
\bsb{a}^\dag\msA^{-1}\msDelta&=\bU_1^\dag \msDelta ,\nonumber\\
\bsb{d}^\dag\msB&=-(\bV_1^\dag)',\nonumber\\
\bsb{b}^\dag&=(\bV_1^\dag +\bsb{a}^\dag\msA^{-1}\mY+\bsb{d}^\dag\msB\mA)=\bV_1^\dag+(\bU_1^\dag)\mY-(\bV_1^\dag)'\mA,\nonumber\\
\bsb{c}^\dag&=-(\bU_1^\dag)'\msDelta+\bsb{a}^\dag\msA^{-1}\mB^{-1}+\bsb{d}^\dag\msB\mY=-(\bU_1^\dag)'\msDelta+(\bU_1^\dag)\mB^{-1}-(\bV_1^\dag)'\mY,\nonumber
\end{align}
whose unique solution corresponds to $\bsb{w}_1\in\mcl{D}(\mL)$ and $\bsb{w}_3=\tilde{\bsb{w}}_1$, i.e. the same domain and action of the original operator.
\subsection{Positivity}
The complete operator (in this case defined in the interval $\mcl{I}=\mR^+$) is a monotone (accretive) operator 
\begin{align}
\langle \mW, \mL\mW\rangle&=\int_{\mR^+} dx\,(\bsb{W}^\dag)'\msSigma\msDelta\bsb{W}'+U^\dag\mB^{-1}U+(V^\dag)'\mA V'\geq0
\end{align}
$\forall \,\mW\in \mcl{D}(\mL)$ as $\msSigma\msDelta$, $\mB^{-1}$ and $\mA$ are real positive symmetric matrices and $\mY$ is a real anti-symmetric matrix.

\subsection{$\mT$ Symmetry}
The new definition for the duality operator $\mT$ implies that applied to eigenvectors $\mW$ of the operator $\mL$, i.e. those that solve the eigenvalue equation ${\cal L}\mW_{\ome}=\omega^2\mW_{\ome}$, 
\begin{align}
\mT\mW_{\ome}&=-i\left(\begin{pmatrix}
\bV'\\ \msDelta\bU'
\end{pmatrix},\begin{pmatrix}
\mA \bV'\\\msDelta\bU'-\mA\msDelta\bU''-\mY\bV'
\end{pmatrix}_0\right)_{\ome}\nonumber\,.
\end{align}
Notice that applied twice again to eigenvectors,
\begin{align}
\mT^2\mW_{\ome}&=-\left(\begin{pmatrix}
\msDelta\bU''\\ \msDelta\bV''
\end{pmatrix},\begin{pmatrix}
\mA\Delta \bU''\\\msDelta\bV''-\mA\msDelta\bV'''-\mY\msDelta\bU''
\end{pmatrix}_0\right)\equiv\mL\mW_{\ome},\nonumber\\
\end{align}
such that extended by linearity $\mL=\mT^2$. 
\subsection{Sum rules and projector operator}
\label{AppSec:Projector_sum_rules}
We prove some useful identities to facilitate the Lagrangian analysis. Analogously to the Appendix \ref{appendix_b}, we find sum rules for the eigenbasis at the boundary that allow to define projector operators for the generalized coordinates by making use of a complete basis of the differential operator. Given that the domain of this operator is dense in the Hilbert space $\mcl{H}=(L^2(\mR_+)\otimes\mathbbm{C}^{2N})\oplus\mathbbm{C}^{2N}$ we can develop the element
\begin{equation}
\mW^{(0,\bone_{i})}=\left(0,\begin{pmatrix}
\bone_{i}\\\bone_{i}
\end{pmatrix}\right),
\end{equation}
where $\bone_{i}=(0,\,...,\, 1,\,...,\,0 )^T$ is the vector with one in the $i$-th position, in an orthonormal basis of the operator
\begin{equation}
\mW^{(0,\bone_{i})}=\int d\Omega \mW_\ome\langle \mW_{\ome},\mW^{(0,\bone_{i})}\rangle. \nonumber
\end{equation}
Doing so for all the rows we find the sum rule identities written in matrix form
\begin{align}
\mone_{N}&=\int d\Omega \mA^{-1}\bU_{\ome 0}\bU_{\ome 0}^\dag,\label{eq:sum_rule_A}\\
\mone_{N}&=\int d\Omega \mB(\bV-\mA \bV' - \mY \bU)_{\ome 0}(\bV-\mA \bV' - \mY \bU)_{\ome 0}^\dag\nonumber\\
&=\int d\Omega \left(\frac{1}{\omega^4}\right)\mB^{-1}\bV'_{\ome 0}(\bV'_{\ome 0})^\dag,\label{eq:sum_rule_B}
\end{align}
where again $\int d\Omega \equiv \sum_{\epsilon=1}^{2N}\int_{\mR_+}d\omega$.
The above identites allow to construct the projector $\mK$ that acting on an eigenbasis of $\mL$ 
\begin{align}
\mK\mW_{\ome}&=\int d\Omega' \mk_{\epsilon\epsilon'}^{\omega\omega'}\mW_{\omep},\label{eq:app_K_oper_def}\\
\mk_{\epsilon\epsilon'}^{\omega\omega'}&=(\bV-\mA \bV' - \mY \bU)_{\ome 0}^\dag\mB(\bV-\mA \bV' - \mY \bU)_{\omep 0}\nonumber\\
&=\left(\frac{1}{\omega\omega'}\right)^2(\bV'_{\ome 0})^\dag\mB^{-1}\bV'_{\omep 0}\nonumber
\end{align}
where $\mk$ is a hermitian (symmetric for real bases) matrix. It can be trivially checked using the sum rules that $\mK[\mK\mW_{\ome}]=\mK\mW_{\ome}$ and thus $\mK^2=\mK$ by linear extension to all elements in the full Hilbert space. 

\subsection{Orthonormal eigenbasis}
Let us find now an orthonormal basis for the spectral decomposition of the differential operator. Given that it acts as a second derivative on an interval, we look for sinusoidal solutions 
\begin{align}
\bW_{\ome}(x)&=\mcl{N}_\ome\left[\cos\left(\omega x\msDelta^{-\frac{1}{2}}\right)\begin{pmatrix}
\msDelta^{-\frac{1}{2}}\be_c\\ \msDelta^{\frac{1}{2}}\br_c
\end{pmatrix}_{\ome} + \sin\left(\omega x\msDelta^{-\frac{1}{2}}\right)\begin{pmatrix}
\be_s\\ \br_s
\end{pmatrix}_{\ome}\right].\notag
\end{align}
Introducing the boundary conditions of the eigenvalue problem $-\tilde{\bsb{w}}_{\ome}=\omega^2 \bsb{w}_{\ome}$, one can reduce $2N$ variables, i.e. 
\begin{align}
\br_c&=-\mC^{-1}\frac{\br_s}{\omega}+\tilde{\mY}\be_c,\notag\\
\be_s&=\mC^{-1}\frac{\be_c}{\omega}+\tilde{\mY}\br_s,\notag
\end{align}
where $\mC^{-1}=\tilde{\mB}^{-1}-\omega^2\tilde{\mA}$, and we have rescaled the inductance, capacitance and admittance matrices with the velocity matrix, i.e.  $\tilde{\mB}^{-1}=\msDelta^{-\frac{1}{2}}\mB^{-1}\msDelta^{-\frac{1}{2}}$, $\tilde{\mA}=\msDelta^{-\frac{1}{2}}\mA\msDelta^{-\frac{1}{2}}$, and $\tilde{\mY}=\msDelta^{-\frac{1}{2}}\mY\msDelta^{-\frac{1}{2}}$. From here on, we will take for granted that  $\br_c$, $\be_r$, and the free $\br_s\equiv \br$ and $\be_c\equiv \be$  depend on frequency and have a degeneracy index $\epsilon$. It must be appreciated that this system, that contains only one boundary, has maximum degeneracy for all eigenvalues. To find an orthonormal basis, we may fix the rest of the free parameters imposing 
\begin{align}
\langle\mW_{\ome},\mW_{\omep}\rangle&=\frac{\pi}{2}\mcl{N}_\ome^2\delta_{\omega\omega'}\begin{pmatrix}
\be\\\br
\end{pmatrix}_\ome^T \mcM\begin{pmatrix}
\be\\\br
\end{pmatrix}_{\omega\epsilon'}\nonumber\\
&=\delta_{\omega\omega'}\delta_{\epsilon\epsilon'},\nonumber
\end{align}
where 
\begin{align}
\mcM&=\begin{pmatrix}
\mM&\mN\\-\mN&\mM
\end{pmatrix}=\mcM^T,\nonumber\\
\mM&= \msDelta^{-\frac{1}{2}}+ \frac{1}{\omega^2}\mC^{-1}\msDelta^{\frac{1}{2}}\mC^{-1}+\mY^T\msDelta^{\frac{1}{2}}\mY,\nonumber\\
\mN&=(\omega\mC)^{-1}\msDelta^{\frac{1}{2}}\mY-\mY^T \msDelta^{\frac{1}{2}}(\omega\mC)^{-1} ,\nonumber
\end{align}
where $\mM$ symmetric and $\mN$ skew-symmetric matrices, and where we have used the above  identities (\ref{eq:identities_int_cossine}). We may diagonalize the symmetric matrix $\mcM$ and find $N$ double-degenerate eigenvalues $m_\lambda$, associated with $N$ pairs of orthogonal eigenvectors $\bc_{\omega\lambda}^T=(\be^T, \br^T)_{\omega\lambda}$ and $\bd_{\omega\lambda}^T=(-\br^T, \be^T)_{\omega\lambda}$. For convenience, we match the first $\bc_{\omega\lambda}$ (second $\bd_{\omega\lambda}$) $N$ eigenvectors with indices $1\leq\epsilon\leq N\rightleftarrows u \lambda$ ($N+1\leq\epsilon\leq 2N\rightleftarrows v \lambda$), where $1\leq \lambda\leq N$, such that we write
\begin{align}
\begin{pmatrix}
\msE^T&\msR^T\\-\msR^T&\msE^T
\end{pmatrix}\mcM\begin{pmatrix}
\msE&-\msR\\\msR&\msE
\end{pmatrix}=\begin{pmatrix}
\mm&0\\0&\mm
\end{pmatrix}.\nonumber
\end{align}
Finally, the orthonormal basis is 
\begin{align}
\label{eq:basis_LUV_ABY}
\bW_{\omega u\lambda}(x)=&\sqrt{\frac{2}{\pi m_\lambda}}\left[\cos\left(\omega x\msDelta^{-\frac{1}{2}}\right)\begin{pmatrix}
\msDelta^{-\frac{1}{2}}\be\\ \msDelta^{\frac{1}{2}}(\tilde{\mY}\be-(\omega\mC)^{-1}\br)
\end{pmatrix} \right.\notag\\
&\left.+ \sin\left(\omega x\msDelta^{-\frac{1}{2}}\right)\begin{pmatrix}
\tilde{\mY}\br+(\omega\mC)^{-1}\be\\ \br
\end{pmatrix}\right]_{\omega\lambda},\notag\\
\bW_{\omega v\lambda}(x)=&\sqrt{\frac{2}{\pi m_\lambda}}\left[\cos\left(\omega x\msDelta^{-\frac{1}{2}}\right)\begin{pmatrix}
-\msDelta^{-\frac{1}{2}}\br\\ -\msDelta^{\frac{1}{2}}(\tilde{\mY}\br+(\omega\mC)^{-1}\be)
\end{pmatrix}\right.\notag\\
&\left.+ \sin\left(\omega x\msDelta^{-\frac{1}{2}}\right)\begin{pmatrix}
\tilde{\mY}\be-(\omega\mC)^{-1}\br\\ \be
\end{pmatrix}\right]_{\omega\lambda}.\notag
\end{align}

\subsubsection*{Projector matrix representation}
In this specific basis, the matrix $\mk_{\epsilon\epsilon'}^{\omega\omega'}$ can be written in two blocks of the degeneracy index
\begin{align}
\mk_{\epsilon\epsilon'}^{\omega\omega'}
&=\frac{1}{(\omega\omega')^2}(\bV'_{\omega\epsilon 0})^T\mB^{-1}\bV'_{\omega'\epsilon'0}=\begin{pmatrix}
\mk_{uu}&\mk_{uv}\\\mk_{vu}&\mk_{vv}
\end{pmatrix}\notag\\
&=\frac{2}{\pi\omega\omega'\sqrt{m_\lambda m'_{\lambda'}}}\begin{pmatrix}
\br^T \\\be^T
\end{pmatrix}_{\omega\lambda}\tilde{\mB}^{-1}\begin{pmatrix}
\br &\be
\end{pmatrix}_{\omega'\lambda'}.\notag
\end{align}
In matrix form with respect to the degeneracy indices, it is written as
\begin{align}
\mk^{\omega\omega'}&\equiv\frac{2}{\pi\omega\omega'}\mm^{-\frac{1}{2}}_\omega\begin{pmatrix}
\msR^T \\\msE^T
\end{pmatrix}_{\omega}\tilde{\mB}^{-1}\begin{pmatrix}
\msR &\msE
\end{pmatrix}_{\omega'}\mm^{-\frac{1}{2}}_{\omega'},\nonumber
\end{align}
where we have defined the $N\times2N$ matrices
\begin{align}
\msR_\omega=\left[
\br_{1}\,\,\dots\,\,\br_{\lambda}\,\,\dots\,\,\br_{N}
\right]_\omega,\notag\\
\msE_\omega=\left[
\be_{1}\,\,\dots\,\,\be_{\lambda}\,\,\dots\,\,\be_{N}
\right]_\omega.\notag
\end{align}
and the square matrix $\mm_\omega=\mathrm{diag}(m_1, ..., m_\lambda,..., m_N)$. For the particular case that $\mY=0$, it can be trivially seen above that $\msN=0$ such that $\msR=0$ and thus 
\begin{align}
	\mk_{\epsilon\epsilon'}^{\omega\omega'}=\begin{pmatrix}
		0&0\\0&\tilde{\mk}_{\lambda\lambda'}^{\omega\omega'}
	\end{pmatrix}.
\end{align}

\subsubsection*{Orthogonal rotation}
In this subsection, we prove the statement that with the basis that we have for the general operator it is possible to find an orthogonal rotation at constant frequency
\begin{equation}
\mO_\omega=\begin{pmatrix}
\mM^{-\frac{1}{2}}\msE&-\mM^{-\frac{1}{2}}\msR\\
(\mH\mM^{-\frac{1}{2}}\msE+\mI\mM^{-\frac{1}{2}}\msR)&(\mI\mM^{-\frac{1}{2}}\msE-\mH\mM^{-\frac{1}{2}}\msR)
\end{pmatrix}\mm^{\frac{1}{2}},\label{eq:ortho_k_matrix_blocks}
\end{equation}
where we have defined the matrices
\begin{align}
\mI&=(1+\mN_{\mM}^2)^{-\frac{1}{2}},\nonumber\\
\mH&=\mI\mN_{\mM}\nonumber,\\
\mN_{\mM}&=\mM^{-\frac{1}{2}}\mN\mM^{-\frac{1}{2}},\nonumber
\end{align}
to put the projector in two $N$ dimensional blocks in the degeneracy subspace, 
\begin{align}
\mO_\omega\mk_{\epsilon\epsilon'}^{\omega\omega'}	\mO_\omega^T=\frac{2}{\pi\omega\omega'}\begin{pmatrix}
0&0\\0&\mI_\omega\mM_\omega^{-\frac{1}{2}}\tilde{\mB}^{-1}\mM_{\omega'}^{-\frac{1}{2}}\mI_{\omega'}^T
\end{pmatrix}\nonumber.
\end{align} 
However, under this rotation of coordinates the telegrapher's matrix becomes
\begin{align}
\mO_\omega (i\sigma_y)\mO_\omega^T=\begin{pmatrix}
-\mN_{\mM}&(1+\mN_{\mM}^2)^{\frac{1}{2}}\\
-(1+\mN_{\mM}^2)^{\frac{1}{2}}&\mN_{\mM}
\end{pmatrix}.\nonumber
\end{align}
This is the reason why the reduction of nondynamical variables in this generic case is going to be nontrivial in the phase space. Further work will be required to bring the Hamiltonian into its simplest form.  

\section{Capacitive and inductive singular networks}
\label{AppSec:Degenerate_cases_AB}
Here, we briefly discuss how the operators should be updated in the case that the capacitance or inductance coupling matrices are not full rank. This section completes the discussion in Sec. \ref{subsec:invertibility_variable_counting} for the invertibility of kinetic and potential infinite-dimensional matrices of two transmission line resonators coupled by capacitors and inductors. The results here relate to the generalized common basis for describing such coupled systems. If the capacitance or inductance matrices are not full rank, we have a separation of Walter-type and common Robin-type boundary conditions. We now correctly describe the new Hilbert sub-space of the operator, its domain and the inner product in which such operator is self-adjoint.

Let $\mcd,\,\mld,\,\msDelta\in M_{N}^{N}(\mathbbm{R})$ be positive square $N$-dimensional diagonal matrices , and $\mA, \,\mB^{-1}\in M_{N}^{D_i}(\mathbbm{R})$ be singular real symmetric semi-positive matrices of rank $D_i$, with $D_i<N$ and $i\in\{\mA,\,\mB\}$. The domain of the new self-adjoint operator is
\begin{align}
&\mcl{D}(\mL)=\left\{\left(\bsb{W},\mcl{P}\bsb{w}\right),\bsb{W}(x)\in AC^1(\mcl{I})\otimes\mathbbm{C}^{2N},\right.\nonumber\\
&\left.\bsb{w}=\begin{pmatrix}
\mA_{\mcl{P}} \mP_{\mA} \bU\\\mP_{\mB}(\bV-\mA \bV' - \mY \bU)
\end{pmatrix}_0\in\mathbbm{C}^{2N-D},\right.\left.\begin{pmatrix}
\mQ_{\mA}(\msDelta\bU'-\mB^{-1}\bU-\mY \bV')\\\mQ_{\mB}(\bV-\mA \bV' - \mY \bU)
\end{pmatrix}_0=0\right\},\nonumber
\end{align}
where $D=D_{\mA}+D_{\mB}$, and  $\mcl{P}=\mcl{P}_{\mA}\otimes\mcl{P}_{\mB}$ is the projector map constructed with maps that exclude the kernels of $\mA$ and $\mB$; through matrices $\msf{P}_i\in M_{D_i,n}^{D_i}(\mathbbm{R})$ of (vertical) eigenvectors expanding the subspace of $\mcl{P}_i$.  $\mA_{\mcl{P}}=\tilde{\mcl{P}}_{\mA}(\mA)$ and $\mB_{\mcl{P}}=\tilde{\mcl{P}}_{\mB}(\mB)$ are the original matrices projected to the subspaces through maps 
\begin{align}
\tilde{\mcl{P}}_i:M_{N}^{N}(\mathbbm{R})&\rightarrow M_{D_i}^{D_i}(\mathbbm{R})\nonumber\\
\msf{X}&\mapsto \tilde{\mcl{P}}_i(\msf{X})=\msf{P}_i^T\msf{X}\msf{P}_i.
\end{align}

The operator acts as 
\begin{equation}
\mL \mW =\left(- \msDelta\bsb{W}'',\tilde{\bsb{w}}=-\begin{pmatrix}
\mP_{\mA}(\msDelta\bU'-\mB^{-1}\bU-\mY \bV')\\\mB_{\mcl{P}}^{-1}\mP_{\mB}\bV'
\end{pmatrix}_0\right).\nonumber
\end{equation}
It can be easily checked that under the inner product
\begin{align}
\langle \mW_1,\mW_2\rangle_{\mcl{P}}&=\int_{\mcl{I}}dx \,\bsb{W}_1^\dag \msSigma \bsb{W}_2 +\bsb{w}_1^\dag\msGamma_{\mcl{P}}\bsb{w}_2 \nonumber
\end{align}
this operator is self-adjoint with $\msSigma=\text{diag}(\mathbbm{1},\msDelta^{-1})$, and $\msGamma_{\mcl{P}}=\text{diag}(\mA_{\mcl{P}}^{-1},\mB_{\mcl{P}})$. A related case of this has been used above in the circuit example of three transmission lines with only one of them being coupled to a Josephson junction, see Sec. \ref{AppSec:example_3TL_Y_JJ}. This generalizes the jump conditions in chapter~\ref{chapter:chapter_3} for the galvanic coupling in the spirit of~\cite{Mortensen:2016}.

\providecommand{\href}[2]{#2}\begingroup\raggedright\endgroup

\cleardoublepage


\end{document}